\documentclass[journal]{IEEEtran}
\IEEEoverridecommandlockouts

\usepackage{cite}
\usepackage{amsmath,amssymb,amsfonts,bm,amsthm}
\usepackage{algorithm,algpseudocode}
\usepackage{standalone}
\usepackage{myPlotStyle}
\usepackage{booktabs,multirow}
\usepackage{url}
\usepackage{graphicx}
\usepackage{textcomp}
\usepackage[T1]{fontenc}
\usepackage{xcolor}
\usepackage{pgfplots}
\usepackage{pgfplotstable}
\usepackage{tikz}
\usepackage{circuitikz}
\usepackage{subcaption}
\usepackage{amsmath}
\usepackage{xcolor}
\usepackage{pgfplots}
\usepackage{pgfplotstable}

\usetikzlibrary{arrows.meta,calc,matrix,positioning,decorations.pathreplacing,shapes.geometric}

\newcommand{\norm}[1]{\left\lVert#1\right\rVert}

\def\BibTeX{{\rm B\kern-.05em{\sc i\kern-.025em b}\kern-.08em
    T\kern-.1667em\lower.7ex\hbox{E}\kern-.125emX}}

\renewcommand{\baselinestretch}{0.945}

\theoremstyle{definition}

\theoremstyle{remark}
\newtheorem{remark}{Remark}

\usepackage{acronym}
\usepackage{bm}
\usepackage{url}
\usepackage{amsmath}
\usepackage[T1]{fontenc}
\usepackage{xcolor}
\usetikzlibrary{arrows.meta,positioning,calc}

\definecolor{myblue}{RGB}{31,119,180}
\definecolor{mygreen}{RGB}{44,130,44}
\definecolor{mypurple}{RGB}{112,48,160}
\definecolor{mygray}{RGB}{110,110,110}

\newacro{ML}[ML]{Machine Learning}
\newacro{NN}[NN]{Neural Network}
\newacro{PIML}[PIML]{Physics-Informed Machine Learning}
\newacro{PINN}[PINN]{Physics-Informed Neural Network}
\newacro{RPNN}[RPNN]{Random-Projection Neural Network}
\newacro{PIRPNN}[PIRPNN]{Physics-Informed Random-Projection Neural Network}
\newacro{MLP}[MLP]{Multi-Layer Perceptron}
\newacro{NTK}[NTK]{Neural Tangent Kernel}
\newacro{KAN}[KAN]{Kolmogorov--Arnold Network}

\newacro{RAPTOR}[RAPTOR]{Random-Projection Physics-Informed Transient Solver}

\newacro{DER}[DER]{Distributed Energy Resource}
\newacro{IBR}[IBR]{Inverter-Based Resource}
\newacro{CBR}[CBR]{Converter-Based Resource}
\newacro{GFL}[GFL]{Grid-Following}
\newacro{GFM}[GFM]{Grid-Forming}
\newacro{PLL}[PLL]{Phase-Locked Loop}
\newacro{PCC}[PCC]{Point of Common Coupling}
\newacro{SCR}[SCR]{Short-Circuit Ratio}

\newacro{ODE}[ODE]{Ordinary Differential Equation}
\newacro{DAE}[DAE]{Differential-Algebraic Equation}
\newacro{RMS}[RMS]{Root-Mean-Square}
\newacro{EMT}[EMT]{Electromagnetic Transient}
\newacro{HIL}[HIL]{Hardware-in-the-Loop}
\newacro{DSA}[DSA]{Dynamic Security Assessment}

\newacro{RBF}[RBF]{Radial Basis Function}
\newacro{BDF}[BDF]{Backward Differentiation Formula}
\newacro{PI}[PI]{Proportional-Integral}
\newacro{NR}[NR]{Newton--Raphson}
\newacro{LU}[LU]{Lower--Upper}
\newacro{WRMS}[WRMS]{Weighted Root-Mean-Square}
\newacro{MAE}[MAE]{Mean Absolute Error}

\definecolor{C1}{HTML}{1F77B4}
\definecolor{C2}{HTML}{D62728}
\definecolor{C3}{HTML}{2CA02C}
\definecolor{C4}{HTML}{9467BD}
\definecolor{C5}{HTML}{FF7F0E}
\usepackage{pgfplots}
\usepgfplotslibrary{groupplots}
\usetikzlibrary{arrows.meta,calc,positioning,decorations.pathreplacing}
\pgfplotsset{compat=1.18}

\definecolor{myblue}{RGB}{31,119,180}
\definecolor{myorange}{RGB}{255,127,14}
\definecolor{mygreen}{RGB}{44,160,44}
\definecolor{mygray}{RGB}{120,120,120}

\pgfplotsset{
  TR/.style={thick, solid, color=sBlue, no marks},
  MS/.style={only marks, mark=*, mark size=1pt, color=sOrange},
  SO/.style={thin, scatter, only marks, color=sOrange, mark=asterisk},
}

\begin{document}

\title{RAPTOR: RAndom-projection Physics-informed Transient sOlveR}

\author{
\IEEEauthorblockN{Petros Ellinas, Benjamin Vilmann, Spyros Chatzivasileiadis, and Johanna Vorwerk \\}
\IEEEauthorblockA{
Department of Wind and Energy Systems, Technical University of Denmark, Lyngby, Denmark\\
Email: \{petrel, bevil, spchatz, vorjo\}@dtu.dk}
\thanks{The work of P. Ellinas and S. Chatzivasileiadis is supported by the European Research Council (ERC) Starting Grant VeriPhIED, Grant Agreement No.~949899.}
}


\maketitle

\begin{abstract}
The complexity of time-domain simulation of modern power systems has increased
significantly because converter-based resources introduce control dynamics that
must be simulated alongside slower system-level and fast electromagnetic
dynamics. The resulting wide range of timescales may force classical
time-domain solvers to use small timesteps, complicate the solution of the nonlinear equations at each timestep, and may result in reduced solver reliability under strongly nonlinear and multi-timescale transient conditions. 
This paper introduces RAPTOR, a first-of-its-kind time-domain simulation framework for power system dynamic simulations. At its core, RAPTOR introduces a new integration technique: It represents the unknown trajectory of hybrid differential-algebraic equations (DAEs) over a time interval using a physics-informed random-projection neural network (PIRPNN) built from fixed Gaussian radial basis functions (RBFs). A nonlinear solver, e.g. a Newton--Raphson, determines how to combine the fixed RBFs so that the resulting trajectory satisfies the governing equations. By combining RBFs with broad and localized shapes, RAPTOR can represent complex multi-timescale behavior over comparatively long time intervals. This enables larger effective simulation advances, fewer overall nonlinear solver iterations, and faster integration over long time horizons. Numerical studies on stiff RMS models, IEEE 9-, 14-, 39-, 57-, and 118-bus RMS simulation benchmarks, and EMT test cases show that RAPTOR can accurately find solutions with substantially fewer sequential simulation advances while offering superior accuracy--runtime performance. Including speedups of over 10x in certain cases, these results showcase RAPTOR's potential to overtake long-standing and widely used integration methods, such as the Radau method and the trapezoidal rule.
\end{abstract}


\section{Introduction}
Modern power systems increasingly contain converter controls and electrical dynamics that evolve much faster than electromechanical states. These dynamics may interact, so that detailed RMS and EMT studies increasingly need to retain several time scales within the same simulation \cite{Hatziargyriou2021Stability,Lara2024TDS}. Simplifying the model can reduce runtime, but it can also remove fast behavior that materially affects the transient response being studied. Besides model reduction, one can target a different accuracy--speed gap: accelerating the numerical solution of detailed stiff, multi-timescale models without changing their physical equations.

Power-system dynamics are commonly written as hybrid \acp{DAE}. Differential
equations describe how components and subsystems evolve, while algebraic equations, such as Kirchhoff's laws \cite{Milano2010},  enforce instantaneous constraints that usually connect these subsystems. These systems are commonly simulated using implicit integration methods. Conventional implicit integrators build the trajectory through successive timesteps using a prescribed local representation of how the states evolve.
For example, the trapezoidal rule assumes a linear representation of the state derivative over a timestep, while the Radau method uses a higher-order polynomial representation determined from several points within the timestep \cite{HairerWanner1996}. 

These classic implicit integration methods perform a \emph{nonlinear solve} at each timestep. This is a repeated calculation that adjusts the unknown state values until the dynamic equations and algebraic constraints are satisfied. For stiff multi-timescale DAEs, this calculation can become more sensitive to its initial estimate, the numerical scaling of the equations, and algebraic consistency \cite{NocedalWright2006,BrenanCampbellPetzold1995}. Preconditioning and multi-rate integration can reduce some of these difficulties, but their benefit depends on the model structure, and they introduce additional computational or modeling requirements \cite{Saad2003,Savcenco2007,Constantinescu2010}. Consequently, when accuracy or convergence deteriorates, the usual solution is to reduce the timestep. Nonetheless, this further increases the number of timesteps, nonlinear solves, and overall computational cost \cite{Gear1971,HairerWanner1996}.

Another promising approach for solving stiff \ac{DAE} systems is provided by spectral integration methods that change how the trajectory is represented over time. These methods aim to accurately cover more simulation time with each nonlinear solve, thereby reducing the total number of nonlinear solves per time-domain simulation. Rather than constructing a new local trajectory at every timestep, as the classic Radau or trapezoidal method does, time-spectral collocation represents each system state with one continuous function over a finite interval.
A nonlinear solve determines this function by enforcing the governing equations at selected internal times, called \emph{collocation points}. As such, time-spectral integration methods are a promising alternative integration approach that, to the best of the authors' knowledge, has not been explored for time-domain simulations on a power system scale.

Classical time-spectral methods typically use a polynomial for representing the interval-wide trajectory. Such a polynomial can represent smooth trajectories accurately over relatively long intervals \cite{trefethen2000spectral}. Multi-timescale dynamics, however, can require more collocation points for an accurate representation. This increases the number of unknown quantities that must be determined together, making the nonlinear solve larger, more complex, and resulting in overall increased computational cost. 

With this paper, we introduce \ac{RAPTOR}, a time-domain solver built around a novel Gaussian \acp{RBF} time-spectral collocation method. To the best of our knowledge, this is the first exploration of time-spectral integration methods in a power system level context. To address the identified drawbacks of some of the aforementioned conventional time-spectral methods that employ polynomials, \ac{RAPTOR} uses a \ac{PIRPNN} that represents the trajectory as a linear combination of Gaussian \ac{RBF}. A Gaussian \ac{RBF} is a smooth bell-shaped function: its center determines where it acts within the interval, while its shape parameter determines whether it is broad or localized. In \ac{RAPTOR}, the centers are distributed uniformly across the time interval, while the shape parameters are sampled before the nonlinear solve from a physics-informed random distribution. These functions are then kept fixed, and the nonlinear solver decides how to combine the \acp{RBF} so that the represented solution trajectory satisfies the governing equations. Broad \acp{RBF} can represent slow evolution across the entire interval, while localized \acp{RBF} represent short fast transients. RAPTOR then turns this representation into a complete time-domain solver by combining advanced initialization and nonlinear solve techniques to reduce the computational burden \cite{Kelley1995Iterative,Brown1989VODE}, algebraic variable enforcement, novel adaptive interval length control, and event handling for discrete events.

We summarize this paper's contributions as:
\begin{itemize}

\item We introduce RAPTOR, a first-of-its-kind time-domain solver built around a novel Gaussian-RBF time-spectral integration method for hybrid, multi-timescale power-system time-domain simulation. 

\item We develop the complete numerical procedure that turns this interval trajectory representation into a practical power systems simulator. RAPTOR automatically constructs and initializes each trajectory interval, solves it while satisfying the algebraic constraints, checks whether the interval is accurate enough to accept, adjusts the following interval length, and restarts the trajectory when a discrete event changes the system.

\item We validate RAPTOR across various power system time-domain simulation tasks including stiff and multi-timescale RMS, classic IEEE standard system-level test cases, and EMT benchmarks. For comparison, we compare RAPTOR against classic integration methods established in the power system domain. The results indicate accurate performance with substantially fewer accepted advances and speedups exceeding \(10\times\) for difficult transient cases.
\end{itemize}

The remainder of this paper is organized as follows: Section~II provides an intuitive introduction to the RAPTOR conceptual framework, while Section~III presents the RAPTOR trajectory representation and adaptive simulation strategy in detail. Section~IV presents the case studies and Section~V concludes the work.

\begin{figure*}[!t]
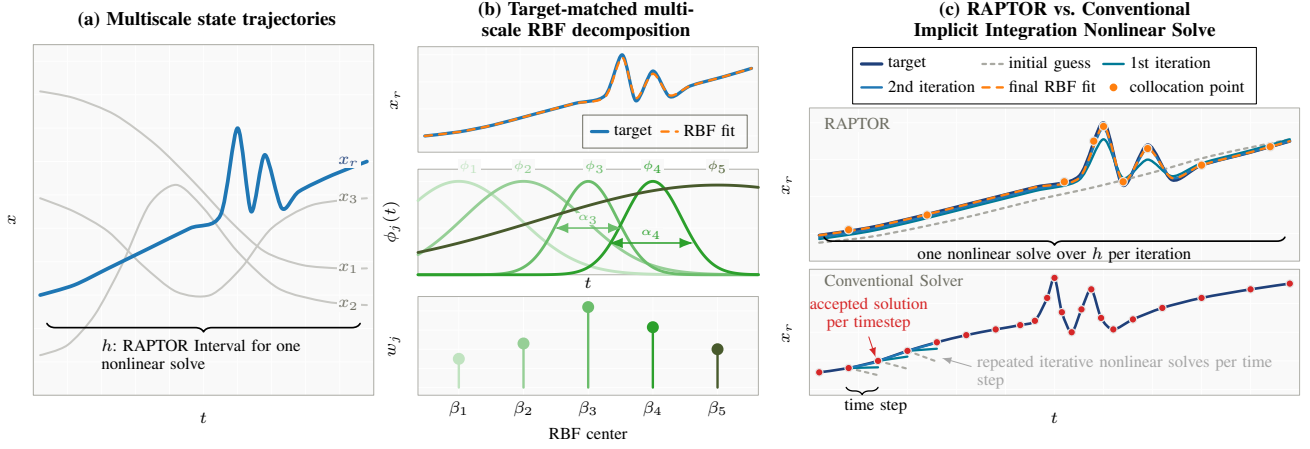

    \centering
    \includestandalone[width=0.95\textwidth]
    {Plots/explanation_raptor}
    \caption{Conceptual illustration of RAPTOR. (a) Multi-timescale state trajectories. (b) RAPTOR represents each trajectory as a weighted combination of overlapping Gaussian \acp{RBF} with broad and localized temporal support. (c) RAPTOR solves for a continuous trajectory over an interval by enforcing the governing equations at collocation points, whereas conventional implicit integrators advance through successive timesteps with separate nonlinear solves. Iteration traces are schematic and do not represent measured iteration counts.
    }
    \label{fig:raptor_conceptual}
\end{figure*}
\section{Conceptual Foundation of RAPTOR}

Power-system dynamic models consist of coupled differential and algebraic equations that describe the dynamic behavior of generators, power electronic converters, control loops, and the transmission network. The differential states represent electrical and control dynamics, while the algebraic states enforce instantaneous physical constraints such as Kirchhoff's current and voltage laws. In addition, known exogenous inputs, references, and disturbances need to be considered. This structure naturally leads to semi-explicit \acp{DAE} of the form
\begin{subequations}\label{eq:dae}
\begin{align}
\dot{\mathbf{x}}(t) &= \mathbf{f}\bigl(t,\mathbf{x}(t),\mathbf{y}(t),\mathbf{u}(t),\mathbf{\sigma}(t)\bigr), \label{eq:dae_diff} \\
\mathbf{0} &= \mathbf{g}\bigl(t,\mathbf{x}(t),\mathbf{y}(t),\mathbf{u}(t),\mathbf{\sigma}(t)\bigr). \label{eq:dae_alg}
\end{align}
\end{subequations}
Here, \(\mathbf{x}\in\mathbb{R}^{n_x}\) contains the differential states,
\(\mathbf{y}\in\mathbb{R}^{n_y}\) the algebraic variables, and \(\mathbf{u}\)
the known inputs and disturbances. The functions \(\mathbf{f}\) and
\(\mathbf{g}\) describe the differential dynamics and algebraic constraints,
respectively. The discrete state \(\mathbf{\sigma}(t)\) identifies the active network or controller configuration \cite{Hiskens2000Hybrid}. We consider hybrid semi-explicit index-1 \acp{DAE} meaning that the algebraic Jacobian
\(\mathbf{g}_{\mathbf{y}}\) is non-singular, so the algebraic variables can be
locally recovered from \(\mathbf{g}=0\).


Power system DAE systems are typically solved with implicit time-domain integrators because they are more robust to stiffness and algebraic constraints than explicit methods. Following \cite{HairerWanner1996}, we call a problem \emph{stiff} when fast,
rapidly decaying dynamics impose a stability restriction that forces standard
nonstiff integration methods to use timesteps much smaller than would be
required by accuracy alone. Classical implicit integrators require solving a nonlinear algebraic system at each timestep and typically employ a Newton-type nonlinear solver. While Newton-type methods have fast local convergence, this behavior is guaranteed \textit{only} when the initial guess is sufficiently close to the consistent solution and the relevant Jacobian is nonsingular and well conditioned \cite{Kelley1995,BrenanCampbellPetzold1996}. In stiff multi-timescale \acp{DAE} systems, both requirements are harder to satisfy: When fast and slow modes coexist within the same solve, the initial guess may be poor, and the Jacobian may become strongly anisotropic. Strong algebraic coupling further redistributes local errors through the network, while switching, protection logic, converter mode changes, and control limiters, all considered through the signal $\mathbf{\sigma}(t)$ in \eqref{eq:dae}, introduce discontinuities that require repeated re-initialization and restoration of algebraic consistency. \emph{Therefore, the central goal of the presented work is to create an integration method that requires fewer simulation advancements while maintaining accuracy and reliable and fast convergence of the nonlinear solver in each simulation advance.}


Fig.~\ref{fig:raptor_conceptual} illustrates the central idea behind RAPTOR. Power-system physical states can contain substantially different temporal behavior caused by the participation of different eigenmodes. As a result, the trajectory of a state changes non-uniformly within the same simulation interval. Fig.~\ref{fig:raptor_conceptual}(a) depicts how some states evolve gradually, whereas others exhibit short and rapidly varying transients such as the bold one in Fig.~\ref{fig:raptor_conceptual}(a). RAPTOR addresses this multi-timescale structure by solving directly for how the differential states evolve throughout a finite time interval, rather than building the trajectory through a sequence of separate local timestep solves.

RAPTOR constructs the trajectory from a set of overlapping Gaussian \acp{RBF}, as depicted by Fig.~\ref{fig:raptor_conceptual}(b). Their centers are distributed uniformly over the interval, while their shape parameters determine whether each function is broad or localized. For example, the broad \ac{RBF} \(\phi_5\) influences a large portion of the interval, whereas the localized \acp{RBF} \(\phi_3\) and \(\phi_4\) mainly affect smaller regions around their centers. In this way, the RBF set provides a simple temporal decomposition of the multi-timescale trajectory into broad functions for slow changes and localized functions for fast transients. 
Together, these simple functions give the nonlinear solver suitable building blocks for constructing complex trajectory shapes that may otherwise require shorter timesteps. The \acp{RBF} are fixed during the nonlinear solve, and each differential state can combine them differently to construct its trajectory.

This representation also changes how the nonlinear computation is organized,
as illustrated in Fig.~\ref{fig:raptor_conceptual}(c). RAPTOR adjusts how the
RBFs are combined so that one continuous candidate trajectory satisfies the
governing equations at multiple collocation points throughout the interval.
A single update, therefore, changes the trajectory simultaneously at many
physical times. In a conventional implicit integrator, each timestep instead
defines a new local nonlinear problem starting from the previous timestep
endpoint, and this process is repeated as the simulation advances. RAPTOR is
designed to reduce this repeated local-solve burden by covering more physical
time within each interval while retaining localized temporal flexibility
through the RBF representation.

\section{RAPTOR: Mathematical Formulation}
This section presents the mathematical formulation of RAPTOR. Generally,
given the differential and algebraic state at \(t_k\), RAPTOR constructs a differential-state trajectory over
\([t_k,t_{k+1}]\) and enforces the initial condition exactly. The trajectory is represented by a \ac{PIRPNN} as a weighted combination
of Gaussian \acp{RBF}. Their centers are placed uniformly over the interval
,and their shape parameters are randomly sampled from a predefined
physics-informed distribution before the nonlinear solve. These \ac{RBF}
parameters are then kept fixed, while the \ac{RBF} coefficients, collected
in \(\mathbf{w}\), are adjusted by RAPTOR by minimizing the \ac{DAE}
residual at specified collocation points. Here, the \ac{DAE} residual measures how well the
represented trajectory satisfies the governing equations: it is based on
the difference between the time derivative of the represented differential
states and the derivative prescribed by
\(\mathbf{f}(t,\mathbf{x},\mathbf{y},\mathbf{u},\mathbf{\sigma})\).

Throughout the paper, we stick to the following terminology: \emph{RAPTOR} denotes the complete time-domain solver,
whereas \emph{PIRPNN} denotes the trajectory representation used within one
RAPTOR interval. An \emph{interval} is the span
\([t_k,t_{k+1}]\) over which RAPTOR computes one candidate trajectory. After
the trajectory is computed, RAPTOR checks its equation residual. If the
residual satisfies the acceptance criterion, the interval is
\emph{accepted} and its endpoint becomes the starting state of the next
interval; otherwise, the interval is \emph{rejected}, shortened, and solved
again. In contrast, we use the terms \emph{timestep} for one accepted step of a comparison, classic time-domain integrator, and
\emph{accepted advance} when referring generically to either one accepted
conventional timestep or one accepted RAPTOR interval.

Although a \ac{PIRPNN} is used to represent the trajectory within each
RAPTOR interval, it is not used as a predictor of future trajectories.
Instead, RAPTOR uses the \ac{PIRPNN} structure as a parameterization of the
unknown trajectory over the current interval. For every new interval, the
\ac{RBF} coefficients are determined again by solving the governing
\ac{DAE} equations. Consequently, the \ac{PIRPNN} fitted on one interval is
valid only for that interval and is not reused to predict the trajectory on
future intervals. The complete power-system trajectory is obtained by
combining the successively solved interval trajectories.


The remainder of this section first details how a single RAPTOR interval is solved, before presenting the workflow on how to simulate an entire time horizon and obtain the power system dynamic trajectories. 

\subsection{Solving a Single RAPTOR Interval}
\label{sec:framework}

This section presents the single-interval solve at the core of RAPTOR, starting from the PIRPNN formulation, the enforcement of algebraic states, the nonlinear RAPTOR interval solving mechanism, and leading to a discussion of the basis function design.

\subsubsection{PIRPNN Trajectory Representation}

Within one RAPTOR interval \([t_k,t_{k+1}]\), one \ac{PIRPNN} is used
for each differential state to parameterize its trajectory using a fixed
set of Gaussian \acp{RBF}. We denote this represented trajectory by
\(\mathbf{x}_{\mathbf{w}}(t)\), where \(\mathbf{w}\) denotes the
\ac{RBF} coefficients that will be determined for the current interval. Let
\begin{equation}
s=\frac{t-t_k}{\Delta t_k},
\qquad
\Delta t_k=t_{k+1}-t_k,
\qquad
s\in[0,1],
\end{equation}
Here, \(t\) denotes physical time, \(\Delta t_k\) is the duration of the
current RAPTOR interval, and \(s\) is the normalized local time that maps
the interval \([t_k,t_{k+1}]\) to \([0,1]\).

Let \(N\) denote the number of Gaussian \acp{RBF} used to represent the
trajectory over one RAPTOR interval. The \(j\)-th basis function,
\(j=1,\ldots,N\), is denoted by \(\psi_j\):
\begin{equation}
\label{eq:gaussian_rbf}
\psi_j(s)
=
\exp\!\left[-\alpha_j(s-\beta_j)^2\right],
\qquad j=1,\ldots,N ,
\end{equation}
where the center \(\beta_j\) determines where the function acts within the
interval and the shape parameter \(\alpha_j>0\) determines its localization.
Small \(\alpha_j\) values give broad functions, while large values give narrow
functions.

For a system with \(n_x\) differential states, each basis function has a
state-specific coefficient vector
\(\mathbf{w}_j\in\mathbb{R}^{n_x}\). RAPTOR represents the differential-state
trajectory over the interval as
\begin{equation}
\label{functional_form_rbf}
\mathbf{x}_{\mathbf{w}}(s)
=
\mathbf{x}(t_k)
+
s\sum_{j=1}^{N}
\mathbf{w}_j
\exp\!\left[-\alpha_j(s-\beta_j)^2\right].
\end{equation}

Because \(s=0\) corresponds to \(t=t_k\), the factor \(s\) lets the PIRPNN correction vanish at the beginning of the interval. Therefore,
$\mathbf{x}_{\mathbf{w}}(0)=\mathbf{x}(t_k)$
for any value of the coefficients. Consequently, the initial condition is satisfied exactly. 

All state-specific coefficients are collected in
\(\mathbf{w}\in\mathbb{R}^{N \times n_x}\), and these coefficients are the only
PIRPNN quantities changed by the nonlinear interval solve.
\subsubsection{Algebraic States Enforcement}
\label{sec:algebraic_consistency}

In RAPTOR, only the differential states are represented by an independent trajectory \(\mathbf{x}_{\mathbf{w}}(t)\). The algebraic variables are treated differently because they do not have their own time dynamics. In power-system \acp{DAE}, variables such as bus voltages and voltage angles are instantaneous constraint variables. At each collocation point, they must satisfy the algebraic equations imposed by the network and component interconnections.

RAPTOR recovers the algebraic variables directly from the algebraic equations. This is justified locally for index-1 \acp{DAE} by the \emph{implicit function theorem} \cite{KrantzParks2002}: If the algebraic equations are smooth and the algebraic Jacobian with respect to \(\mathbf{y}\) is nonsingular, then the algebraic variables are locally determined by the differential states, inputs, and time. In other words, there exists a local function
\begin{equation}
\mathbf{y}=\varphi(t,\mathbf{x},\mathbf{u},\sigma)
\end{equation}
such that the algebraic constraints are satisfied. In practice, at each nonlinear-solver iteration, the current coefficient
vector \(\mathbf{w}\) defines a represented differential-state trajectory
\(\mathbf{x}_{\mathbf{w}}(t)\). For this current trajectory, RAPTOR treats
\(\mathbf{x}_{\mathbf{w}}(t)\), \(\mathbf{u}(t)\), and
\(\mathbf{\sigma}(t)\) as fixed and solves
\(\mathbf{g}(t,\mathbf{x}_{\mathbf{w}},\mathbf{y},
\mathbf{u},\mathbf{\sigma})=\mathbf{0}\)
for \(\mathbf{y}\). The resulting algebraic state is denoted by
\(\mathbf{y}_{\mathbf{w}}(t)=
\varphi(t,\mathbf{x}_{\mathbf{w}}(t),\mathbf{u}(t),\mathbf{\sigma}(t))\).
After the nonlinear solver updates \(\mathbf{w}\), the represented trajectory
\(\mathbf{x}_{\mathbf{w}}(t)\) changes and the algebraic variables are
recomputed accordingly.

Accordingly, RAPTOR optimizes only the differential trajectory \(\mathbf{x}_{\mathbf{w}}(t)\). At each collocation point $m$ at time $t_m \in [t_k,t_{k+1}]$, the algebraic variables are recovered by solving the algebraic constraints, and the resulting values are used to evaluate the differential equation mismatch.

For compactness, we write
\begin{equation}
\mathbf{F}(t,\mathbf{x},\mathbf{u},\mathbf{\sigma})
:=
\mathbf{f}\!\left(
t,\mathbf{x},
\varphi(t,\mathbf{x},\mathbf{u},\mathbf{\sigma}),
\mathbf{u},\mathbf{\sigma}
\right),
\end{equation}
so that the differential dynamics are
\begin{equation}
\label{eq:reduced_ode}
\dot{\mathbf{x}}(t)
=
\mathbf{F}\bigl(
t,\mathbf{x}(t),\mathbf{u}(t),\mathbf{\sigma}(t)
\bigr).
\end{equation}
 In practice, each evaluation of the equation mismatch first evaluates \(\mathbf{x}_{\mathbf{w}}(t_m)\), then solves the algebraic equations for \(\mathbf{y}_{\mathbf{w}}(t_m)\), and finally evaluates the differential equation mismatch.

\subsubsection{Nonlinear Interval Solve}
\label{sec:interval_optimization}

The remaining task is to choose the coefficients \(\mathbf{w}\) so that the
trajectory follows the governing \acp{DAE} over the interval. For this
purpose, we evaluate the equations at a finite set of \(M\) collocation
points. These collocation points are independent of the \(N\) \acp{RBF}
and their centers, and may be chosen separately over the interval. Let
\(
s_m\in[0,1],\) \(m\in \{1,\ldots,M\}
\),
be the normalized collocation points, and let
\(
t_{k,m}=t_k+s_m\Delta t_k
\)
be the corresponding physical times. At each collocation point, the algebraic variables are recovered from the algebraic equations as
\begin{equation}
\mathbf{y}_{\mathbf{w}}(t_{k,m})
=
\varphi\bigl(
t_{k,m},
\mathbf{x}_{\mathbf{w}}(t_{k,m}),
\mathbf{u}(t_{k,m}),
\sigma(t_{k,m})
\bigr).
\end{equation}
The equation mismatch at the collocation points becomes
\begin{equation}
\label{eq:collocation_residual}
\mathbf{r}_m(\mathbf{w})
=
\dot{\mathbf{x}}_{\mathbf{w}}(t_{k,m})
-
\mathbf{F}\bigl(
t_{k,m},
\mathbf{x}_{\mathbf{w}}(t_{k,m}),
\mathbf{u}(t_{k,m}),
\mathbf{\sigma}(t_{k,m})
\bigr),
\end{equation}
where the derivative \(\dot{\mathbf{x}}_{\mathbf{w}}(t)\) can be computed analytically. The interval solve is therefore the nonlinear problem
\begin{equation}
\label{eq:w_star}
\mathbf{w}^{\star}
=
\arg\min_{\mathbf{w}}
\sum_{m=1}^{M}
q_m
\left\|
\mathbf{r}_m(\mathbf{w})
\right\|^2,
\end{equation}
where \(q_m>0\) weights the residual contribution of each collocation point. The collocation points are not additional accepted timesteps. Instead, they are internal points used to check whether the candidate trajectory satisfies the equations over the interval. The interval problem is solved using a modified Newton (Chord) method,
where the residual Jacobian and its factorization are reused across
successive coefficient updates and refreshed when convergence deteriorates
\cite{Brown1989VODE}.

Because the \acp{RBF} overlap, different coefficient values can sometimes produce nearly the same trajectory. The coefficients therefore are not unique either. Nonetheless, only the resulting trajectory and its \ac{DAE} residual matter. In practice, this is handled using standard rank-aware least-squares methods, such as QR/SVD-based solves or regularization \cite{Bjorck1996LeastSquares}.

\begin{figure*}[t]
    \centering
    \includegraphics[scale=0.8]{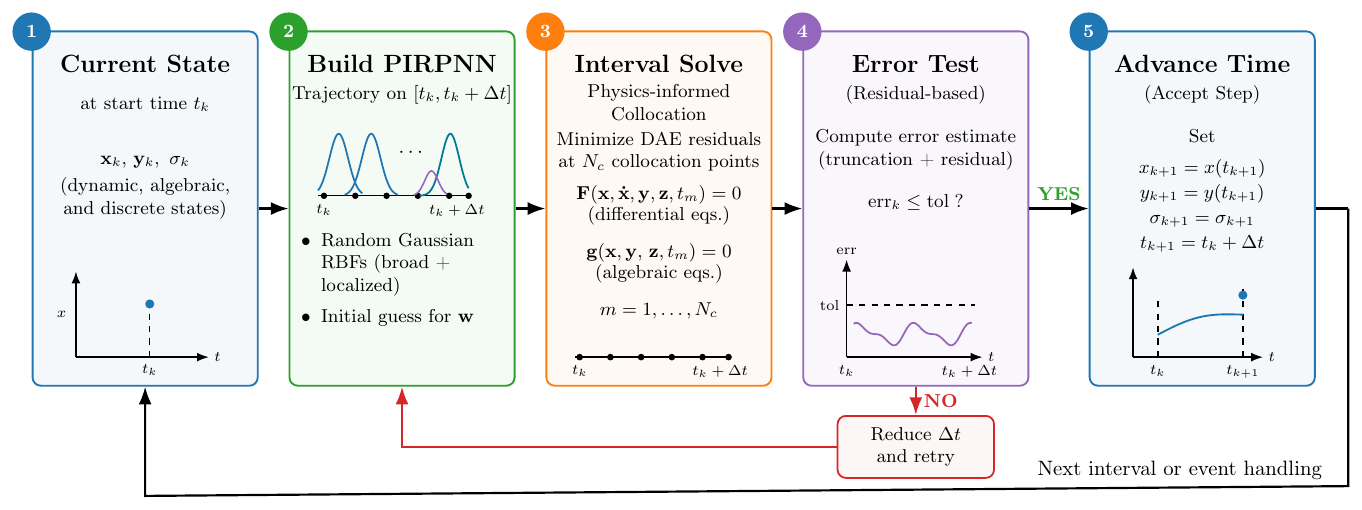}
    \caption{Workflow of the RAPTOR solver, including interval proposal, PIRPNN coefficient initialization, nonlinear residual minimization, residual-based acceptance, and PI-based interval-length update.}
\label{fig:raptor_workflow}
\end{figure*}

\subsubsection{PIRPNN Basis Design}
\label{sec:rpnn_design}

The RBF parameters are defined on the normalized coordinate \(s\in[0,1]\).
This provides a common dimensionless time coordinate for every RAPTOR
interval, while the basis density and localization determine the available
temporal resolution.

Centers $\beta_j$ are placed uniformly over the normalized interval through
\begin{equation}
\beta_j = N^{-1}\left(j-\tfrac{1}{2}\right), \qquad j=1,\dots,N
\end{equation}
to ensure consistent coverage of each interval and separates the density of the basis pool from the local shape characteristics.


The temporal resolution of the Gaussian basis is controlled by its shape
parameters. Following the interval-dependent scaling used for Gaussian
\acp{PIRPNN} \cite{Fabiani2023PIRPNN}, we relate the admissible shape range to the
length of the current interval. In \ac{RAPTOR}, we additionally use information
from the local power-system dynamics so that this range reflects the time
scales that must be represented on that interval.

Let
$A_k
=
\left.
\frac{\partial \mathbf{F}}
{\partial \mathbf{x}}
\right|_{t_k,\mathbf{x}(t_k)}$
denote the local Jacobian of the reduced differential dynamics. The
eigenvalues \(\lambda_i(A_k)\) characterize the local rates of its linearized
dynamic modes. Define
$\rho_k
=
\max_i |\lambda_i(A_k)|.$
The quantity \(1/\rho_k\) provides a local estimate of the fastest dynamic
time scale. Matching the localization of the Gaussian basis to this scale
motivates
\begin{equation}
\label{eq:spectral_c_heuristic}
\alpha_j
\sim
\mathcal{U}(0,c_k),
\qquad
c_k
=
\frac{1}{2}
\bigl(\rho_k\Delta t_k\bigr)^2 .
\end{equation}

Consequently, the available \ac{RBF} localization changes with both the proposed
interval length and the local system dynamics. Faster local dynamics increase
the available localization, whereas shorter intervals reduce the required
range. Equation~\eqref{eq:spectral_c_heuristic} is used as a physics-informed
design rule for the RBF basis.

\subsection{Building the Simulation Over the Full Time Horizon}

To simulate over a longer time horizon, RAPTOR repeats the single-interval solve. On each interval, it chooses a tentative interval length, initializes the coefficients, solves the residual-minimization problem, and accepts or rejects the result using a residual-based error indicator. Fig.~\ref{fig:raptor_workflow} summarizes the workflow.

Let the simulation horizon be partitioned into intervals
\(
[t_0,t_f]=[t_0,t_1]\cup\cdots\cup[t_{K-1},t_K]\), where $t_K=t_f$,
and define \(\Delta t_k=t_{k+1}-t_k\). While initially only the state at $t_0$ is known, every interval provides the initial state for the next interval. In addition, we assume that all these initial conditions are algebraically consistent, and that the input signal is prescribed. At the beginning of interval \(k\), RAPTOR starts from a state \((\mathbf{x}(t_k),\mathbf{y}(t_k))\).

\subsubsection{Coefficient Initialization} 
On each interval \([t_k,t_{k+1}]\), the Gaussian \ac{RBF} basis functions are fixed once a tentative interval length \(\Delta t_k\) is chosen. The nonlinear interval solve then adjusts only the stacked output coefficients \(\mathbf{w}\in\mathbb{R}^{N \times n_x}\).

To start the nonlinear solve, RAPTOR needs an initial coefficient vector $\mathbf{w}$. Ideally, such initialization provides a physically reasonable starting point for the nonlinear solver. We build this initial guess from three pieces of information: the known derivative at the start of the interval, a predicted derivative at the end of the interval, and, when appropriate, the coefficients of the previous interval. The following paragraphs introduce these three components and then combine them into one small regularized least-squares problem.

\paragraph*{Derivative at the Start of the Interval}
At the start time \(t_k\), the state \(\mathbf{x}(t_k)\) is known from the previous interval, and the reduced \acp{DAE} provides its time derivative:
\begin{equation}
\label{eq:left_slope_true}
\dot{\mathbf{x}}(t_k) =
\mathbf{F}\bigl(t_k,\mathbf{x}(t_k),\mathbf{u}(t_k),\mathbf{\sigma}(t_k)\bigr),
\end{equation}
The derivative of the \ac{PIRPNN} model at the same point remains linear in the coefficient vector \(\mathbf{w}\), because the basis functions are fixed and only their evaluated derivatives appear. Thus,
\begin{equation}
\label{eq:left_slope_model}
\dot{\mathbf{x}}_{\mathbf{w}}(t_k)=\Phi(t_k)\mathbf{w},
\end{equation}
where \(\Phi(t_k)\) is the basis-derivative matrix that maps the coefficient vector \(\mathbf{w}\) to the trajectory derivative at \(t_k\).

We therefore enforce the slope-matching condition
\begin{equation}
\label{eq:left_ls}
\Phi(t_k)\mathbf{w} \approx \dot{\mathbf{x}}(t_k).
\end{equation}

\paragraph*{Estimated Derivative at the End of the Interval}
To incorporate knowledge about the expected trend over the interval, we compute a forward Euler prediction of the state:
\begin{equation}
\label{eq:euler_predict}
\tilde{\mathbf{x}}(t_{k+1})=
\mathbf{x}(t_k)+\Delta t_k \,\dot{\mathbf{x}}(t_k).
\end{equation}
This predicted state yields a predicted derivative:
\begin{equation}
\label{eq:right_slope_true}
\dot{\mathbf{x}}_{\mathrm{pred}}
=
\mathbf{F}\bigl(t_{k+1},\tilde{\mathbf{x}}(t_{k+1}),\mathbf{u}(t_{k+1}),\mathbf{\sigma}(t_{k+1})\bigr),
\end{equation}
and the \ac{PIRPNN} derivative at the tentative end of the interval is
\begin{equation}
\label{eq:right_slope_model}
\dot{\mathbf{x}}_{\mathbf{w}}(t_{k+1})=\Phi(t_{k+1})\mathbf{w}.
\end{equation}

The corresponding matching condition is
\begin{equation}
\label{eq:right_ls}
\Phi(t_{k+1})\mathbf{w} \approx \dot{\mathbf{x}}_{\mathrm{pred}}.
\end{equation}

\paragraph*{Coefficients from Previous Intervals}
When two consecutive intervals belong to the same smooth operating segment, the accepted trajectory coefficients from the previous interval provide a useful initialization for the next one. RAPTOR therefore initializes the new coefficients close to the previously accepted coefficients, except after faults, switching actions, limiter activations, or other discrete events, where the trajectory may change abruptly and the continuation bias should be reset or weakened. We add the quadratic penalty:
\begin{equation}
\label{eq:regularization}
\norm{\mathbf{w}-\mathbf{w}_{\mathrm{prev}}}^2.
\end{equation}
This term is used only after the first accepted interval and only when continuation from the previous interval is appropriate. 

\paragraph*{Linear System for the Initial Coefficients}
Finally, combining \eqref{eq:left_ls}, \eqref{eq:right_ls}, and \eqref{eq:regularization} yields the linear system
\begin{align}
\label{eq:continuation_linear_system}
&\Bigl(
\theta I
+\Phi(t_k)^\top\Phi(t_k)
+\Phi(t_{k+1})^\top\Phi(t_{k+1})
\Bigr)\mathbf{w}
\\
&=
\theta\mathbf{w}_{\mathrm{prev}}
+\Phi(t_k)^\top \dot{\mathbf{x}}(t_k)
+\Phi(t_{k+1})^\top \dot{\mathbf{x}}_{\mathrm{pred}}.
\notag
\end{align}
 Here, \(I\) is the identity matrix with the same dimension as \(\mathbf{w}\), and \(\theta I\) regularizes the linear system while biasing the initialization toward the previous accepted coefficients. Note that we use the parameter \(\theta\) to control the strength of bias for coefficients from previous intervals as discussed in the previous paragraph. All terms on the right-hand side are known before the nonlinear interval solve begins. The resulting \(\mathbf{w}\) is used as the initial guess for the nonlinear solver.

\subsubsection{Residual-Based Interval Acceptance}
After solving the interval problem, we evaluate a residual-based error indicator at the right endpoint:

\begin{equation}
\label{eq:err_estimate}
\mathrm{err}_k
=
\frac{
\norm{
\dot{\mathbf{x}}_{\mathbf{w}}(t_{k+1})
-
\mathbf{F}\bigl(
t_{k+1},
\mathbf{x}_{\mathbf{w}}(t_{k+1}),
\mathbf{u}(t_{k+1}),
\mathbf{\sigma}(t_{k+1})
\bigr)
}
}{
\mathrm{atol}+\mathrm{rtol}\,\norm{\mathbf{x}_{\mathbf{w}}(t_{k+1})}
}
\end{equation}
The numerator measures the differential-equation mismatch at the end of the interval. The denominator normalizes this mismatch using the user-defined absolute and relative tolerances, \(\mathrm{atol}\) and \(\mathrm{rtol}\). If \(\mathrm{err}_k < 1\), the interval is accepted and \(\mathbf{x}_{\mathbf{w}}(t_{k+1})\) is used as the initial condition for the next interval. If \(\mathrm{err}_k > 1\), the interval is rejected, the interval length is reduced, and the solve is repeated. This is analogous to step acceptance in classical adaptive ODE solvers, except that the decision is based on the DAE residual rather than on a local truncation-error estimate. 

\subsubsection{Adaptive Interval-Length Selection}

After an interval is accepted, RAPTOR updates the next interval length \(\Delta t_{k+1}\). The goal is simple: if the error indicator is much smaller than one, the next interval can be larger; if the error is close to one, the interval length should remain similar; and if the error is too large, the interval is rejected and retried with a smaller length.

We use a \ac{PI}-type controller, as in standard adaptive time integration. After each accepted interval, the proposed growth factor is
\begin{equation}
\label{eq:pi_growth_factor}
G_k
=
\left(
\mathrm{err}_k^{k_P}
\mathrm{err}_{k-1}^{k_I}
\right)^{-1/r},
\end{equation}
where \(r>0\) is a prescribed controller parameter, and \(k_P\) and \(k_I\) are the proportional and integral controller gains. The current error \(\mathrm{err}_k\) makes the controller react to the present interval, while the previous error \(\mathrm{err}_{k-1}\) smooths the interval-length changes across successive intervals. For the first accepted interval, where no previous error is available, we set \(\mathrm{err}_{k-1}=\mathrm{err}_k\).

The next interval length is then computed as
\begin{equation}
\label{eq:pi_stepsize_simplified}
\Delta t_{k+1}
=
\Delta t_k \,
\eta_{\mathrm{saf}}\,
\max\!\Bigl(
\gamma_{\min},
\min(G_k,\gamma_{\max})
\Bigr),
\end{equation}
where \(\eta_{\mathrm{saf}}\in(0,1]\) is a safety factor that makes the update slightly conservative. The constants \(\gamma_{\min}\) and \(\gamma_{\max}\) limit how much the interval length can shrink or grow after one accepted interval. The interval length is also constrained by
\begin{equation}
    \Delta t_{\min} \le \Delta t_k \le t_f-t_k .
\end{equation}

If an interval is rejected, the state is not advanced. RAPTOR retries the same interval with a reduced interval length, for example
\begin{equation}
    \Delta t_k
\leftarrow
\max(\Delta t_{\min},\gamma_{\mathrm{rej}}\Delta t_k),
\qquad
0<\gamma_{\mathrm{rej}}<1 ,
\end{equation}

where \(\gamma_{\mathrm{rej}}\) is the rejection shrink factor. Thus, accepted intervals allow controlled growth or reduction of the next interval, while rejected intervals force an immediate reduction and retry.

Before the PI controller has any previous error information, RAPTOR chooses the first interval length using the standard starting-step heuristic of adaptive ODE solvers \cite{GladwellShampineBrankin1987,Hairer2000}. 

\begin{remark}[Index-1 DAEs and discrete events]

When a fault, breaker action, limiter activation, or other discrete event occurs, RAPTOR does not represent the discontinuity inside one smooth trajectory. Instead, the current interval is terminated at the event time, the active state \(\mathbf{\sigma}(t)\) and algebraic equations are updated, a new algebraically consistent post-event state is computed, and \ac{RAPTOR} starts a new trajectory interval on the updated system. Thus, discontinuities are handled by interval termination, reinitialization, and continuation on the updated index-1 \ac{DAE}, consistent with standard event-driven time-domain simulation practice.
\end{remark}

\section{Results and Performance Analysis}
\label{sec:results}

We demonstrate and evaluate RAPTOR on a set of RMS and EMT test cases covering both differential-equation and index-1 DAE simulations. The benchmarks are selected to span a wide spectrum of numerical characteristics encountered in power system dynamics, including stiffness, multi-timescale behaviour, severe oscillations, and fast electromagnetic transients. They comprise converter and transmission-line models, IEEE benchmark systems, and EMT models of a synchronous machine and a grid-following inverter. Because no individual test case captures all relevant numerical challenges, a diverse benchmark suite is required to evaluate RAPTOR's robustness and efficiency across fundamentally different simulation regimes. The presented analysis therefore focuses on testing consistent solver performance across diverse numerical regimes. Even though one larger benchmark system is included, scalability tests for very large systems are left for future work. For all benchmarks, RAPTOR is compared against established time-domain solvers that represent the current state of practice in power-system simulation, thereby providing a direct assessment against widely used industry and research tools rather than alternative spectral or frequency-domain approaches.

\subsection{Experimental and Evaluation Setup}
RAPTOR is implemented in Python using JAX. The nonlinear interval problem
is solved using the modified Newton (Chord) method described in
Section~\ref{sec:interval_optimization}, where the residual Jacobian and its
factorization are reused across coefficient updates. All reported experiments
are executed on a single CPU core, and one-time JAX compilation and warm-up
costs are excluded from the reported runtimes. The complete implementation
and benchmark configurations are available at \cite{RAPTORSupplementary}.

Unless stated otherwise, RMS results are averaged over \(10\)
initial-condition instances and EMT results over \(10\) disturbance events.
A run is marked as failed if it terminates early, returns NaNs/Infs, or
produces an incomplete trajectory. RMS component-model errors are evaluated against a high-accuracy
Kvaerno--5 reference computed with Diffrax
\cite{Kidger2021Diffrax}, using
\(\mathrm{atol}=\mathrm{rtol}=10^{-9}\).
For evaluation, each solver is represented by two configurations: the configuration with the shortest execution time, denoted \emph{Best Time} and the configuration with the lowest mean absolute error, denoted \emph{Best MAE}. For the conventional solvers, these use \((\mathrm{atol},\mathrm{rtol})=(10^{-3},10^{-5})\) and
\((10^{-5},10^{-8})\), respectively. RAPTOR similarly uses
\((\pi_{\mathrm{atol}},\pi_{\mathrm{rtol}})=(30,10^{-1})\) and
\((10^{-1},10^{-2})\), respectively. These RAPTOR parameters scale the
differential-equation residual used for interval acceptance and are therefore
not directly comparable in magnitude to the state-error tolerances
\(\mathrm{atol}\) and \(\mathrm{rtol}\) used by conventional integrators.
The reported configurations were selected empirically to represent the
lowest-error and lowest-runtime operating points of the tested solver settings. A fixed random seed is used for the sampled RAPTOR RBF parameters in all
reported comparisons. A configuration denoted RAPTOR-\(N\) uses \(N\) Gaussian RBFs per
differential state. The number of RBFs \(N\) and the number of collocation
points \(M\) are selected empirically from a small preliminary configuration
sweep, retaining the combination that provides the best accuracy--runtime
trade-off for the corresponding benchmark. This preliminary selection is performed once for each benchmark configuration
and is not repeated for every simulation run.

We compare RAPTOR against representative stiff and nonstiff integration
families: BDF, Radau, LSODA, DOP853, and RK45 use the standard SciPy
implementations \cite{Virtanen2020SciPy}, while Kvaerno3--5, Tsit5, Dopri5,
Dopri8, and implicit Euler use Diffrax \cite{Kidger2021Diffrax}. RAPTOR and
the fixed-step EMT trapezoidal solvers are custom implementations, explained in \cite{RAPTORSupplementary}. Trapezoidal
integration is included because it is widely used in EMTDC/PSCAD
\cite{HairerWanner1996,PSCADTrapezoidal}.


All benchmark-specific physical parameters, controller gains, initialization settings, and RAPTOR internal numerical parameters required to reproduce the experiments are reported in Supplementary Material \cite{RAPTORSupplementary}. Note that all physical system and control parameters are kept the same per benchmark. In other words, each presented method solves the exact same system of \acp{DAE}.

\subsection{RMS Benchmarks: Semi-Explicit ODE/DAE Models}
This section evaluates RAPTOR on RMS benchmarks with increasingly demanding
oscillatory, multi-timescale, and high-dimensional \ac{DAE} dynamics. We first consider a transmission-line model represented by \(M\) cascaded lumped \(LC\) sections and then the IEEE 9-, 14-, 39-, 57-, and
118-bus systems. The supplementary material \cite{RAPTORSupplementary} provides additional insights and results for a standalone grid-following inverter and the impact on simulation performance under progressively weaker grid conditions.

\subsubsection{Cascaded M-Section Line Benchmark}
\begin{figure}[t]
\begin{minipage}[t]{\columnwidth}
    \centering
    \begin{tikzpicture}

\begin{axis}[
    title={},
    xlabel={Average overall MAE},
    ylabel={Average runtime (s)},
    ylabel style={xshift=-10pt},
    xmode=log,
    ymode=log,
    grid=major,
    xmin=5e-7, xmax=3e-1,
    ymin=3e-3, ymax=10,
    clip=false,
    legend columns=5, 
    legend style={
        at={(axis description cs:1,1.1)}, 
        anchor=south east,
        nodes={scale=1, anchor=west},
        draw=none,
        column sep=0.1cm
    }
]

\tikzset{
  paretoLine/.style={line width=1.1pt, opacity=0.8}
}

\node[anchor=south west, draw=none, font=\scriptsize, fill = white!80, fill opacity = 0.4, text opacity =1] at (axis description cs:0.05,0.05)
{%
\begin{tabular}{@{}c@{\hspace{4pt}}l@{}}
$\bullet$ & Best MAE \\
$\circ$   & Best Time
\end{tabular}%
};

\addplot[only marks, mark=square*, color=red, mark size=3pt] coordinates {(2.82232e-4, 0.20802)};
\addlegendentry{RAPTOR--6}
\addplot[only marks, mark=square, color=red, mark size=3pt, forget plot] coordinates {(6.96550e-4, 0.16333)};
\addplot[paretoLine, color=red, forget plot] coordinates {(2.82232e-4, 0.20802) (6.96550e-4, 0.16333)};

\addplot[only marks, mark=triangle*, color=magenta, mark size=3pt] coordinates {(5.61817e-7, 0.24907)};
\addlegendentry{RAPTOR--8}
\addplot[only marks, mark=triangle, color=magenta, mark size=3pt, forget plot] coordinates {(1.19912e-6, 0.22399)};
\addplot[paretoLine, color=magenta, forget plot] coordinates {(5.61817e-7, 0.24907) (1.19912e-6, 0.22399)};

\addplot[only marks, mark=diamond*, color=red!70!black, mark size=3pt] coordinates {(1.58129e-6, 3.77371)};
\addlegendentry{RAPTOR--10}
\addplot[only marks, mark=diamond, color=red!70!black, mark size=3pt, forget plot] coordinates {(3.42324e-6, 2.96721)};
\addplot[paretoLine, color=red!70!black, forget plot] coordinates {(1.58129e-6, 3.77371) (3.42324e-6, 2.96721)};

\addplot[only marks, mark=*, color=gray, mark size=2pt] coordinates {(2.99916e-3, 1.19996)};
\addlegendentry{BDF}
\addplot[only marks, mark=o, color=gray, mark size=2pt, forget plot] coordinates {(4.51695e-2, 0.85855)};

\addplot[only marks, mark=*, color=orange, mark size=2pt] coordinates {(1.23890e-4, 0.66115)};
\addlegendentry{DOP853}
\addplot[only marks, mark=o, color=orange, mark size=2pt, forget plot] coordinates {(4.56419e-3, 0.22495)};

\addplot[only marks, mark=*, color=blue!50!black, mark size=2pt] coordinates {(1.26973e-1, 0.05540)};
\addlegendentry{Dopri5}
\addplot[only marks, mark=o, color=blue!50!black, mark size=2pt, forget plot] coordinates {(1.48554e-1, 0.04547)};

\addplot[only marks, mark=*, color=brown, mark size=2pt] coordinates {(3.64117e-4, 0.05651)};
\addlegendentry{Dopri8}
\addplot[only marks, mark=o, color=brown, mark size=2pt, forget plot] coordinates {(1.73803e-2, 0.02688)};

\addplot[only marks, mark=*, color=black!30!red, mark size=2pt] coordinates {(7.16869e-3, 0.00912)};
\addlegendentry{ImpEuler}
\addplot[only marks, mark=o, color=black!30!red, mark size=2pt, forget plot] coordinates {(7.16869e-3, 0.00492)};

\addplot[only marks, mark=*, color=green, mark size=2pt] coordinates {(6.48125e-3, 0.03810)};
\addlegendentry{Kvaerno3}
\addplot[only marks, mark=o, color=green, mark size=2pt, forget plot] coordinates {(6.48125e-3, 0.02518)};

\addplot[only marks, mark=*, color=cyan, mark size=2pt] coordinates {(6.77124e-3, 0.04029)};
\addlegendentry{Kvaerno4}
\addplot[only marks, mark=o, color=cyan, mark size=2pt, forget plot] coordinates {(6.77124e-3, 0.01954)};

\addplot[only marks, mark=*, color=blue, mark size=2pt] coordinates {(2.22424e-3, 0.32906)};
\addlegendentry{Kvaerno5}
\addplot[only marks, mark=o, color=blue, mark size=2pt, forget plot] coordinates {(7.56975e-3, 0.01258)};

\addplot[only marks, mark=*, color=purple, mark size=2pt] coordinates {(1.35973e-3, 0.57484)};
\addlegendentry{LSODA}
\addplot[only marks, mark=o, color=purple, mark size=2pt, forget plot] coordinates {(1.21922e-2, 0.18283)};

\addplot[only marks, mark=*, color=black, mark size=2pt] coordinates {(1.90466e-5, 2.03270)};
\addlegendentry{Radau}
\addplot[only marks, mark=o, color=black, mark size=2pt, forget plot] coordinates {(4.25717e-3, 0.16535)};

\addplot[only marks, mark=*, color=black!50!green, mark size=2pt] coordinates {(1.83984e-1, 0.07141)};
\addlegendentry{Tsit5}
\addplot[only marks, mark=o, color=black!50!green, mark size=2pt, forget plot] coordinates {(2.28356e-1, 0.04202)};

\end{axis}
\end{tikzpicture}
    \captionof{figure}{Speed--accuracy trade-off for the cascaded M-section line benchmark.}
    \label{fig:line_speed_accuracy}
    \vspace{2pt}
    
    \begin{tikzpicture}

\begin{axis}[
    title={},
    xlabel={Average number of advances},
    ylabel={Average runtime (s)},
    ylabel style={xshift=-10pt},
    ymode=log,
    grid=both,
    xmin=0, xmax=4100,
    xtick distance={1000},
    ymin=3e-3, ymax=10,
    clip=false,
    legend columns=5, 
    legend style={
        at={(axis description cs:1,1.1)}, 
        anchor=south east,
        nodes={scale=1, anchor=west},
        draw=none,
        column sep=0.1cm
    }
]

\tikzset{
  paretoLine/.style={line width=1.1pt, opacity=0.8}
}

\node[anchor=north east, draw=none, font=\scriptsize, fill = white!80, fill opacity = 0.4, text opacity =1] at (axis description cs:0.99,0.99)
{%
\begin{tabular}{@{}c@{\hspace{4pt}}l@{}}
$\bullet$ & Best MAE \\
$\circ$   & Best Time
\end{tabular}%
};


\addplot[only marks, mark=square*, color=red, mark size=3pt] coordinates {(1844, 0.20802)};
\addlegendentry{RAPTOR--6}
\addplot[only marks, mark=square, color=red, mark size=3pt, forget plot] coordinates {(1724, 0.16333)};
\addplot[paretoLine, color=red, forget plot] coordinates {(1844, 0.20802) (1724, 0.16333)};

\addplot[only marks, mark=triangle*, color=magenta, mark size=3pt] coordinates {(1987, 0.24907)};
\addlegendentry{RAPTOR--8}
\addplot[only marks, mark=triangle, color=magenta, mark size=3pt, forget plot] coordinates {(2002, 0.22399)};
\addplot[paretoLine, color=magenta, forget plot] coordinates {(1987, 0.24907) (2002, 0.22399)};

\addplot[only marks, mark=diamond*, color=red!70!black, mark size=3pt] coordinates {(1680, 3.77371)};
\addlegendentry{RAPTOR-10}
\addplot[only marks, mark=diamond, color=red!70!black, mark size=3pt, forget plot] coordinates {(1295, 2.96721)};
\addplot[paretoLine, color=red!70!black, forget plot] coordinates {(1680, 3.77371) (1295, 2.96721)};

\addplot[only marks, mark=*, color=gray, mark size=2pt] coordinates {(1999, 1.19996)};
\addlegendentry{BDF}
\addplot[only marks, mark=o, color=gray, mark size=2pt, forget plot] coordinates {(1999, 0.85855)};

\addplot[only marks, mark=*, color=orange, mark size=2pt] coordinates {(1999, 0.66115)};
\addlegendentry{DOP853}
\addplot[only marks, mark=o, color=orange, mark size=2pt, forget plot] coordinates {(1999, 0.22495)};

\addplot[only marks, mark=*, color=blue!50!black, mark size=2pt] coordinates {(3795, 0.05540)};
\addlegendentry{Dopri5}
\addplot[only marks, mark=o, color=blue!50!black, mark size=2pt, forget plot] coordinates {(3396, 0.04547)};

\addplot[only marks, mark=*, color=brown, mark size=2pt] coordinates {(1461, 0.05651)};
\addlegendentry{Dopri8}
\addplot[only marks, mark=o, color=brown, mark size=2pt, forget plot] coordinates {(1138, 0.02688)};

\addplot[only marks, mark=*, color=black!30!red, mark size=2pt] coordinates {(142, 0.00912)};
\addlegendentry{ImpEuler}
\addplot[only marks, mark=o, color=black!30!red, mark size=2pt, forget plot] coordinates {(142, 0.00492)};

\addplot[only marks, mark=*, color=green, mark size=2pt] coordinates {(366, 0.03810)};
\addlegendentry{Kvaerno3}
\addplot[only marks, mark=o, color=green, mark size=2pt, forget plot] coordinates {(366, 0.02518)};

\addplot[only marks, mark=*, color=cyan, mark size=2pt] coordinates {(218, 0.04029)};
\addlegendentry{Kvaerno4}
\addplot[only marks, mark=o, color=cyan, mark size=2pt, forget plot] coordinates {(218, 0.019544)};

\addplot[only marks, mark=*, color=blue, mark size=2pt] coordinates {(3051, 0.32906)};
\addlegendentry{Kvaerno5}
\addplot[only marks, mark=o, color=blue, mark size=2pt, forget plot] coordinates {(88, 0.01258)};

\addplot[only marks, mark=*, color=purple, mark size=2pt] coordinates {(1999, 0.57484)};
\addlegendentry{LSODA}
\addplot[only marks, mark=o, color=purple, mark size=2pt, forget plot] coordinates {(1999, 0.18283)};

\addplot[only marks, mark=*, color=black, mark size=2pt] coordinates {(1999, 2.03270)};
\addlegendentry{Radau}
\addplot[only marks, mark=o, color=black, mark size=2pt, forget plot] coordinates {(1999, 0.16535)};

\addplot[only marks, mark=*, color=black!50!green, mark size=2pt] coordinates {(3968, 0.07141)};
\addlegendentry{Tsit5}
\addplot[only marks, mark=o, color=black!50!green, mark size=2pt, forget plot] coordinates {(3419, 0.04202)};

\end{axis}
\end{tikzpicture}
    \captionof{figure}{Accepted intervals/timesteps versus runtime for the cascaded M-section line benchmark.}
    \label{fig:line_steps_time}
    \end{minipage}
\end{figure}

\begin{figure}[t]
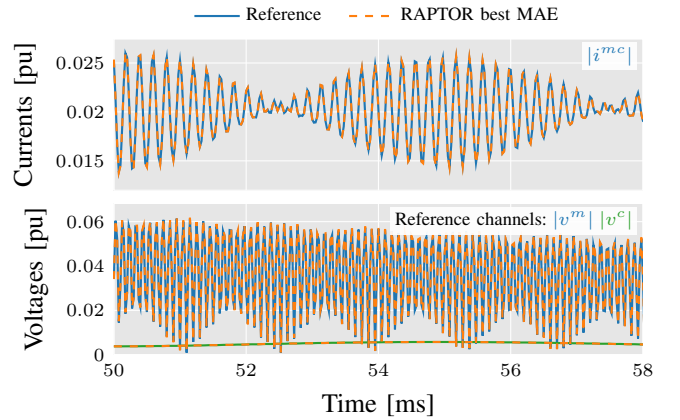

    \centering
    \includestandalone{Plots/Mline_trajectory}
    \caption{Example results for representative M-section trajectories over the 50--58~ms
    interval. Solid curves denote the reference solution and green dotted
    curves denote the RAPTOR best-MAE configuration.}
    \label{fig:mline_trajectory}
\end{figure}
Here, we consider a cascaded M-section model of a single transmission line driven by a sinusoidal current injection at the sending end. This benchmark is considerably more demanding because the distributed $LC$ dynamics introduce fast oscillatory modes of up to approximately $7$~kHz together with strong multi-timescale behavior. This is a representative case in which conventional time-domain solvers face a difficult trade-off between numerical stability, accuracy, and computational cost. Figure~\ref{fig:mline_trajectory} further shows that
RAPTOR closely reproduces the fast oscillatory current and voltage trajectories
of the high-accuracy reference solution.

As shown in Fig.~\ref{fig:line_speed_accuracy}, where the lower-left region again corresponds to low error and runtime, the RAPTOR configurations depict significantly lower error than the other variants. We additionally include several explicit methods in this benchmark to illustrate the numerical difficulty introduced by the fast oscillatory dynamics. In particular, Tsit5 and Dopri5 fail in \(80\%\) of the tested initial conditions, while the completed explicit runs generally exhibit substantially larger errors. Among the runs that complete, several conventional configurations still produce substantially larger errors despite additional timestep refinement.

Considering the different RAPTOR variants, the results in Fig.~\ref{fig:line_speed_accuracy} highlight the importance of carefully selecting the number of Gaussian RBFs used to represent the trajectory per RAPTOR interval solve. Here, RAPTOR-8 achieves the highest accuracy while runtime is similar to using 6 RBFs. Interestingly, increasing to 10 RBFs increases runtime, while reducing accuracy compared to RAPTOR-6. This illustrates the trade-off in the
RAPTOR representation: too few RBFs can limit the temporal flexibility of the
trajectory, whereas additional RBFs increase the size of the nonlinear
coefficient problem without necessarily providing useful additional
resolution.

Fig.~\ref{fig:line_steps_time} provides the complementary view in terms of accepted advances and runtime. While RAPTOR does not achieve the lowest runtime or the lowest number of advances, it completes all trials and keeps its accepted-advance count within a comparatively narrow range. However, read together Figs.~\ref{fig:line_speed_accuracy} and~\ref{fig:line_steps_time} highlight that short runtime and low number of advances do not imply an accurate solution: several conventional configurations with low runtime and low number of advances in Fig.~\ref{fig:line_steps_time} depict substantial errors in Fig.~\ref{fig:line_speed_accuracy}. Conventional configurations with similar runtime to RAPTOR produce significantly larger errors than RAPTOR. Hence, RAPTOR achieves robust accuracy at competitive runtime.

\subsection{Classic RMS Power System Benchmarks: IEEE 9-, 14-, 39-, 57- and 118-Bus Systems}
\label{subsec:dae_ieee_9_14_39_57_118}
We evaluate RAPTOR over a $1$~s simulation horizon on the IEEE 9-, 14-, 39-, 57-, and 118-bus networks. The tested systems span \(71\) states for the 9-bus to \(794\) states for the 118-bus case, testing whether RAPTOR remains effective across a substantial increase in system dimension. Each system is initialized from an AC power-flow operating point and includes sixth-order synchronous machines with AVR and governor dynamics, grid-following inverter aggregates with PLL and active/reactive-power control, fifth-order induction motors, and static loads. We use the network topology and operating points from the original IEEE case data, while the device dynamic parameters are taken from a common physically consistent benchmark rather than historical IEEE dynamic data. Details are provided in the supplementary material \cite{RAPTORSupplementary}.

We apply the same disturbance sequence to every test case and to all devices of the corresponding type: a $5~\%$ increase in the mechanical torque of all induction motors at $t=0.20$~s, a $3\%$ increase in the active-power reference of all inverters at $t=0.55$~s, and a $0.01$~pu increase in the AVR voltage reference of all synchronous machines at $t=0.80$~s.

We evaluate the accuracy against a high-accuracy Radau solver reference at
\(1001\) time steps, providing an independent reference and reducing reliance
on a single solver for validation. 
Runtime comparisons are performed under uniform accuracy requirements. A solver configuration is classified as \emph{accuracy-qualified} when its normalized mean absolute state error (NMAE) is at most \(3\times10^{-5}\), its bus-voltage-magnitude MAE is at most \(10^{-6}\), and its maximum Kirchhoff's Current Law (KCL) residual is at most \(10^{-10}\). The KCL residual measures the current-balance error at each bus, and the reported maximum is the largest mismatch over all buses and all audited time points. For every solver and network, the fastest tested configuration satisfying all three requirements is compared in the following for the runtime comparison.

Table~\ref{tab:dae_performance} depicts the accuracy and speed-up potentials for all test cases and different time-domain solvers. For accuracy comparison, it provides the \ac{MAE} of the differential states and bus voltage magnitudes for the selected RAPTOR configuration. Despite the increase in test system size, RAPTOR consistently requires only 22-32 intervals, while staying highly accurate in the order of \(10^{-7}\)--\(10^{-8}\). While accepted advances do not generally increase with system size, RAPTOR runtime increases with \ac{DAE} order. The larger the test case, the longer the runtime. Nonetheless, even the largest test case successfully solves within 0.312~s.

\begin{table}[t] 
\centering 
\caption{IEEE RMS DAE benchmark performance for \(t_f=1\)~s. Errors are RAPTOR MAEs, and speedup of RAPTOR is relative to the fastest accuracy-qualified configuration of each solver.} \label{tab:dae_performance} 
\footnotesize 
\setlength{\tabcolsep}{3.2pt} 
\begin{tabular}{lccccc} \toprule \textbf{Metric} & \textbf{IEEE 9} & \textbf{IEEE 14} & \textbf{IEEE 39} & \textbf{IEEE 57} & \textbf{IEEE 118} \\ \midrule State MAE ($10^{-8}$) & 5.44 & 8.54 & 18.7 & 9.70 & 2.42 \\ $|V|$ MAE ($10^{-8}$) & 1.65 & 2.52 & 6.60 & 1.81 & 0.671 \\ 
Time [ms] & \textbf{26.0} & \textbf{39.0} & \textbf{80.9} & \textbf{62.0} & \textbf{311.5} \\
Intervals & 28 & 22 & 27 & 23 & 32 \\  \midrule 
\multicolumn{6}{l}{\textbf{Relative Speed Up}} \\ \midrule
BDF ($\times$) & \textbf{2.79} & \textbf{2.45} & \textbf{1.91} & \textbf{1.96} & \textbf{2.19} \\ Radau ($\times$) & \textbf{6.16} & \textbf{6.76} & \textbf{5.33} & \textbf{5.94} & \textbf{16.17} \\ LSODA ($\times$) & \textbf{2.00} & \textbf{1.88} & \textbf{1.83} & \textbf{1.57} & \textbf{1.12} \\ DOP853 ($\times$) & \textbf{4.81} & \textbf{5.08} & \textbf{3.95} & \textbf{4.20} & \textbf{2.89} \\ RK45 ($\times$) & \textbf{2.58} & \textbf{2.80} & \textbf{2.44} & \textbf{2.36} & \textbf{1.69} \\ \bottomrule \end{tabular}
\end{table}

In addition, Table~\ref{tab:dae_performance} compares runtime between configurations that first meet the same prescribed accuracy requirements. Note that RAPTOR provides a speed-up compared to all tested classic \ac{DAE} solver benchmarks. Generally, the reported speedup depends on how much computational work each benchmark solver can avoid as the dynamics change during the simulation. On the IEEE 118-bus case, RAPTOR is \(16.17\times\) faster than the accuracy-qualified Radau configuration, whereas the speedup relative to LSODA is \(1.12\times\). LSODA automatically detects changes in stiffness and switches between a non-stiff Adams method and a stiff BDF method \cite{Petzold1983AutoSwitch}. It can therefore reduce its computational work when the trajectory becomes less oscillatory. RAPTOR follows a different strategy: It reduces the number of accepted advances while requiring only \(1.6\)--\(2.1\) nonlinear coefficient corrections per accepted interval. The corresponding solver-effort statistics are reported in the supplementary material \cite{RAPTORSupplementary}.

Generally, the provided comparison illustrates two different ways of reducing runtime: LSODA changes the integration method when the local problem permits it, and thus is one of the fastest benchmarks. In contrast, RAPTOR reduces the number of accepted advances by sectionalizing a trajectory into intervals rather than computing only the next state.





\subsection{EMT Benchmarks: Trapezoidal and Delayed-Control Comparisons}\label{sec:res_emt}
In the final experiments, we evaluate RAPTOR on an EMT synchronous-machine model and an EMT grid-following inverter model against two custom fixed-step implicit trapezoidal solvers, described in the supplementary material \cite{RAPTORSupplementary}. The first trapezoidal solver is a fine reference in which the controller states, electromagnetic states, and algebraic network variables are evaluated using current-step quantities within the same timestep. The second, used for the grid-following inverter, is a coarser delayed-control implementation motivated by the sequential calculation used in EMTDC/PSCAD: the electromagnetic equations use the controller output from the previous timestep, introducing a one-step controller delay.

The reported runtimes come from our custom implementations, not the PSCAD executable. EMTDC uses fixed timesteps and trapezoidal integration \cite{PSCADTrapezoidal}. At the same time, its sequential component evaluation can introduce one-timestep signal delays when a required signal is evaluated later in the calculation order \cite{PSCADTimeStepDelay}.

\subsubsection{EMT Synchronous Machine Benchmark}
We first consider an EMT synchronous-machine benchmark with 8 differential states, 11 algebraic states, and one switching event at $t=0.05$~s, where the infinite-bus voltage source is switched off. The reference solution is generated using the implicit trapezoidal rule with a full Newton solve at each micro-step, using fixed timesteps of $\Delta t=10^{-6}$~s and $\Delta t=10^{-7}$~s.

\begin{figure}[t]
    \centering
    \begin{tikzpicture}

\begin{groupplot}[
  group style={
    group size=1 by 2, 
    vertical sep=10pt,
  },
  grid=both,
  xtick distance={0.05},
  legend columns = 2,
  legend style={at={(0.5,1.05)},anchor=south,draw=none,fill=none,
  /tikz/every even column/.append style={column sep=5pt}},
  xmin = 0, xmax = 0.1
]

\nextgroupplot[
  ylabel={Currents [pu]},
  xlabel={}, 
  xticklabels=\empty
]

\addplot[thick, solid, color=sBlue] table[col sep=comma, x=t, y=ia] {Plots/data/emt_traditional_latex_solution.csv};
\addlegendentry{Trapezoidal Rule}
\addplot[thick, solid, color=sRed, forget plot] table[col sep=comma, x=t, y=ib] {Plots/data/emt_traditional_latex_solution.csv};
\addplot[thick, solid, color=sGreen, forget plot] table[col sep=comma, x=t, y=ic] {Plots/data/emt_traditional_latex_solution.csv};

\addplot[only marks, mark=*, mark size=1pt, color=sBlue, each nth point={12}] table[col sep=comma, x=t, y=ia] {Plots/data/emt_jax_solution.csv};
\addlegendentry{RAPTOR}
\addplot[only marks, mark=*, mark size=1pt, color=sRed, each nth point={12}, forget plot] table[col sep=comma, x=t, y=ib] {Plots/data/emt_jax_solution.csv};
\addplot[only marks, mark=*, mark size=1pt, color=sGreen, each nth point={12}, forget plot] table[col sep=comma, x=t, y=ic] {Plots/data/emt_jax_solution.csv};

\nextgroupplot[
  ylabel={Voltages [pu]},
  xlabel={Time [ms]}, 
  xticklabel={\pgfmathparse{\tick*1000}\pgfmathprintnumber{\pgfmathresult}},
]

\addplot[thick, solid, color=sBlue, forget plot] table[col sep=comma, x=t, y=va] {Plots/data/emt_traditional_latex_solution.csv};
\addplot[thick, solid, color=sRed, forget plot] table[col sep=comma, x=t, y=vb] {Plots/data/emt_traditional_latex_solution.csv};
\addplot[thick, solid, color=sGreen, forget plot] table[col sep=comma, x=t, y=vc] {Plots/data/emt_traditional_latex_solution.csv};

\addplot[only marks, mark=*, mark size=1pt, color=sBlue, each nth point={12}, forget plot] table[col sep=comma, x=t, y=va] {Plots/data/emt_jax_solution.csv};
\addplot[only marks, mark=*, mark size=1pt, color=sRed, each nth point={12}, forget plot] table[col sep=comma, x=t, y=vb] {Plots/data/emt_jax_solution.csv};
\addplot[only marks, mark=*, mark size=1pt, color=sGreen, each nth point={12}, forget plot] table[col sep=comma, x=t, y=vc] {Plots/data/emt_jax_solution.csv};

\end{groupplot}

\end{tikzpicture}
    \caption{EMT synchronous-machine trajectories: fine simultaneous
    trapezoidal reference versus RAPTOR at common evaluation times.}
    \label{fig:emt-trajectory}
\end{figure}
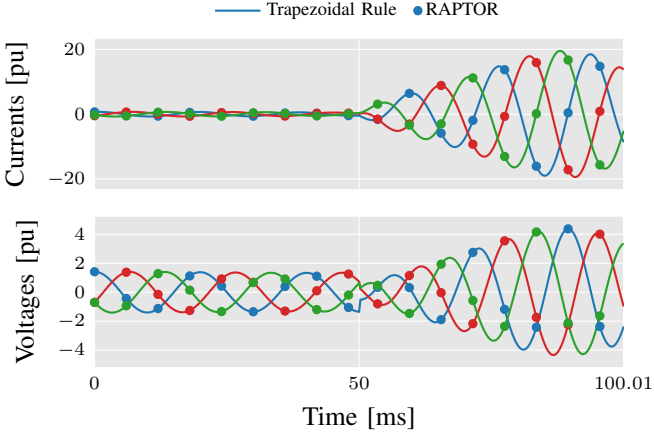

Figure~\ref{fig:emt-trajectory} shows time-domain trajectories for one operating condition subjected to the switching event at \(t=50\)~ms. Despite using only a small number of accepted intervals, RAPTOR accurately reproduces the trajectory obtained with the fine simultaneous trapezoidal reference solution. The interval endpoints shown in Fig.~\ref{fig:emt-trajectory} are widely spaced, highlighting the substantial reduction in time advances achieved by RAPTOR. However, these endpoints merely define the validity ranges of the underlying PIRPNNs and do not represent the only times at which the solution is available. Within each interval, the corresponding PIRPNN provides a continuous approximation of the system states. As a result, the state trajectory can be evaluated at any desired time instant throughout the simulation horizon without requiring additional numerical integration or nonlinear solves.

The best tested RAPTOR configuration reaches a last-point maximum error of \(2.3\times10^{-6}\) with runtime below \(0.5\)~s. Relative to the finer \(\Delta t=10^{-7}\)~s trapezoidal reference, the measured runtime gain exceeds one order of magnitude. This comparison establishes that RAPTOR can reproduce the current and voltage waveforms of this benchmark while advancing over substantially longer effective intervals.

\subsubsection{EMT Grid-Following Inverter}
We adapt a grid-following converter EMT model from \cite{VenturaNadal2025EMT}. It combines high-bandwidth controls, PLL dynamics, and disturbance-driven transient behavior. We evaluate the method over 10 disturbance events, including reference changes and voltage-sag scenarios.

\begin{figure}[t]
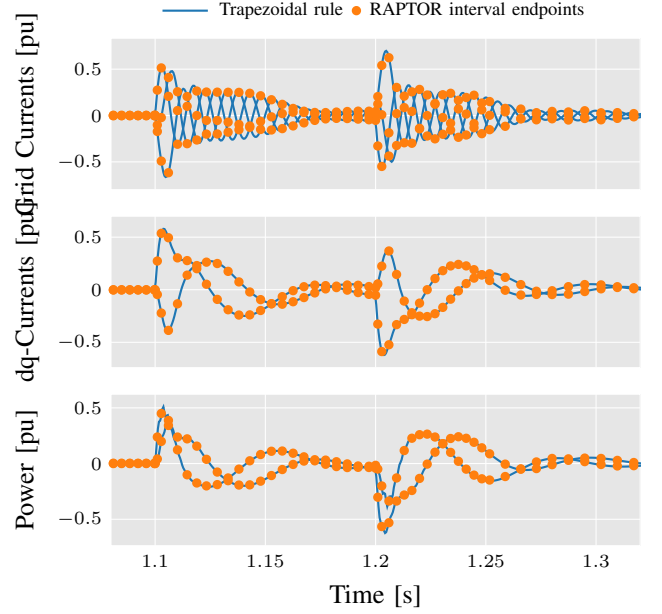

    \centering
    \includestandalone{Plots/emt_pll1}
    \caption{Grid-following EMT inverter trajectories over the
\(1.1\)--\(1.4\)~s disturbance window.}
    \label{fig:emt-pll}
\end{figure}

Fig.~\ref{fig:emt-pll} depicts time-domain trajectories for one test case including two consecutive disturbances.  Like for the synchronous-machine case, close agreement between RAPTOR and the fine fully coupled implicit-trapezoidal reference at common evaluation times is achieved. RAPTOR captures both the fast electrical response and the slower controller dynamics with high accuracy.

Due to space limitations, we only present the results for one example trajectory and selected states in this manuscript. The interested reader can find additional evaluation and tests in the supplementary material \cite{RAPTORSupplementary}.

Across all test cases, the custom PSCAD/EMTDC-style delayed-control trapezoidal baseline with $\Delta t=10^{-5}\,\mathrm{s}$ requires an average runtime of 3.5~s for the 2~s simulation. The fine simultaneous trapezoidal reference, using $\Delta t=10^{-6}\,\mathrm{s}$, increases the average runtime to 29.1~s. In contrast, RAPTOR evaluates the electromagnetic and network equations using the current controller values and therefore avoids the one-step controller delay introduced by the delayed-control formulation. Using only 360 accepted intervals on average, RAPTOR completes the same simulation in 1.6~s. Consequently, RAPTOR is approximately $2.2\times$ faster than the PSCAD/EMTDC-style delayed-control baseline and $18.2\times$ faster than the fine fully coupled reference, while simultaneously eliminating the controller--electromagnetic interface delay.

\section{Conclusion}

This paper introduces RAPTOR, a novel high-performance time-domain solver whose numerical core is a new time-spectral integration method for stiff power-system ODEs and index-1 DAEs. RAPTOR represents the unknown solutions of the \acp{DAE} with a set of fixed Gaussian \acp{RBF}. These are bell-shaped functions whose broad and localized shapes allow one interval trajectory to represent slow evolution and fast changes. The nonlinear solver determines how to linearly combine the \acp{RBF} so that the governing equations are satisfied at collocation points, while the algebraic variables are recovered from their constraint equations. Physics-guided random RBF shape selection, coefficient initialization, residual-based interval acceptance, interval-length control, advanced linear-algebra techniques, and event-aligned restart complete the solver.

The results identify the numerical gap targeted by RAPTOR more clearly. Stiffness alone is not sufficient to make RAPTOR preferable: mature implicit methods are already highly effective for small stiff systems whose trajectories remain smooth. The harder regime occurs when short fast transients are embedded within slower dynamics. This occurs naturally after faults, switching actions, or reference changes, when fast electrical and control states respond while slower system dynamics remain active. Conventional implicit methods may remain numerically stable in this regime, but accuracy and nonlinear convergence can still require many short sequential timesteps. RAPTOR addresses this difficulty by giving the nonlinear solver both broad and localized temporal functions and solving for a trajectory over each interval rather than only the next state.

The numerical results support this interpretation across the tested RMS,
networked DAE, and EMT benchmarks. RAPTOR reaches a \(16.17\times\) speedup
over the accuracy-qualified Radau configuration on IEEE~118, approximately
\(2.2\times\) over the custom delayed-control trapezoidal implementation, and
\(18.2\times\) over the fine fully coupled trapezoidal reference in the
grid-following EMT benchmark, while satisfying the prescribed accuracy
requirements. These results are obtained with a custom research implementation
of RAPTOR, whereas established time-domain solvers have benefited from decades
of algorithmic and software optimization. Further work will therefore focus on reducing RAPTOR's per-interval
computational cost to improve its computational advantage on larger systems.

\bibliographystyle{IEEEtran}
\bibliography{bibliography}

\clearpage

\end{document}


\definecolor{raptorred}{RGB}{190,30,45}
\definecolor{eventgray}{RGB}{120,120,120}
\pgfplotstableread[col sep=comma]{%
t,delta_ref,delta_rap,omega_ref,omega_rap,vmag_ref,vmag_rap,slip_ref,slip_rap,pf_ref,pf_rap,imag_ref,imag_rap
0.000000000,0,0,0,0,0.943,0.943,0.04,0.04,2.04,2.04,2.28774253205,2.28774253205
0.001000000,0,0,0,0,0.943,0.943,0.04,0.04,2.04,2.04,2.28774253205,2.28774253205
0.002000000,0,0,0,0,0.943,0.943,0.04,0.04,2.04,2.04,2.28774253205,2.28774253205
0.003000000,0,0,0,0,0.943,0.943,0.04,0.04,2.04,2.04,2.28774253205,2.28774253205
0.004000000,0,0,0,0,0.943,0.943,0.04,0.04,2.04,2.04,2.28774253205,2.28774253205
0.005000000,0,0,0,0,0.943,0.943,0.04,0.04,2.04,2.04,2.28774253205,2.28774253205
0.006000000,0,0,0,0,0.943,0.943,0.04,0.04,2.04,2.04,2.28774253205,2.28774253205
0.007000000,0,0,0,0,0.943,0.943,0.04,0.04,2.04,2.04,2.28774253205,2.28774253205
0.008000000,0,0,0,0,0.943,0.943,0.04,0.04,2.04,2.04,2.28774253205,2.28774253205
0.009000000,0,0,0,0,0.943,0.943,0.04,0.04,2.04,2.04,2.28774253205,2.28774253205
0.010000000,0,0,0,0,0.943,0.943,0.04,0.04,2.04,2.04,2.28774253205,2.28774253205
0.011000000,0,0,0,0,0.943,0.943,0.04,0.04,2.04,2.04,2.28774253205,2.28774253205
0.012000000,0,0,0,0,0.943,0.943,0.04,0.04,2.04,2.04,2.28774253205,2.28774253205
0.013000000,0,0,0,0,0.943,0.943,0.04,0.04,2.04,2.04,2.28774253205,2.28774253205
0.014000000,0,0,0,0,0.943,0.943,0.04,0.04,2.04,2.04,2.28774253205,2.28774253205
0.015000000,0,0,0,0,0.943,0.943,0.04,0.04,2.04,2.04,2.28774253205,2.28774253205
0.016000000,0,0,0,0,0.943,0.943,0.04,0.04,2.04,2.04,2.28774253205,2.28774253205
0.017000000,0,0,0,0,0.943,0.943,0.04,0.04,2.04,2.04,2.28774253205,2.28774253205
0.018000000,0,0,0,0,0.943,0.943,0.04,0.04,2.04,2.04,2.28774253205,2.28774253205
0.019000000,0,0,0,0,0.943,0.943,0.04,0.04,2.04,2.04,2.28774253205,2.28774253205
0.020000000,0,0,0,0,0.943,0.943,0.04,0.04,2.04,2.04,2.28774253205,2.28774253205
0.021000000,0,0,0,0,0.943,0.943,0.04,0.04,2.04,2.04,2.28774253205,2.28774253205
0.022000000,0,0,0,0,0.943,0.943,0.04,0.04,2.04,2.04,2.28774253205,2.28774253205
0.023000000,0,0,0,0,0.943,0.943,0.04,0.04,2.04,2.04,2.28774253205,2.28774253205
0.024000000,0,0,0,0,0.943,0.943,0.04,0.04,2.04,2.04,2.28774253205,2.28774253205
0.025000000,0,0,0,0,0.943,0.943,0.04,0.04,2.04,2.04,2.28774253205,2.28774253205
0.026000000,0,0,0,0,0.943,0.943,0.04,0.04,2.04,2.04,2.28774253205,2.28774253205
0.027000000,0,0,0,0,0.943,0.943,0.04,0.04,2.04,2.04,2.28774253205,2.28774253205
0.028000000,0,0,0,0,0.943,0.943,0.04,0.04,2.04,2.04,2.28774253205,2.28774253205
0.029000000,0,0,0,0,0.943,0.943,0.04,0.04,2.04,2.04,2.28774253205,2.28774253205
0.030000000,0,0,0,0,0.943,0.943,0.04,0.04,2.04,2.04,2.28774253205,2.28774253205
0.031000000,0,0,0,-2.22044604925e-14,0.943,0.943,0.04,0.04,2.04,2.04,2.28774253205,2.28774253205
0.032000000,0,0,0,-1.11022302463e-13,0.943,0.943,0.04,0.04,2.04,2.04,2.28774253205,2.28774253205
0.033000000,0,0,0,-1.11022302463e-13,0.943,0.943,0.04,0.04,2.04,2.04,2.28774253205,2.28774253205
0.034000000,0,0,0,-1.11022302463e-13,0.943,0.943,0.04,0.04,2.04,2.04,2.28774253205,2.28774253205
0.035000000,0,0,0,-1.11022302463e-13,0.943,0.943,0.04,0.04,2.04,2.04,2.28774253205,2.28774253205
0.036000000,0,0,0,-1.11022302463e-13,0.943,0.943,0.04,0.04,2.04,2.04,2.28774253205,2.28774253205
0.037000000,0,0,0,-1.11022302463e-13,0.943,0.943,0.04,0.04,2.04,2.04,2.28774253205,2.28774253205
0.038000000,0,0,0,-1.11022302463e-13,0.943,0.943,0.04,0.04,2.04,2.04,2.28774253205,2.28774253205
0.039000000,0,0,0,-1.11022302463e-13,0.943,0.943,0.04,0.04,2.04,2.04,2.28774253205,2.28774253205
0.040000000,0,0,0,-1.11022302463e-13,0.943,0.943,0.04,0.04,2.04,2.04,2.28774253205,2.28774253205
0.041000000,0,0,0,-1.11022302463e-13,0.943,0.943,0.04,0.04,2.04,2.04,2.28774253205,2.28774253205
0.042000000,0,0,0,-1.11022302463e-13,0.943,0.943,0.04,0.04,2.04,2.04,2.28774253205,2.28774253205
0.043000000,0,0,0,-1.11022302463e-13,0.943,0.943,0.04,0.04,2.04,2.04,2.28774253205,2.28774253205
0.044000000,0,0,0,-1.11022302463e-13,0.943,0.943,0.04,0.04,2.04,2.04,2.28774253205,2.28774253205
0.045000000,0,0,0,-1.11022302463e-13,0.943,0.943,0.04,0.04,2.04,2.04,2.28774253205,2.28774253205
0.046000000,0,0,0,-1.11022302463e-13,0.943,0.943,0.04,0.04,2.04,2.04,2.28774253205,2.28774253205
0.047000000,0,0,0,-1.11022302463e-13,0.943,0.943,0.04,0.04,2.04,2.04,2.28774253205,2.28774253205
0.048000000,0,0,0,-1.11022302463e-13,0.943,0.943,0.04,0.04,2.04,2.04,2.28774253205,2.28774253205
0.049000000,0,0,0,-1.11022302463e-13,0.943,0.943,0.04,0.04,2.04,2.04,2.28774253205,2.28774253205
0.050000000,0,0,0,-1.11022302463e-13,0.943,0.943,0.04,0.04,2.04,2.04,2.28774253205,2.28774253205
0.051000000,0,0,0,-1.25825276124e-13,0.943,0.943,0.04,0.04,2.04,2.04,2.28774253205,2.28774253205
0.052000000,0,0,0,-2.14643118094e-13,0.943,0.943,0.04,0.04,2.04,2.04,2.28774253205,2.28774253205
0.053000000,0,0,0,-2.22044604925e-13,0.943,0.943,0.04,0.04,2.04,2.04,2.28774253205,2.28774253205
0.054000000,0,0,0,-2.22044604925e-13,0.943,0.943,0.04,0.04,2.04,2.04,2.28774253205,2.28774253205
0.055000000,0,0,0,-2.22044604925e-13,0.943,0.943,0.04,0.04,2.04,2.04,2.28774253205,2.28774253205
0.056000000,0,0,0,-2.22044604925e-13,0.943,0.943,0.04,0.04,2.04,2.04,2.28774253205,2.28774253205
0.057000000,0,0,0,-2.22044604925e-13,0.943,0.943,0.04,0.04,2.04,2.04,2.28774253205,2.28774253205
0.058000000,0,0,0,-2.22044604925e-13,0.943,0.943,0.04,0.04,2.04,2.04,2.28774253205,2.28774253205
0.059000000,0,0,0,-2.22044604925e-13,0.943,0.943,0.04,0.04,2.04,2.04,2.28774253205,2.28774253205
0.060000000,0,0,0,-2.22044604925e-13,0.943,0.943,0.04,0.04,2.04,2.04,2.28774253205,2.28774253205
0.061000000,0,0,0,-2.22044604925e-13,0.943,0.943,0.04,0.04,2.04,2.04,2.28774253205,2.28774253205
0.062000000,0,0,0,-2.22044604925e-13,0.943,0.943,0.04,0.04,2.04,2.04,2.28774253205,2.28774253205
0.063000000,0,0,0,-2.22044604925e-13,0.943,0.943,0.04,0.04,2.04,2.04,2.28774253205,2.28774253205
0.064000000,0,0,0,-2.22044604925e-13,0.943,0.943,0.04,0.04,2.04,2.04,2.28774253205,2.28774253205
0.065000000,0,0,0,-2.59052039079e-13,0.943,0.943,0.04,0.04,2.04,2.04,2.28774253205,2.28774253205
0.066000000,0,0,0,-3.33066907388e-13,0.943,0.943,0.04,0.04,2.04,2.04,2.28774253205,2.28774253205
0.067000000,0,0,0,-3.33066907388e-13,0.943,0.943,0.04,0.04,2.04,2.04,2.28774253205,2.28774253205
0.068000000,0,0,0,-3.33066907388e-13,0.943,0.943,0.04,0.04,2.04,2.04,2.28774253205,2.28774253205
0.069000000,0,0,0,-3.33066907388e-13,0.943,0.943,0.04,0.04,2.04,2.04,2.28774253205,2.28774253205
0.070000000,0,0,0,-3.33066907388e-13,0.943,0.943,0.04,0.04,2.04,2.04,2.28774253205,2.28774253205
0.071000000,0,0,0,-3.33066907388e-13,0.943,0.943,0.04,0.04,2.04,2.04,2.28774253205,2.28774253205
0.072000000,0,0,0,-3.33066907388e-13,0.943,0.943,0.04,0.04,2.04,2.04,2.28774253205,2.28774253205
0.073000000,0,0,0,-3.33066907388e-13,0.943,0.943,0.04,0.04,2.04,2.04,2.28774253205,2.28774253205
0.074000000,0,0,0,-3.33066907388e-13,0.943,0.943,0.04,0.04,2.04,2.04,2.28774253205,2.28774253205
0.075000000,0,0,0,-3.33066907388e-13,0.943,0.943,0.04,0.04,2.04,2.04,2.28774253205,2.28774253205
0.076000000,0,0,0,-3.33066907388e-13,0.943,0.943,0.04,0.04,2.04,2.04,2.28774253205,2.28774253205
0.077000000,0,0,0,-3.33066907388e-13,0.943,0.943,0.04,0.04,2.04,2.04,2.28774253205,2.28774253205
0.078000000,0,0,0,-3.33066907388e-13,0.943,0.943,0.04,0.04,2.04,2.04,2.28774253205,2.28774253205
0.079000000,0,0,0,-3.33066907388e-13,0.943,0.943,0.04,0.04,2.04,2.04,2.28774253205,2.28774253205
0.080000000,0,0,0,-3.33066907388e-13,0.943,0.943,0.04,0.04,2.04,2.04,2.28774253205,2.28774253205
0.081000000,0,0,0,-3.33066907388e-13,0.943,0.943,0.04,0.04,2.04,2.04,2.28774253205,2.28774253205
0.082000000,0,0,0,-3.33066907388e-13,0.943,0.943,0.04,0.04,2.04,2.04,2.28774253205,2.28774253205
0.083000000,0,0,0,-3.33066907388e-13,0.943,0.943,0.04,0.04,2.04,2.04,2.28774253205,2.28774253205
0.084000000,0,0,0,-3.33066907388e-13,0.943,0.943,0.04,0.04,2.04,2.04,2.28774253205,2.28774253205
0.085000000,0,0,0,-3.70074341542e-13,0.943,0.943,0.04,0.04,2.04,2.04,2.28774253205,2.28774253205
0.086000000,0,0,0,-4.4408920985e-13,0.943,0.943,0.04,0.04,2.04,2.04,2.28774253205,2.28774253205
0.087000000,0,0,0,-4.4408920985e-13,0.943,0.943,0.04,0.04,2.04,2.04,2.28774253205,2.28774253205
0.088000000,0,0,0,-4.4408920985e-13,0.943,0.943,0.04,0.04,2.04,2.04,2.28774253205,2.28774253205
0.089000000,0,0,0,-4.4408920985e-13,0.943,0.943,0.04,0.04,2.04,2.04,2.28774253205,2.28774253205
0.090000000,0,0,0,-4.4408920985e-13,0.943,0.943,0.04,0.04,2.04,2.04,2.28774253205,2.28774253205
0.091000000,0,0,0,-4.4408920985e-13,0.943,0.943,0.04,0.04,2.04,2.04,2.28774253205,2.28774253205
0.092000000,0,0,0,-4.4408920985e-13,0.943,0.943,0.04,0.04,2.04,2.04,2.28774253205,2.28774253205
0.093000000,0,0,0,-4.4408920985e-13,0.943,0.943,0.04,0.04,2.04,2.04,2.28774253205,2.28774253205
0.094000000,0,0,0,-4.4408920985e-13,0.943,0.943,0.04,0.04,2.04,2.04,2.28774253205,2.28774253205
0.095000000,0,0,0,-4.4408920985e-13,0.943,0.943,0.04,0.04,2.04,2.04,2.28774253205,2.28774253205
0.096000000,0,0,0,-4.4408920985e-13,0.943,0.943,0.04,0.04,2.04,2.04,2.28774253205,2.28774253205
0.097000000,0,0,0,-4.4408920985e-13,0.943,0.943,0.04,0.04,2.04,2.04,2.28774253205,2.28774253205
0.098000000,0,0,0,-4.4408920985e-13,0.943,0.943,0.04,0.04,2.04,2.04,2.28774253205,2.28774253205
0.099000000,0,0,0,-4.4408920985e-13,0.943,0.943,0.04,0.04,2.04,2.04,2.28774253205,2.28774253205
0.100000000,0,0,0,-4.4408920985e-13,0.943,0.943,0.04,0.04,2.04,2.04,2.28774253205,2.28774253205
0.101000000,0,0,0,-4.58892183512e-13,0.943,0.943,0.04,0.04,2.04,2.04,2.28774253205,2.28774253205
0.102000000,0,0,0,-5.47710025482e-13,0.943,0.943,0.04,0.04,2.04,2.04,2.28774253205,2.28774253205
0.103000000,0,0,0,-5.55111512313e-13,0.943,0.943,0.04,0.04,2.04,2.04,2.28774253205,2.28774253205
0.104000000,0,0,0,-5.55111512313e-13,0.943,0.943,0.04,0.04,2.04,2.04,2.28774253205,2.28774253205
0.105000000,0,0,0,-5.55111512313e-13,0.943,0.943,0.04,0.04,2.04,2.04,2.28774253205,2.28774253205
0.106000000,0,0,0,-5.55111512313e-13,0.943,0.943,0.04,0.04,2.04,2.04,2.28774253205,2.28774253205
0.107000000,0,0,0,-5.55111512313e-13,0.943,0.943,0.04,0.04,2.04,2.04,2.28774253205,2.28774253205
0.108000000,0,0,0,-5.55111512313e-13,0.943,0.943,0.04,0.04,2.04,2.04,2.28774253205,2.28774253205
0.109000000,0,0,0,-5.55111512313e-13,0.943,0.943,0.04,0.04,2.04,2.04,2.28774253205,2.28774253205
0.110000000,0,0,0,-5.55111512313e-13,0.943,0.943,0.04,0.04,2.04,2.04,2.28774253205,2.28774253205
0.111000000,0,0,0,-5.55111512313e-13,0.943,0.943,0.04,0.04,2.04,2.04,2.28774253205,2.28774253205
0.112000000,0,0,0,-5.55111512313e-13,0.943,0.943,0.04,0.04,2.04,2.04,2.28774253205,2.28774253205
0.113000000,0,0,0,-5.55111512313e-13,0.943,0.943,0.04,0.04,2.04,2.04,2.28774253205,2.28774253205
0.114000000,0,0,0,-6.14323406959e-13,0.943,0.943,0.04,0.04,2.04,2.04,2.28774253205,2.28774253205
0.115000000,0,0,0,-6.66133814775e-13,0.943,0.943,0.04,0.04,2.04,2.04,2.28774253205,2.28774253205
0.116000000,0,0,0,-6.66133814775e-13,0.943,0.943,0.04,0.04,2.04,2.04,2.28774253205,2.28774253205
0.117000000,0,0,0,-6.66133814775e-13,0.943,0.943,0.04,0.04,2.04,2.04,2.28774253205,2.28774253205
0.118000000,0,0,0,-6.66133814775e-13,0.943,0.943,0.04,0.04,2.04,2.04,2.28774253205,2.28774253205
0.119000000,0,0,0,-6.66133814775e-13,0.943,0.943,0.04,0.04,2.04,2.04,2.28774253205,2.28774253205
0.120000000,0,0,0,-6.66133814775e-13,0.943,0.943,0.04,0.04,2.04,2.04,2.28774253205,2.28774253205
0.121000000,0,0,0,-6.66133814775e-13,0.943,0.943,0.04,0.04,2.04,2.04,2.28774253205,2.28774253205
0.122000000,0,0,0,-6.66133814775e-13,0.943,0.943,0.04,0.04,2.04,2.04,2.28774253205,2.28774253205
0.123000000,0,0,0,-6.66133814775e-13,0.943,0.943,0.04,0.04,2.04,2.04,2.28774253205,2.28774253205
0.124000000,0,0,0,-6.66133814775e-13,0.943,0.943,0.04,0.04,2.04,2.04,2.28774253205,2.28774253205
0.125000000,0,0,0,-6.66133814775e-13,0.943,0.943,0.04,0.04,2.04,2.04,2.28774253205,2.28774253205
0.126000000,0,0,0,-6.66133814775e-13,0.943,0.943,0.04,0.04,2.04,2.04,2.28774253205,2.28774253205
0.127000000,0,0,0,-6.66133814775e-13,0.943,0.943,0.04,0.04,2.04,2.04,2.28774253205,2.28774253205
0.128000000,0,0,0,-6.66133814775e-13,0.943,0.943,0.04,0.04,2.04,2.04,2.28774253205,2.28774253205
0.129000000,0,0,0,-6.66133814775e-13,0.943,0.943,0.04,0.04,2.04,2.04,2.28774253205,2.28774253205
0.130000000,0,0,0,-6.66133814775e-13,0.943,0.943,0.04,0.04,2.04,2.04,2.28774253205,2.28774253205
0.131000000,0,0,0,-6.66133814775e-13,0.943,0.943,0.04,0.04,2.04,2.04,2.28774253205,2.28774253205
0.132000000,0,0,0,-6.66133814775e-13,0.943,0.943,0.04,0.04,2.04,2.04,2.28774253205,2.28774253205
0.133000000,0,0,0,-6.66133814775e-13,0.943,0.943,0.04,0.04,2.04,2.04,2.28774253205,2.28774253205
0.134000000,0,0,0,-7.25345709422e-13,0.943,0.943,0.04,0.04,2.04,2.04,2.28774253205,2.28774253205
0.135000000,0,0,0,-7.77156117238e-13,0.943,0.943,0.04,0.04,2.04,2.04,2.28774253205,2.28774253205
0.136000000,0,0,0,-7.77156117238e-13,0.943,0.943,0.04,0.04,2.04,2.04,2.28774253205,2.28774253205
0.137000000,0,0,0,-7.77156117238e-13,0.943,0.943,0.04,0.04,2.04,2.04,2.28774253205,2.28774253205
0.138000000,0,0,0,-7.77156117238e-13,0.943,0.943,0.04,0.04,2.04,2.04,2.28774253205,2.28774253205
0.139000000,0,0,0,-7.77156117238e-13,0.943,0.943,0.04,0.04,2.04,2.04,2.28774253205,2.28774253205
0.140000000,0,0,0,-7.77156117238e-13,0.943,0.943,0.04,0.04,2.04,2.04,2.28774253205,2.28774253205
0.141000000,0,0,0,-7.77156117238e-13,0.943,0.943,0.04,0.04,2.04,2.04,2.28774253205,2.28774253205
0.142000000,0,0,0,-7.77156117238e-13,0.943,0.943,0.04,0.04,2.04,2.04,2.28774253205,2.28774253205
0.143000000,0,0,0,-7.77156117238e-13,0.943,0.943,0.04,0.04,2.04,2.04,2.28774253205,2.28774253205
0.144000000,0,3.39259166023e-15,0,-7.77156117238e-13,0.943,0.943,0.04,0.04,2.04,2.04,2.28774253205,2.28774253205
0.145000000,0,7.42129425675e-15,0,-7.77156117238e-13,0.943,0.943,0.04,0.04,2.04,2.04,2.28774253205,2.28774253205
0.146000000,0,1.03898119594e-14,0,-7.77156117238e-13,0.943,0.943,0.04,0.04,2.04,2.04,2.28774253205,2.28774253205
0.147000000,0,1.54786994498e-14,0,-7.77156117238e-13,0.943,0.943,0.04,0.04,2.04,2.04,2.28774253205,2.28774253205
0.148000000,0,1.82351801737e-14,0,-7.77156117238e-13,0.943,0.943,0.04,0.04,2.04,2.04,2.28774253205,2.28774253205
0.149000000,0,2.2475919749e-14,0,-7.77156117238e-13,0.943,0.943,0.04,0.04,2.04,2.04,2.28774253205,2.28774253205
0.150000000,0,2.75648072394e-14,0,-7.77156117238e-13,0.943,0.943,0.04,0.04,2.04,2.04,2.28774253205,2.28774253205
0.151000000,0,3.26536947297e-14,0,-7.77156117238e-13,0.943,0.943,0.04,0.04,2.04,2.04,2.28774253205,2.28774253205
0.152000000,0,3.774258222e-14,0,-7.77156117238e-13,0.943,0.943,0.04,0.04,2.04,2.04,2.28774253205,2.28774253205
0.153000000,0,4.0499062944e-14,0,-7.77156117238e-13,0.943,0.943,0.04,0.04,2.04,2.04,2.28774253205,2.28774253205
0.154000000,0,4.47398025193e-14,0,-7.77156117238e-13,0.943,0.943,0.04,0.04,2.04,2.04,2.28774253205,2.28774253205
0.155000000,0,4.98286900096e-14,0,-7.77156117238e-13,0.943,0.943,0.04,0.04,2.04,2.04,2.28774253205,2.28774253205
0.156000000,0,5.49175774999e-14,0,-7.77156117238e-13,0.943,0.943,0.04,0.04,2.04,2.04,2.28774253205,2.28774253205
0.157000000,0,6.00064649903e-14,0,-7.77156117238e-13,0.943,0.943,0.04,0.04,2.04,2.04,2.28774253205,2.28774253205
0.158000000,0,6.50953524806e-14,0,-7.77156117238e-13,0.943,0.943,0.04,0.04,2.04,2.04,2.28774253205,2.28774253205
0.159000000,0,7.0184239971e-14,0,-7.77156117238e-13,0.943,0.943,0.04,0.04,2.04,2.04,2.28774253205,2.28774253205
0.160000000,0,7.42129425675e-14,0,-7.77156117238e-13,0.943,0.943,0.04,0.04,2.04,2.04,2.28774253205,2.28774253205
0.161000000,0,7.71814602702e-14,0,-7.77156117238e-13,0.943,0.943,0.04,0.04,2.04,2.04,2.28774253205,2.28774253205
0.162000000,0,8.22703477605e-14,0,-7.77156117238e-13,0.943,0.943,0.04,0.04,2.04,2.04,2.28774253205,2.28774253205
0.163000000,0,8.73592352509e-14,0,-7.77156117238e-13,0.943,0.943,0.04,0.04,2.04,2.04,2.28774253205,2.28774253205
0.164000000,0,9.24481227412e-14,0,-7.77156117238e-13,0.943,0.943,0.04,0.04,2.04,2.04,2.28774253205,2.28774253205
0.165000000,0,9.75370102315e-14,0,-7.77156117238e-13,0.943,0.943,0.04,0.04,2.04,2.04,2.28774253205,2.28774253205
0.166000000,0,1.02201823764e-13,0,-7.77156117238e-13,0.943,0.943,0.04,0.04,2.04,2.04,2.28774253205,2.28774253205
0.167000000,0,1.0474626751e-13,0,-7.77156117238e-13,0.943,0.943,0.04,0.04,2.04,2.04,2.28774253205,2.28774253205
0.168000000,0,1.09623118021e-13,0,-7.77156117238e-13,0.943,0.943,0.04,0.04,2.04,2.04,2.28774253205,2.28774253205
0.169000000,0,1.13015709681e-13,0,-7.77156117238e-13,0.943,0.943,0.04,0.04,2.04,2.04,2.28774253205,2.28774253205
0.170000000,0,1.1662033832e-13,0,-7.77156117238e-13,0.943,0.943,0.04,0.04,2.04,2.04,2.28774253205,2.28774253205
0.171000000,0,1.21285151853e-13,0,-7.77156117238e-13,0.943,0.943,0.04,0.04,2.04,2.04,2.28774253205,2.28774253205
0.172000000,0,1.23829595598e-13,0,-7.77156117238e-13,0.943,0.943,0.04,0.04,2.04,2.04,2.28774253205,2.28774253205
0.173000000,0,1.2870644611e-13,0,-7.77156117238e-13,0.943,0.943,0.04,0.04,2.04,2.04,2.28774253205,2.28774253205
0.174000000,0,1.3209903777e-13,0,-7.77156117238e-13,0.943,0.943,0.04,0.04,2.04,2.04,2.28774253205,2.28774253205
0.175000000,0,1.35703666409e-13,0,-7.77156117238e-13,0.943,0.943,0.04,0.04,2.04,2.04,2.28774253205,2.28774253205
0.176000000,0,1.40368479942e-13,0,-7.77156117238e-13,0.943,0.943,0.04,0.04,2.04,2.04,2.28774253205,2.28774253205
0.177000000,0,1.42912923687e-13,0,-7.77156117238e-13,0.943,0.943,0.04,0.04,2.04,2.04,2.28774253205,2.28774253205
0.178000000,0,1.45457367432e-13,0,-7.77156117238e-13,0.943,0.943,0.04,0.04,2.04,2.04,2.28774253205,2.28774253205
0.179000000,0,1.48001811177e-13,0,-7.77156117238e-13,0.943,0.943,0.04,0.04,2.04,2.04,2.28774253205,2.28774253205
0.180000000,0,1.50546254923e-13,0,-7.77156117238e-13,0.943,0.943,0.04,0.04,2.04,2.04,2.28774253205,2.28774253205
0.181000000,0,1.53090698668e-13,0,-7.77156117238e-13,0.943,0.943,0.04,0.04,2.04,2.04,2.28774253205,2.28774253205
0.182000000,0,1.55635142413e-13,0,-7.77156117238e-13,0.943,0.943,0.04,0.04,2.04,2.04,2.28774253205,2.28774253206
0.183000000,0,1.55847179392e-13,0,-7.77156117238e-13,0.943,0.943,0.04,0.04,2.04,2.04,2.28774253205,2.28774253206
0.184000000,0,1.57543475222e-13,0,-7.77156117238e-13,0.943,0.943,0.04,0.04,2.04,2.04,2.28774253205,2.28774253206
0.185000000,0,1.60087918967e-13,0,-7.77156117238e-13,0.943,0.943,0.04,0.04,2.04,2.04,2.28774253205,2.28774253206
0.186000000,0,1.62208288755e-13,0,-7.77156117238e-13,0.943,0.943,0.04,0.04,2.04,2.04,2.28774253205,2.28774253206
0.187000000,0,1.62208288755e-13,0,-7.77156117238e-13,0.943,0.943,0.04,0.04,2.04,2.04,2.28774253205,2.28774253206
0.188000000,0,1.64540695521e-13,0,-7.77156117238e-13,0.943,0.943,0.04,0.04,2.04,2.04,2.28774253205,2.28774253206
0.189000000,0,1.65388843436e-13,0,-7.77156117238e-13,0.943,0.943,0.04,0.04,2.04,2.04,2.28774253205,2.28774253206
0.190000000,0,1.65388843436e-13,0,-7.77156117238e-13,0.943,0.943,0.04,0.04,2.04,2.04,2.28774253205,2.28774253206
0.191000000,0,1.65388843436e-13,0,-7.77156117238e-13,0.943,0.943,0.04,0.04,2.04,2.04,2.28774253205,2.28774253206
0.192000000,0,1.65388843436e-13,0,-7.77156117238e-13,0.943,0.943,0.04,0.04,2.04,2.04,2.28774253205,2.28774253206
0.193000000,0,1.65388843436e-13,0,-7.77156117238e-13,0.943,0.943,0.04,0.04,2.04,2.04,2.28774253205,2.28774253206
0.194000000,0,1.65388843436e-13,0,-7.77156117238e-13,0.943,0.943,0.04,0.04,2.04,2.04,2.28774253205,2.28774253205
0.195000000,0,1.62208288755e-13,0,-7.77156117238e-13,0.943,0.943,0.04,0.04,2.04,2.04,2.28774253205,2.28774253205
0.196000000,0,1.62208288755e-13,0,-7.77156117238e-13,0.943,0.943,0.04,0.04,2.04,2.04,2.28774253205,2.28774253205
0.197000000,0,1.59027734073e-13,0,-7.77156117238e-13,0.943,0.943,0.04,0.04,2.04,2.04,2.28774253205,2.28774253205
0.198000000,0,1.59027734073e-13,0,-7.77156117238e-13,0.943,0.943,0.04,0.04,2.04,2.04,2.28774253205,2.28774253205
0.199000000,0,1.59027734073e-13,0,-7.77156117238e-13,0.943,0.943,0.04,0.04,2.04,2.04,2.28774253205,2.28774253205
0.200000000,0,1.59027734073e-13,0,-7.77156117238e-13,0.943,0.943,0.0400000005769,0.04,2.04,2.04,2.28774253205,2.28774253205
0.201000000,-1.35082928154e-09,-2.87073621383e-09,-8.14145417749e-08,-8.98242369374e-08,0.942999940123,0.942999936945,0.0400398779739,0.0400398732098,2.04000000833,2.04000000937,2.28774259959,2.28774260634
0.202000000,-2.09047554215e-08,-2.09171627654e-08,-6.17544571035e-07,-6.51914433547e-07,0.942999746604,0.942999740848,0.0400795798981,0.040079570332,2.04000006558,2.04000006902,2.28774304116,2.28774306967
0.203000000,-1.02960575829e-07,-1.04559225026e-07,-1.97623084564e-06,-2.02315111331e-06,0.94299940315,0.942999396923,0.040119041021,0.0401190291777,2.04000021724,2.04000022237,2.28774415133,2.28774418969
0.204000000,-3.16562906987e-07,-3.20916695774e-07,-4.44347048045e-06,-4.47797785519e-06,0.942998898621,0.94299889444,0.0401582032248,0.0401581939467,2.04000050409,2.04000050849,2.28774615124,2.28774617864
0.205000000,-7.51951390943e-07,-7.53940007313e-07,-8.2354448816e-06,-8.23103252223e-06,0.942998226105,0.942998226358,0.0401970150665,0.0401970148308,2.0400009616,2.04000096165,2.28774920085,2.28774919664
0.206000000,-1.51740040076e-06,-1.53253038462e-06,-1.35089250897e-05,-1.3540729582e-05,0.94299738214,0.942997378242,0.0402354312771,0.0402354205092,2.04000161956,2.04000162503,2.28775340952,2.28775343292
0.207000000,-2.73627794026e-06,-2.76221683437e-06,-2.03706940205e-05,-2.04259240189e-05,0.94299636603,0.942996359962,0.0402734122439,0.0402733946632,2.04000250203,2.04000251092,2.2877588452,2.28775888637
0.208000000,-4.54449880848e-06,-4.57405723553e-06,-2.88854781205e-05,-2.89461552061e-05,0.942995179283,0.942995173119,0.0403109235445,0.0403109044729,2.04000362742,2.04000363678,2.28776554221,2.28776558776
0.209000000,-7.08826215031e-06,-7.10941945895e-06,-3.90828041041e-05,-3.91277319212e-05,0.942993825151,0.942993820941,0.0403479355106,0.0403479215529,2.04000500874,2.04000501514,2.28777350778,2.28777354205
0.210000000,-1.05220377678e-05,-1.05177488311e-05,-5.09629327627e-05,-5.09698838691e-05,0.942992308227,0.942992307889,0.0403844228199,0.0403844215607,2.04000665391,2.04000665382,2.28778272755,2.28778273422
0.211000000,-1.50067487829e-05,-1.50308310657e-05,-6.4502107433e-05,-6.45493576812e-05,0.942990634118,0.942990630075,0.0404203640906,0.0404203488965,2.04000856615,2.04000857273,2.28779317038,2.28779320647
0.212000000,-2.07082149788e-05,-2.07549180427e-05,-7.96568954131e-05,-7.97189356971e-05,0.942988809178,0.942988803712,0.0404557415321,0.0404557190932,2.04001074443,2.04001075419,2.28780479218,2.28780483796
0.213000000,-2.77957550891e-05,-2.78522381418e-05,-9.63680109045e-05,-9.64224655009e-05,0.942986840277,0.942986835334,0.0404905406035,0.0404905182511,2.04001318391,2.04001319339,2.28781753923,2.28781757809
0.214000000,-3.64409669312e-05,-3.64862890886e-05,-0.00011456357063,-0.000114593052336,0.942984734624,0.942984731754,0.0405247496987,0.0405247350186,2.04001587638,2.04001588242,2.28783135089,2.28783137065
0.215000000,-4.68166622192e-05,-4.68213584672e-05,-0.000134161849408,-0.000134154639286,0.942982499613,0.942982499981,0.0405583598611,0.0405583604041,2.04001881071,2.04001881076,2.2878461618,2.28784615498
0.216000000,-5.90959343436e-05,-5.91515052488e-05,-0.000155073649033,-0.000155098011567,0.942980142705,0.942980140127,0.0405913645146,0.0405913499367,2.0400219733,2.04002197855,2.28786190379,2.28786191966
0.217000000,-7.34513510192e-05,-7.35366579498e-05,-0.000177204298013,-0.000177243799104,0.942977671332,0.942977667422,0.0406237592168,0.0406237364532,2.04002534845,2.04002535557,2.2878785073,2.28787853486
0.218000000,-9.00542543244e-05,-9.01425088748e-05,-0.00020045533855,-0.000200495275404,0.94297509282,0.942975089002,0.0406555414342,0.040655517849,2.04002891877,2.04002892509,2.28789590261,2.28789593168
0.219000000,-0.000109074166424,-0.000109134159478,-0.000224725948184,-0.000224752960709,0.942972414332,0.942972411857,0.0406867103321,0.0406866938051,2.04003266554,2.04003266917,2.28791402079,2.28791404138
0.220000000,-0.000130678280377,-0.000130674629389,-0.000249914121397,-0.000249917010864,0.942969642821,0.942969642737,0.0407172665848,0.0407172654146,2.04003656902,2.04003656893,2.28793279442,2.28793279729
0.221000000,-0.000155031008852,-0.00015510293143,-0.000275917657944,-0.000275933693383,0.942966785001,0.942966783141,0.0407472122092,0.0407471970444,2.04004060874,2.04004061167,2.28795215819,2.28795216967
0.222000000,-0.000182293628008,-0.000182407714561,-0.000302634981297,-0.000302656173767,0.94296384732,0.942963844686,0.0407765504013,0.0407765279151,2.04004476382,2.04004476732,2.2879720493,2.28797206519
0.223000000,-0.000212623977151,-0.000212644077768,-0.00032996580035,-0.000329968498358,0.942960835949,0.942960835571,0.0408052853924,0.0408052811936,2.04004901313,2.04004901348,2.28799240777,2.28799241
0.224000000,-0.000246176200846,-0.000246210508251,-0.000357811653084,-0.000357815739926,0.94295775677,0.94295775606,0.0408334223211,0.0408334157114,2.04005333556,2.04005333598,2.28801317667,2.2880131802
0.225000000,-0.000283100543408,-0.000283160242363,-0.000386076341674,-0.00038608127495,0.942954615374,0.942954613988,0.0408609671149,0.0408609567862,2.0400577102,2.04005771042,2.2880343022,2.28803430763
0.226000000,-0.000323543218847,-0.000323586806778,-0.000414666250781,-0.000414667550053,0.942951417062,0.942951415296,0.0408879263804,0.0408879194371,2.04006211648,2.04006211598,2.28805573377,2.28805573821
0.227000000,-0.000367646253484,-0.000367674420109,-0.000443490632995,-0.000443488381252,0.942948166851,0.942948164148,0.0409143073123,0.0409143031442,2.04006653431,2.04006653314,2.288077424,2.28807742926
0.228000000,-0.00041554734127,-0.00041561502627,-0.000472461847201,-0.000472455422451,0.942944869471,0.942944864701,0.0409401176099,0.0409401078916,2.04007094425,2.04007094181,2.28809932877,2.28809933891
0.229000000,-0.000467379883169,-0.000467439432218,-0.000501495425964,-0.000501480775528,0.942941529385,0.942941522625,0.0409653653885,0.0409653571385,2.04007532754,2.0400753236,2.28812140699,2.28812142135
0.230000000,-0.000523272965925,-0.000523275024549,-0.000530510153252,-0.00053048624693,0.942938150796,0.942938141949,0.0409900591129,0.0409900589221,2.04007966619,2.04007966123,2.28814362058,2.28814364066
0.231000000,-0.000583351164994,-0.000583420492636,-0.000559428260605,-0.0005593890128,0.942934737652,0.942934725314,0.041014207544,0.0410141996932,2.04008394311,2.04008393488,2.28816593446,2.28816596587
0.232000000,-0.000647734686199,-0.000647821407667,-0.000588175344207,-0.000588117690348,0.942931293657,0.942931278135,0.0410378196742,0.0410378110335,2.0400881421,2.04008813076,2.28818831628,2.28818835964
0.233000000,-0.000716539387575,-0.000716584796892,-0.000616680358889,-0.000616603599268,0.942927822294,0.942927804064,0.0410609046777,0.041060901716,2.0400922479,2.04009223423,2.28821073634,2.28821079252
0.234000000,-0.000789876866518,-0.000789943910013,-0.000644875552513,-0.000644771396252,0.94292432683,0.942924305972,0.0410834718682,0.0410834666538,2.04009624619,2.04009622825,2.28823316739,2.28823323713
0.235000000,-0.000867854085308,-0.000867950791778,-0.00067269674453,-0.000672561244197,0.942920810313,0.942920787476,0.0411055306647,0.0411055226653,2.04010012374,2.04010010094,2.28825558475,2.28825566784
0.236000000,-0.000950574186383,-0.000950648219083,-0.000700082686933,-0.000699921011882,0.942917275616,0.942917251769,0.0411270905461,0.0411270862512,2.04010386818,2.04010384226,2.28827796569,2.2882780628
0.237000000,-0.00103813571815,-0.00103819045001,-0.000726975636356,-0.000726787867422,0.942913725415,0.942913701756,0.0411481610295,0.0411481595389,2.04010746824,2.0401074393,2.2883002898,2.28830039875
0.238000000,-0.00113063330221,-0.0011308100041,-0.000753320823166,-0.00075308990195,0.942910162221,0.942910140607,0.0411687516391,0.0411687386526,2.04011091361,2.04011088063,2.28832253848,2.28832264695
0.239000000,-0.00122815747771,-0.00122837009314,-0.000779066545165,-0.000778803714681,0.942906588384,0.942906568867,0.0411888718837,0.0411888560237,2.04011419493,2.04011416117,2.28834469498,2.28834480079
0.240000000,-0.00133079461191,-0.00133098385944,-0.000804164205448,-0.000803879060725,0.942903006098,0.942902988536,0.0412085312335,0.0412085179464,2.04011730385,2.04011727177,2.28836674427,2.28836684655
0.241000000,-0.0014386275003,-0.00143873377453,-0.000828567842226,-0.000828271399023,0.942899417424,0.942899401606,0.0412277391098,0.0412277332981,2.04012023287,2.04012020504,2.28838867279,2.28838877199
0.242000000,-0.0015517346897,-0.00155180269069,-0.000852234595117,-0.000851923447017,0.942895824279,0.942895809908,0.0412465048556,0.0412465024381,2.04012297551,2.04012295066,2.28841046854,2.28841056303
0.243000000,-0.00167019130045,-0.00167037431025,-0.000875124079425,-0.000874777165401,0.942892228466,0.94289221532,0.0412648377372,0.041264826544,2.04012552608,2.04012549874,2.28843212074,2.28843220586
0.244000000,-0.00179406862172,-0.00179429664792,-0.000897198649596,-0.000896826860094,0.942888631665,0.942888619682,0.0412827469215,0.0412827327713,2.0401278798,2.04012785239,2.28845361992,2.28845369715
0.245000000,-0.00192343438126,-0.00192363414797,-0.000918423179841,-0.000918038756274,0.942885035449,0.942885024524,0.0413002414708,0.0413002297806,2.04013003268,2.04013000778,2.2884749577,2.28847502905
0.246000000,-0.00205835284848,-0.00205845057701,-0.000938764962433,-0.000938381120904,0.94288144129,0.942881431267,0.0413173303338,0.0413173260245,2.04013198152,2.04013196175,2.28849612674,2.28849619464
0.247000000,-0.00219888487436,-0.00219892492791,-0.00095819365753,-0.000957804918955,0.942877850565,0.942877841327,0.041334022339,0.0413340218376,2.04013372386,2.04013370754,2.28851712064,2.28851718415
0.248000000,-0.00234508793473,-0.0023452383709,-0.000976681234111,-0.00097626086546,0.942874264559,0.942874256145,0.0413503261854,0.0413503182665,2.04013525798,2.04013523897,2.28853793388,2.28853798819
0.249000000,-0.00249701637717,-0.00249721373894,-0.000994201781235,-0.000993761528704,0.942870684478,0.942870676732,0.0413662504452,0.0413662398746,2.0401365828,2.04013656352,2.28855856174,2.28855860952
0.250000000,-0.00265472122864,-0.0026548994361,-0.00101073161107,-0.00101028412854,0.942867111444,0.94286710418,0.0413818035472,0.0413817948152,2.04013769791,2.04013768082,2.28857900021,2.28857904434
0.251000000,-0.00281825066949,-0.00281833879451,-0.00102624892429,-0.00102580788685,0.942863546513,0.942863539528,0.0413969937921,0.0413969911528,2.04013860349,2.04013859107,2.28859924593,2.28859928934
0.252000000,-0.00298764961792,-0.00298769259276,-0.00104073405882,-0.0010402922915,0.942859990664,0.94285998397,0.041411829326,0.0414118296905,2.04013930031,2.04013929076,2.28861929618,2.28861933766
0.253000000,-0.00316296041262,-0.00316312262576,-0.00105416902851,-0.00105369708634,0.94285644482,0.94285643872,0.0414263181686,0.0414263117712,2.04013978963,2.04013977702,2.28863914875,2.28863918277
0.254000000,-0.00334422219622,-0.00334442413644,-0.00106653789811,-0.001066049372,0.942852909834,0.942852904243,0.041440468176,0.0414404595273,2.04014007327,2.04014006022,2.28865880194,2.28865883101
0.255000000,-0.00353147177932,-0.00353163504069,-0.00107782621628,-0.00107733526241,0.942849386514,0.942849381286,0.0414542870819,0.0414542803035,2.04014015343,2.04014014249,2.28867825448,2.28867828126
0.256000000,-0.00372474285299,-0.00372479429921,-0.0010880214929,-0.00108754225858,0.942845875603,0.94284587052,0.0414677824502,0.041467781317,2.04014003285,2.04014002634,2.28869750555,2.28869753291
0.257000000,-0.00392406700032,-0.00392406477765,-0.00109711255258,-0.00109663639956,0.942842377803,0.942842372836,0.0414809617272,0.0414809633498,2.04013971457,2.04013971064,2.28871655464,2.28871658157
0.258000000,-0.00412947277406,-0.00412960288325,-0.00110509008699,-0.00110458471706,0.94283889376,0.942838889196,0.041493832185,0.0414938277656,2.04013920208,2.04013919499,2.28873540164,2.28873542297
0.259000000,-0.00434098681712,-0.00434116468965,-0.00111194595531,-0.00111142577916,0.942835424085,0.942835419839,0.0415064009837,0.0415063944482,2.04013849912,2.04013849143,2.28875404665,2.28875406468
0.260000000,-0.00455863284528,-0.00455878695318,-0.00111767378541,-0.0011171517604,0.942831969338,0.942831965271,0.0415186751071,0.0415186695886,2.04013760984,2.04013760367,2.28877249014,2.28877250689
0.261000000,-0.00478243283533,-0.00478253888416,-0.00112226824744,-0.00112174930528,0.942828530048,0.94282852603,0.0415306614324,0.041530658294,2.04013653859,2.04013653458,2.28879073273,2.28879074946
0.262000000,-0.00501240590482,-0.00501276198888,-0.00112572570454,-0.00112515493099,0.942825106697,0.942825103497,0.0415423666568,0.0415423524706,2.04013529003,2.0401352796,2.28880877534,2.28880878257
0.263000000,-0.00524856980207,-0.00524900842318,-0.00112804332808,-0.00112745212933,0.942821699743,0.942821696921,0.0415537973908,0.0415537798609,2.04013386899,2.04013385709,2.28882661899,2.2888266224
0.264000000,-0.00549093923797,-0.00549129594917,-0.00112922005213,-0.00112863917163,0.942818309599,0.942818306693,0.0415649600463,0.0415649463015,2.04013228058,2.04013227169,2.28884426498,2.28884426999
0.265000000,-0.00573952756922,-0.005739636601,-0.00112925558904,-0.00112871593002,0.942814936653,0.942814933171,0.0415758609448,0.0415758576805,2.04013053003,2.04013052828,2.28886171467,2.28886172665
0.266000000,-0.00599434604084,-0.00599442347453,-0.00112815085984,-0.00112761331717,0.942811581265,0.942811577783,0.0415865062665,0.0415865044184,2.04012862275,2.04012862253,2.28887896961,2.28887898193
0.267000000,-0.00625540356987,-0.00625575866888,-0.00112590809498,-0.001125316589,0.942808243758,0.942808241161,0.0415969020304,0.0415968894271,2.04012656431,2.04012655825,2.28889603147,2.28889603473
0.268000000,-0.00652270799379,-0.00652316429051,-0.00112253013251,-0.00112191713542,0.94280492444,0.942804922224,0.0416070541813,0.041607037966,2.04012436034,2.04012435281,2.28891290199,2.28891290175
0.269000000,-0.00679626446303,-0.00679664298687,-0.00111802129532,-0.00111741906492,0.942801623581,0.942801621228,0.0416169684728,0.0416169554482,2.04012201664,2.04012201158,2.28892958305,2.28892958472
0.270000000,-0.00707607697039,-0.00707619719518,-0.00111238653855,-0.00111182713961,0.942798341436,0.942798338409,0.041626650575,0.0416266471288,2.04011953905,2.04011953998,2.28894607657,2.28894608544
0.271000000,-0.00736214773565,-0.00736221531057,-0.00110563178257,-0.00110507804414,0.942795078235,0.942795075165,0.0416361060289,0.0416361044381,2.04011693346,2.04011693592,2.28896238458,2.28896239439
0.272000000,-0.0076544769529,-0.00765479630997,-0.00109776402979,-0.00109716181803,0.94279183418,0.942791832021,0.0416453402247,0.0416453298011,2.04011420583,2.04011420398,2.28897850914,2.28897851098
0.273000000,-0.00795306391319,-0.00795345921372,-0.00108879076577,-0.00108817163036,0.942788609464,0.942788607683,0.0416543584851,0.0416543454705,2.04011136217,2.04011135951,2.28899445238,2.28899445142
0.274000000,-0.00825790561937,-0.00825820440505,-0.00107872067467,-0.00107811503052,0.942785404249,0.942785402305,0.0416631659575,0.0416631561641,2.04010840846,2.04010840801,2.28901021647,2.28901021769
0.275000000,-0.0085689980648,-0.00856903077355,-0.00106756296314,-0.0010670002783,0.942782218684,0.942782216021,0.0416717677109,0.0416717665157,2.04010535072,2.04010535504,2.28902580361,2.28902581182
0.276000000,-0.00888633576284,-0.00888632086137,-0.00105532760042,-0.00105477041382,0.942779052901,0.942779050177,0.0416801686986,0.0416801689578,2.04010219496,2.04010220039,2.28904121605,2.28904122522
0.277000000,-0.00920991147307,-0.00921016441532,-0.0010420254406,-0.00104142057933,0.942775907009,0.942775905201,0.0416883737333,0.0416883657441,2.04009894715,2.04009894918,2.28905645604,2.28905645769
0.278000000,-0.00953971720121,-0.00954007066171,-0.00102766771859,-0.00102704459964,0.942772781111,0.942772779654,0.041696387567,0.0416963767371,2.04009561324,2.04009561446,2.28907152585,2.28907152483
0.279000000,-0.00987574301692,-0.00987602883389,-0.00101226662452,-0.00101165386093,0.942769675283,0.942769673615,0.0417042147931,0.0417042062556,2.04009219915,2.04009220172,2.2890864278,2.28908642877
0.280000000,-0.0102179781116,-0.0102180261038,-0.000995834775308,-0.000995260379604,0.942766589596,0.942766587149,0.0417118599312,0.041711858536,2.04008871073,2.04008871642,2.28910116416,2.28910117164
0.281000000,-0.010566410449,-0.0105664294898,-0.000978385382044,-0.000977814750976,0.942763524104,0.942763521563,0.041719327398,0.0417193270087,2.04008515379,2.04008516008,2.28911573726,2.28911574564
0.282000000,-0.0109210264832,-0.0109213160025,-0.000959932368372,-0.000959316967417,0.942760478845,0.942760477222,0.0417266214805,0.0417266137531,2.04008153404,2.04008153803,2.28913014939,2.28913015078
0.283000000,-0.0112818120403,-0.0112821919205,-0.00094048995436,-0.000939858973075,0.942757453849,0.942757452599,0.0417337464114,0.0417337363494,2.04007785716,2.04007786078,2.28914440287,2.28914440177
0.284000000,-0.0116487513204,-0.0116490419408,-0.000920073107147,-0.000919454609686,0.942754449131,0.94275444771,0.0417407062802,0.0417406985953,2.04007412871,2.04007413348,2.28915849998,2.28915850072
0.285000000,-0.012021827716,-0.0120218517288,-0.000898697156915,-0.000898117948147,0.942751464697,0.942751462561,0.0417475051034,0.0417475041814,2.04007035417,2.0400703612,2.28917244302,2.28917244973
0.286000000,-0.0124010238246,-0.0124009848487,-0.000876377795223,-0.00087580595764,0.942748500543,0.942748498391,0.041754146827,0.0417541472161,2.04006653896,2.0400665467,2.28918623427,2.2891862417
0.287000000,-0.0127863205936,-0.0127865211165,-0.000853131429945,-0.000852521467586,0.94274555665,0.942745555478,0.0417606352441,0.0417606294451,2.04006268834,2.04006269513,2.28919987599,2.2891998768
0.288000000,-0.0131776985587,-0.0131779764705,-0.000828974642597,-0.000828352005312,0.942742632994,0.942742632217,0.0417669741087,0.0417669663638,2.04005880749,2.04005881434,2.28921337044,2.28921336886
0.289000000,-0.013575137385,-0.013575336595,-0.000803924391501,-0.000803312929174,0.942739729543,0.942739728561,0.0417731670941,0.0417731613222,2.04005490155,2.04005490897,2.28922671984,2.28922671999
0.290000000,-0.0139786152662,-0.0139785825875,-0.000777998235502,-0.000777420286739,0.942736846251,0.942736844466,0.0417792177307,0.0417792176365,2.04005097541,2.04005098362,2.2892399264,2.28923993225
0.291000000,-0.0143881100271,-0.0143880567325,-0.000751213879435,-0.00075063924682,0.94273398307,0.942733981144,0.0417851295126,0.0417851302365,2.04004703395,2.04004704219,2.28925299232,2.28925299908
0.292000000,-0.01480359873,-0.0148038010597,-0.000723589339113,-0.000722978901403,0.942731139941,0.942731138923,0.0417909058597,0.0417909009221,2.04004308193,2.04004308995,2.28926591977,2.28926592046
0.293000000,-0.0152250571044,-0.0152258402342,-0.000695143137275,-0.000694454162598,0.942728316797,0.94272831788,0.0417965500563,0.0417965323926,2.04003912391,2.0400391323,2.28927871088,2.2892786975
0.294000000,-0.0156524605455,-0.0156527400702,-0.000665893919005,-0.000665274007379,0.942725513566,0.94272551298,0.0418020653522,0.0418020586802,2.04003516439,2.04003517331,2.28929136779,2.28929136502
0.295000000,-0.0160857837837,-0.0160858567939,-0.000635860582521,-0.000635266674849,0.942722730171,0.942722728904,0.0418074549296,0.0418074525564,2.04003120776,2.04003121661,2.28930389259,2.2893038938
0.296000000,-0.0165250003471,-0.0165256478466,-0.000605062445813,-0.000604394442916,0.942719966525,0.942719967198,0.0418127218423,0.0418127077817,2.04002725821,2.04002726837,2.28931628733,2.28931627562
0.297000000,-0.0169700834611,-0.0169709349737,-0.000573518928015,-0.000572826822721,0.942717222539,0.942717223911,0.041817869111,0.0418178513141,2.04002331985,2.0400233313,2.28932855407,2.28932853785
0.298000000,-0.0174210057738,-0.0174216998458,-0.000541249650321,-0.000540578987041,0.942714498118,0.942714498977,0.0418228996947,0.0418228853685,2.04001939669,2.04001940846,2.28934069483,2.28934068197
0.299000000,-0.0178777388541,-0.0178779226602,-0.00050827457343,-0.00050766630121,0.942711793159,0.942711792324,0.0418278164314,0.041827812155,2.04001549251,2.04001550288,2.28935271156,2.28935270948
0.300000000,-0.0183402539965,-0.0183402484214,-0.000474613739532,-0.000474027465671,0.942709107558,0.942709106101,0.0418326221273,0.0418326213563,2.04001161105,2.04001162059,2.28936460624,2.28936460801
0.301000000,-0.0188085219972,-0.0188091036639,-0.000440287348247,-0.000439635444767,0.942706441207,0.942706441718,0.0418373195325,0.0418373078461,2.04000775592,2.04000776847,2.28937638078,2.28937637037
0.302000000,-0.0192825126882,-0.0192833227864,-0.000405315864538,-0.000404642047564,0.94270379399,0.942703795262,0.041841911283,0.0418418956818,2.04000393051,2.04000394481,2.28938803708,2.28938802217
0.303000000,-0.0197621956542,-0.0197628800301,-0.000369719816762,-0.000369063487591,0.94270116579,0.942701166636,0.0418463999843,0.0418463869203,2.04000013815,2.04000015218,2.289399577,2.28939956489
0.304000000,-0.0202475400509,-0.020247749337,-0.000333519851958,-0.000332916040375,0.942698556489,0.942698555743,0.0418507881912,0.0418507835859,2.03999638207,2.03999639308,2.28941100238,2.28941099998
0.305000000,-0.0207385141788,-0.0207385514213,-0.000296736814565,-0.000296151241703,0.94269596596,0.942695964638,0.0418550783507,0.0418550766368,2.03999266526,2.03999267487,2.28942231501,2.28942231607
0.306000000,-0.021235086115,-0.0212356946119,-0.000259391595092,-0.000258751406086,0.942693394078,0.942693394661,0.0418592728807,0.0418592615656,2.03998899066,2.03998900485,2.28943351665,2.28943350655
0.307000000,-0.0217372235706,-0.0217380442583,-0.000221505169651,-0.000220848347735,0.942690840714,0.942690842012,0.0418633741535,0.0418633595453,2.03998536112,2.03998537752,2.28944460905,2.289444595
0.308000000,-0.0222448935025,-0.0222455741777,-0.00018309864902,-0.000182458561784,0.942688305735,0.942688306589,0.0418673844397,0.0418673723592,2.03998177923,2.03998179492,2.28945559391,2.28945558273
0.309000000,-0.0227580626672,-0.0227582583045,-0.000144193174734,-0.000143598555665,0.942685789006,0.942685788287,0.0418713059823,0.0418713017579,2.03997824756,2.03997825905,2.28946647291,2.28946647105
0.310000000,-0.0232766975106,-0.0232766984746,-0.000104809944612,-0.000104232276185,0.942683290393,0.942683289082,0.0418751409826,0.0418751397157,2.03997476856,2.03997477822,2.28947724769,2.28947724938
0.311000000,-0.0238007638193,-0.0238012907843,-6.49702348587e-05,-6.43507925119e-05,0.942680809757,0.94268081025,0.0418788915471,0.0418788821881,2.03997134446,2.03997135993,2.28948791986,2.28948791154
0.312000000,-0.0243302272009,-0.0243309342834,-2.46953366645e-05,-2.40652414917e-05,0.942678346959,0.942678348117,0.0418825597553,0.0418825478324,2.03996797746,2.03996799552,2.28949849099,2.28949847925
0.313000000,-0.0248650530013,-0.0248656032185,1.5993426139e-05,1.66078099174e-05,0.942675901858,0.942675902574,0.0418861476495,0.0418861382001,2.03996466963,2.03996468657,2.28950896264,2.28950895372
0.314000000,-0.0254052059957,-0.0254052717223,5.70747031858e-05,5.76517857256e-05,0.942673474312,0.942673473511,0.0418896571827,0.0418896548189,2.03996142285,2.03996143458,2.28951933632,2.2895193361
0.315000000,-0.0259506507753,-0.0259505223111,9.85271459975e-05,9.90903908063e-05,0.942671064177,0.942671062817,0.0418930902778,0.0418930905765,2.03995823894,2.03995824852,2.28952961351,2.28952961658
0.316000000,-0.026501351809,-0.0265017361854,0.000140329385223,0.000140923839113,0.94266867131,0.9426686717,0.0418964488412,0.0418964418917,2.03995511965,2.03995513586,2.28953979568,2.28953978935
0.317000000,-0.0270572729727,-0.0270578431369,0.000182460079712,0.000183061774184,0.942666295566,0.942666296616,0.0418997346775,0.0418997252729,2.03995206653,2.03995208547,2.28954988425,2.28954987456
0.318000000,-0.0276183779861,-0.0276188153459,0.000224897901635,0.000225487554531,0.942663936799,0.94266393745,0.0419029495638,0.0419029421042,2.03994908101,2.03994909848,2.2895598806,2.28955987328
0.319000000,-0.0281846304227,-0.0281846245923,0.000267621542038,0.000268184532896,0.942661594864,0.942661594089,0.0419060952501,0.0419060937514,2.03994616449,2.03994617597,2.2895697861,2.28956978659
0.320000000,-0.0287559936328,-0.0287558297153,0.000310609695298,0.000311164016512,0.942659269614,0.942659268329,0.0419091734495,0.0419091739523,2.03994331827,2.03994332724,2.2895796021,2.28957960535
0.321000000,-0.0293324305209,-0.0293327920666,0.000353841153933,0.000354417517523,0.942656960904,0.942656961316,0.041912185786,0.0419121795782,2.03994054342,2.0399405596,2.28958932988,2.28958932415
0.322000000,-0.0299139039438,-0.029914473065,0.000397294727339,0.000397875630992,0.942654668586,0.942654669647,0.0419151338729,0.0419151252324,2.03993784102,2.03993786006,2.28959897072,2.28959896164
0.323000000,-0.0305003766429,-0.0305008419104,0.000440949237568,0.00044152175077,0.942652392516,0.942652393207,0.0419180193026,0.0419180121516,2.03993521204,2.03993522933,2.28960852588,2.28960851881
0.324000000,-0.0310918110344,-0.031091867583,0.000484783612809,0.000485339294635,0.942650132545,0.94265013188,0.0419208435972,0.0419208415537,2.03993265734,2.03993266809,2.28961799657,2.28961799663
0.325000000,-0.0316881693423,-0.0316880833704,0.000528776865183,0.000529327397395,0.94264788853,0.942647887382,0.0419236082327,0.0419236079235,2.03993017761,2.03993018574,2.28962738397,2.28962738654
0.326000000,-0.0322894137377,-0.0322898316049,0.000572908039453,0.000573469296818,0.942645660325,0.942645660802,0.0419263146717,0.0419263085012,2.03992777355,2.03992778931,2.28963668925,2.28963668345
0.327000000,-0.032895506271,-0.0328961116091,0.000617156209248,0.000617716795956,0.942643447783,0.942643448882,0.0419289643519,0.0419289561593,2.03992544577,2.03992546445,2.28964591356,2.28964590465
0.328000000,-0.0335064087109,-0.0335068926186,0.000661500641597,0.000662053627986,0.942641250762,0.942641251512,0.0419315586327,0.041931551974,2.0399231947,2.03992321152,2.28965505799,2.28965505101
0.329000000,-0.0341220827763,-0.0341221441876,0.000705920641497,0.000706463592104,0.942639069117,0.94263906858,0.0419340988592,0.0419340970013,2.03992102074,2.03992103085,2.28966412362,2.28966412336
0.330000000,-0.0347424901067,-0.0347423783604,0.000750395584115,0.000750934271687,0.942636902706,0.94263690172,0.0419365863514,0.0419365863395,2.03991892422,2.0399189317,2.28967311151,2.28967311368
0.331000000,-0.0353675922303,-0.0353679256689,0.000794904912782,0.000795441378615,0.942634751386,0.942634751968,0.0419390224002,0.0419390174983,2.0399169054,2.03991692068,2.2896820227,2.28968201709
0.332000000,-0.0359973505014,-0.0359978274343,0.000839428290433,0.000839957815486,0.942632615016,0.94263261621,0.0419414082305,0.0419414018198,2.03991496438,2.03991498263,2.28969085817,2.2896908497
0.333000000,-0.036631726261,-0.0366320561291,0.000883945409091,0.000884467897934,0.942630493457,0.942630494335,0.0419437450604,0.0419437402272,2.03991310124,2.03991311759,2.28969961892,2.28969961226
0.334000000,-0.0372706807943,-0.0372705844891,0.000928436002523,0.000928956008908,0.942628386567,0.942628386233,0.0419460340917,0.0419460336277,2.03991131604,2.03991132557,2.2897083059,2.28970830551
0.335000000,-0.0379141752625,-0.0379139073878,0.000972880034755,0.000973398758478,0.942626294209,0.942626293457,0.0419482764671,0.041948277646,2.03990960867,2.03990961552,2.28971692005,2.28971692187
0.336000000,-0.0385621707801,-0.0385623441564,0.00101725758594,0.00101776464542,0.942624216247,0.942624216987,0.0419504733015,0.0419504700497,2.03990797898,2.03990799356,2.28972546226,2.28972545672
0.337000000,-0.0392146284455,-0.0392149702185,0.00106154879376,0.0010620467698,0.942622152545,0.942622153852,0.0419526256975,0.0419526209032,2.03990642679,2.03990644418,2.28973393342,2.28973392511
0.338000000,-0.0398715093061,-0.0398717549731,0.00110573387069,0.00110622972659,0.942620102966,0.942620103947,0.0419547347367,0.0419547310426,2.03990495184,2.03990496716,2.28974233441,2.28974232771
0.339000000,-0.0405327743556,-0.0405326659872,0.00114979332033,0.00115029807472,0.942618067379,0.942618067166,0.0419568014417,0.041956801301,2.03990355378,2.03990356226,2.28975066605,2.28975066519
0.340000000,-0.0411983845826,-0.0411981655631,0.00119370769402,0.00119421701148,0.942616045652,0.942616044999,0.0419588268261,0.0419588278602,2.03990223222,2.03990223801,2.28975892917,2.28975893037
0.341000000,-0.0418683009623,-0.041868530736,0.00123745762393,0.00123794652146,0.942614037653,0.942614038401,0.0419608118875,0.0419608088722,2.03990098672,2.039901,2.28976712457,2.2897671188
0.342000000,-0.0425424844483,-0.0425428505726,0.00128102390562,0.00128149837401,0.942612043253,0.942612044567,0.0419627575915,0.0419627533484,2.03989981679,2.0398998328,2.28977525302,2.28977524457
0.343000000,-0.0432208960037,-0.0432211337707,0.00132438758271,0.00132485897276,0.942610062324,0.94261006335,0.0419646648621,0.0419646617912,2.03989872184,2.03989873605,2.28978331528,2.28978330779
0.344000000,-0.0439034965983,-0.0439033756402,0.00136752974167,0.00136801507129,0.942608094741,0.942608094605,0.041966534617,0.0419665347883,2.03989770128,2.03989770923,2.28979131209,2.28979130896
0.345000000,-0.0445902471905,-0.0445899806204,0.00141043153379,0.00141092318924,0.942606140377,0.942606139745,0.0419683677606,0.0419683692185,2.03989675447,2.03989675994,2.28979924417,2.28979924248
0.346000000,-0.0452811087869,-0.0452812391887,0.00145307442634,0.00145353635019,0.94260419911,0.942604199711,0.0419701651499,0.0419701632193,2.0398958807,2.0398958933,2.28980711222,2.28980710382
0.347000000,-0.0459760424159,-0.0459763089678,0.00149543999672,0.00149588616072,0.942602270818,0.942602271826,0.0419719276257,0.0419719246182,2.03989507921,2.03989509429,2.2898149169,2.28981490604
0.348000000,-0.0466750091268,-0.0466751619061,0.00153750990672,0.00153795807624,0.942600355381,0.942600356002,0.0419736560177,0.0419736540356,2.03989434922,2.0398943623,2.2898226589,2.2898226496
0.349000000,-0.0473779699797,-0.0473777692208,0.00157926592492,0.00157973773109,0.942598452679,0.94259845215,0.0419753511381,0.0419753520908,2.03989368991,2.03989369672,2.28983033886,2.28983033496
0.350000000,-0.0480848861405,-0.048084554854,0.00162069016252,0.00162117021591,0.942596562596,0.942596561627,0.0419770137558,0.0419770157309,2.0398931004,2.03989310487,2.2898379574,2.28983795554
0.351000000,-0.0487957188004,-0.048795784458,0.00166176479399,0.00166220364356,0.942594685018,0.94259468529,0.0419786446318,0.041978643405,2.03989257979,2.03989259113,2.28984551513,2.28984550722
0.352000000,-0.049510429158,-0.0495106458948,0.0017024720429,0.00170288927368,0.942592819828,0.942592820555,0.0419802445202,0.041980242126,2.03989212715,2.03989214096,2.28985301265,2.28985300256
0.353000000,-0.050228978478,-0.0502291075579,0.00174279433152,0.00174321358475,0.942590966915,0.942590967333,0.0419818141502,0.0419818124678,2.03989174152,2.03989175359,2.28986045055,2.28986044202
0.354000000,-0.0509513281875,-0.0509511377956,0.00178271445295,0.0017831631333,0.942589126169,0.942589125532,0.0419833542099,0.0419833549949,2.03989142188,2.03989142823,2.28986782939,2.28986782605
0.355000000,-0.0516774397135,-0.0516771350539,0.00182221520162,0.00182267388512,0.942587297482,0.942587296447,0.0419848653877,0.0419848670067,2.03989116719,2.03989117145,2.2898751497,2.2898751484
0.356000000,-0.0524072745547,-0.0524073476077,0.00186127951296,0.00186168810396,0.942585480746,0.942585480899,0.0419863483573,0.0419863471187,2.03989097642,2.03989098702,2.28988241203,2.28988240507
0.357000000,-0.0531407941887,-0.053141003656,0.00189989033572,0.00190027470507,0.942583675854,0.942583676433,0.0419878037877,0.0419878015354,2.0398908485,2.03989086127,2.28988961691,2.289889608
0.358000000,-0.0538779603668,-0.0538780728224,0.00193803112536,0.00193842069011,0.942581882703,0.942581882967,0.041989232299,0.041989230751,2.0398907823,2.03989079333,2.28989676483,2.28989675758
0.359000000,-0.0546187348764,-0.0546185250389,0.00197568540283,0.00197611312162,0.942580101193,0.942580100422,0.041990634505,0.0419906352506,2.0398907767,2.03989078231,2.28990385629,2.28990385419
0.360000000,-0.055363079549,-0.0553627385042,0.00201283675927,0.00201327889164,0.942578331222,0.942578330043,0.0419920110136,0.0419920126117,2.03989083055,2.03989083408,2.28991089177,2.28991089178
0.361000000,-0.0561109562281,-0.0561109498306,0.00204946883531,0.00204985440175,0.94257657269,0.942576572628,0.0419933624262,0.0419933615729,2.03989094271,2.03989095181,2.28991787174,2.28991786646
0.362000000,-0.0568623269752,-0.0568624279625,0.00208556563353,0.00208592550691,0.942574825501,0.942574825832,0.0419946893119,0.041994687639,2.03989111198,2.03989112287,2.28992479666,2.2899247896
0.363000000,-0.0576171540188,-0.0576171446181,0.00212111141629,0.00212147975656,0.942573089561,0.942573089582,0.0419959922184,0.041995991238,2.03989133716,2.03989134632,2.28993166696,2.28993166154
0.364000000,-0.0583753995873,-0.0583750716295,0.00215609044618,0.0021565047818,0.942571364775,0.942571363804,0.0419972716935,0.0419972727915,2.03989161705,2.03989162123,2.28993848307,2.2899384826
0.365000000,-0.0591370260352,-0.0591365679792,0.00219048717143,0.00219091893013,0.942569651051,0.942569649698,0.04199852827,0.0419985301272,2.03989195043,2.0398919527,2.28994524543,2.28994524695
0.366000000,-0.059901995706,-0.0599018572711,0.00222428605423,0.00222465329591,0.942567948298,0.942567948033,0.0419997624774,0.0419997621097,2.03989233607,2.03989234334,2.28995195444,2.28995195079
0.367000000,-0.0606702713276,-0.0606702443688,0.00225747209104,0.00225781016879,0.942566256428,0.942566256561,0.0420009748056,0.0420009736496,2.03989277272,2.03989278161,2.2899586105,2.28995860502
0.368000000,-0.0614418156393,-0.0614417007244,0.00229003029384,0.00229037781647,0.942564575355,0.942564575214,0.0420021657433,0.0420021651282,2.03989325912,2.03989326654,2.28996521399,2.28996520995
0.369000000,-0.0622165914582,-0.0622161975956,0.00232194577543,0.00232234460335,0.942562904994,0.942562903923,0.0420033357725,0.0420033369216,2.03989379403,2.03989379715,2.2899717653,2.28997176587
0.370000000,-0.0629945616059,-0.0629940716991,0.00235320367636,0.00235362103315,0.942561245258,0.942561243843,0.042004485371,0.0420044870913,2.03989437617,2.03989437776,2.28997826479,2.28997826714
0.371000000,-0.0637756892577,-0.0637755293275,0.00238378959239,0.00238413347662,0.942559596068,0.942559595718,0.0420056149863,0.0420056146256,2.03989500429,2.03989501041,2.28998471284,2.28998471005
0.372000000,-0.0645599377239,-0.0645599095723,0.00241368928888,0.00241399984,0.942557957344,0.942557957391,0.0420067250545,0.04200672391,2.03989567711,2.03989568475,2.28999110978,2.2899911051
0.373000000,-0.0653472703352,-0.0653471823394,0.00244288855455,0.00244320915,0.942556329006,0.942556328801,0.0420078160106,0.0420078152848,2.03989639336,2.0398963998,2.28999745595,2.28999745255
0.374000000,-0.0661376505337,-0.0661373174588,0.00247137331888,0.00247175052108,0.942554710978,0.942554709885,0.0420088882799,0.0420088890851,2.03989715176,2.03989715454,2.2900037517,2.29000375268
0.375000000,-0.0669310418725,-0.066930628063,0.00249912965788,0.00249952715434,0.942553103182,0.942553101761,0.042009942278,0.0420099435773,2.03989795104,2.03989795253,2.29000999735,2.29000999998
0.376000000,-0.0677274082841,-0.0677273038704,0.00252614408747,0.0025264609757,0.942551505547,0.942551505153,0.0420109783943,0.042010977851,2.03989878993,2.03989879529,2.29001619322,2.29001619085
0.377000000,-0.0685267137007,-0.0685267221877,0.00255240312397,0.0025526844407,0.942549918,0.94254991799,0.0420119970185,0.0420119958146,2.03989966716,2.03989967373,2.29002233962,2.29002233539
0.378000000,-0.069328922171,-0.0693288539982,0.00257789341251,0.00257818720923,0.942548340471,0.942548340218,0.042012998533,0.0420129977643,2.03990058145,2.03990058687,2.29002843684,2.29002843383
0.379000000,-0.0701339977484,-0.0701336707281,0.0026026016211,0.00260295900945,0.942546772888,0.942546771784,0.0420139833174,0.0420139839903,2.03990153155,2.03990153372,2.29003448519,2.29003448641
0.380000000,-0.0709419049433,-0.0709414666957,0.00262651491711,0.0026268960435,0.942545215187,0.94254521377,0.0420149517249,0.042014952928,2.03990251619,2.03990251712,2.29004048496,2.29004048777
0.381000000,-0.0717526083535,-0.0717524225164,0.00264962056207,0.00264991585839,0.942543667301,0.942543666885,0.0420159041033,0.0420159037407,2.03990353413,2.03990353813,2.29004643642,2.29004643436
0.382000000,-0.072566072631,-0.0725659564367,0.00267190587344,0.00267216452005,0.942542129165,0.942542129132,0.0420168407981,0.0420168399169,2.03990458412,2.03990458894,2.29005233986,2.29005233596
0.383000000,-0.0733822624891,-0.0733820437426,0.00269335824132,0.00269363221931,0.942540600717,0.942540600462,0.0420177621502,0.0420177617107,2.03990566491,2.03990566861,2.29005819554,2.29005819279
0.384000000,-0.0742011429277,-0.0742006601236,0.00271396535956,0.00271430921919,0.942539081894,0.942539080826,0.0420186684853,0.0420186693723,2.03990677528,2.03990677619,2.29006400373,2.29006400505
0.385000000,-0.0750226792471,-0.075022086049,0.00273371522685,0.00273408507375,0.942537572638,0.942537571274,0.0420195601143,0.0420195614826,2.039907914,2.03990791387,2.29006976468,2.29006976751
0.386000000,-0.0758468367476,-0.0758464943273,0.0027525958417,0.0027528732884,0.94253607289,0.942536072497,0.0420204373484,0.0420204372779,2.03990907987,2.03990908226,2.29007547864,2.29007547672
0.387000000,-0.076673580941,-0.0766733348704,0.00277059541065,0.00277083300604,0.942534582595,0.942534582564,0.0420213004897,0.0420212998953,2.03991027169,2.0399102748,2.29008114587,2.29008114215
0.388000000,-0.0775028771766,-0.0775025807697,0.00278770199902,0.00278795515386,0.942533101693,0.942533101433,0.0420221498445,0.0420221495656,2.03991148824,2.03991149058,2.2900867666,2.29008676401
0.389000000,-0.0783346914915,-0.0783342032959,0.00280390433494,0.00280423079188,0.942531630134,0.942531629061,0.0420229856912,0.0420229865189,2.03991272837,2.03991272866,2.29009234108,2.29009234248
0.390000000,-0.0791689899785,-0.0791684536535,0.00281919120049,0.0028195437673,0.942530167865,0.942530166478,0.0420238083057,0.0420238094905,2.0399139909,2.03991399055,2.29009786952,2.29009787244
0.391000000,-0.0800057387305,-0.0800054643309,0.00283355137687,0.00283380476981,0.942528714835,0.942528714377,0.0420246179637,0.0420246178213,2.03991527469,2.03991527637,2.29010335217,2.29010335045
0.392000000,-0.0808449040811,-0.0808446935018,0.00284697386976,0.00284718575653,0.942527270994,0.942527270906,0.0420254149322,0.0420254143556,2.03991657859,2.03991658071,2.29010878923,2.29010878569
0.393000000,-0.0816864522355,-0.0816861537576,0.00285944758338,0.00285967590375,0.942525836292,0.942525836009,0.0420261994807,0.0420261992366,2.03991790146,2.03991790286,2.29011418095,2.29011417816
0.394000000,-0.0825303500919,-0.0825298467393,0.00287096205254,0.00287126636264,0.942524410684,0.942524409626,0.0420269718545,0.0420269726317,2.03991924218,2.03991924205,2.29011952751,2.29011952803
0.395000000,-0.0833765646056,-0.0833759786934,0.00288150686334,0.00288184185637,0.942522994126,0.942522992735,0.0420277322971,0.0420277334729,2.03992059969,2.03992059902,2.29012482914,2.2901248309
0.396000000,-0.0842250627321,-0.0842246944404,0.00289107160234,0.00289130611305,0.942521586572,0.942521586041,0.0420284810517,0.0420284810951,2.03992197288,2.03992197362,2.29013008604,2.29013008324
0.397000000,-0.0850758116744,-0.0850755145805,0.00289964607347,0.00289983996302,0.942520187981,0.942520187745,0.042029218354,0.0420292180156,2.03992336069,2.03992336174,2.2901352984,2.29013529393
0.398000000,-0.0859287785064,-0.0859284150477,0.00290721998453,0.00290743494229,0.942518798309,0.942518797821,0.0420299444417,0.0420299444062,2.03992476204,2.03992476256,2.29014046645,2.29014046309
0.399000000,-0.0867839310128,-0.0867833715341,0.00291378365613,0.00291408290795,0.942517417517,0.94251741624,0.0420306595314,0.042030660439,2.03992617591,2.03992617527,2.29014559036,2.29014559084
0.400000000,-0.0876412370371,-0.0876406050288,0.00291932745977,0.00291965662882,0.942516045567,0.942516043992,0.0420313638381,0.0420313650637,2.0399276013,2.03992760028,2.29015067032,2.29015067227
0.401000000,-0.0885006644229,-0.0885002450527,0.00292384176692,0.00292406018249,0.942514682423,0.942514681737,0.0420320575766,0.0420320577098,2.03992903719,2.03992903711,2.29015570653,2.29015570406
0.402000000,-0.0893621812664,-0.0893618406306,0.00292731716045,0.00292748732321,0.942513328047,0.942513327702,0.0420327409551,0.0420327406722,2.03993048261,2.0399304827,2.29016069917,2.29016069503
0.403000000,-0.0902257555347,-0.0902253661444,0.00292974413107,0.00292993118979,0.942511982403,0.942511981858,0.0420334141835,0.0420334141122,2.03993193654,2.03993193628,2.29016564842,2.29016564533
0.404000000,-0.0910913559193,-0.0910907959642,0.00293111376393,0.00293138499647,0.942510645459,0.942510644174,0.0420340774533,0.0420340781885,2.03993339807,2.03993339711,2.29017055447,2.29017055511
0.405000000,-0.0919589511715,-0.0919583311133,0.00293141719454,0.00293171759033,0.942509317183,0.942509315622,0.0420347309545,0.0420347319513,2.03993486626,2.0399348651,2.29017541748,2.29017541955
0.406000000,-0.0928285100428,-0.0928280875592,0.00293064555756,0.00293083069253,0.942507997544,0.942507996858,0.0420353748772,0.0420353748784,2.0399363402,2.0399363395,2.29018023764,2.29018023543
0.407000000,-0.0937000015415,-0.0936996472194,0.00292879019348,0.00292892718604,0.942506686512,0.942506686153,0.0420360094053,0.0420360090385,2.039937819,2.03993781828,2.2901850151,2.29018501135
0.408000000,-0.0945733945464,-0.0945729856221,0.00292584235329,0.002926000509,0.942505384057,0.942505383485,0.0420366347245,0.0420366345709,2.03993930173,2.03993930075,2.29018975005,2.29018974743
0.409000000,-0.0954486586711,-0.0954480785849,0.00292179386485,0.00292204411715,0.942504090152,0.942504088833,0.0420372510044,0.0420372516126,2.03994078756,2.03994078623,2.29019444265,2.2901944438
0.410000000,-0.0963257635897,-0.0963251110679,0.00291663660401,0.00291692220555,0.942502804772,0.94250280316,0.042037858413,0.0420378592927,2.03994227566,2.03994227423,2.29019909305,2.29019909573
0.411000000,-0.0972046789764,-0.0972041902036,0.00291036244615,0.00291053338836,0.942501527892,0.942501527123,0.0420384571185,0.042038457125,2.0399437652,2.03994376375,2.29020370142,2.29020370002
0.412000000,-0.0980853747653,-0.0980849311878,0.00290296346583,0.00290308904342,0.942500259488,0.942500259028,0.0420390472839,0.0420390469816,2.03994525538,2.03994525373,2.29020826792,2.29020826507
0.413000000,-0.0989678207612,-0.0989673117195,0.00289443165147,0.00289458288023,0.942498999535,0.942498998858,0.0420396290736,0.042039628984,2.03994674537,2.03994674355,2.2902127927,2.29021279102
0.414000000,-0.0998519875105,-0.0998513096622,0.00288475954946,0.00288500867451,0.942497748013,0.942497746597,0.0420402026378,0.042040203252,2.03994823443,2.0399482326,2.29021727593,2.29021727797
0.415000000,-0.100737845621,-0.100737095472,0.00287393975196,0.00287422641302,0.942496504901,0.942496503202,0.0420407681255,0.0420407689874,2.03994972182,2.03994972002,2.29022171774,2.29022172127
0.416000000,-0.1016253657,-0.10162476705,0.00286196485133,0.00286213236818,0.942495270182,0.942495269326,0.0420413256857,0.0420413257413,2.03995120681,2.03995120464,2.29022611829,2.29022611775
0.417000000,-0.102514518618,-0.102513968499,0.00284882763157,0.00284894652998,0.942494043835,0.942494043301,0.0420418754631,0.0420418752185,2.03995268869,2.03995268622,2.29023047773,2.2902304757
0.418000000,-0.103405275114,-0.103404677781,0.00283452079475,0.00283466332033,0.942492825843,0.942492825112,0.0420424176032,0.0420424175284,2.03995416672,2.03995416423,2.29023479621,2.29023479522
0.419000000,-0.104297606677,-0.104296872738,0.00281903758048,0.00281927727065,0.94249161619,0.942491614747,0.0420429522395,0.0420429527792,2.03995564026,2.03995563813,2.29023907387,2.29023907643
0.420000000,-0.105191484853,-0.105190707019,0.00280237127326,0.00280264514849,0.942490414862,0.942490413155,0.0420434795041,0.0420434802407,2.03995710867,2.03995710673,2.29024331084,2.29024331475
0.421000000,-0.106086881192,-0.106086265563,0.0027845151569,0.00278466184862,0.942489221845,0.942489220989,0.0420439995295,0.042043999502,2.03995857129,2.03995856872,2.29024750727,2.29024750709
0.422000000,-0.106983767504,-0.106983219043,0.00276546270039,0.00276555532416,0.942488037126,0.942488036601,0.0420445124439,0.0420445121183,2.03996002752,2.0399600246,2.29025166331,2.29025166156
0.423000000,-0.107882115474,-0.107881544307,0.00274520729437,0.00274532082816,0.942486860691,0.942486859981,0.0420450183768,0.0420450181881,2.03996147672,2.03996147391,2.29025577908,2.29025577828
0.424000000,-0.108781897531,-0.108781218148,0.00272374284682,0.00272395369442,0.942485692531,0.94248569112,0.0420455174465,0.0420455178087,2.03996291834,2.03996291619,2.29025985472,2.29025985736
0.425000000,-0.109683086169,-0.109682376116,0.00270106330835,0.00270130801495,0.942484532635,0.942484530965,0.0420460097709,0.0420460103107,2.03996435181,2.03996434997,2.29026389036,2.29026389429
0.426000000,-0.110585653882,-0.110585090522,0.0026771626298,0.00267727736426,0.942483380996,0.942483380172,0.0420464954674,0.0420464953139,2.03996577659,2.03996577382,2.29026788614,2.29026788602
0.427000000,-0.111489573425,-0.111489061618,0.00265203493965,0.0026520962405,0.942482237605,0.942482237111,0.0420469746502,0.0420469742363,2.03996719216,2.03996718892,2.29027184219,2.29027184053
0.428000000,-0.112394817428,-0.112394267348,0.0026256742911,0.00262576030122,0.942481102454,0.942481101775,0.0420474474344,0.0420474471646,2.03996859797,2.03996859488,2.29027575863,2.29027575791
0.429000000,-0.113301359267,-0.113300686084,0.00259807523517,0.00259826521465,0.942479975538,0.942479974161,0.0420479139254,0.0420479141839,2.03996999356,2.03996999131,2.29027963559,2.29027963826
0.430000000,-0.114209172383,-0.114208439645,0.0025692323633,0.00256946211803,0.942478856852,0.942478855216,0.0420483742282,0.0420483746754,2.03997137846,2.03997137653,2.2902834732,2.29028347715
0.431000000,-0.115118230214,-0.115117594673,0.0025391402676,0.00253924253488,0.942477746392,0.942477745603,0.0420488284474,0.0420488282824,2.0399727522,2.03997274904,2.29028727158,2.29028727156
0.432000000,-0.116028506463,-0.116027883498,0.00250779371003,0.00250784647887,0.942476644154,0.942476643699,0.0420492766846,0.0420492763012,2.03997411436,2.03997411061,2.29029103085,2.29029102931
0.433000000,-0.116939974707,-0.116939288306,0.00247518738128,0.00247526962614,0.942475550136,0.942475549502,0.0420497190425,0.0420497188073,2.03997546448,2.0399754609,2.29029475113,2.29029475052
0.434000000,-0.117852609267,-0.117851791705,0.002441316449,0.00244150767703,0.942474464337,0.942474463012,0.0420501556151,0.0420501558751,2.03997680219,2.03997679961,2.29029843255,2.29029843527
0.435000000,-0.118766384529,-0.118765506811,0.00240617612035,0.00240640865947,0.942473386757,0.942473385177,0.0420505864959,0.0420505869307,2.03997812709,2.03997812491,2.29030207522,2.29030207918
0.436000000,-0.119681274878,-0.119680496277,0.00236976160206,0.0023698622758,0.942472317396,0.942472316662,0.0420510117785,0.0420510116412,2.03997943882,2.03997943524,2.29030567925,2.29030567928
0.437000000,-0.120597254961,-0.12059651568,0.0023320682645,0.00233211456154,0.942471256255,0.942471255851,0.0420514315535,0.0420514311996,2.03998073701,2.03998073285,2.29030924477,2.29030924329
0.438000000,-0.1215142993,-0.121513545361,0.00229309141009,0.00229316203711,0.942470203336,0.942470202744,0.0420518459124,0.0420518456753,2.03998202132,2.03998201747,2.29031277189,2.2903127713
0.439000000,-0.122432383162,-0.122431563908,0.00225282679733,0.00225300150332,0.942469158641,0.942469157345,0.0420522549395,0.0420522551375,2.03998329142,2.03998328884,2.29031626071,2.29031626339
0.440000000,-0.123351481874,-0.123350661077,0.00221127022249,0.00221148014399,0.942468122177,0.94246812061,0.0420526587182,0.0420526590588,2.03998454703,2.03998454499,2.29031971135,2.29031971525
0.441000000,-0.124271570766,-0.124270863909,0.00216841748246,0.00216848987288,0.942467093946,0.94246709322,0.0420530573321,0.0420530571374,2.03998578786,2.03998578431,2.29032312392,2.29032312393
0.442000000,-0.125192625425,-0.125191927693,0.00212426452983,0.00212428458368,0.942466073955,0.942466073569,0.0420534508625,0.0420534504767,2.03998701361,2.03998700939,2.29032649852,2.29032649701
0.443000000,-0.126114621318,-0.126113869535,0.00207880725211,0.00207885523624,0.942465062209,0.942465061662,0.0420538393911,0.042053839123,2.03998822402,2.03998822009,2.29032983526,2.29032983448
0.444000000,-0.127037534651,-0.127036697939,0.0020320419738,0.00203219590822,0.942464058716,0.94246405749,0.0420542229933,0.042054223131,2.03998941885,2.03998941627,2.29033313426,2.29033313645
0.445000000,-0.12796134169,-0.127960470987,0.00198396505513,0.00198416139443,0.942463063483,0.942463061975,0.0420546017439,0.0420546020383,2.03999059788,2.03999059588,2.2903363956,2.29033639897
0.446000000,-0.128886018704,-0.128885227158,0.00193457285613,0.0019346365378,0.942462076521,0.942462075828,0.0420549757179,0.0420549755412,2.0399917609,2.0399917573,2.2903396194,2.290339619
0.447000000,-0.129811542219,-0.129810772451,0.00188386188671,0.00188387444501,0.942461097837,0.942461097448,0.042055344988,0.0420553446449,2.03999290769,2.03999290348,2.29034280575,2.29034280396
0.448000000,-0.130737888641,-0.130737089345,0.00183182859437,0.00183187166855,0.942460127441,0.942460126847,0.0420557096275,0.042055709402,2.03999403804,2.03999403425,2.29034595475,2.2903459539
0.449000000,-0.131665035109,-0.131664160395,0.00177846984339,0.00177862511387,0.942459165345,0.942459164037,0.0420560697041,0.0420560698658,2.03999515181,2.03999514948,2.29034906651,2.29034906887
0.450000000,-0.132592958826,-0.132592055499,0.00172378253249,0.00172397768524,0.942458211562,0.942458209993,0.042056425285,0.0420564255773,2.03999624884,2.03999624707,2.29035214111,2.29035214466
0.451000000,-0.133521636993,-0.133520806122,0.00166776356036,0.00166781762293,0.942457266103,0.942457265403,0.0420567764374,0.0420567762674,2.03999732897,2.03999732547,2.29035517865,2.29035517839
0.452000000,-0.134451047071,-0.134450240475,0.0016104099676,0.0016104048221,0.942456328981,0.942456328637,0.0420571232267,0.0420571228759,2.03999839207,2.03999838787,2.29035817924,2.29035817746
0.453000000,-0.135381166397,-0.135380340669,0.00155171873706,0.00155173752427,0.942455400209,0.942455399704,0.0420574657187,0.0420574654539,2.039999438,2.03999943418,2.29036114295,2.29036114197
0.454000000,-0.136311973041,-0.136311088791,0.0014916872475,0.00149181402757,0.942454479803,0.942454478613,0.0420578039743,0.0420578040519,2.04000046667,2.04000046432,2.29036406989,2.29036407198
0.455000000,-0.137243445133,-0.137242540982,0.00143031291122,0.00143047704923,0.942453567778,0.942453566343,0.042058138054,0.0420581382442,2.04000147798,2.0400014762,2.29036696013,2.29036696335
0.456000000,-0.138175560804,-0.13817471917,0.00136759313962,0.00136761487679,0.942452664151,0.942452663593,0.0420584680183,0.0420584677786,2.04000247185,2.04000246828,2.29036981378,2.29036981324
0.457000000,-0.139108298439,-0.139107474458,0.00130352548022,0.00130348987353,0.942451768937,0.942451768728,0.042058793926,0.0420587935236,2.0400034482,2.04000344394,2.2903726309,2.29037262894
0.458000000,-0.140041636305,-0.140040789726,0.00123810742481,0.00123810031145,0.942450882153,0.942450881762,0.0420591158364,0.0420591155241,2.04000440695,2.04000440312,2.29037541161,2.29037541051
0.459000000,-0.140975553393,-0.140974648088,0.00117133684263,0.00117144443535,0.942450003817,0.942450002708,0.0420594338044,0.0420594338241,2.04000534806,2.04000534579,2.29037815597,2.29037815803
0.460000000,-0.141910028756,-0.141909094557,0.00110321163427,0.00110336352484,0.942449133949,0.942449132565,0.0420597478846,0.0420597480241,2.04000627151,2.04000626986,2.29038086406,2.29038086736
0.461000000,-0.142845041444,-0.142844145282,0.00103372969962,0.00103374465238,0.942448272568,0.942448272045,0.0420600581315,0.0420600578842,2.04000717726,2.04000717384,2.29038353597,2.29038353568
0.462000000,-0.143780570765,-0.143779675006,0.000962889068035,0.000962852213199,0.942447419693,0.942447419502,0.0420603645982,0.0420603642108,2.04000806529,2.04000806122,2.29038617179,2.29038617017
0.463000000,-0.144716595905,-0.144715668523,0.000890687715804,0.000890684423352,0.942446575344,0.942446574954,0.0420606673383,0.0420606670435,2.04000893557,2.04000893197,2.29038877158,2.29038877089
0.464000000,-0.14565309677,-0.145652110792,0.000817123978036,0.00081723953608,0.942445739542,0.942445738418,0.0420609664013,0.0420609664215,2.04000978813,2.0400097861,2.29039133543,2.2903913379
0.465000000,-0.146590053327,-0.146589037458,0.000742196219372,0.000742357642683,0.94244491231,0.942444910907,0.0420612618366,0.0420612619699,2.04001062297,2.04001062156,2.29039386341,2.2903938671
0.466000000,-0.147527445543,-0.14752645914,0.000665902804009,0.000665925393916,0.942444093669,0.942444093148,0.0420615536936,0.0420615534628,2.04001144011,2.04001143695,2.29039635559,2.29039635568
0.467000000,-0.148465253635,-0.148464270678,0.000588242218269,0.000588209920185,0.942443283643,0.942443283473,0.0420618420201,0.0420618416525,2.04001223958,2.04001223577,2.29039881205,2.29039881077
0.468000000,-0.149403457705,-0.149402457306,0.000509212899402,0.000509210009421,0.942442482253,0.942442481902,0.0420621268648,0.0420621265756,2.04001302142,2.04001301805,2.29040123285,2.29040123241
0.469000000,-0.150342038565,-0.150341004167,0.000428813624387,0.000428924554182,0.942441689525,0.942441688451,0.0420624082725,0.0420624082684,2.04001378566,2.04001378384,2.29040361808,2.29040362066
0.470000000,-0.15128097709,-0.151279935753,0.000347043197735,0.00034719356616,0.942440905483,0.942440904149,0.0420626862879,0.0420626863798,2.04001453239,2.04001453114,2.2904059678,2.29040597147
0.471000000,-0.152220254153,-0.152219253932,0.000263900423958,0.000263904493646,0.942440130151,0.942440129734,0.0420629609558,0.0420629606979,2.04001526164,2.04001525864,2.29040828206,2.29040828207
0.472000000,-0.153159850877,-0.153158871206,0.000179384223253,0.000179326890703,0.942439363556,0.942439363515,0.0420632323196,0.0420632319259,2.04001597351,2.04001596984,2.29041056095,2.29041055949
0.473000000,-0.154099748268,-0.154098771875,9.34934696328e-05,9.34603381846e-05,0.942438605723,0.942438605512,0.0420635004234,0.0420635000976,2.04001666805,2.04001666481,2.29041280453,2.29041280381
0.474000000,-0.15503992804,-0.155038940188,6.22735818467e-06,6.30448399575e-06,0.942437856678,0.942437855745,0.042063765308,0.0420637652463,2.04001734536,2.04001734362,2.29041501285,2.29041501506
0.475000000,-0.155980371965,-0.155979388158,-8.24148894685e-05,-8.23002069663e-05,0.94243711645,0.942437115257,0.0420640270139,0.0420640270423,2.04001800555,2.04001800436,2.29041718597,2.29041718924
0.476000000,-0.156921061815,-0.15692010889,-0.000172434051926,-0.000172465563523,0.942436385064,0.9424363848,0.0420642855819,0.0420642852857,2.04001864869,2.0400186458,2.29041932397,2.29041932359
0.477000000,-0.157861979611,-0.157861034972,-0.00026383079843,-0.000263921129774,0.94243566255,0.942435662663,0.0420645410515,0.0420645406334,2.04001927492,2.04001927139,2.29042142689,2.29042142508
0.478000000,-0.158803107257,-0.158802151631,-0.000356605841745,-0.000356667156695,0.942434948935,0.942434948866,0.0420647934626,0.0420647931156,2.04001988432,2.04001988123,2.29042349479,2.29042349376
0.479000000,-0.159744427357,-0.159743444476,-0.000450759591097,-0.000450703937463,0.942434244249,0.942434243434,0.0420650428524,0.0420650427622,2.04002047703,2.0400204754,2.29042552774,2.29042552966
0.480000000,-0.160685922575,-0.160684916879,-0.000546292430847,-0.000546191307919,0.942433548521,0.942433547428,0.042065289258,0.042065289262,2.04002105317,2.04002105211,2.29042752577,2.29042752881
0.481000000,-0.161627575575,-0.161626559471,-0.000643204745354,-0.000643241750263,0.94243286178,0.942432861618,0.0420655327163,0.0420655324242,2.04002161288,2.0400216102,2.29042948895,2.29042948843
0.482000000,-0.162569369265,-0.162568327378,-0.000741496815948,-0.000741584946518,0.942432184057,0.942432184267,0.0420657732634,0.0420657728642,2.04002215629,2.04002215304,2.29043141733,2.29043141544
0.483000000,-0.163511286439,-0.163510209744,-0.000841168964705,-0.000841221606945,0.942431515383,0.9424315154,0.0420660109356,0.0420660106091,2.04002268355,2.04002268073,2.29043331096,2.29043330987
0.484000000,-0.164453310587,-0.164452196114,-0.000942221227485,-0.000942152466177,0.942430855788,0.942430855041,0.0420662457671,0.0420662456854,2.04002319479,2.04002319339,2.29043516988,2.29043517175
0.485000000,-0.165395425253,-0.165394286041,-0.00104465361672,-0.00104453829719,0.942430205304,0.942430204275,0.0420664777915,0.0420664777991,2.04002369018,2.04002368933,2.29043699414,2.29043699711
0.486000000,-0.166337613986,-0.166336469557,-0.00114846614496,-0.00114849216444,0.942429563963,0.942429563884,0.0420667070426,0.0420667067691,2.04002416988,2.04002416747,2.29043878379,2.29043878319
0.487000000,-0.167279860573,-0.167278716338,-0.00125365872761,-0.00125374247194,0.942428931797,0.942428932103,0.0420669335534,0.0420669331746,2.04002463405,2.04002463107,2.29044053887,2.29044053685
0.488000000,-0.168222148692,-0.168221013908,-0.00136023131803,-0.00136028905108,0.942428308838,0.942428308956,0.0420671573571,0.0420671570408,2.04002508283,2.04002508027,2.29044225943,2.29044225813
0.489000000,-0.169164462705,-0.169163348153,-0.00146818360058,-0.00146813132829,0.94242769512,0.942427694469,0.0420673784848,0.0420673783932,2.04002551643,2.04002551518,2.2904439455,2.29044394707
0.490000000,-0.170106787034,-0.170105701211,-0.00157751523722,-0.00157742764571,0.942427090676,0.942427089744,0.0420675969673,0.0420675969548,2.040025935,2.04002593424,2.29044559713,2.29044559971
0.491000000,-0.171049106101,-0.171048032756,-0.00168822589008,-0.00168828515605,0.94242649554,0.942426495578,0.0420678128357,0.0420678125561,2.04002633872,2.04002633649,2.29044721436,2.29044721334
0.492000000,-0.171991404568,-0.171990304423,-0.00180031512997,-0.00180042715844,0.942425909745,0.942425910173,0.0420680261202,0.0420680257431,2.04002672778,2.04002672503,2.29044879721,2.29044879479
0.493000000,-0.172933666983,-0.172932537173,-0.00191378256342,-0.00191386213256,0.942425333326,0.942425333574,0.042068236851,0.0420682365337,2.04002710236,2.04002709999,2.29045034573,2.290450344
0.494000000,-0.173875878578,-0.173874745198,-0.00202862754373,-0.00202859477761,0.942424766317,0.942424765806,0.0420684450568,0.0420684449498,2.04002746264,2.04002746153,2.29045185996,2.29045186096
0.495000000,-0.174818024642,-0.174816889834,-0.00214484940353,-0.00214477206966,0.942424208754,0.942424207934,0.0420686507658,0.0420686507425,2.04002780883,2.04002780823,2.29045333992,2.29045334201
0.496000000,-0.175760090462,-0.175758941906,-0.00226244747559,-0.00226250831719,0.942423660672,0.942423660805,0.0420688540065,0.0420688537413,2.04002814111,2.04002813919,2.29045478564,2.29045478429
0.497000000,-0.176702061565,-0.17670090207,-0.00238142100695,-0.00238153417658,0.942423122106,0.942423122607,0.0420690548066,0.042069054456,2.04002845967,2.04002845732,2.29045619716,2.29045619452
0.498000000,-0.177643923366,-0.177642759329,-0.00250176927763,-0.0025018498552,0.942422593092,0.942422593373,0.0420692531939,0.0420692529063,2.04002876472,2.04002876277,2.29045757451,2.29045757269
0.499000000,-0.178585661956,-0.178584502956,-0.00262349133073,-0.0026234552331,0.942422073666,0.942422073136,0.0420694491947,0.0420694491126,2.04002905646,2.0400290557,2.29045891772,2.2904589188
0.500000000,-0.179527263481,-0.179526104956,-0.00274658618948,-0.00274650883223,0.942421563865,0.942421563046,0.042069642835,0.0420696428251,2.04002933509,2.04002933477,2.2904602268,2.29046022892
0.501000000,-0.180468714089,-0.180467535459,-0.00287105287755,-0.00287112031816,0.942421063725,0.942421063931,0.0420698341408,0.0420698338897,2.04002960081,2.04002959922,2.29046150179,2.29046150034
0.502000000,-0.181410000163,-0.181408810325,-0.002996890338,-0.00299701781691,0.942420573283,0.942420573921,0.0420700231377,0.0420700227929,2.04002985384,2.04002985178,2.2904627427,2.29046273979
0.503000000,-0.182351107975,-0.182349918896,-0.00312409754499,-0.00312419996622,0.942420092575,0.942420093042,0.0420702098514,0.0420702095547,2.04003009437,2.04003009264,2.29046394957,2.29046394732
0.504000000,-0.183292024468,-0.183290850474,-0.00325267325063,-0.0032526653613,0.94241962164,0.942419621321,0.0420703943059,0.0420703941953,2.04003032262,2.04003032195,2.29046512241,2.29046512294
0.505000000,-0.184232736639,-0.184231568922,-0.00338261618849,-0.0033825705154,0.942419160516,0.942419159922,0.0420705765253,0.0420705764789,2.04003053879,2.04003053852,2.29046626125,2.29046626275
0.506000000,-0.185173231487,-0.185172038501,-0.00351392509257,-0.00351402363599,0.942418709238,0.942418709689,0.0420707565336,0.0420707562589,2.04003074311,2.04003074166,2.29046736609,2.29046736404
0.507000000,-0.186113496243,-0.18611228863,-0.00364659862151,-0.00364675447075,0.942418267846,0.942418268723,0.0420709343542,0.0420709339937,2.04003093578,2.04003093393,2.29046843696,2.29046843351
0.508000000,-0.187053518031,-0.187052309112,-0.0037806354628,-0.00378076179879,0.942417836378,0.942417837052,0.0420711100108,0.0420711097014,2.04003111702,2.04003111549,2.29046947388,2.29046947118
0.509000000,-0.187993284635,-0.187992089932,-0.00391603409644,-0.00391604445675,0.942417414872,0.942417414708,0.0420712835256,0.0420712833998,2.04003128703,2.04003128652,2.29047047686,2.29047047706
0.510000000,-0.188932783897,-0.188931588539,-0.00405279298543,-0.00405275844914,0.942417003366,0.942417002882,0.0420714549209,0.0420714548638,2.04003144604,2.04003144592,2.29047144592,2.29047144723
0.511000000,-0.189872003658,-0.189870766362,-0.00419091059256,-0.00419101208492,0.942416601899,0.942416602441,0.0420716242187,0.0420716239523,2.04003159426,2.04003159311,2.29047238106,2.29047237895
0.512000000,-0.190810931991,-0.190809667407,-0.00433038531111,-0.00433053706164,0.942416210509,0.942416211452,0.0420717914407,0.0420717910991,2.0400317319,2.04003173043,2.2904732823,2.29047327891
0.513000000,-0.191749556861,-0.191748283132,-0.00447121556035,-0.00447133245949,0.942415829237,0.942415829947,0.0420719566088,0.0420719563203,2.04003185919,2.04003185804,2.29047414965,2.29047414709
0.514000000,-0.192687866889,-0.192686605158,-0.00461339956637,-0.0046133973474,0.94241545812,0.942415457959,0.0420721197435,0.0420721196322,2.04003197632,2.0400319761,2.29047498311,2.2904749835
0.515000000,-0.193625850753,-0.193624586337,-0.00475693553958,-0.004756887174,0.942415097197,0.942415096707,0.0420722808653,0.0420722808202,2.04003208353,2.04003208367,2.29047578271,2.29047578422
0.516000000,-0.194563497129,-0.19456218592,-0.00490182169,-0.00490190963295,0.942414746509,0.942414747073,0.0420724399947,0.0420724397494,2.04003218103,2.04003218024,2.29047654843,2.2904765465
0.517000000,-0.195500794924,-0.195499459633,-0.00504805616297,-0.00504819728172,0.942414406095,0.942414407085,0.0420725971517,0.042072596832,2.04003226903,2.04003226793,2.2904772803,2.29047727702
0.518000000,-0.196437732937,-0.19643639945,-0.00519563712775,-0.00519574877374,0.942414075993,0.942414076773,0.0420727523564,0.0420727520834,2.04003234774,2.04003234691,2.29047797831,2.29047797576
0.519000000,-0.197374300617,-0.197372997276,-0.00534456257417,-0.00534456266585,0.942413756244,0.942413756169,0.0420729056281,0.0420729055189,2.04003241738,2.04003241735,2.29047864247,2.29047864275
0.520000000,-0.198310487471,-0.198309199948,-0.00549483047707,-0.00549479270493,0.942413446888,0.942413446511,0.0420730569855,0.0420730569343,2.04003247817,2.04003247842,2.29047927278,2.29047927405
0.521000000,-0.199246283002,-0.199244961736,-0.00564643881129,-0.00564654458748,0.942413147963,0.942413148697,0.0420732064476,0.0420732062015,2.04003253031,2.0400325297,2.29047986924,2.29047986694
0.522000000,-0.200181676943,-0.200180347698,-0.00579938549128,-0.0057995517297,0.94241285951,0.942412860711,0.042073354033,0.0420733537123,2.04003257401,2.04003257311,2.29048043185,2.29048042808
0.523000000,-0.201116658921,-0.201115348997,-0.0059536684538,-0.00595381204002,0.942412581568,0.942412582581,0.0420734997604,0.0420734994813,2.0400326095,2.04003260881,2.29048096062,2.29048095749
0.524000000,-0.202051219206,-0.202049956742,-0.00610928546863,-0.00610932336967,0.942412314178,0.942412314338,0.0420736436474,0.0420736435228,2.04003263697,2.04003263697,2.29048145555,2.29048145516
0.525000000,-0.202985348125,-0.202984110446,-0.00626623429212,-0.00626623721834,0.942412057379,0.942412057242,0.0420737857117,0.0420737856424,2.04003265664,2.04003265688,2.29048191662,2.29048191718
0.526000000,-0.203919036002,-0.203917759156,-0.00642451268007,-0.00642465740679,0.942411811211,0.942411812202,0.0420739259706,0.0420739257174,2.04003266871,2.0400326682,2.29048234384,2.29048234083
0.527000000,-0.204852273389,-0.204850979277,-0.00658411833276,-0.00658431974942,0.942411575714,0.942411577168,0.0420740644415,0.0420740641201,2.04003267339,2.04003267264,2.29048273721,2.29048273276
0.528000000,-0.205785050732,-0.205783762746,-0.00674504897069,-0.00674522215425,0.942411350929,0.942411352169,0.0420742011417,0.0420742008637,2.04003267089,2.04003267035,2.29048309671,2.29048309296
0.529000000,-0.206717359115,-0.206716101839,-0.00690730215991,-0.00690736261032,0.942411136895,0.942411137237,0.0420743360876,0.0420743359614,2.04003266141,2.04003266149,2.29048342235,2.29048342145
0.530000000,-0.207649189677,-0.207647932118,-0.0070708754536,-0.0070708914551,0.942410933652,0.942410933655,0.0420744692954,0.0420744692269,2.04003264514,2.04003264546,2.29048371412,2.29048371428
0.531000000,-0.208580533558,-0.20857920309,-0.00723576640493,-0.00723591247994,0.942410741239,0.942410742357,0.0420746007814,0.042074600542,2.0400326223,2.040032622,2.29048397201,2.2904839687
0.532000000,-0.209511382119,-0.209510004885,-0.00740197251525,-0.00740216496485,0.942410559698,0.942410561253,0.0420747305617,0.042074730261,2.04003259308,2.04003259259,2.290484196,2.29048419137
0.533000000,-0.210441726619,-0.210440333023,-0.00756949130487,-0.00756964764215,0.942410389068,0.942410390378,0.0420748586522,0.042074858396,2.04003255767,2.04003255739,2.29048438611,2.29048438225
0.534000000,-0.211371558949,-0.211370183404,-0.00773832015077,-0.0077383593079,0.942410229388,0.942410229766,0.0420749850683,0.0420749849588,2.04003251628,2.04003251655,2.29048454231,2.29048454135
0.535000000,-0.212300871054,-0.212299492186,-0.00790845641863,-0.0079084501677,0.942410080698,0.942410080727,0.042075109825,0.0420751097709,2.04003246909,2.04003246956,2.29048466459,2.2904846647
0.536000000,-0.213229654879,-0.213228211289,-0.00807989747365,-0.00808002408094,0.942409943039,0.942409944213,0.0420752329375,0.0420752327188,2.0400324163,2.04003241623,2.29048475295,2.29048474954
0.537000000,-0.21415790259,-0.214156437371,-0.00825264063342,-0.00825282170324,0.942409816449,0.942409818093,0.0420753544206,0.0420753541411,2.0400323581,2.04003235785,2.29048480736,2.29048480251
0.538000000,-0.21508560625,-0.215084164522,-0.00842668323264,-0.00842684087022,0.942409700969,0.942409702397,0.0420754742893,0.0420754740492,2.04003229467,2.04003229456,2.29048482783,2.29048482363
0.539000000,-0.216012758551,-0.216011385312,-0.00860202247421,-0.00860207891188,0.942409596637,0.942409597156,0.0420755925576,0.0420755924545,2.0400322262,2.04003222651,2.29048481434,2.29048481287
0.540000000,-0.216939352235,-0.216938023989,-0.00877865554993,-0.00877868166972,0.942409503493,0.942409503695,0.0420757092397,0.0420757091868,2.04003215288,2.04003215332,2.29048476687,2.29048476632
0.541000000,-0.217865380047,-0.21786400719,-0.00895657965183,-0.00895674458587,0.942409421577,0.942409422963,0.0420758243497,0.042075824138,2.04003207488,2.04003207485,2.2904846854,2.29048468124
0.542000000,-0.21879083495,-0.218789416784,-0.00913579192752,-0.00913600420031,0.942409350928,0.942409352774,0.0420759379012,0.0420759376319,2.04003199238,2.04003199222,2.29048456994,2.29048456438
0.543000000,-0.219715709804,-0.219714378358,-0.00931628954082,-0.00931625110943,0.942409291585,0.942409291264,0.0420760499083,0.0420760499364,2.04003190557,2.04003190625,2.29048442045,2.29048442141
0.544000000,-0.220639998093,-0.220638666593,-0.00949806953343,-0.00949803406982,0.942409243586,0.942409243266,0.042076160384,0.0420761604059,2.04003181462,2.04003181536,2.29048423693,2.29048423769
0.545000000,-0.221563693352,-0.221562372397,-0.00968112893762,-0.0096810879644,0.94240920697,0.942409206511,0.0420762693414,0.0420762693471,2.04003171969,2.04003172071,2.29048401935,2.29048401962
0.546000000,-0.222486789115,-0.222485470479,-0.00986546478521,-0.00986542360972,0.942409181777,0.942409181182,0.0420763767937,0.0420763767499,2.04003162097,2.04003162238,2.29048376771,2.29048376685
0.547000000,-0.223409279028,-0.223407987731,-0.0100510740871,-0.01005102794,0.942409168044,0.94240916723,0.0420764827537,0.0420764826626,2.04003151861,2.04003152042,2.29048348198,2.29048348046
0.548000000,-0.224331157052,-0.224329903285,-0.0102379537933,-0.0102379059048,0.942409165811,0.942409164862,0.0420765872342,0.0420765871142,2.04003141278,2.04003141468,2.29048316215,2.29048316113
0.549000000,-0.225252417251,-0.225251181571,-0.0104261008349,-0.0104260505619,0.942409175114,0.94240917423,0.0420766902474,0.0420766901534,2.04003130365,2.04003130505,2.2904828082,2.29048280989
0.550000000,-0.226173053844,-0.226171870528,-0.0106155121143,-0.0106154650517,0.942409196032,0.942409195431,0.0420767918057,0.0420767918504,2.04003119138,2.04003119141,2.29048243224,2.29048242769
0.551000000,-0.227091507582,-0.227090184515,-0.010801617449,-0.0108014788777,0.942463667557,0.942463539023,0.0420724984766,0.0420724009938,2.04050735627,2.040518428,2.30669184659,2.30663975345
0.552000000,-0.228000647445,-0.227999168021,-0.0109811859841,-0.0109810726304,0.942509156751,0.942509051354,0.0420613061549,0.042061245138,2.04176266408,2.04177063157,2.31927694744,2.31923459954
0.553000000,-0.228893382776,-0.22889205656,-0.0111559516324,-0.0111558900145,0.942547525331,0.942547488602,0.0420455293638,0.0420455098624,2.04358095465,2.043583563,2.32900507162,2.32899058558
0.554000000,-0.22976405383,-0.22976258893,-0.0113273377275,-0.0113272661742,0.942580236957,0.942580172284,0.042026913202,0.0420268789845,2.04579461073,2.04579911481,2.33648448345,2.33645900171
0.555000000,-0.23060815532,-0.230606444318,-0.0114965118291,-0.0114964493479,0.942608441622,0.942608370435,0.0420067598533,0.0420067398908,2.04827418518,2.04827797308,2.34219657842,2.34216906698
0.556000000,-0.231422115022,-0.231420527389,-0.0116644304411,-0.0116643955345,0.942633042603,0.94263301676,0.0419860281531,0.0419860374637,2.05092020412,2.05092048407,2.34652187473,2.34651248553
0.557000000,-0.232203116462,-0.23220166025,-0.0118318755425,-0.0118318530291,0.942654749362,0.942654749585,0.0419654117418,0.0419654336835,2.05365670453,2.05365527132,2.34976082827,2.34976168501
0.558000000,-0.232948956324,-0.232947119797,-0.0119994845994,-0.011999471989,0.942674119333,0.942674098081,0.0419454004432,0.0419454278287,2.0564261323,2.05642486423,2.35215046772,2.35214377779
0.559000000,-0.233657929691,-0.233656173562,-0.0121677752475,-0.0121677634466,0.942691590839,0.942691564983,0.0419263283273,0.0419263400221,2.05918532264,2.05918511663,2.3538776296,2.35386891534
0.560000000,-0.234328737498,-0.234327545096,-0.012337165644,-0.0123371363633,0.942707508962,0.942707495477,0.0419084112765,0.041908394394,2.06190233508,2.06190383325,2.35508946552,2.35508404326
0.561000000,-0.234960411712,-0.234958628005,-0.0125079912854,-0.0125079915704,0.942722145858,0.942722121522,0.0418917762826,0.041891784552,2.06455396294,2.0645537661,2.35590177043,2.35589371863
0.562000000,-0.235552254544,-0.235550346135,-0.0126805189783,-0.0126805341064,0.942735716718,0.942735706422,0.0418764842668,0.0418765116118,2.06712377155,2.06712172933,2.3564055894,2.35640359837
0.563000000,-0.236103789248,-0.236102280329,-0.012854958384,-0.0128549587064,0.942748392219,0.942748397166,0.0418625477244,0.0418625667285,2.06960056145,2.06959887059,2.35667242622,2.35667536004
0.564000000,-0.236614719813,-0.236613041197,-0.013031471678,-0.013031482482,0.942760308305,0.942760298848,0.0418499443921,0.0418499538703,2.07197715735,2.07197632514,2.35675837902,2.35675593902
0.565000000,-0.237084898178,-0.237083416688,-0.0132101815585,-0.0132101854395,0.942771573795,0.942771570895,0.0418386277184,0.0418386332024,2.07424946187,2.07424886445,2.35670739538,2.35670693027
0.566000000,-0.237514297318,-0.237512861709,-0.0133911779435,-0.0133911820729,0.942782276334,0.942782274493,0.0418285348689,0.0418285385494,2.07641571352,2.0764152426,2.35655384711,2.35655361941
0.567000000,-0.237902989123,-0.237901576317,-0.0135745235691,-0.0135745295276,0.942792487019,0.942792485705,0.0418195927944,0.0418195953049,2.07847590641,2.07847552418,2.35632456408,2.3563244136
0.568000000,-0.238251126178,-0.238249748157,-0.0137602586729,-0.0137602656018,0.94280226401,0.942802263266,0.0418117227818,0.041811723986,2.08043133784,2.08043106899,2.35604043873,2.35604035792
0.569000000,-0.238558926751,-0.238557614011,-0.013948404907,-0.0139484087977,0.942811655317,0.942811655359,0.0418048438283,0.0418048432339,2.08228425646,2.08228414942,2.3557176902,2.35571772903
0.570000000,-0.238826662349,-0.238825410441,-0.0141389686213,-0.0141389705334,0.942820700984,0.942820701603,0.0417988751041,0.0417988730212,2.08403758901,2.08403763681,2.3553688629,2.35536894377
0.571000000,-0.23905464734,-0.239053319549,-0.0143319436301,-0.0143319555986,0.942829434768,0.942829434879,0.0417937377064,0.0417937369991,2.08569472862,2.08569463053,2.35500361711,2.35500361358
0.572000000,-0.239243230426,-0.239241841793,-0.0145273135063,-0.0145273332084,0.942837885447,0.94283788527,0.0417893558731,0.0417893560534,2.08725937273,2.08725915884,2.35462934679,2.35462932378
0.573000000,-0.239392787513,-0.239391365001,-0.0147250535099,-0.0147250737383,0.942846077824,0.942846077538,0.041785657793,0.0417856582006,2.08873539855,2.08873509889,2.35425166816,2.35425171505
0.574000000,-0.23950371569,-0.239502250328,-0.01492513223,-0.0149251589362,0.942854033517,0.942854033119,0.0417825760798,0.0417825767955,2.09012676804,2.09012640596,2.35387480997,2.35387485197
0.575000000,-0.239576428284,-0.239574771793,-0.0151275129346,-0.0151275538718,0.942861771564,0.942861770267,0.0417800480193,0.0417800509655,2.09143745726,2.0914367296,2.35350191595,2.35350198812
0.576000000,-0.239611350765,-0.239609436838,-0.0153321546892,-0.0153322134922,0.942869308882,0.94286930651,0.0417780156699,0.0417780211216,2.09267140297,2.09267020766,2.35313528122,2.35313543014
0.577000000,-0.239608917236,-0.239607190123,-0.0155390133175,-0.0155390634699,0.942876660642,0.942876659396,0.0417764257978,0.0417764284612,2.0938324639,2.09383167632,2.35277654385,2.35277666411
0.578000000,-0.239569567564,-0.239568104372,-0.0157480421864,-0.0157480800979,0.942883840545,0.942883840612,0.0417752297425,0.0417752295094,2.09492439297,2.09492412113,2.35242683243,2.3524268724
0.579000000,-0.239493744979,-0.239491840693,-0.0159591928585,-0.0159592594086,0.942890861046,0.942890859403,0.0417743832172,0.0417743866469,2.09595081795,2.09594980872,2.35208688258,2.35208703935
0.580000000,-0.239381894078,-0.239380031012,-0.0161724156336,-0.0161724813538,0.942897733527,0.942897732235,0.0417738460707,0.0417738485486,2.09691522871,2.09691433546,2.35175712853,2.35175728554
0.581000000,-0.239234459176,-0.239233088028,-0.0163876599931,-0.0163876983279,0.942904468433,0.942904469061,0.0417735820233,0.0417735805664,2.09782096955,2.0978208873,2.35143777493,2.35143780262
0.582000000,-0.239051882952,-0.239050137666,-0.0166048749652,-0.0166049352534,0.942911075386,0.942911074744,0.0417735583838,0.0417735591119,2.09867123571,2.09867058989,2.35112885357,2.35112899733
0.583000000,-0.238834605333,-0.238832732316,-0.0168240094187,-0.0168240802881,0.942917563269,0.94291756236,0.0417737457836,0.0417737469477,2.09946907192,2.09946828105,2.35083026673,2.35083043473
0.584000000,-0.238583062563,-0.238581462902,-0.0170450123045,-0.0170450742615,0.942923940297,0.94292394043,0.04177411791,0.0417741175784,2.10021737317,2.10021701132,2.35054182169,2.35054188664
0.585000000,-0.2382976865,-0.238296012065,-0.0172678328447,-0.0172679037417,0.942930214083,0.942930214126,0.0417746512262,0.0417746511744,2.1009188883,2.10091845223,2.35026325783,2.35026333206
0.586000000,-0.237978904023,-0.23797699522,-0.0174924206826,-0.0174925042111,0.942936391686,0.942936391145,0.0417753247269,0.0417753252577,2.10157622349,2.10157550968,2.34999426721,2.3499944143
0.587000000,-0.237627136577,-0.237625420356,-0.0177187259981,-0.017718797014,0.942942479654,0.942942479623,0.0417761197001,0.0417761191418,2.10219184696,2.10219138895,2.34973451054,2.34973461866
0.588000000,-0.237242799831,-0.237241346226,-0.0179466995943,-0.0179467592239,0.942948484063,0.942948484727,0.041777019504,0.0417770180385,2.10276809423,2.10276795988,2.34948362957,2.34948366419
0.589000000,-0.236826303422,-0.236824444622,-0.0181762929613,-0.0181763811238,0.942954410552,0.942954410461,0.0417780093603,0.0417780090005,2.10330717366,2.10330661768,2.34924125645,2.34924137183
0.590000000,-0.236378050758,-0.236376206094,-0.018407458321,-0.0184075453167,0.942960264351,0.942960264316,0.0417790761672,0.0417790755681,2.1038111719,2.10381064869,2.34900702084,2.34900713635
0.591000000,-0.235898438992,-0.235896071614,-0.0186401486445,-0.0186402631379,0.942966050316,0.94296604932,0.0417802083077,0.0417802084641,2.10428206104,2.10428103702,2.34878055513,2.34878079156
0.592000000,-0.235387858736,-0.235385366014,-0.0188743176927,-0.0188744399311,0.942971772939,0.942971771869,0.0417813955316,0.0417813955441,2.1047217007,2.10472060875,2.34856149871,2.34856175477
0.593000000,-0.234846694259,-0.234844416044,-0.0191099199938,-0.01911003416,0.942977436386,0.942977435873,0.0417826287675,0.0417826282469,2.10513184724,2.10513100753,2.34834950051,2.3483496977
0.594000000,-0.234275323554,-0.234273590873,-0.0193469108328,-0.019347002983,0.942983044521,0.942983045079,0.0417838999895,0.0417838988094,2.1055141597,2.10551383807,2.34814422079,2.34814429196
0.595000000,-0.23367411813,-0.233672264532,-0.0195852462693,-0.0195853472903,0.942988600913,0.942988601339,0.0417852021366,0.0417852011164,2.10587020132,2.10586978877,2.34794533302,2.3479454251
0.596000000,-0.233043443258,-0.233041054581,-0.0198248831,-0.0198250101779,0.942994108865,0.94299410854,0.0417865289786,0.0417865282755,2.10620144709,2.10620062488,2.34775252436,2.34775271761
0.597000000,-0.232383658093,-0.232381133387,-0.0200657788334,-0.0200659129502,0.94299957143,0.942999571015,0.0417878750177,0.0417878741875,2.10650928906,2.10650840448,2.34756549603,2.34756570723
0.598000000,-0.231695115805,-0.231692833821,-0.0203078916627,-0.0203080156807,0.943004991433,0.943004991451,0.0417892354043,0.0417892342446,2.106795041,2.1067943788,2.3473839634,2.34738412265
0.599000000,-0.230978163479,-0.230976472768,-0.020551180464,-0.0205512801499,0.943010371475,0.943010372333,0.041790605879,0.0417906043893,2.10705994018,2.10705972433,2.34720765644,2.34720770666
0.600000000,-0.23023314248,-0.230231406984,-0.0207956047408,-0.0207957069189,0.943015713962,0.943015714769,0.0417919826819,0.0417919811307,2.10730515447,2.10730491727,2.34703631893,2.34703637619
0.601000000,-0.229460388618,-0.229458169044,-0.0210411245923,-0.021041247717,0.943021021112,0.943021021373,0.0417933624925,0.0417933609444,2.10753178645,2.10753125135,2.34686970805,2.34686984148
0.602000000,-0.228660232108,-0.228657844709,-0.0212877007035,-0.02128783406,0.94302629497,0.943026295145,0.0417947423891,0.0417947408593,2.10774087514,2.10774026392,2.34670759437,2.34670774609
0.603000000,-0.227832997785,-0.227830756591,-0.0215352943068,-0.0215354277521,0.943031537418,0.943031537899,0.0417961197945,0.0417961183642,2.10793340035,2.10793290444,2.34654976119,2.34654988054
0.604000000,-0.226979005412,-0.226977231759,-0.0217838671364,-0.0217839898208,0.943036750193,0.943036751306,0.0417974924258,0.0417974912073,2.10811028778,2.10811007548,2.34639600364,2.34639604602
0.605000000,-0.226098569618,-0.22609670067,-0.0220333814207,-0.0220335119707,0.943041934894,0.943041935984,0.0417988582686,0.0417988570829,2.10827240992,2.10827215494,2.34624612863,2.34624618088
0.606000000,-0.225192000039,-0.22518968701,-0.0222837998526,-0.0222839461929,0.943047092988,0.943047093661,0.0418002155428,0.0418002141253,2.10842058915,2.10842012162,2.34609995426,2.34610006508
0.607000000,-0.224259601908,-0.224257202408,-0.0225350855234,-0.0225352318471,0.943052225834,0.943052226394,0.041801562657,0.0418015610408,2.10855560472,2.10855511854,2.34595730832,2.34595742883
0.608000000,-0.223301675352,-0.223299509491,-0.0227872019701,-0.0227873372222,0.943057334671,0.943057335395,0.0418028982149,0.04180289655,2.10867818755,2.10867783491,2.34581802956,2.34581811859
0.609000000,-0.222318516513,-0.222316855541,-0.0230401130658,-0.0230402327953,0.943062420646,0.943062421811,0.0418042209583,0.041804219529,2.10878903177,2.10878891976,2.34568196506,2.3456819893
0.610000000,-0.22131041688,-0.221308648659,-0.0232937830629,-0.0232939130776,0.94306748481,0.943067485995,0.041805529775,0.0418055284295,2.10888878982,2.10888863632,2.34554897141,2.34554900308
0.611000000,-0.220277663863,-0.220275407536,-0.0235481765344,-0.0235483292662,0.943072528129,0.943072529071,0.0418068236675,0.0418068221765,2.10897807862,2.10897773562,2.34541891328,2.34541899364
0.612000000,-0.219220540931,-0.219217272822,-0.0238032583517,-0.0238034355557,0.943077551491,0.943077551587,0.0418081017396,0.0418080994411,2.10905748171,2.10905676549,2.34529166294,2.34529185517
0.613000000,-0.218139327316,-0.218135813907,-0.0240589936954,-0.0240591777314,0.94308255571,0.943082555673,0.0418093631932,0.0418093606875,2.10912754781,2.10912676857,2.34516710061,2.34516731488
0.614000000,-0.217034298829,-0.21703125595,-0.0243153479859,-0.0243155215431,0.943087541536,0.943087541937,0.0418106073019,0.0418106051074,2.10918879825,2.10918822116,2.34504511273,2.34504527107
0.615000000,-0.215905727201,-0.215903797556,-0.0245722869193,-0.0245724354049,0.943092509652,0.943092510943,0.0418118334176,0.0418118320705,2.10924172269,2.10924155307,2.34492559294,2.34492563241
0.616000000,-0.214753880827,-0.214751643022,-0.0248297764077,-0.0248299337565,0.94309746069,0.943097461803,0.0418130409499,0.0418130393457,2.10928678567,2.10928653081,2.34480844056,2.34480850795
0.617000000,-0.213579024534,-0.21357591174,-0.0250877825861,-0.0250879604787,0.943102395226,0.943102395763,0.0418142293657,0.041814227022,2.10932442555,2.10932391136,2.34469356084,2.34469371021
0.618000000,-0.21238141971,-0.212378051169,-0.0253462717954,-0.0253464561818,0.943107313787,0.943107314198,0.0418153981818,0.0418153956162,2.1093550563,2.10935448455,2.34458086452,2.34458103424
0.619000000,-0.211161324577,-0.211158313316,-0.0256052105582,-0.0256053895986,0.94311221686,0.943112217569,0.0418165469578,0.0418165447268,2.10937907008,2.10937861843,2.34447026724,2.34447040051
0.620000000,-0.20991899407,-0.209916948398,-0.0258645655793,-0.0258647294327,0.943117104885,0.943117106295,0.0418176752944,0.0418176739869,2.10939683706,2.10939666211,2.34436168957,2.34436173381
0.621000000,-0.208654680053,-0.208652300932,-0.0261243037265,-0.0261244753108,0.943121978269,0.943121979506,0.0418187828276,0.041818781216,2.10940870755,2.10940845689,2.3442550565,2.34425512715
0.622000000,-0.207368631397,-0.207365438917,-0.0263843920205,-0.0263845772974,0.943126837381,0.943126838124,0.0418198692261,0.0418198668075,2.10941501306,2.1094145752,2.34415029722,2.34415043377
0.623000000,-0.206061093943,-0.20605771945,-0.0266447976311,-0.0266449852666,0.943131682559,0.943131683208,0.0418209341893,0.0418209315566,2.10941606674,2.10941560707,2.34404734498,2.34404749289
0.624000000,-0.204732310817,-0.204729363089,-0.0269054878564,-0.0269056690421,0.943136514112,0.943136515037,0.0418219774428,0.04182197521,2.10941216576,2.10941182464,2.34394613654,2.34394624675
0.625000000,-0.203382522142,-0.203380584094,-0.0271664301317,-0.0271665988154,0.943141332318,0.943141333869,0.0418229987397,0.0418229975357,2.10940359025,2.10940348535,2.34384661244,2.34384664081
0.626000000,-0.202011965586,-0.202009793426,-0.0274275920006,-0.0274277618022,0.943146137435,0.943146138815,0.0418239978547,0.0418239963664,2.10939060683,2.10939046722,2.34374871614,2.34374876128
0.627000000,-0.200620875833,-0.200617980488,-0.0276889411338,-0.0276891158767,0.943150929692,0.943150930603,0.0418249745872,0.0418249722799,2.10937346634,2.10937319801,2.34365239458,2.34365249282
0.628000000,-0.199209485348,-0.199206419076,-0.0279504452938,-0.0279506206298,0.943155709302,0.943155710105,0.0418259287551,0.0418259262504,2.10935240819,2.10935212157,2.34355759716,2.34355770601
0.629000000,-0.19777802363,-0.197775316746,-0.0282120723617,-0.0282122467455,0.943160476453,0.9431604775,0.0418268601994,0.0418268581285,2.1093276572,2.1093274461,2.34346427645,2.34346435684
0.630000000,-0.196326718165,-0.196324883389,-0.0284737902979,-0.0284739647084,0.943165231318,0.943165232955,0.0418277687766,0.0418277677819,2.10929942855,2.10929937085,2.34337238711,2.34337240323
0.631000000,-0.194855793505,-0.194853632158,-0.028735567174,-0.0287357474931,0.943169974049,0.943169975566,0.0418286543652,0.0418286530762,2.10926792397,2.10926781498,2.34328188666,2.34328192291
0.632000000,-0.193365472359,-0.193362521734,-0.0289973711305,-0.0289975577404,0.943174704788,0.943174705916,0.0418295168573,0.0418295147256,2.109233337,2.10923310803,2.3431927344,2.34319282151
0.633000000,-0.191855974482,-0.191852777726,-0.0292591704162,-0.0292593621174,0.943179423657,0.943179424735,0.0418303561664,0.0418303538334,2.10919584866,2.10919558937,2.34310489229,2.34310499193
0.634000000,-0.190327518194,-0.190324614734,-0.0295209333319,-0.0295211302616,0.943184130769,0.943184132113,0.0418311722163,0.0418311703338,2.10915563395,2.10915542193,2.34301832356,2.34301840005
0.635000000,-0.18878031854,-0.188778243659,-0.0297826282943,-0.0297828319557,0.94318882622,0.943188828132,0.0418319649538,0.0418319641746,2.10911285504,2.10911276019,2.34293299416,2.34293301355
0.636000000,-0.187214589152,-0.187212268459,-0.0300442237708,-0.030044425925,0.943193510097,0.943193511852,0.0418327343355,0.0418327332727,2.10906766878,2.10906755183,2.34284887117,2.34284890457
0.637000000,-0.18563054132,-0.185627582921,-0.0303056883107,-0.030305881775,0.943198182478,0.943198183772,0.0418334803343,0.0418334784406,2.10902022336,2.10902004237,2.34276592347,2.34276599654
0.638000000,-0.184028383774,-0.184025321489,-0.0305669905531,-0.0305671765835,0.943202843427,0.943202844554,0.0418342029403,0.0418342008694,2.1089706581,2.10897048164,2.3426841218,2.34268420104
0.639000000,-0.182408324043,-0.182405658617,-0.0308280991789,-0.0308282828343,0.943207493004,0.943207494268,0.0418349021519,0.041834900561,2.1089191082,2.10891898921,2.34260343777,2.34260349276
0.640000000,-0.18077056657,-0.180768761918,-0.0310889829782,-0.0310891736022,0.943212131256,0.943212132983,0.0418355779882,0.0418355775291,2.10886569887,2.10886567838,2.34252384502,2.34252384756
0.641000000,-0.179115314463,-0.179113277959,-0.0313496107884,-0.0313497977699,0.943216758226,0.943216759789,0.0418362304767,0.0418362297603,2.1088105513,2.1088105135,2.34244531806,2.3424453334
0.642000000,-0.177442768773,-0.177440042288,-0.0316099515205,-0.0316101321305,0.943221373948,0.943221375148,0.0418368596587,0.04183685814,2.10875378039,2.10875367998,2.34236783267,2.34236788585
0.643000000,-0.175753128079,-0.175750155191,-0.0318699741759,-0.0318701597437,0.94322597845,0.94322597965,0.0418374655913,0.041837463919,2.10869549408,2.10869537138,2.34229136606,2.34229142846
0.644000000,-0.174046589902,-0.174043854669,-0.0321296477942,-0.0321298466432,0.943230571754,0.943230573248,0.0418380483383,0.0418380471565,2.10863579736,2.10863569396,2.34221589603,2.34221593977
0.645000000,-0.172323348922,-0.172321438728,-0.0323889415225,-0.0323891526773,0.943235153878,0.9432351558,0.0418386079814,0.0418386079196,2.10857478775,2.10857475889,2.34214140189,2.34214139858
0.646000000,-0.170583598495,-0.17057965116,-0.0326478245589,-0.0326479801437,0.943239724833,0.943239725355,0.0418391446104,0.0418391417338,2.10851255948,2.1085123874,2.34206786363,2.34206798033
0.647000000,-0.168827530167,-0.168822856462,-0.0329062661717,-0.0329063970442,0.943244284627,0.94324428465,0.0418396583261,0.0418396544367,2.1084492023,2.10844899491,2.34199526211,2.3419954209
0.648000000,-0.167055333075,-0.16705098059,-0.033164235727,-0.033164372581,0.943248833263,0.943248833516,0.0418401492436,0.0418401458052,2.10838480049,2.10838462979,2.3419235793,2.34192371953
0.649000000,-0.165267195401,-0.165264149456,-0.0334217026294,-0.0334218811909,0.943253370743,0.943253371928,0.0418406174837,0.0418406159091,2.1083194361,2.10831935293,2.34185279759,2.3418528643
0.650000000,-0.163463302707,-0.163461290797,-0.0336786363949,-0.0336788541713,0.943257897063,0.943257899007,0.0418410631823,0.041841063101,2.10825318565,2.10825316162,2.34178290047,2.34178291
0.651000000,-0.161643839222,-0.161640161516,-0.0339350065979,-0.0339351626716,0.943262412216,0.94326241297,0.0418414864823,0.0418414840049,2.10818612295,2.10818602019,2.34171387195,2.34171397418
0.652000000,-0.159808987571,-0.159804620699,-0.0341907828838,-0.0341909074021,0.943266916196,0.943266916442,0.041841887536,0.0418418840612,2.10811831861,2.1081181941,2.34164569667,2.34164583673
0.653000000,-0.157958928029,-0.157954810578,-0.0344459350073,-0.0344460649054,0.94327140899,0.943271409417,0.0418422665077,0.0418422634016,2.10804983885,2.1080497353,2.3415783601,2.34157848613
0.654000000,-0.156093839979,-0.156090873973,-0.0347004327652,-0.034700611812,0.943275890586,0.94327589189,0.0418426235666,0.0418426221634,2.10798074828,2.10798069438,2.34151184803,2.34151191108
0.655000000,-0.154213900402,-0.154211822479,-0.0349542460719,-0.0349544695603,0.943280360972,0.943280362987,0.041842958894,0.0418429588278,2.10791110734,2.10791108285,2.34144614702,2.34144616189
0.656000000,-0.152319284815,-0.152315567653,-0.0352073449177,-0.0352074871445,0.94328482013,0.943284820918,0.0418432726782,0.0418432702356,2.10784097411,2.10784089834,2.3413812441,2.34138134735
0.657000000,-0.150410167721,-0.150405755617,-0.0354596993424,-0.0354598012154,0.943289268045,0.94328926831,0.0418435651125,0.041843561674,2.10777040491,2.10777031569,2.3413171266,2.3413172666
0.658000000,-0.148486720714,-0.148482525142,-0.0357112795454,-0.0357113885801,0.943293704699,0.943293705148,0.0418438364035,0.0418438333075,2.10769945173,2.10769937612,2.34125378266,2.34125391014
0.659000000,-0.146549114617,-0.146546013497,-0.0359620557701,-0.0359622260526,0.943298130076,0.943298131418,0.041844086761,0.0418440853032,2.10762816544,2.10762811961,2.34119120062,2.34119126872
0.660000000,-0.144597519048,-0.144595292101,-0.036211998325,-0.036212223733,0.943302544155,0.943302546232,0.0418443164001,0.0418443162519,2.10755659511,2.10755656683,2.34112936915,2.34112939021
0.661000000,-0.142632100785,-0.142628393221,-0.0364610776905,-0.0364612072809,0.943306946919,0.943306947774,0.0418445255476,0.0418445232011,2.10748478603,2.10748473946,2.3410682776,2.34106837747
0.662000000,-0.140653025816,-0.140648739815,-0.036709264396,-0.0367093484251,0.94331133835,0.943311338686,0.0418447144331,0.0418447112138,2.1074127825,2.10741273938,2.34100791547,2.34100804604
0.663000000,-0.138660459047,-0.138656457543,-0.0369565290353,-0.0369566241684,0.943315718429,0.943315718952,0.0418448832907,0.0418448804733,2.10734062737,2.10734059885,2.34094827253,2.34094838823
0.664000000,-0.136654562599,-0.136651670632,-0.0372028423866,-0.0372030115675,0.943320087139,0.943320088557,0.0418450323642,0.0418450311655,2.10726836019,2.10726834921,2.34088933912,2.3408893965
0.665000000,-0.134635497756,-0.134633498864,-0.0374481752857,-0.0374484096951,0.943324444461,0.943324446608,0.0418451619002,0.0418451619916,2.10719601957,2.10719601817,2.34083110573,2.34083111663
0.666000000,-0.132603424837,-0.132600081435,-0.0376924986314,-0.0376926216481,0.94332879038,0.94332879128,0.0418452721487,0.0418452702174,2.10712364299,2.10712364653,2.34077356304,2.34077364619
0.667000000,-0.130558501444,-0.130554641026,-0.0379357835188,-0.0379358543194,0.943333124878,0.943333125242,0.0418453633679,0.0418453606501,2.10705126503,2.10705127958,2.34071670223,2.34071681364
0.668000000,-0.128500884281,-0.128497295634,-0.0381780011111,-0.0381780852197,0.943337447942,0.943337448484,0.0418454358185,0.0418454334922,2.10697891938,2.10697894306,2.34066051458,2.34066061233
0.669000000,-0.126430729195,-0.126428163022,-0.0384191226326,-0.0384192919252,0.943341759557,0.943341760998,0.0418454897638,0.041845488948,2.10690663881,2.10690666206,2.34060499156,2.34060503568
0.670000000,-0.124348189398,-0.12434641599,-0.0386591195151,-0.0386593629734,0.943346059711,0.943346061891,0.0418455254739,0.041845525839,2.1068344536,2.10683447068,2.34055012502,2.34055012718
0.671000000,-0.122253417141,-0.122250308854,-0.0388979632702,-0.0388980796209,0.943350348391,0.943350349337,0.0418455432209,0.0418455416687,2.10676239316,2.10676242617,2.34049590692,2.34049597904
0.672000000,-0.1201465639,-0.120142885605,-0.0391356254695,-0.0391356831961,0.943354625587,0.943354626025,0.041845543279,0.041845540961,2.10669048615,2.10669052929,2.34044232937,2.34044243043
0.673000000,-0.118027778629,-0.1180242647,-0.0393720779012,-0.0393721516679,0.943358891291,0.943358891948,0.0418455259273,0.0418455239354,2.10661875908,2.10661880102,2.34038938478,2.34038947531
0.674000000,-0.115897209241,-0.115894564752,-0.0396072924468,-0.0396074630509,0.943363145495,0.943363147099,0.041845491447,0.0418454908133,2.1065472376,2.10654726182,2.34033706569,2.34033710774
0.675000000,-0.113755002943,-0.113753018342,-0.0398412410475,-0.0398414954464,0.943367388194,0.943367390586,0.0418454401207,0.0418454405414,2.10647594677,2.10647595136,2.34028536474,2.34028536885
0.676000000,-0.111601304555,-0.111598001339,-0.0400738958669,-0.0400740084815,0.943371619383,0.943371620578,0.0418453722346,0.0418453708704,2.10640490979,2.10640494054,2.3402342748,2.3402343458
0.677000000,-0.109436257764,-0.109432379034,-0.0403052291781,-0.0403052770928,0.943375839061,0.943375839776,0.0418452880764,0.0418452860068,2.10633414901,2.10633419368,2.34018378888,2.34018388746
0.678000000,-0.107260005609,-0.107256268018,-0.0405352133147,-0.0405352796304,0.943380047225,0.943380048176,0.041845187935,0.0418451861796,2.10626368629,2.10626372765,2.34013390007,2.3401339884
0.679000000,-0.105072688895,-0.105069783801,-0.0407638208344,-0.0407639945086,0.943384243879,0.943384245776,0.0418450721015,0.0418450716186,2.10619354192,2.10619355881,2.34008460163,2.34008464326
0.680000000,-0.102874447197,-0.102872211285,-0.0409910244235,-0.0409912896016,0.943388429025,0.943388431682,0.0418449408682,0.0418449413904,2.10612373535,2.10612373073,2.34003588696,2.34003589118
0.681000000,-0.10066541948,-0.100662029852,-0.0412167968303,-0.0412169035288,0.943392602667,0.943392604076,0.0418447945278,0.0418447934842,2.10605428558,2.10605432416,2.33998774951,2.33998781498
0.682000000,-0.09844574265,-0.0984419167328,-0.0414411110253,-0.0414411445206,0.943396764813,0.943396765674,0.0418446333747,0.0418446317629,2.10598521032,2.10598527338,2.33994018289,2.33994027234
0.683000000,-0.0962155522846,-0.096211974977,-0.0416639401286,-0.0416639916255,0.943400915472,0.943400916485,0.0418444577035,0.0418444564551,2.10591652643,2.10591659128,2.3398931808,2.33989325846
0.684000000,-0.093974983365,-0.0939723054167,-0.0418852573264,-0.0418854240103,0.943405054655,0.943405056521,0.0418442678093,0.0418442677888,2.10584825039,2.10584829026,2.33984673701,2.33984676854
0.685000000,-0.0917241690161,-0.0917222266412,-0.0421050360192,-0.0421053000888,0.943409182373,0.94340918491,0.0418440639877,0.0418440649412,2.10578039759,2.10578041608,2.3398008454,2.33980083997
0.686000000,-0.0894632409264,-0.0894602925403,-0.0423232497812,-0.0423233388885,0.943413298643,0.943413299873,0.041843846534,0.0418438461268,2.10571298258,2.10571305596,2.33975549994,2.33975555154
0.687000000,-0.0871923301856,-0.0871890078176,-0.0425398722589,-0.0425398812428,0.943417403481,0.943417404135,0.0418436157438,0.0418436148607,2.10564601957,2.10564612092,2.33971069466,2.33971076867
0.688000000,-0.0849115662515,-0.0849084633824,-0.0427548772998,-0.0427549070753,0.943421496905,0.94342149772,0.0418433719123,0.0418433713692,2.10557952189,2.10557962075,2.33966642366,2.33966648684
0.689000000,-0.0826210770334,-0.0826187510461,-0.0429682389517,-0.0429683963515,0.943425578937,0.943425580653,0.041843115334,0.0418431158794,2.10551350202,2.10551356505,2.33962268115,2.33962270154
0.690000000,-0.080320989819,-0.0803192298449,-0.0431799313435,-0.0431801982892,0.943429649599,0.943429652086,0.0418428463033,0.0418428476805,2.10544797213,2.10544800176,2.33957946136,2.33957944812
0.691000000,-0.0780114305124,-0.0780085589548,-0.0433899287855,-0.0433900121249,0.943433708916,0.943433710262,0.0418425651137,0.0418425652239,2.10538294367,2.10538302486,2.3395367586,2.33953680098
0.692000000,-0.0756925233268,-0.0756891310849,-0.0435982058212,-0.0435982086695,0.943437756915,0.943437757851,0.0418422720574,0.0418422716938,2.10531842727,2.10531852948,2.33949456726,2.33949463225
0.693000000,-0.0733643919145,-0.0733610763219,-0.0438047370731,-0.0438047669496,0.943441793625,0.943441794843,0.0418419674261,0.0418419673311,2.10525443327,2.10525452525,2.33945288176,2.33945293733
0.694000000,-0.07102715866,-0.0710245377317,-0.0440094973397,-0.0440096655365,0.943445819075,0.943445821214,0.0418416515103,0.0418416523806,2.10519097141,2.10519102212,2.33941169657,2.3394117116
0.695000000,-0.0686809442057,-0.0686790122762,-0.0442124616763,-0.044212741013,0.9434498333,0.943449836039,0.0418413245986,0.0418413262754,2.10512805063,2.10512807399,2.33937100624,2.33937098798
0.696000000,-0.0663258686862,-0.0663229687972,-0.0444136052139,-0.0444136762764,0.943453836332,0.943453837821,0.0418409869788,0.0418409875263,2.10506567964,2.10506576495,2.33933080534,2.33933084491
0.697000000,-0.0639620510473,-0.0639587661252,-0.0446129032594,-0.0446128846881,0.943457828208,0.943457829201,0.0418406389373,0.0418406390752,2.10500386662,2.10500397851,2.33929108849,2.33929115286
0.698000000,-0.0615896084769,-0.0615864977901,-0.0448103313948,-0.0448103443484,0.943461808967,0.943461810185,0.0418402807576,0.041840281141,2.10494261898,2.10494272151,2.33925185038,2.33925190784
0.699000000,-0.0592086575672,-0.0592062456631,-0.0450058652982,-0.0450060326952,0.943465778648,0.943465780785,0.0418399127226,0.0418399139333,2.1048819439,2.10488200028,2.33921308572,2.33921310653
0.700000000,-0.0568193143163,-0.0568174533478,-0.0451994807341,-0.0451997780701,0.943469737292,0.943469740171,0.041839535114,0.0418395369468,2.10482184827,2.10482186593,2.33917478923,2.33917478107
0.701000000,-0.0544216922329,-0.0544189484077,-0.0453911538825,-0.045391240498,0.943473684943,0.943473686651,0.0418391482091,0.0418391490601,2.10476233801,2.10476241943,2.33913695574,2.33913699847
0.702000000,-0.0520159045319,-0.0520127987641,-0.0455808609739,-0.0455808523433,0.943477621646,0.943477622865,0.0418387522843,0.0418387527573,2.10470341891,2.1047035289,2.33909958006,2.33909964437
0.703000000,-0.0496020638396,-0.0495990920432,-0.0457685783354,-0.045768595538,0.943481547448,0.943481548854,0.0418383476152,0.0418383482539,2.10464509652,2.10464519835,2.33906265705,2.33906271472
0.704000000,-0.047180281175,-0.0471779163036,-0.0459542825699,-0.0459544522641,0.943485462395,0.943485464657,0.0418379344737,0.0418379357647,2.10458737576,2.10458743156,2.33902618161,2.33902620546
0.705000000,-0.0447506659641,-0.0447487644217,-0.0461379505678,-0.046138245919,0.943489366538,0.943489369487,0.0418375131283,0.041837514893,2.10453026099,2.10453027922,2.33899014868,2.33899014624
0.706000000,-0.0423133273142,-0.0423105590176,-0.0463195592791,-0.0463196223845,0.943493259929,0.943493261711,0.0418370838465,0.0418370847294,2.10447375643,2.10447384298,2.33895455322,2.33895459961
0.707000000,-0.0398683737872,-0.0398652375431,-0.0464990857447,-0.0464990429807,0.943497142619,0.943497143918,0.0418366468947,0.0418366474263,2.10441786609,2.104417983,2.3389193902,2.33891945724
0.708000000,-0.0374159117322,-0.037412886537,-0.0466765074233,-0.0466764904999,0.943501014662,0.943501016149,0.0418362025335,0.04183620319,2.10436259321,2.10436270179,2.33888465466,2.33888471518
0.709000000,-0.0349560468107,-0.0349535916582,-0.0468518019067,-0.0468519477309,0.943504876113,0.943504878448,0.0418357510219,0.0418357522251,2.10430794078,2.10430800172,2.33885034167,2.33885036952
0.710000000,-0.0324888841385,-0.0324868841199,-0.0470249468946,-0.0470252299259,0.94350872703,0.943508730048,0.0418352926168,0.0418352942236,2.10425391163,2.10425393328,2.33881644629,2.33881644817
0.711000000,-0.030014528098,-0.0300117627264,-0.0471959202206,-0.0471959656785,0.943512567468,0.943512569363,0.0418348275739,0.0418348284618,2.10420050831,2.10420059846,2.33878296362,2.33878301011
0.712000000,-0.0275330809139,-0.0275300258212,-0.0473647001589,-0.0473646422638,0.943516397487,0.94351639892,0.0418343561419,0.0418343567651,2.10414773267,2.10414785227,2.33874988883,2.33874995322
0.713000000,-0.0250446445394,-0.0250417513554,-0.0475312650402,-0.0475312425331,0.943520217147,0.943520218766,0.0418338785688,0.0418338793258,2.10409558647,2.10409569607,2.33871721706,2.33871727379
0.714000000,-0.0225493203803,-0.0225470162414,-0.047695593306,-0.0476957493926,0.943524026509,0.943524028947,0.0418333951011,0.041833396334,2.10404407131,2.10404413107,2.33868494352,2.33868496811
0.715000000,-0.0200472085132,-0.0200453807294,-0.0478576636744,-0.0478579698547,0.943527825634,0.943527828725,0.0418329059807,0.0418329075617,2.10399318834,2.10399320719,2.3386530634,2.33865306247
0.716000000,-0.0175384075164,-0.0175359080149,-0.0480174551903,-0.0480175158906,0.943531614584,0.943531616574,0.041832411445,0.0418324124548,2.1039429383,2.10394302579,2.33862157198,2.33862161238
0.717000000,-0.0150230157031,-0.015020271039,-0.0481749469592,-0.048174901288,0.943535393424,0.943535394956,0.0418319117299,0.0418319125351,2.1038933218,2.10389343853,2.33859046451,2.33859052143
0.718000000,-0.0125011309145,-0.0124985421659,-0.0483301181819,-0.0483301096811,0.943539162218,0.943539163925,0.0418314070704,0.0418314079821,2.10384433935,2.10384444566,2.33855973629,2.33855978598
0.719000000,-0.00997284906435,-0.00997079364797,-0.0484829484981,-0.0484831248516,0.94354292103,0.943542923532,0.041830897694,0.0418308989741,2.10379599087,2.10379604726,2.33852938265,2.33852940242
0.720000000,-0.00743826540817,-0.00743661782072,-0.0486334176999,-0.0486337467949,0.943546669926,0.943546673071,0.0418303838261,0.0418303853591,2.10374827611,2.10374829218,2.33849939893,2.33849939529
0.721000000,-0.00489747474491,-0.00489514880236,-0.0487815056882,-0.0487815729323,0.943550408972,0.943550411072,0.0418298656904,0.0418298667456,2.10370119469,2.10370128008,2.33846978052,2.33846981649
0.722000000,-0.00235057124172,-0.00234795072872,-0.048927192503,-0.0489271430648,0.943554138235,0.943554139924,0.0418293435082,0.0418293443724,2.10365474604,2.10365486107,2.33844052279,2.33844057511
0.723000000,0.000202352840365,0.000204902587273,-0.0490704586493,-0.049070442056,0.943557857783,0.943557859679,0.0418288174934,0.0418288184088,2.1036089291,2.10360903436,2.3384116212,2.33841166754
0.724000000,0.0027612054907,0.00276333698805,-0.0492112846945,-0.04921145491,0.943561567681,0.943561570389,0.0418282878589,0.0418282890224,2.10356374272,2.10356379906,2.33838307119,2.33838309019
0.725000000,0.00532589520535,0.00532772114525,-0.0493496513291,-0.0493499754427,0.943565268,0.943565271372,0.0418277548154,0.0418277561335,2.10351918565,2.10351920258,2.33835486823,2.33835486589
0.726000000,0.00789633155646,0.00789884194777,-0.0494855395099,-0.0494855877541,0.943568958806,0.943568961208,0.0418272185696,0.041827219502,2.10347525632,2.10347534215,2.33832700783,2.338327043
0.727000000,0.0104724255079,0.0104752446475,-0.0496189305542,-0.0496188556099,0.943572640168,0.943572642202,0.0418266793221,0.0418266800988,2.10343195286,2.10343206809,2.33829948553,2.33829953613
0.728000000,0.0130540882414,0.0130568551302,-0.0497498058385,-0.0497497647777,0.943576312154,0.943576314404,0.0418261372726,0.0418261380806,2.10338927331,2.10338937881,2.33827229688,2.33827234178
0.729000000,0.0156412313446,0.0156435999754,-0.0498781468373,-0.0498783010949,0.943579974834,0.943579977865,0.0418255926191,0.0418255936022,2.10334721562,2.10334727261,2.33824543746,2.33824545647
0.730000000,0.0182337680618,0.0182358136879,-0.0500039354712,-0.0500042521439,0.943583628275,0.943583631925,0.0418250455523,0.0418250466466,2.10330577737,2.10330579529,2.33821890288,2.33821890144
0.731000000,0.0208316122593,0.0208342173722,-0.0501271538309,-0.0501271886718,0.943587272546,0.943587275215,0.0418244962602,0.0418244971059,2.10326495597,2.1032650412,2.3381926888,2.33819272179
0.732000000,0.0234346781815,0.0234374767996,-0.0502477841134,-0.0502476964511,0.943590907716,0.943590909969,0.0418239449291,0.0418239456972,2.10322474877,2.1032248623,2.33816679087,2.33816683756
0.733000000,0.0260428806103,0.0260455281767,-0.0503658086564,-0.0503657616745,0.943594533853,0.94359453624,0.0418233917428,0.0418233925608,2.10318515297,2.10318525632,2.33814120477,2.33814124542
0.734000000,0.0286561359931,0.0286583095631,-0.0504812102744,-0.0504813705856,0.943598151025,0.943598154085,0.0418228368775,0.0418228378347,2.10314616545,2.10314622093,2.33811592624,2.33811594206
0.735000000,0.0312743610061,0.0312761377806,-0.050593971849,-0.0505943043728,0.943601759299,0.943601762879,0.0418222805081,0.0418222815527,2.103107783,2.10310780012,2.33809095102,2.33809094722
0.736000000,0.0338974727922,0.0338996940206,-0.0507040763962,-0.0507041205769,0.943605358744,0.943605361322,0.0418217228074,0.0418217237184,2.10307000234,2.10307008487,2.3380662749,2.33806630289
0.737000000,0.0365253893451,0.036527754566,-0.0508115071778,-0.0508114263829,0.943608949426,0.943608951575,0.0418211639442,0.0418211648189,2.10303282001,2.10303292982,2.33804189367,2.33804193426
0.738000000,0.0391580299375,0.0391602656575,-0.0509162478433,-0.0509162086689,0.943612531412,0.943612533699,0.041820604081,0.0418206049791,2.1029962323,2.10299633211,2.33801780318,2.33801783811
0.739000000,0.0417953140188,0.041797172541,-0.0510182820976,-0.0510184544777,0.943616104768,0.943616107751,0.0418200433793,0.0418200443225,2.10296023547,2.10296028881,2.33799399929,2.3379940112
0.740000000,0.0444371613852,0.04443876836,-0.0511175937453,-0.0511179395568,0.943619669559,0.943619673139,0.0418194819992,0.041819482931,2.1029248257,2.10292484189,2.33797047788,2.33797047178
0.741000000,0.0470834932401,0.0470856650828,-0.0512141670315,-0.051214210464,0.943623225851,0.943623228604,0.0418189200933,0.0418189209179,2.10288999894,2.10289007841,2.33794723489,2.33794725865
0.742000000,0.0497342313679,0.0497366944921,-0.0513079863865,-0.0513078959944,0.943626773707,0.94362677619,0.0418183578117,0.0418183585764,2.102855751,2.10285585678,2.33792426629,2.33792430193
0.743000000,0.052389297863,0.0523917634848,-0.0513990363424,-0.051398984994,0.943630313192,0.943630315917,0.0418177953025,0.0418177960311,2.10282207767,2.1028221737,2.33790156804,2.33790159847
0.744000000,0.0550486152723,0.0550507659418,-0.0514873015686,-0.0514874666806,0.943633844367,0.94363384779,0.0418172327116,0.0418172334088,2.10278897465,2.10278902584,2.33787913615,2.33787914511
0.745000000,0.0577121075839,0.0577138752314,-0.0515727672132,-0.0515731139333,0.943637367295,0.943637371149,0.0418166701776,0.0418166708607,2.10275643741,2.1027564532,2.33785696668,2.33785695864
0.746000000,0.0603796990118,0.0603819016707,-0.0516554185009,-0.0516554469455,0.943640882037,0.943640884973,0.0418161078376,0.0418161085366,2.10272446137,2.10272453961,2.33783505573,2.33783507691
0.747000000,0.0630513142737,0.0630536875475,-0.051735240827,-0.051735135119,0.943644388653,0.943644391233,0.0418155458261,0.0418155465433,2.1026930419,2.10269314526,2.33781339938,2.33781343227
0.748000000,0.0657268782583,0.0657291678942,-0.0518122196328,-0.05181216417,0.943647887202,0.943647889955,0.0418149842768,0.0418149849841,2.10266217431,2.10266226698,2.33779199377,2.33779202193
0.749000000,0.0684063176938,0.0684082868202,-0.0518863410052,-0.0518865183225,0.943651377742,0.943651381167,0.0418144233134,0.0418144239599,2.10263185368,2.10263190174,2.33777083508,2.33777084328
0.750000000,0.0710895593083,0.0710912773005,-0.051957591031,-0.0519579601857,0.943654860329,0.943654864293,0.0418138630595,0.0418138636342,2.10260207507,2.1026020877,2.33774991955,2.33774991209
0.751000000,0.0737765302138,0.0737786410597,-0.0520259559378,-0.052026012815,0.94365833502,0.94365833815,0.0418133036367,0.0418133042937,2.10257283352,2.10257290387,2.33772924339,2.33772926163
0.752000000,0.0764671576391,0.0764694252838,-0.0520914219783,-0.0520913414092,0.943661801871,0.943661804657,0.041812745166,0.0418127458663,2.10254412399,2.10254421834,2.33770880285,2.33770883159
0.753000000,0.0791613704225,0.0791635746178,-0.0521539759973,-0.052153936006,0.943665260935,0.943665263859,0.0418121877595,0.0418121884407,2.10251594132,2.102516027,2.33768859427,2.33768861907
0.754000000,0.0818590978832,0.0818610333457,-0.0522136050184,-0.052213786988,0.943668712262,0.943668715799,0.0418116315264,0.0418116321038,2.10248828023,2.10248832568,2.33766861399,2.33766862115
0.755000000,0.0845602694218,0.0845620055612,-0.052270296097,-0.0522706573032,0.943672155904,0.943672159931,0.0418110765756,0.0418110770487,2.10246113549,2.10246114954,2.33764885838,2.33764885197
0.756000000,0.0872648149051,0.0872669328579,-0.0523240364606,-0.0523240655328,0.943675591913,0.943675595115,0.0418105230135,0.0418105236294,2.10243450177,2.10243457238,2.33762932385,2.33762934181
0.757000000,0.0899726646967,0.089974943829,-0.0523748135171,-0.0523746942372,0.943679020335,0.943679023199,0.0418099709434,0.0418099716266,2.10240837371,2.10240846802,2.3376100068,2.3376100348
0.758000000,0.0926837505487,0.0926859824389,-0.0524226152165,-0.0524225345833,0.943682441218,0.943682444218,0.0418094204609,0.041809421117,2.10238274582,2.10238283213,2.33759090376,2.33759092817
0.759000000,0.095398004213,0.0953999930916,-0.0524674295093,-0.052467577701,0.943685854607,0.943685858209,0.0418088716616,0.0418088721756,2.10235761262,2.10235766035,2.33757201124,2.33757201916
0.760000000,0.0981153578165,0.0981171535504,-0.0525092444967,-0.0525095825112,0.943689260545,0.943689264634,0.0418083246387,0.0418083250207,2.10233296857,2.10233298558,2.33755332578,2.33755332074
0.761000000,0.10083574358,0.100837854187,-0.0525480483029,-0.0525480584068,0.943692659077,0.943692662373,0.0418077794853,0.0418077800627,2.10230880814,2.10230887716,2.33753484392,2.33753486072
0.762000000,0.103559095198,0.103561312909,-0.0525838296517,-0.0525837009411,0.943696050243,0.943696053213,0.0418072362856,0.0418072369535,2.10228512569,2.10228521543,2.33751656231,2.33751658768
0.763000000,0.106285346707,0.106287478985,-0.0526165774076,-0.0526165009736,0.943699434082,0.943699437182,0.0418066951221,0.0418066957579,2.10226191556,2.10226199606,2.33749847762,2.33749849903
0.764000000,0.109014432251,0.10901630242,-0.0526462804815,-0.0526464493795,0.943702810631,0.94370281431,0.0418061560763,0.0418061565393,2.10223917207,2.10223921467,2.33748058654,2.33748059221
0.765000000,0.111746286304,0.111747942232,-0.0526729279162,-0.0526733004008,0.943706179927,0.943706184069,0.0418056192282,0.0418056195352,2.10221688954,2.10221690199,2.33746288576,2.3374628791
0.766000000,0.114480843966,0.114482747535,-0.052696509012,-0.0526965543695,0.943709542007,0.943709545363,0.0418050846539,0.0418050852011,2.10219506225,2.10219512283,2.33744537203,2.33744538521
0.767000000,0.117218041418,0.117220018259,-0.0527170135325,-0.052716921043,0.943712896901,0.943712899926,0.0418045524233,0.0418045530784,2.1021736844,2.10217376405,2.3374280422,2.33742806315
0.768000000,0.119957814838,0.11995970756,-0.0527344312413,-0.0527343919526,0.943716244639,0.943716247785,0.0418040226063,0.0418040232205,2.10215275025,2.10215282118,2.33741089309,2.33741091048
0.769000000,0.12270010077,0.122701768601,-0.0527487520628,-0.052748958803,0.943719585253,0.943719588964,0.0418034952705,0.04180349568,2.10213225401,2.10213228974,2.33739392156,2.33739392474
0.770000000,0.125444835836,0.125446339546,-0.0527599659419,-0.0527603727644,0.943722918772,0.943722922943,0.0418029704833,0.0418029707093,2.10211218988,2.10211219812,2.33737712448,2.33737711675
0.771000000,0.128191957993,0.128193722683,-0.0527680634193,-0.0527681279293,0.94372624522,0.943726248639,0.0418024483042,0.0418024487995,2.10209255204,2.10209260638,2.33736049884,2.33736050971
0.772000000,0.130941405433,0.130943287319,-0.052773035141,-0.0527729484769,0.943729564619,0.943729567741,0.0418019287911,0.0418019293996,2.10207333467,2.10207340793,2.33734404162,2.33734406005
0.773000000,0.133693116476,0.133694984363,-0.0527748718123,-0.0527748275042,0.943732876994,0.943732880266,0.0418014120015,0.0418014125535,2.10205453196,2.10205459816,2.33732774983,2.33732776542
0.774000000,0.136447029672,0.13644876425,-0.0527735642375,-0.052773758295,0.943736182366,0.943736186229,0.0418008979913,0.0418008983038,2.10203613807,2.1020361725,2.33731162051,2.33731162351
0.775000000,0.139203084271,0.13920473743,-0.0527691035372,-0.0527694903057,0.943739480753,0.943739485108,0.0418003868124,0.0418003869144,2.10201814718,2.10201815694,2.33729565074,2.33729564408
0.776000000,0.141961220355,0.141963150375,-0.0527614812214,-0.0527615132598,0.94374277217,0.943742775818,0.0417998785119,0.0417998789044,2.10200055346,2.10200060678,2.33727983768,2.3372798482
0.777000000,0.144721378009,0.14472344312,-0.0527506888002,-0.0527505632052,0.943746056634,0.94374606001,0.0417993731369,0.0417993736463,2.10198335109,2.10198342228,2.33726417849,2.33726419598
0.778000000,0.14748349767,0.147485564786,-0.0527367179517,-0.0527366343032,0.943749334156,0.94374933769,0.0417988707324,0.0417988711747,2.10196653426,2.10196659886,2.33724867038,2.33724868523
0.779000000,0.15024751984,0.150249464924,-0.0527195603738,-0.05271972075,0.94375260475,0.943752608864,0.0417983713429,0.0417983715233,2.10195009715,2.10195013194,2.33723331054,2.33723331378
0.780000000,0.153013386226,0.153015229342,-0.0526992083453,-0.0526995698745,0.943755868424,0.943755873003,0.0417978750062,0.0417978749632,2.10193403398,2.10193404533,2.33721809628,2.33721809049
0.781000000,0.155781038691,0.155783058549,-0.0526756542196,-0.0526756660998,0.943759125183,0.943759129015,0.041797381759,0.041797382032,2.10191833898,2.10191838981,2.33720302493,2.33720303452
0.782000000,0.158550419237,0.158552475542,-0.0526488904194,-0.0526487543459,0.943762375034,0.943762378537,0.0417968916374,0.0417968920392,2.10190300637,2.10190307271,2.33718809385,2.33718810946
0.783000000,0.161321470027,0.161323437416,-0.052618909441,-0.0526188285938,0.943765617981,0.94376562157,0.0417964046765,0.0417964050099,2.1018880304,2.10188808957,2.33717330038,2.3371733133
0.784000000,0.164094133956,0.164095902833,-0.0525857041379,-0.0525858827631,0.943768854026,0.943768858118,0.0417959209071,0.0417959209681,2.10187340534,2.10187343598,2.33715864195,2.33715864403
0.785000000,0.166868354561,0.166869947697,-0.0525492676859,-0.0525496608035,0.943772083166,0.943772087652,0.0417954403566,0.0417954401869,2.10185912549,2.1018591337,2.33714411607,2.3371441097
0.786000000,0.169644075382,0.169645749796,-0.0525095932605,-0.0525096403295,0.943775305399,0.943775309086,0.0417949630524,0.0417949632125,2.10184518518,2.10184522928,2.33712972022,2.33712972769
0.787000000,0.172421240392,0.172422908949,-0.0524666742583,-0.0524665750869,0.943778520719,0.94377852405,0.0417944890194,0.0417944893057,2.10183157876,2.1018316369,2.33711545195,2.3371154648
0.788000000,0.175199793307,0.175201390667,-0.0524205039346,-0.0524204591685,0.943781729125,0.943781732546,0.0417940182837,0.0417940184827,2.10181830052,2.10181835223,2.33710130877,2.33710131916
0.789000000,0.177979679224,0.177981159438,-0.0523710762524,-0.0523712869079,0.943784930605,0.943784934574,0.0417935508643,0.0417935507592,2.10180534492,2.10180537093,2.33708728833,2.33708728889
0.790000000,0.180760843373,0.18076227592,-0.0523183852423,-0.0523188002818,0.943788125147,0.943788129599,0.041793086779,0.041793086408,2.10179270638,2.10179271267,2.33707338832,2.33707338122
0.791000000,0.183543230982,0.183544867608,-0.0522624249353,-0.052262474336,0.943791312739,0.943791316511,0.0417926260461,0.0417926259806,2.10178037937,2.10178041964,2.33705960639,2.33705961186
0.792000000,0.186326787729,0.186328561154,-0.0522031895945,-0.0522030752922,0.943794493366,0.943794496905,0.0417921686813,0.0417921687085,2.10176835837,2.1017684125,2.33704594029,2.33704595058
0.793000000,0.189111459117,0.189113283714,-0.0521406733851,-0.0521406018107,0.943797667016,0.943797670741,0.0417917147008,0.0417917146061,2.10175663784,2.10175668681,2.3370323877,2.33703239566
0.794000000,0.191897191914,0.191898949773,-0.052074871137,-0.052075053756,0.943800833668,0.943800837969,0.0417912641148,0.0417912636885,2.10174521239,2.10174523808,2.33701894646,2.33701894536
0.795000000,0.194683933015,0.194685512217,-0.0520057777452,-0.0520061798569,0.943803993299,0.943803997959,0.0417908169326,0.0417908162381,2.10173407664,2.10173408362,2.33700561444,2.33700560626
0.796000000,0.19747162931,0.197473174739,-0.0519333881048,-0.0519339049476,0.943807145888,0.943807150792,0.0417903731634,0.0417903723233,2.10172322521,2.10172322289,2.33699238948,2.33699237483
0.797000000,0.200260227902,0.20026180384,-0.0518576972207,-0.0518582033992,0.943810291411,0.943810296254,0.0417899328156,0.0417899319422,2.10171265275,2.10171265252,2.33697926947,2.3369792458
0.798000000,0.203049676544,0.203051321775,-0.0517787004418,-0.051779185881,0.943813429843,0.943813434584,0.0417894958941,0.0417894949799,2.10170235398,2.10170235654,2.33696625237,2.33696621574
0.799000000,0.20583992319,0.205841670769,-0.0516963932257,-0.0516968638399,0.943816561153,0.943816565888,0.0417890624028,0.0417890614359,2.10169232366,2.10169232786,2.33695333614,2.33695328951
0.800000000,0.208630916102,0.208632795738,-0.0516107711926,-0.0516112613704,0.943819685313,0.943819690375,0.0417886323444,0.0417886313035,2.10168255658,2.10168255887,2.3369405188,2.33694047395
0.801000000,0.211422603849,0.211424482945,-0.0515218309315,-0.0515223164686,0.943822804603,0.943822809673,0.0417882057188,0.0417882046832,2.10167304764,2.10167304888,2.3369277981,2.33692775813
0.802000000,0.214214935285,0.214216826754,-0.0514295758608,-0.0514300508202,0.943825930408,0.943825935479,0.0417877825165,0.0417877814847,2.10166379222,2.10166379321,2.33691516925,2.33691513135
0.803000000,0.217007859597,0.217009776855,-0.05133402532,-0.0513344802417,0.943829075741,0.943829080827,0.0417873627108,0.0417873616828,2.10165478655,2.10165478807,2.33690262249,2.33690258374
0.804000000,0.21980132621,0.219803248788,-0.0512352101931,-0.0512356399107,0.943832253091,0.943832258059,0.0417869462647,0.0417869452411,2.10164602737,2.10164602993,2.33689014502,2.33689010298
0.805000000,0.222595284885,0.222597224118,-0.0511331772104,-0.0511335736262,0.943835474205,0.943835479053,0.0417865331306,0.0417865321091,2.10163751189,2.1016375157,2.33687772103,2.33687767459
0.806000000,0.225389685596,0.225391705288,-0.0510279832351,-0.0510283404464,0.94383875035,0.943838755151,0.0417861232559,0.0417861222256,2.10162923752,2.10162924229,2.33686533318,2.33686528274
0.807000000,0.228184478628,0.228186588842,-0.0509196992468,-0.0509200217801,0.943842092108,0.943842096955,0.0417857165828,0.0417857155357,2.10162120183,2.10162120662,2.33685296274,2.33685291118
0.808000000,0.230979614465,0.230981744432,-0.0508084055032,-0.0508087132454,0.943845509602,0.943845514482,0.0417853130508,0.0417853119857,2.10161340243,2.10161340574,2.33684059034,2.33684054313
0.809000000,0.233775043864,0.233777241076,-0.0506941945341,-0.0506945103479,0.943849012335,0.943849017537,0.0417849125967,0.0417849114968,2.10160583695,2.10160583663,2.33682819618,2.33682816032
0.810000000,0.236570717769,0.23657300822,-0.0505771675132,-0.0505774983396,0.943852609364,0.943852615934,0.0417845151559,0.0417845140333,2.10159850297,2.10159849916,2.33681576041,2.336815741
0.811000000,0.239366587356,0.239368874534,-0.0504574360773,-0.0504578011979,0.94385630919,0.943856314417,0.0417841206622,0.0417841194923,2.10159139799,2.1015913914,2.3368032633,2.33680324406
0.812000000,0.242162603974,0.242164898658,-0.0503351199588,-0.0503354488811,0.943860119879,0.943860126358,0.0417837290485,0.0417837279087,2.10158451943,2.10158451571,2.33679068541,2.3367906638
0.813000000,0.244958719163,0.244961037144,-0.0502103476839,-0.0502106469004,0.943864049007,0.943864056005,0.0417833402459,0.0417833391236,2.1015778646,2.10157786248,2.33677800775,2.33677798293
0.814000000,0.247754884619,0.247757237886,-0.0500832553182,-0.0500835366473,0.943868103719,0.943868109541,0.0417829541848,0.0417829530552,2.10157143073,2.10157142771,2.33676521184,2.33676518451
0.815000000,0.250551052188,0.250553433557,-0.0499539862591,-0.0499542096843,0.943872290735,0.943872296461,0.0417825707935,0.041782569695,2.1015652149,2.10156521314,2.33675227982,2.33675224897
0.816000000,0.253347173852,0.253349572098,-0.0498226908199,-0.0498228306774,0.943876616353,0.943876622666,0.0417821899988,0.0417821889559,2.10155921413,2.10155921436,2.33673919447,2.33673916073
0.817000000,0.256143201709,0.256145632653,-0.0496895254251,-0.0496895911256,0.943881086496,0.943881091583,0.0417818117257,0.0417818107142,2.1015534253,2.10155342401,2.33672593924,2.33672590796
0.818000000,0.258939087968,0.258941597918,-0.0495546527415,-0.049554635997,0.943885706686,0.943885710132,0.0417814358968,0.0417814349141,2.10154784522,2.10154784039,2.3367124983,2.33671247544
0.819000000,0.26173478492,0.261737417492,-0.0494182403962,-0.049418108874,0.943890482115,0.943890487178,0.0417810624324,0.0417810615277,2.1015424706,2.10154246415,2.3366988565,2.33669884562
0.820000000,0.264530244943,0.264533008177,-0.049280461738,-0.0492802439983,0.943895417584,0.943895422459,0.0417806912501,0.0417806904019,2.10153729808,2.10153728478,2.33668499947,2.33668501288
0.821000000,0.267325420467,0.267328276451,-0.0491414940422,-0.0491412260854,0.943900517614,0.943900520202,0.0417803222649,0.0417803214618,2.10153232423,2.1015322984,2.33667091346,2.33667096646
0.822000000,0.270120263982,0.270123164185,-0.0490015196449,-0.0490012142902,0.943905786349,0.943905790877,0.0417799553883,0.0417799547001,2.10152754554,2.10152750802,2.33665658554,2.33665668698
0.823000000,0.272914728004,0.272917722866,-0.0488607240723,-0.0488604366722,0.943911227688,0.943911233383,0.0417795905296,0.0417795899517,2.10152295846,2.10152290354,2.33664200335,2.33664217037
0.824000000,0.275708765078,0.275711965887,-0.04871929713,-0.0487190907568,0.94391684518,0.943916850653,0.0417792275938,0.0417792271153,2.10151855938,2.10151848031,2.33662715531,2.33662740742
0.825000000,0.278502327756,0.278505578534,-0.0485774312744,-0.0485773185127,0.943922642147,0.943922654214,0.0417788664837,0.0417788661164,2.10151434467,2.10151426046,2.33661203042,2.33661231637
0.826000000,0.281295368586,0.281298613791,-0.0484353224307,-0.0484352142386,0.943928621598,0.943928633149,0.041778507098,0.0417785065965,2.10151031064,2.10151024759,2.33659661841,2.33659683187
0.827000000,0.284087840105,0.284091088707,-0.0482931687444,-0.0482930605984,0.943934786329,0.943934794879,0.0417781493324,0.0417781487237,2.10150645361,2.10150640483,2.33658090951,2.33658107023
0.828000000,0.286879694818,0.286882947342,-0.0481511710551,-0.0481510584396,0.943941138855,0.943941144825,0.0417777930793,0.0417777924046,2.10150276985,2.10150273215,2.33656489468,2.33656501473
0.829000000,0.289670885197,0.289674124434,-0.0480095320413,-0.0480094319108,0.943947681489,0.943947691927,0.0417774382277,0.0417774375616,2.10149925563,2.10149923217,2.33654856537,2.33654864316
0.830000000,0.29246136366,0.292464601939,-0.0478684563767,-0.0478683697095,0.943954416302,0.943954428982,0.0417770846636,0.0417770840035,2.10149590723,2.10149589362,2.33653191363,2.33653196025
0.831000000,0.29525108257,0.29525432565,-0.0477281502061,-0.0477280688948,0.943961345164,0.943961357607,0.0417767322697,0.0417767316132,2.1014927209,2.10149271327,2.33651493202,2.33651495757
0.832000000,0.298039994214,0.298043240761,-0.0475888210653,-0.0475887282704,0.943968469741,0.943968479307,0.0417763809258,0.0417763802729,2.10148969291,2.10148968788,2.33649761364,2.33649762741
0.833000000,0.300828050802,0.300831285272,-0.0474506775874,-0.0474505673952,0.943975791506,0.943975798231,0.0417760305089,0.0417760298733,2.10148681954,2.10148681624,2.33647995209,2.33647995789
0.834000000,0.303615204454,0.303618384113,-0.0473139292918,-0.0473138534803,0.943983311752,0.943983322641,0.0417756808931,0.041775680308,2.10148409708,2.10148409991,2.33646194141,2.33646192992
0.835000000,0.306401407186,0.306404546281,-0.0471787864138,-0.0471787328677,0.943991031594,0.943991044285,0.0417753319502,0.0417753314051,2.10148152185,2.10148152823,2.33644357612,2.33644355372
0.836000000,0.309186610913,0.309189739546,-0.0470454596547,-0.0470454081999,0.94399895199,0.943998964189,0.0417749835494,0.0417749830361,2.10147909017,2.10147909784,2.33642485116,2.33642482374
0.837000000,0.311970767426,0.311973925593,-0.0469141600585,-0.046914082181,0.944007073732,0.944007083165,0.0417746355578,0.0417746350713,2.1014767984,2.10147680537,2.33640576191,2.33640573507
0.838000000,0.314753828389,0.31475703942,-0.0467850987733,-0.0467849911328,0.944015397466,0.944015404574,0.0417742878405,0.0417742873827,2.10147464292,2.10147464922,2.33638630412,2.33638627825
0.839000000,0.317535745349,0.317538973864,-0.0466584869571,-0.046658439445,0.944023923708,0.944023935742,0.0417739402608,0.0417739398333,2.10147262015,2.10147263023,2.33636647391,2.3363664365
0.840000000,0.320316469677,0.320319709711,-0.046534535499,-0.0465345228386,0.944032652811,0.94403266714,0.0417735926803,0.0417735922794,2.10147072654,2.10147073843,2.33634626781,2.33634622424
0.841000000,0.323095952648,0.323099181157,-0.046413455091,-0.0464134434125,0.944041585045,0.944041598876,0.0417732449594,0.0417732445886,2.10146895857,2.10146897048,2.33632568261,2.33632563856
0.842000000,0.325874145316,0.325877338675,-0.0462954557297,-0.0462954028931,0.944050720509,0.944050731069,0.041772896957,0.0417728966259,2.10146731276,2.10146732308,2.33630471553,2.33630467664
0.843000000,0.328650998638,0.328654138663,-0.0461807470417,-0.0461806505317,0.944060059235,0.944060066682,0.0417725485313,0.041772548244,2.10146578568,2.10146579458,2.336283364,2.33628333035
0.844000000,0.331426463348,0.331429532865,-0.0460695376755,-0.0460695243781,0.944069601112,0.944069612901,0.0417721995396,0.0417721992676,2.10146437393,2.10146438548,2.33626162581,2.33626158344
0.845000000,0.334200490036,0.334203578331,-0.0459620355584,-0.0459620667734,0.944079345949,0.944079360005,0.0417718498387,0.0417718495729,2.10146307415,2.10146308691,2.33623949901,2.33623945195
0.846000000,0.336973029106,0.336976195904,-0.0458584476247,-0.0458584766021,0.944089293452,0.944089307509,0.0417714992849,0.0417714990253,2.10146188304,2.10146189568,2.33621698191,2.33621693458
0.847000000,0.339744030754,0.339747155031,-0.0457589795892,-0.0457590066222,0.94409944323,0.944099456565,0.0417711477345,0.0417711474922,2.10146079734,2.1014608097,2.3361940731,2.33619402694
0.848000000,0.342513445027,0.34251648751,-0.0456638362634,-0.0456640672531,0.944109794825,0.944109816772,0.0417707950436,0.0417707947532,2.10145981383,2.10145983024,2.33617077135,2.33617070836
0.849000000,0.345281221708,0.34528423893,-0.0455732207338,-0.0455735197631,0.944120347673,0.944120372078,0.0417704410687,0.0417704407643,2.10145892935,2.10145894664,2.33614707573,2.33614700795
0.850000000,0.348047310458,0.348050348459,-0.0454873352586,-0.0454875678564,0.944131101163,0.944131122364,0.0417700856667,0.0417700853881,2.10145814076,2.10145815625,2.33612298546,2.33612292409
0.851000000,0.350811660648,0.350814766986,-0.0454063800837,-0.0454064057879,0.944142054589,0.944142067237,0.0417697286951,0.0417697284894,2.10145744502,2.10145745632,2.33609850002,2.33609845576
0.852000000,0.353574221494,0.353577318055,-0.045330554316,-0.0453306035985,0.944153207194,0.944153220389,0.041769370012,0.0417693698026,2.10145683909,2.10145685038,2.33607361903,2.33607357391
0.853000000,0.356334941968,0.356337958186,-0.0452600552685,-0.0452603496933,0.944164558153,0.944164579848,0.0417690094766,0.0417690091826,2.10145632002,2.10145633467,2.33604834233,2.33604828044
0.854000000,0.359093770821,0.359096760378,-0.0451950785554,-0.0451954770194,0.944176106582,0.944176131353,0.0417686469491,0.0417686466171,2.10145588487,2.10145590046,2.33602266992,2.33602260205
0.855000000,0.361850656602,0.361853675261,-0.0451358182563,-0.0451361613969,0.944187851551,0.944187873919,0.041768282291,0.04176828198,2.1014555308,2.10145554505,2.33599660193,2.33599653914
0.856000000,0.364605547575,0.364608651732,-0.0450824662004,-0.0450825776299,0.944199792066,0.94419980653,0.0417679153652,0.0417679151464,2.10145525498,2.1014552658,2.33597013869,2.33597009218
0.857000000,0.36735839185,0.367361486635,-0.04503521299,-0.0450353511307,0.944211927103,0.944211942081,0.0417675460358,0.0417675458108,2.10145505466,2.10145506539,2.33594328063,2.33594323338
0.858000000,0.370109137204,0.370112130191,-0.0449942465548,-0.0449946565158,0.944224255576,0.94422427837,0.0417671741693,0.0417671738396,2.10145492713,2.10145494046,2.33591602837,2.33591596545
0.859000000,0.372857731263,0.372860670904,-0.0449597535707,-0.0449602586165,0.944236776376,0.9442368014,0.0417667996332,0.0417667992716,2.10145486973,2.10145488353,2.33588838258,2.33588831538
0.860000000,0.375604121335,0.375607061774,-0.0449319181427,-0.0449323299129,0.944249488346,0.944249510168,0.0417664222975,0.0417664219856,2.10145487986,2.10145489224,2.33586034412,2.33586028379
0.861000000,0.378348254519,0.378351257661,-0.0449109226222,-0.0449110416546,0.944262390298,0.944262403666,0.0417660420338,0.0417660418612,2.10145495497,2.10145496424,2.33583191391,2.33583187136
0.862000000,0.381090077661,0.381093045379,-0.0448969472593,-0.0448970809867,0.944275481015,0.944275494387,0.0417656587159,0.0417656585515,2.10145509256,2.10145510158,2.33580309299,2.33580305077
0.863000000,0.383829537315,0.383832384887,-0.0448901698883,-0.0448906096148,0.944288759247,0.944288780089,0.0417652722203,0.0417652719343,2.10145529019,2.1014553013,2.33577388251,2.33577382506
0.864000000,0.386566579842,0.386569385511,-0.0448907667272,-0.0448913209109,0.944302223724,0.94430224702,0.041764882425,0.0417648821009,2.10145554545,2.10145555702,2.33574428368,2.33574422139
0.865000000,0.389301151244,0.389303999573,-0.0448989109735,-0.0448993759249,0.944315873144,0.944315893994,0.0417644892112,0.0417644889406,2.10145585602,2.10145586657,2.33571429783,2.33571424095
0.866000000,0.392033197389,0.392036176164,-0.0449147747119,-0.0449149327372,0.944329706194,0.944329719759,0.0417640924618,0.0417640923456,2.1014562196,2.10145622783,2.33568392634,2.33568388507
0.867000000,0.394762663744,0.394765671721,-0.0449385266865,-0.0449387256899,0.944343721533,0.944343736051,0.0417636920629,0.041763691941,2.10145663395,2.10145664221,2.33565317066,2.33565312767
0.868000000,0.39748949555,0.397492430623,-0.0449703337202,-0.0449708944074,0.944357917807,0.944357940325,0.0417632879032,0.0417632876256,2.1014570969,2.10145710703,2.33562203234,2.33562197265
0.869000000,0.400213637871,0.400216564581,-0.0450103611808,-0.0450110597938,0.944372293651,0.944372319003,0.0417628798731,0.0417628795462,2.10145760629,2.10145761691,2.33559051295,2.33559044742
0.870000000,0.402935035318,0.402938013118,-0.0450587706571,-0.045059369701,0.944386847679,0.944386870703,0.0417624678671,0.0417624676066,2.10145816005,2.10145816989,2.33555861414,2.33555855374
0.871000000,0.40565363234,0.405656717747,-0.0451157223836,-0.0451159707644,0.9444015785,0.944401594064,0.0417620517818,0.0417620517126,2.10145875615,2.10145876409,2.3355263376,2.33552629331
0.872000000,0.408369373132,0.408372413018,-0.0451813745207,-0.0451816451905,0.944416484712,0.944416500292,0.0417616315159,0.041761631464,2.10145939259,2.10145940044,2.33549368507,2.33549364087
0.873000000,0.411082201462,0.411085057713,-0.045255881688,-0.0452565203143,0.944431564904,0.944431587015,0.0417612069726,0.0417612067729,2.10146006745,2.10146007664,2.33546065835,2.33546060013
0.874000000,0.413792060945,0.413794805524,-0.0453393972708,-0.0453401585452,0.944446817661,0.944446841366,0.0417607780565,0.0417607778277,2.10146077884,2.10146078829,2.33542725925,2.33542719757
0.875000000,0.416498894878,0.416501622454,-0.0454320720542,-0.0454327034663,0.944462241562,0.944462262227,0.0417603446755,0.0417603445381,2.10146152491,2.10146153375,2.33539348963,2.3353934345
0.876000000,0.419202646168,0.419205476258,-0.0455340536907,-0.0455342967692,0.944477835184,0.944477848483,0.0417599067414,0.0417599068161,2.10146230389,2.10146231138,2.33535935138,2.33535931223
0.877000000,0.421903257592,0.421906103282,-0.0456454887723,-0.0456457677881,0.9444935971,0.944493611023,0.0417594641671,0.0417594642366,2.10146311403,2.10146312173,2.33532484642,2.33532480584
0.878000000,0.424600671488,0.424603450888,-0.045766520399,-0.0457672239079,0.944509525885,0.944509547346,0.0417590168699,0.041759016733,2.10146395363,2.10146396278,2.33528997669,2.3352899195
0.879000000,0.427294829912,0.427297626951,-0.04589728944,-0.0458981529377,0.944525620114,0.944525644407,0.0417585647701,0.0417585645531,2.10146482106,2.10146483084,2.33525474415,2.33525468055
0.880000000,0.429985674747,0.429988472646,-0.0460379354068,-0.0460386634281,0.944541878361,0.944541899902,0.0417581077894,0.0417581076493,2.1014657147,2.10146572426,2.33521915078,2.33521909295
0.881000000,0.432673147375,0.432675500321,-0.0461885939147,-0.0461905238703,0.944558299206,0.944558339337,0.0417576458546,0.0417576451248,2.10146663301,2.10146664502,2.33518319858,2.33518309935
0.882000000,0.435357188978,0.435359514263,-0.0463493990632,-0.0463513802975,0.944574881233,0.944574921349,0.0417571788942,0.0417571781371,2.10146757447,2.10146758591,2.33514688955,2.33514679003
0.883000000,0.43803774049,0.438040279428,-0.0465204830192,-0.0465218294293,0.944591623028,0.944591652831,0.0417567068393,0.0417567063918,2.10146853761,2.10146854718,2.3351102257,2.33511014893
0.884000000,0.440714742349,0.440717736002,-0.0467019742464,-0.04670199247,0.944608523185,0.944608532928,0.0417562296253,0.0417562298298,2.10146952101,2.10146952775,2.33507320907,2.33507317706
0.885000000,0.443388134831,0.443390690261,-0.0468939999754,-0.0468952880807,0.944625580304,0.9446256087,0.0417557471896,0.0417557467591,2.1014705233,2.10147053189,2.33503584169,2.33503576798
0.886000000,0.446057857877,0.446060170435,-0.0470966849647,-0.0470986590955,0.944642792992,0.944642830772,0.0417552594721,0.0417552586983,2.10147154313,2.10147155227,2.33499812558,2.33499803072
0.887000000,0.448723850985,0.448726143993,-0.0473101507331,-0.0473121609286,0.944660159865,0.944660197479,0.0417547664173,0.0417547656238,2.10147257922,2.10147258784,2.33496006277,2.33495996818
0.888000000,0.451386053499,0.451388564676,-0.0475345176988,-0.0475358870621,0.944677679545,0.944677707749,0.0417542679705,0.0417542674937,2.1014736303,2.1014736376,2.33492165531,2.33492158195
0.889000000,0.454044404324,0.454047386441,-0.047769903132,-0.0477699283873,0.944695350667,0.944695360491,0.0417537640813,0.0417537642679,2.10147469517,2.10147470059,2.33488290521,2.33488287364
0.890000000,0.456698842012,0.456701332107,-0.0480164218217,-0.0480178535875,0.944713171877,0.944713200415,0.041753254702,0.041753254197,2.10147577265,2.10147577933,2.33484381451,2.33484374055
0.891000000,0.459349304947,0.459351522837,-0.0482741873734,-0.0482763936163,0.944731141823,0.944731180076,0.0417527397866,0.0417527389052,2.10147686161,2.10147686863,2.33480438522,2.33480428914
0.892000000,0.461995730972,0.461997925255,-0.0485433096766,-0.0485455875393,0.944749259176,0.944749297798,0.0417522192935,0.0417522183831,2.10147796096,2.10147796757,2.33476461934,2.33476452242
0.893000000,0.464638057686,0.464640490433,-0.0488238969915,-0.0488255166707,0.944767522612,0.944767552501,0.0417516931828,0.0417516926016,2.10147906964,2.10147907534,2.33472451889,2.33472444205
0.894000000,0.467276222446,0.467279168991,-0.0491160559001,-0.0491162615879,0.944785930816,0.94478594311,0.0417511614171,0.0417511615326,2.10148018662,2.10148019112,2.33468408584,2.3346840497
0.895000000,0.469910162053,0.469912582017,-0.0494198893496,-0.0494215408267,0.944804482495,0.944804512296,0.0417506239628,0.0417506233841,2.10148131094,2.10148131602,2.33464332218,2.33464324585
0.896000000,0.472539813122,0.472541941429,-0.0497354990759,-0.0497379177783,0.944823176361,0.944823214856,0.041750080788,0.0417500798514,2.10148244164,2.10148244669,2.33460222987,2.3346021335
0.897000000,0.47516511193,0.475167213003,-0.050062984612,-0.0500654249404,0.944842011141,0.944842049256,0.0417495318634,0.0417495309347,2.10148357782,2.10148358241,2.33456081087,2.33456071541
0.898000000,0.477785994246,0.477788346252,-0.0504024422983,-0.0504041392278,0.944860985579,0.944861014531,0.0417489771632,0.0417489766129,2.10148471859,2.1014847225,2.33451906709,2.33451899305
0.899000000,0.480402395642,0.480405290735,-0.0507539672352,-0.0507541369744,0.94488009843,0.944880109729,0.0417484166631,0.041748416866,2.10148586313,2.10148586631,2.33447700047,2.33447696788
0.900000000,0.483014251367,0.483016567617,-0.051117652512,-0.051119281648,0.944899348458,0.94489937579,0.0417478503412,0.0417478498722,2.1014870106,2.10148701379,2.33443461291,2.33443454296
0.901000000,0.485621496046,0.485623487607,-0.0514935874527,-0.0514959787562,0.944918734456,0.944918769562,0.0417472781797,0.04174727738,2.10148816027,2.10148816316,2.33439190627,2.33439181814
0.902000000,0.48822406418,0.488226018162,-0.0518818606697,-0.0518842524178,0.944938255219,0.944938289629,0.0417467001615,0.0417466993983,2.10148931136,2.10148931383,2.33434888244,2.33434879598
0.903000000,0.49082189001,0.490824109106,-0.0522825591732,-0.0522841724127,0.944957909555,0.944957935108,0.0417461162715,0.0417461159146,2.10149046317,2.10149046527,2.33430554325,2.33430547783
0.904000000,0.493414907041,0.493417710168,-0.0526957657432,-0.0526958072879,0.944977696299,0.94497770512,0.0417455264991,0.0417455269177,2.10149161502,2.10149161692,2.33426189053,2.33426186502
0.905000000,0.496003048604,0.496005242513,-0.0531215622097,-0.0531231444766,0.944997614297,0.944997638925,0.0417449308345,0.0417449305752,2.10149276627,2.10149276786,2.33421792607,2.33421786321
0.906000000,0.498586247859,0.498588112998,-0.053560029362,-0.0535624381548,0.945017662398,0.945017695145,0.0417443292693,0.0417443286747,2.10149391628,2.10149391754,2.33417365168,2.33417356959
0.907000000,0.501164437276,0.501166288035,-0.054011244209,-0.0540136995013,0.945037839483,0.945037872392,0.041743721799,0.0417437212354,2.10149506447,2.10149506551,2.33412906909,2.33412898673
0.908000000,0.503737548986,0.503739714681,-0.0544752819199,-0.0544769867967,0.945058144445,0.945058169787,0.0417431084209,0.0417431082551,2.10149621028,2.10149621132,2.33408418005,2.33408411602
0.909000000,0.506305514899,0.506308339435,-0.0549522165076,-0.0549523568099,0.945078576184,0.945078586448,0.0417424891335,0.0417424897332,2.10149735316,2.10149735454,2.33403898627,2.33403895885
0.910000000,0.508868266566,0.508870472831,-0.0554421200162,-0.0554438944396,0.945099133619,0.945099159886,0.0417418639374,0.0417418638485,2.1014984926,2.10149849353,2.33399348946,2.3339934237
0.911000000,0.511425734834,0.511427618459,-0.0559450608978,-0.0559477226551,0.945119815694,0.945119850429,0.0417412328367,0.0417412324079,2.10149962814,2.10149962875,2.33394769126,2.33394760517
0.912000000,0.513977850423,0.513979739654,-0.0564611069668,-0.0564638404626,0.94514062136,0.94514065672,0.0417405958359,0.0417405954402,2.10150075929,2.10150075986,2.33390159333,2.33390150583
0.913000000,0.516524543795,0.516526779359,-0.056990324686,-0.0569922960306,0.945161549577,0.945161577898,0.0417399529412,0.0417399529525,2.10150188563,2.10150188651,2.3338551973,2.33385512707
0.914000000,0.519065744612,0.51906868022,-0.0575327766096,-0.0575331366362,0.945182599337,0.945182613108,0.0417393041622,0.0417393049525,2.10150300676,2.10150300838,2.33380850475,2.33380847026
0.915000000,0.52160138233,0.521603636243,-0.0580885243077,-0.0580905309414,0.94520376964,0.945203798245,0.0417386495094,0.0417386496441,2.10150412228,2.10150412329,2.33376151725,2.33376144653
0.916000000,0.524131386214,0.524133259921,-0.0586576284143,-0.0586604981657,0.945225059494,0.945225095444,0.0417379889945,0.0417379888303,2.10150523183,2.10150523252,2.33371423636,2.33371414752
0.917000000,0.526655684847,0.526657516448,-0.0592401463452,-0.0592430317058,0.945246467933,0.945246503488,0.0417373226317,0.0417373225449,2.10150633506,2.10150633582,2.33366666361,2.33366657555
0.918000000,0.52917420637,0.529176350151,-0.0598361335267,-0.0598381760966,0.945267994007,0.945268021647,0.0417366504371,0.0417366507994,2.10150743167,2.10150743296,2.33361880047,2.33361873177
0.919000000,0.531686878723,0.531689705983,-0.0604456444664,-0.0604459757596,0.945289636775,0.945289649215,0.0417359724276,0.0417359736055,2.10150852135,2.1015085237,2.33357064844,2.33357061725
0.920000000,0.534193629471,0.534195674118,-0.0610687319533,-0.0610706848587,0.945311395309,0.945311420902,0.0417352886215,0.0417352891938,2.10150960381,2.10150960552,2.33352220897,2.33352214483
0.921000000,0.5366943854,0.536695982767,-0.0617054454015,-0.0617082335891,0.945333268709,0.94533330053,0.0417345990401,0.0417345993488,2.10151067881,2.1015106803,2.33347348346,2.33347340342
0.922000000,0.539189073147,0.539190603336,-0.0623558336188,-0.0623586123156,0.945355256081,0.945355287065,0.0417339037049,0.0417339041067,2.10151174611,2.10151174791,2.33342447334,2.33342439498
0.923000000,0.541677619103,0.541679484151,-0.063019944292,-0.0630218627015,0.945377356539,0.945377379923,0.0417332026385,0.0417332034818,2.10151280546,2.10151280819,2.33337517997,2.33337512034
0.924000000,0.544159948798,0.544162573326,-0.0636978215776,-0.0636980253542,0.94539956923,0.945399578525,0.0417324958662,0.0417324974894,2.10151385669,2.10151386097,2.33332560471,2.33332558035
0.925000000,0.54663598751,0.54663785653,-0.0643895086175,-0.0643914247153,0.945421893306,0.945421916488,0.0417317834141,0.0417317843975,2.1015148996,2.1015149035,2.33327574887,2.33327568943
0.926000000,0.54910566031,0.549107153582,-0.0650950476875,-0.0650978921688,0.945444327929,0.945444358719,0.0417310653087,0.0417310659748,2.10151593402,2.10151593811,2.33322561378,2.33322553501
0.927000000,0.551568891601,0.55157042027,-0.0658144784327,-0.0658173972652,0.94546687228,0.945466903994,0.0417303415789,0.0417303422705,2.10151695979,2.10151696468,2.33317520071,2.33317511951
0.928000000,0.554025605239,0.554027581194,-0.0665478383911,-0.0665499570811,0.945489525562,0.945489551423,0.0417296122547,0.0417296133156,2.10151797678,2.10151798309,2.3331245109,2.33312444453
0.929000000,0.556475724884,0.556478553935,-0.0672951643774,-0.0672955828499,0.94551228698,0.945512300018,0.0417288773665,0.041728879145,2.10151898487,2.10151899322,2.33307354559,2.33307351185
0.930000000,0.558919173822,0.558921097726,-0.0680564917441,-0.0680585677277,0.945535155756,0.945535180576,0.0417281369461,0.0417281381319,2.10151998394,2.10151999207,2.33302230601,2.33302224145
0.931000000,0.561355874478,0.56135730257,-0.0688318527144,-0.0688348502521,0.945558131134,0.945558162488,0.0417273910271,0.0417273918656,2.1015209739,2.10152098126,2.33297079332,2.33297071083
0.932000000,0.563785749107,0.563787139298,-0.0696212789241,-0.0696243120253,0.945581212365,0.945581243729,0.0417266396435,0.0417266404615,2.10152195467,2.10152196149,2.33291900869,2.33291892504
0.933000000,0.566208719721,0.566210539545,-0.0704248010854,-0.0704269452434,0.945604398709,0.945604423298,0.0417258828297,0.041725883976,2.10152292618,2.10152293305,2.33286695327,2.33286688626
0.934000000,0.568624707426,0.568627421783,-0.071242446786,-0.0712427963805,0.94562768945,0.94562770078,0.0417251206222,0.04172512243,2.10152388837,2.10152389589,2.33281462816,2.33281459492
0.935000000,0.571033633024,0.571035497368,-0.0720742425748,-0.0720762878856,0.945651083882,0.945651107516,0.0417243530581,0.0417243542055,2.10152484119,2.10152484705,2.33276203447,2.33276196897
0.936000000,0.573435417071,0.573436821118,-0.0729202141674,-0.0729231456312,0.945674581305,0.94567461117,0.0417235801748,0.0417235809771,2.10152578462,2.10152578936,2.33270917327,2.33270909125
0.937000000,0.575829979514,0.575831357872,-0.073780385234,-0.0737833348031,0.945698181039,0.945698210834,0.0417228020108,0.0417228027953,2.10152671863,2.10152672284,2.33265604562,2.33265596353
0.938000000,0.578217239612,0.578219045882,-0.0746547772466,-0.074656872608,0.945721882418,0.94572190599,0.041722018606,0.0417220196909,2.1015276432,2.10152764747,2.33260265253,2.33260258659
0.939000000,0.580597116363,0.580599824083,-0.0755434108641,-0.0755437736584,0.945745684786,0.945745696112,0.0417212300004,0.041721231696,2.10152855834,2.10152856327,2.33254899503,2.33254896124
0.940000000,0.582969528712,0.582971296448,-0.076446306475,-0.0764484592109,0.945769587488,0.945769611402,0.0417204362343,0.041720437267,2.10152946404,2.10152946732,2.33249507411,2.33249500736
0.941000000,0.585334394298,0.585335645894,-0.0773634805168,-0.0773665925829,0.945793589901,0.94579362044,0.0417196373498,0.0417196380297,2.10153036033,2.10153036252,2.33244089073,2.3324408066
0.942000000,0.587691630583,0.587692838188,-0.0782949488866,-0.0782981261531,0.945817691407,0.945817722313,0.0417188333892,0.0417188340393,2.10153124722,2.10153124893,2.33238644584,2.33238636082
0.943000000,0.590041154741,0.590042809634,-0.0792407266599,-0.0792430671354,0.945841891395,0.945841916503,0.041718024395,0.0417180253312,2.10153212476,2.10153212657,2.33233174038,2.33233167085
0.944000000,0.592382883768,0.592385495876,-0.0802008282749,-0.0802014225173,0.945866189258,0.945866202503,0.0417172104104,0.041717211941,2.10153299297,2.10153299548,2.33227677526,2.33227673748
0.945000000,0.594716733378,0.59471836832,-0.0811752645509,-0.0811776225201,0.945890584419,0.945890609572,0.0417163914796,0.0417163923979,2.1015338519,2.10153385284,2.33222155137,2.33222148211
0.946000000,0.597042619053,0.597043727008,-0.0821640456541,-0.0821673136776,0.945915076307,0.945915107407,0.0417155676471,0.0417155682602,2.1015347016,2.10153470154,2.33216606958,2.33216598457
0.947000000,0.599360456006,0.599361535042,-0.0831671810417,-0.0831704483291,0.945939664353,0.945939695199,0.0417147389577,0.0417147395821,2.10153554214,2.10153554165,2.33211033075,2.33211024648
0.948000000,0.60167015928,0.601671724905,-0.0841846795887,-0.0841870340952,0.945964347997,0.945964372529,0.0417139054564,0.0417139063988,2.10153637357,2.10153637324,2.33205433573,2.33205426849
0.949000000,0.603971642628,0.603974228841,-0.085216546773,-0.0852170786306,0.945989126703,0.94598913899,0.0417130671892,0.0417130687452,2.10153719597,2.10153719636,2.33199808533,2.33199805118
0.950000000,0.606264819506,0.606266381425,-0.0862627872927,-0.0862650274431,0.946013999941,0.946014023161,0.0417122242024,0.0417122252191,2.10153800942,2.10153800832,2.33194158035,2.33194151684
0.951000000,0.608549603123,0.608550611708,-0.0873234052371,-0.0873265114537,0.946038967185,0.946038995701,0.0417113765424,0.0417113773094,2.10153881398,2.10153881195,2.33188482159,2.3318847438
0.952000000,0.610825906518,0.610826883416,-0.0883984041582,-0.0884014806091,0.946064027916,0.946064055888,0.0417105242556,0.0417105250694,2.10153960975,2.10153960736,2.33182780983,2.33182773352
0.953000000,0.61309364145,0.613095128138,-0.0894877844475,-0.0894899394172,0.946089181635,0.946089203371,0.0417096673894,0.0417096685344,2.10154039681,2.1015403946,2.33177054581,2.3317704865
0.954000000,0.615352719304,0.615355277366,-0.0905915455882,-0.090591891516,0.946114427851,0.9461144378,0.0417088059913,0.0417088077401,2.10154117526,2.10154117376,2.33171303028,2.33171300323
0.955000000,0.617603051242,0.617604529276,-0.091709686544,-0.0917117821015,0.946139766078,0.946139787103,0.0417079401086,0.041707941356,2.10154194519,2.10154194232,2.33165526397,2.33165520689
0.956000000,0.61984454825,0.619845444787,-0.0928422057747,-0.0928452222576,0.946165195833,0.946165222649,0.0417070697892,0.0417070708047,2.10154270669,2.10154270299,2.3315972476,2.33159717485
0.957000000,0.622077120061,0.622077987678,-0.0939890988296,-0.0939921528,0.946190716652,0.946190743717,0.0417061950811,0.0417061961412,2.10154345987,2.10154345589,2.33153898187,2.33153890861
0.958000000,0.624300675949,0.624302087577,-0.0951503602201,-0.0951525687611,0.94621632808,0.94621634994,0.0417053160325,0.0417053174036,2.10154420484,2.10154420111,2.33148046744,2.33148040874
0.959000000,0.626515124986,0.626517673698,-0.0963259840185,-0.0963264638535,0.946242029665,0.946242040942,0.0417044326915,0.0417044346302,2.10154494168,2.10154493871,2.33142170501,2.33142167582
0.960000000,0.628720376062,0.628721800748,-0.0975159638127,-0.0975182583206,0.946267820958,0.946267843874,0.0417035451067,0.0417035465702,2.10154567052,2.10154566638,2.33136269524,2.3313626343
0.961000000,0.630916336848,0.63091716192,-0.0987202905441,-0.0987235579985,0.946293701527,0.946293730776,0.0417026533265,0.0417026545764,2.10154639147,2.10154638668,2.33130343878,2.33130336062
0.962000000,0.633102914472,0.633103719064,-0.099938953985,-0.0999422950656,0.946319670949,0.946319700912,0.0417017573994,0.0417017587049,2.10154710462,2.10154709973,2.33124393625,2.33124385635
0.963000000,0.635280015875,0.635281398172,-0.101171943542,-0.101174456649,0.946345728803,0.946345753897,0.041700857374,0.0417008589959,2.10154781009,2.10154780561,2.33118418828,2.33118412213
0.964000000,0.637447547804,0.637450124845,-0.102419248141,-0.102420029406,0.946371874671,0.946371889351,0.0416999532991,0.0416999554891,2.10154850799,2.10154850441,2.3311241955,2.33112415857
0.965000000,0.639605415847,0.639606803002,-0.10368085434,-0.10368340155,0.94639810815,0.9463981337,0.0416990452231,0.0416990470129,2.10154919844,2.10154919399,2.33106395851,2.33106389132
0.966000000,0.641753524943,0.641754266291,-0.104956747395,-0.104960199828,0.946424428843,0.9464244598,0.0416981331947,0.041698134841,2.10154988154,2.10154987674,2.3310034779,2.33100339547
0.967000000,0.643891779871,0.643892476956,-0.106246912261,-0.106250356953,0.946450836359,0.946450867005,0.0416972172624,0.0416972190267,2.10155055741,2.10155055278,2.33094275426,2.33094267243
0.968000000,0.646020085187,0.646021360135,-0.107551333407,-0.107553861422,0.946477330306,0.946477355029,0.0416962974749,0.0416962996069,2.10155122615,2.10155122219,2.33088178816,2.3308817226
0.969000000,0.648138344375,0.64814084117,-0.108869993217,-0.108870702409,0.946503910306,0.946503923603,0.0416953738805,0.0416953766184,2.10155188789,2.10155188507,2.33082058017,2.33082054636
0.970000000,0.650246460148,0.650247677616,-0.110202872643,-0.110205251804,0.946530575989,0.94653059879,0.0416944465276,0.0416944489558,2.10155254273,2.10155253945,2.33075913086,2.33075906973
0.971000000,0.652344335093,0.652344851429,-0.111549952401,-0.111553165775,0.946557326987,0.946557354221,0.0416935154643,0.0416935178136,2.10155319078,2.10155318753,2.33069744076,2.33069736651
0.972000000,0.654431871532,0.654432328569,-0.112911212687,-0.112914381178,0.946584162932,0.946584189394,0.041692580739,0.0416925832395,2.10155383214,2.10155382943,2.33063551043,2.33063543778
0.973000000,0.656508970821,0.656510035647,-0.114286631919,-0.114288888938,0.946611083469,0.946611104137,0.0416916423996,0.0416916452669,2.10155446694,2.10155446525,2.3305733404,2.33057328371
0.974000000,0.658575533422,0.658577899074,-0.115676186947,-0.11567667922,0.946638088249,0.946638098276,0.0416907004938,0.0416907039289,2.10155509526,2.10155509506,2.33051093121,2.33051090447
0.975000000,0.660631459696,0.660632528878,-0.117079854443,-0.117082097761,0.946665176924,0.946665197544,0.0416897550693,0.0416897581864,2.10155571722,2.10155571706,2.33044828337,2.33044822641
0.976000000,0.662676649693,0.662677043402,-0.118497610517,-0.118500801816,0.946692349149,0.946692375799,0.0416888061736,0.0416888091728,2.10155633293,2.10155633329,2.33038539741,2.3303853233
0.977000000,0.664711002622,0.664711399026,-0.119929429812,-0.119932704826,0.946719604587,0.946719632288,0.0416878538542,0.0416878569443,2.10155694248,2.10155694385,2.33032227384,2.33032219686
0.978000000,0.666734416667,0.666735506461,-0.121375285265,-0.121377766603,0.946746942908,0.946746966476,0.041686898158,0.0416869015473,2.10155754596,2.10155754883,2.33025891316,2.33025884813
0.979000000,0.668746789937,0.668749271717,-0.122835149699,-0.122835938744,0.946774363781,0.946774377717,0.0416859391317,0.0416859430321,2.10155814349,2.1015581483,2.33019531588,2.33019527844
0.980000000,0.67074802017,0.670749076396,-0.124308995306,-0.124311396865,0.94680186688,0.946801889481,0.0416849768222,0.0416849804938,2.10155873515,2.10155874059,2.3301314825,2.33013141911
0.981000000,0.672738004418,0.672740077709,-0.12579679314,-0.12579787732,0.946829451884,0.946829466494,0.0416840112759,0.0416840152628,2.10155932103,2.10155932702,2.33006741352,2.33006737223
0.982000000,0.674716638558,0.674718719443,-0.127298512409,-0.127299583813,0.946857118477,0.94685713297,0.0416830425387,0.041683046487,2.10155990123,2.10155990679,2.33000310942,2.33000306788
0.983000000,0.676683818431,0.676685899434,-0.128814122265,-0.128815188164,0.946884866345,0.946884880756,0.0416820706566,0.0416820745612,2.10156047583,2.10156048095,2.3299385707,2.32993852899
0.984000000,0.678639439421,0.678641512531,-0.130343591148,-0.130344669871,0.946912695175,0.946912709664,0.0416810956752,0.0416810995274,2.10156104492,2.10156104959,2.32987379785,2.3298737557
0.985000000,0.680583396415,0.680585463248,-0.131886886695,-0.131887984338,0.946940604661,0.946940619313,0.04168011764,0.0416801214338,2.10156160858,2.1015616128,2.32980879134,2.32980874867
0.986000000,0.682515582964,0.682517649098,-0.13344397456,-0.133445079078,0.946968594498,0.94696860921,0.0416791365957,0.0416791403327,2.10156216689,2.10156217066,2.32974355166,2.32974350882
0.987000000,0.68443589262,0.684437960474,-0.135014820395,-0.135015907989,0.946996664384,0.946996678908,0.0416781525871,0.0416781562745,2.10156271993,2.10156272325,2.32967807929,2.32967803699
0.988000000,0.686344218381,0.686346286637,-0.136599389046,-0.136600438428,0.947024814021,0.947024828115,0.0416771656586,0.0416771693035,2.10156326776,2.10156327066,2.32961237471,2.3296123336
0.989000000,0.688240452962,0.688242520607,-0.138197644896,-0.138198649568,0.947053043109,0.947053056705,0.0416761758546,0.0416761794584,2.10156381047,2.10156381293,2.32954643839,2.32954639869
0.990000000,0.690124487571,0.690126559593,-0.13980955021,-0.139810523906,0.947081351356,0.947081364628,0.0416751832186,0.0416751867747,2.10156434811,2.10156435015,2.32948027083,2.32948023213
0.991000000,0.69199621342,0.69199830109,-0.141435067252,-0.141436036253,0.947109738469,0.947109751765,0.041674187794,0.0416741912901,2.10156488075,2.10156488238,2.3294138725,2.32941383398
0.992000000,0.693855521203,0.693857637374,-0.143074157575,-0.143075145481,0.947138204158,0.947138217809,0.041673189624,0.0416731930488,2.10156540845,2.10156540969,2.32934724387,2.32934720474
0.993000000,0.695702301346,0.695704452089,-0.144726782312,-0.144727793476,0.947166748131,0.947166762228,0.0416721887517,0.0416721921022,2.10156593127,2.10156593214,2.32928038543,2.32928034548
0.994000000,0.697536442741,0.697538622005,-0.146392900545,-0.14639391331,0.947195370104,0.947195384369,0.0416711852191,0.0416711885058,2.10156644926,2.10156644979,2.32921329765,2.32921325757
0.995000000,0.69935783424,0.699360024643,-0.148072471302,-0.148073444795,0.947224069791,0.947224083671,0.0416701790683,0.0416701823116,2.10156696247,2.10156696268,2.32914598103,2.32914594215
0.996000000,0.70116636423,0.701168548982,-0.149765452996,-0.149766351118,0.947252846906,0.947252859905,0.0416691703412,0.0416691735594,2.10156747096,2.10156747088,2.32907843604,2.32907839955
0.997000000,0.702961920824,0.702964102372,-0.151471803645,-0.151472625834,0.947281701166,0.947281713297,0.0416681590793,0.0416681622728,2.10156797475,2.10156797443,2.32901066317,2.32901062904
0.998000000,0.704744390597,0.704746602988,-0.153191479299,-0.153192276079,0.94731063229,0.947310644293,0.0416671453234,0.0416671484672,2.10156847391,2.1015684734,2.32894266291,2.32894262935
0.999000000,0.706513660035,0.706515944517,-0.154924435897,-0.154925266169,0.947339639995,0.947339652745,0.041666129114,0.0416661321788,2.10156896845,2.10156896784,2.32887443575,2.32887440066
1.000000000,0.708269615368,0.708271918944,-0.15667062905,-0.15667140646,0.947368724001,0.947368736247,0.0416651104918,0.0416651135221,2.10156945841,2.10156945775,2.32880598219,2.32880594854
}\curvedata
\pgfplotstableread[col sep=comma]{%
interval,t_end,h,delta_interval,omega_interval,vmag_interval,slip_interval,pf_interval,imag_interval
1,0.01,0.01,0,0,0.943,0.04,2.04,2.28774253205
2,0.0433333333333,0.0333333333333,0,-1.11022302463e-13,0.943,0.04,2.04,2.28774253205
3,0.0933333333333,0.05,0,-4.4408920985e-13,0.943,0.04,2.04,2.28774253205
4,0.143333333333,0.05,0,-7.77156117238e-13,0.943,0.04,2.04,2.28774253205
5,0.193333333333,0.05,1.65388843436e-13,-7.77156117238e-13,0.943,0.04,2.04,2.28774253206
6,0.2,0.00666666666667,1.59027734073e-13,-7.77156117238e-13,0.943,0.04,2.04,2.28774253205
7,0.222222222222,0.0222222222222,-0.000188762660502,-0.00030865854117,0.94296318433,0.0407829875024,2.04004570065,2.28797653476
8,0.237037037037,0.0148148148148,-0.00104149713551,-0.000727775993137,0.942913570113,0.0411489336195,2.04010757053,2.28830122504
9,0.259259259259,0.0222222222222,-0.00439676116476,-0.00111306145067,0.94283452222,0.0415096159774,2.040138286,2.28875887281
10,0.292592592593,0.0333333333333,-0.0150523912403,-0.000706279418283,0.942729462201,0.0417942708644,2.0400407454,2.2892735261
11,0.342592592593,0.05,-0.0429435154518,0.00130725798497,0.942610866487,0.041963895783,2.03989916028,2.28978004291
12,0.392592592593,0.05,-0.0813426126152,0.00285487661689,0.94252641785,0.0420258831942,2.03991735861,2.290111995
13,0.442592592593,0.05,-0.125737974304,0.00209775992799,0.942465471428,0.0420536822872,2.0399877315,2.29032848623
14,0.492592592593,0.05,-0.17254866982,-0.00186722915385,0.94242556564,0.0420681514238,2.04002695157,2.29044972324
15,0.542592592593,0.05,-0.21933765117,-0.00924255566714,0.942409314035,0.0420760044936,2.04003194213,2.29048448653
16,0.55,0.00740740740741,-0.226171870528,-0.0106154650517,0.942409195431,0.0420767918504,2.04003119141,2.29048242769
17,0.564814814815,0.0148148148148,-0.236999663035,-0.0131769070533,0.942769532839,0.041840627385,2.07383665981,2.3567256356
18,0.574691358025,0.00987654320988,-0.239556754653,-0.0150648245424,0.94285940613,0.0417807708218,2.09104132185,2.35361644851
19,0.58950617284,0.0148148148148,-0.236602012089,-0.0182931705986,0.942957383411,0.0417785387879,2.10356655232,2.34912168978
20,0.611728395062,0.0222222222222,-0.219508637027,-0.0237340375838,0.943076190608,0.0418077551945,2.10903685984,2.34532594027
21,0.645061728395,0.0333333333333,-0.172214923172,-0.0324051544666,0.943235438539,0.0418386422633,2.10857098268,2.34213680871
22,0.695061728395,0.05,-0.0685342267543,-0.0442252765362,0.943450083868,0.0418413061455,2.10512418831,2.33936847419
23,0.745061728395,0.05,0.0578782646937,-0.0515784008007,0.94363758864,0.0418166361355,2.10275444254,2.33785558911
24,0.795061728395,0.05,0.194857522244,-0.0520019283816,0.94380419302,0.0417907886177,2.10173339508,2.33700478286
25,0.8,0.00493827160494,0.208632795738,-0.0516112613704,0.943819690375,0.0417886313035,2.10168255887,2.33694047395
26,0.80987654321,0.00987654320988,0.236227849937,-0.0505921361186,0.943852164872,0.0417845628952,2.10159938859,2.33681728024
27,0.824691358025,0.0148148148148,0.277643445309,-0.0486211489474,0.943920838688,0.0417789773625,2.10151552726,2.33661705123
28,0.846913580247,0.0222222222222,0.339507800757,-0.0457672923676,0.944098564896,0.0417711779533,2.10146089616,2.33619603498
29,0.88024691358,0.0333333333333,0.430652580036,-0.0460741998914,0.944545927373,0.0417579944186,2.10146594702,2.33521027555
30,0.93024691358,0.05,0.559524194958,-0.0682469590555,0.945540830096,0.0417279551657,2.1015202387,2.3330095821
31,0.98024691358,0.05,0.671242855329,-0.124675707512,0.946808682509,0.04168474283,2.10155888683,2.33011565138
32,1,0.0197530864198,0.708271918944,-0.15667140646,0.947368736247,0.0416651135221,2.10156945775,2.32880594854
}\intervaldata

\begin{tikzpicture}
\begin{groupplot}[
  group style={group size=2 by 3, horizontal sep=1.90cm, vertical sep=1.52cm},
  width=7cm,
  height=3cm,
  xmin=0, xmax=1,
  xlabel={Time [s]},
  xtick={0,0.2,0.4,0.6,0.8,1.0},
  scaled y ticks=false,
  grid=major,
  major grid style={draw=gray!18},
  tick label style={font=\footnotesize,/pgf/number format/fixed,/pgf/number format/precision=3},
  label style={font=\footnotesize},
  title style={font=\footnotesize, yshift=-1pt},
  ylabel style={at={(axis description cs:-0.16,0.5)}, anchor=south},
  every axis plot/.append style={line join=round},
  clip mode=individual,
  extra x ticks={0.20,0.55,0.80},
  extra x tick labels={,,},
  extra x tick style={grid=major, major grid style={draw=eventgray, densely dashed, line width=0.6pt}},
]

\nextgroupplot[title={Rotor angle (gen @ bus 55)}, ylabel={$\Delta\delta$ [deg]}, ymin=-0.315442, ymax=0.7841026]
\addplot[black, line width=1.1pt] table[x=t, y=delta_ref] {\curvedata};
\addplot[raptorred, dashed, line width=1.15pt] table[x=t, y=delta_rap] {\curvedata};
\addplot[only marks, mark=*, mark size=1.35pt, raptorred] table[x=t_end, y=delta_interval] {\intervaldata};

\nextgroupplot[title={Rotor speed (gen @ bus 89)}, ylabel={$10^{-3}(\omega-1)$ [pu]}, ymin=-0.1694397, ymax=0.01569997]
\addplot[black, line width=1.1pt] table[x=t, y=omega_ref] {\curvedata};
\addplot[raptorred, dashed, line width=1.15pt] table[x=t, y=omega_rap] {\curvedata};
\addplot[only marks, mark=*, mark size=1.35pt, raptorred] table[x=t_end, y=omega_interval] {\intervaldata};

\nextgroupplot[title={Voltage magnitude (bus 76)}, ylabel={$|V|$ [pu]}, ymin=0.9419132, ymax=0.9478647]
\addplot[black, line width=1.1pt] table[x=t, y=vmag_ref] {\curvedata};
\addplot[raptorred, dashed, line width=1.15pt] table[x=t, y=vmag_rap] {\curvedata};
\addplot[only marks, mark=*, mark size=1.35pt, raptorred] table[x=t_end, y=vmag_interval] {\intervaldata};

\nextgroupplot[title={Motor slip (bus 59)}, ylabel={Slip [pu]}, ymin=0.03983386, ymax=0.04224294]
\addplot[black, line width=1.1pt] table[x=t, y=slip_ref] {\curvedata};
\addplot[raptorred, dashed, line width=1.15pt] table[x=t, y=slip_rap] {\curvedata};
\addplot[only marks, mark=*, mark size=1.35pt, raptorred] table[x=t_end, y=slip_interval] {\intervaldata};

\nextgroupplot[title={IBR active power (bus 49)}, ylabel={$P_f$ [pu]}, ymin=2.034329, ymax=2.114978]
\addplot[black, line width=1.1pt] table[x=t, y=pf_ref] {\curvedata};
\addplot[raptorred, dashed, line width=1.15pt] table[x=t, y=pf_rap] {\curvedata};
\addplot[only marks, mark=*, mark size=1.35pt, raptorred] table[x=t_end, y=pf_interval] {\intervaldata};

\nextgroupplot[title={IBR current magnitude (bus 49)}, ylabel={$|I|$ [pu]}, ymin=2.282221, ymax=2.36228]
\addplot[black, line width=1.1pt] table[x=t, y=imag_ref] {\curvedata};
\addplot[raptorred, dashed, line width=1.15pt] table[x=t, y=imag_rap] {\curvedata};
\addplot[only marks, mark=*, mark size=1.35pt, raptorred] table[x=t_end, y=imag_interval] {\intervaldata};

\end{groupplot}

\node[draw=eventgray, fill=white, rounded corners=1pt, inner sep=3pt, font=\scriptsize, anchor=south]
  at ($(group c1r1.north)!0.5!(group c2r1.north)+(0,0.76cm)$)
  {Events: $t{=}0.20$ s motor torque $+5\%$, $t{=}0.55$ s IBR $P_{\mathrm{ref}}$ $+3\%$, $t{=}0.80$ s AVR $V_{\mathrm{ref}}$ $+0.01$ pu};

\node[draw=black, fill=white, rounded corners=1pt, inner sep=4pt, anchor=north]
  at ($(group c1r3.south)!0.5!(group c2r3.south)+(0,-0.94cm)$) {%
  \begin{tikzpicture}[baseline=-0.6ex]
    \draw[black, line width=1.1pt] (0,0) -- (0.8,0) node[right, black, font=\footnotesize] {Reference};
    \draw[raptorred, dashed, line width=1.15pt] (3.1,0) -- (3.9,0) node[right, black, font=\footnotesize] {RAPTOR trajectory};
    \fill[raptorred] (7.1,0) circle (1.35pt) node[right, black, font=\footnotesize] {Interval endpoints};
  \end{tikzpicture}%
  };
\end{tikzpicture}


\begin{tikzpicture}
\begin{axis}[
    width=11.5cm,
    height=7cm,
    xmode=log,
    ymode=log,
    xlabel={Runtime [s]},
    ylabel={Mean differential-state NRMSE},
    xmin=0.9,
    xmax=16,
    ymin=1.3e-5,
    ymax=4.0e-5,
    xtick={1,2,5,10},
    xticklabels={1,2,5,10},
    ytick={1.5e-5,2e-5,3e-5,4e-5},
    scaled y ticks=false,
    yticklabel style={
        /pgf/number format/.cd,
        sci,
        sci zerofill,
        precision=1
    },
    grid=both,
    title={EMT PLL inverter: runtime--accuracy trade-off}
]

\addplot[
    only marks,
    color=pDarkBlue,
    mark=*,
    mark size=3pt,
    mark options={fill=pDarkBlue,draw=pDarkBlue}
] coordinates {
    (1.10358,2.12063e-5)
};
\node[
    anchor=south west,
    text=pDarkBlue
] at (axis cs:1.10358,2.12063e-5) {RBF};

\addplot[
    only marks,
    color=pPurple,
    mark=square*,
    mark size=3pt,
    mark options={fill=pPurple,draw=pPurple}
] coordinates {
    (1.62510,1.74953e-5)
};
\node[
    anchor=north west,
    text=pPurple
] at (axis cs:1.62510,1.74953e-5) {CGL};

\addplot[
    only marks,
    color=pGreen,
    mark=triangle*,
    mark size=3.5pt,
    mark options={fill=pGreen,draw=pGreen}
] coordinates {
    (1.85356,2.54449e-5)
};
\node[
    anchor=south west,
    text=pGreen
] at (axis cs:1.85356,2.54449e-5) {LGL};

\addplot[
    only marks,
    color=pRed,
    mark=diamond*,
    mark size=3pt,
    mark options={fill=pRed,draw=pRed}
] coordinates {
    (1.32835,2.43477e-5)
};
\node[
    anchor=south east,
    text=pRed
] at (axis cs:1.32835,2.43477e-5) {Radau};

\addplot[
    only marks,
    color=pBrown,
    mark=pentagon*,
    mark size=3pt,
    mark options={fill=pBrown,draw=pBrown}
] coordinates {
    (2.33065,2.19401e-5)
};
\node[
    anchor=north west,
    text=pBrown
] at (axis cs:2.33065,2.19401e-5) {BDF};

\addplot[
    only marks,
    color=pGrey,
    mark=x,
    very thick,
    mark size=4pt
] coordinates {
    (3.57659,3.3e-5)
};
\node[
    anchor=south west,
    text=pGrey
] at (axis cs:3.57659,3.3e-5) {Trap.};


\end{axis}
\end{tikzpicture}


\begin{tikzpicture}

\begin{axis}[
    width=11.5cm,
    height=7cm,
    ybar,
    bar width=10pt,
    bar shift=0pt,
    ymode=log,
    ymin=0.7,
    ymax=250,
    ylabel={Runtime [ms]},
    xlabel={},
    symbolic x coords={RBF,CGL,LGL,Radau,BDF,Trap.},
    xtick={RBF,CGL,LGL,Radau,BDF,Trap.},
    enlarge x limits=0.12,
    x tick label style={
        rotate=30,
        anchor=east
    },
    ytick={1,10,100},
    nodes near coords,
    point meta=explicit symbolic,
    nodes near coords style={
        font=\scriptsize
    },
    grid=major,
    title={Cascaded line: target-matched runtime}
]

\addplot[
    fill=pDarkBlue,
    draw=pDarkBlue
] coordinates {
    (RBF,1.1177) [1.1177]
};

\addplot[
    fill=pRed,
    draw=pRed
] coordinates {
    (CGL,1.6315) [1.6315]
};

\addplot[
    fill=pPurple,
    draw=pPurple
] coordinates {
    (LGL,1.6419) [1.6419]
};

\addplot[
    fill=pGreen,
    draw=pGreen
] coordinates {
    (Radau,28.5512) [28.5512]
};

\addplot[
    fill=pBrown,
    draw=pBrown
] coordinates {
    (BDF,56.7247) [56.7247]
};

\addplot[
    fill=pGrey,
    draw=pGrey
] coordinates {
    (Trap.,181.2914) [181.2914]
};

\end{axis}

\end{tikzpicture}


\begin{tikzpicture}
\begin{axis}[
    width=11.5cm,
    height=7cm,
    xlabel={Runtime [ms]},
    ylabel={Scaled RMSE},
    xmin=55,
    xmax=78,
    ymin=3.0e-5,
    ymax=6.8e-5,
    xtick={60,65,70,75},
    ytick={3e-5,4e-5,5e-5,6e-5},
    scaled y ticks=false,
    yticklabel style={
        /pgf/number format/.cd,
        sci,
        sci zerofill,
        precision=1
    },
    grid=major,
    title={RMS inverter: equal-size representation comparison}
]

\addplot[
    only marks,
    color=pDarkBlue,
    mark=*,
    mark size=3pt,
    mark options={fill=pDarkBlue,draw=pDarkBlue}
] coordinates {
    (58.6153,3.523211e-5)
};
\node[
    anchor=south west,
    text=pDarkBlue
] at (axis cs:58.6153,3.523211e-5) {RBF};

\addplot[
    only marks,
    color=pRed,
    mark=square*,
    mark size=3pt,
    mark options={fill=pRed,draw=pRed}
] coordinates {
    (74.6934,4.100549e-5)
};
\node[
    anchor=south east,
    text=pRed
] at (axis cs:74.6934,4.100549e-5) {CGL};

\addplot[
    only marks,
    color=pPurple,
    mark=triangle*,
    mark size=3.5pt,
    mark options={fill=pPurple,draw=pPurple}
] coordinates {
    (71.7254,6.293404e-5)
};
\node[
    anchor=south east,
    text=pPurple
] at (axis cs:71.7254,6.293404e-5) {LGL};

\end{axis}
\end{tikzpicture}


\appendix


This appendix provides the model definitions, numerical implementations, benchmark construction details, and reproducibility information supporting the experiments in the main paper. Section~\ref{app:emt_trapezoidal} defines the fixed-step fully coupled and PSCAD/EMTDC-style delayed-control trapezoidal implementations used in the EMT comparisons, while Section~\ref{app:rms_inverter_results} presents the standalone RMS inverter results and weak-grid sensitivity analysis. The cascaded M-section line, EMT grid-following inverter, and synchronous-machine--infinite-bus EMT benchmark models are given in Sections~\ref{app:msection_model}, \ref{app:emt_gfl_model}, and \ref{app:emt_sm_model}, respectively. Section~\ref{app:ieee_rms_models} describes the construction, initialization, dynamic-device models, algebraic coupling, disturbances, and additional solver-effort results for the IEEE 9-, 14-, 39-, 57-, and 118-bus RMS benchmarks. Finally, Section~\ref{app:reproducibility} collects the benchmark-specific physical parameters, controller gains, initial conditions, disturbance settings, and internal RAPTOR numerical parameters required to reproduce the reported numerical studies.

\subsection{Fixed-Step EMT Trapezoidal Implementations}
\label{app:emt_trapezoidal}

For the EMT benchmarks, we implement two fixed-step implicit trapezoidal
solvers. The first is a fully coupled solver used to obtain the high-accuracy
reference trajectory. The second is a lower-cost delayed-control implementation
used for the grid-following inverter and is motivated by the sequential
calculation used in EMTDC/PSCAD. Both implementations solve the same physical
model; they differ in how the controller and electromagnetic equations are
coupled within one timestep.

\paragraph{Fully coupled trapezoidal solver.}
Consider the DAE
\begin{align}
\dot{\mathbf{x}} &= \mathbf{f}
(t,\mathbf{x},\mathbf{y}),\\
\mathbf{0} &= \mathbf{g}
(t,\mathbf{x},\mathbf{y}),
\end{align}
where $\mathbf{x}$ contains the differential states, including the controller
and electromagnetic states, and $\mathbf{y}$ contains the algebraic network
variables. For a fixed timestep $\Delta t$, the trapezoidal rule advances the
differential states according to
\begin{equation}
\mathbf{x}_{n+1}
=
\mathbf{x}_{n}
+
\frac{\Delta t}{2}
\left[
\mathbf{f}(t_n,\mathbf{x}_n,\mathbf{y}_n)
+
\mathbf{f}(t_{n+1},\mathbf{x}_{n+1},\mathbf{y}_{n+1})
\right],
\end{equation}
while the algebraic equations must simultaneously satisfy
\begin{equation}
\mathbf{g}
(t_{n+1},\mathbf{x}_{n+1},\mathbf{y}_{n+1})
=
\mathbf{0}.
\end{equation}

The new differential and algebraic states are therefore unknown together.
At every timestep, we solve these equations with a Newton iteration until a
consistent solution at $t_{n+1}$ is obtained. In the grid-following inverter,
this means that the new controller states, controller voltage commands,
electromagnetic states, and network variables all use quantities from the same
new timestep. Consequently, a change in the controller can immediately affect
the electromagnetic solution within that timestep.

This fully coupled implementation is used as the waveform reference. For the
grid-following inverter, a fixed timestep of
$\Delta t=10^{-6}$~s is used. For the synchronous-machine EMT benchmark,
calculations with $\Delta t=10^{-6}$~s and $\Delta t=10^{-7}$~s are used to
verify convergence of the reference solution.

\paragraph{PSCAD/EMTDC-style delayed-control solver.}
For the grid-following inverter, we additionally implement a sequential
trapezoidal calculation with a fixed timestep of
$\Delta t=10^{-5}$~s. The main difference is the treatment of the controller
output. Let $\mathbf{u}_{c,n}$ denote the converter voltage command available
from the controller at time $t_n$. During the electromagnetic calculation from
$t_n$ to $t_{n+1}$, this value is kept as the controller input to the
electromagnetic subsystem. The electromagnetic and algebraic equations are
therefore solved as
\begin{align}
\mathbf{x}^{\mathrm{em}}_{n+1}
&=
\mathbf{x}^{\mathrm{em}}_n
+
\frac{\Delta t}{2}
\left[
\mathbf{f}_{\mathrm{em}}
(t_n,\mathbf{x}^{\mathrm{em}}_n,\mathbf{y}_n,\mathbf{u}_{c,n})
\right.
\nonumber\\
&\hspace{2.6cm}\left.
+
\mathbf{f}_{\mathrm{em}}
(t_{n+1},\mathbf{x}^{\mathrm{em}}_{n+1},
\mathbf{y}_{n+1},\mathbf{u}_{c,n})
\right],
\\
\mathbf{0}
&=
\mathbf{g}
(t_{n+1},\mathbf{x}^{\mathrm{em}}_{n+1},
\mathbf{y}_{n+1},\mathbf{u}_{c,n}).
\end{align}

After the electromagnetic and network quantities at $t_{n+1}$ have been
obtained, these quantities are used by the controller to compute its updated
states and the new controller output $\mathbf{u}_{c,n+1}$. This new controller
output is then applied during the following electromagnetic timestep.

The calculation can therefore be summarized as
\begin{equation}
\mathbf{u}_{c,n}
\;\longrightarrow\;
\left(
\mathbf{x}^{\mathrm{em}}_{n+1},
\mathbf{y}_{n+1}
\right)
\;\longrightarrow\;
\mathbf{u}_{c,n+1}.
\end{equation}

This sequential calculation avoids solving the controller and electromagnetic
equations together in one nonlinear system and is therefore less
computationally expensive. However, the electromagnetic equations at
$t_{n+1}$ use the controller output computed at $t_n$. The controller action
therefore enters the electromagnetic system one timestep later, introducing a
delay of $\Delta t$ at the controller--electromagnetic interface.

This implementation is referred to as
\emph{PSCAD/EMTDC-style} because fixed-step trapezoidal integration and
sequential component evaluation are characteristic of EMTDC calculations.
It is a custom implementation used for the numerical comparison and is not a
runtime measurement of the PSCAD executable itself. The comparison therefore
illustrates the computational effect of a representative sequential
controller--network calculation rather than claiming that every PSCAD model
necessarily introduces such a delay.

\subsection{RMS Single-Inverter}

\subsection{Additional RMS Inverter Results}
\label{app:rms_inverter_results}

\subsubsection{Single-Inverter Benchmark}

\begin{figure}[h]
\begin{minipage}[t]{\columnwidth}
    \centering
    \begin{tikzpicture}

\begin{axis}[
    xlabel={Average overall MAE},
    ylabel={Average runtime (s)},
    ylabel style={xshift=-10pt},
    xmode=log,
    ymode=log,
    grid=major,
    xmin=3e-8, xmax=3e-3,
    ymin=5e-3, ymax=1,
    clip=false,
    legend columns=3, 
    legend style={
        at={(1,1.05)}, 
        anchor=south east,
        nodes={scale=1, anchor=west},
        draw=none,
        column sep=0.2cm
    }
]

\tikzset{
  baselinePoint/.style={only marks, mark size=2.0pt, opacity=0.6},
  hiPoint/.style={only marks, mark size=3.0pt, opacity=1, line width=0.55pt},
  paretoLine/.style={line width=1.1pt, opacity=0.9}
}

\addplot[hiPoint, mark=square*, color=sRed] coordinates {(4.09318e-8, 0.0952671)};
\addlegendentry{RAPTOR--6}
\addplot[hiPoint, mark=square, color=sRed, mark options={fill=white}, forget plot] coordinates {(1.55793e-6, 0.0146212)};

\addplot[hiPoint, mark=triangle*, color=sPurple] coordinates {(5.5198e-8, 0.132425)};
\addlegendentry{RAPTOR--8}
\addplot[hiPoint, mark=triangle, color=sPurple, mark options={fill=white}, forget plot] coordinates {(7.5074e-8, 0.0370395)};

\addplot[baselinePoint, mark=diamond*, color=black] coordinates {(1.63365e-7, 0.0649912)};
\addlegendentry{Radau}
\addplot[baselinePoint, mark=diamond, color=black, forget plot] coordinates {(1.06973e-5, 0.026045)};

\addplot[baselinePoint, mark=*, color=gray] coordinates {(2.30982e-6, 0.0514887)};
\addlegendentry{BDF}
\addplot[baselinePoint, mark=o, color=gray, forget plot] coordinates {(1.62557e-4, 0.0280677)};

\addplot[baselinePoint, mark=*, color=sBlue] coordinates {(6.2725e-6, 0.0112367)};
\addlegendentry{Kvaerno 5}
\addplot[baselinePoint, mark=o, color=sBlue, forget plot] coordinates {(1.88664e-4, 0.00726472)};

\addplot[baselinePoint, mark=*, color=cyan] coordinates {(1.69302e-6, 0.0180287)};
\addlegendentry{Kvaerno 4}
\addplot[baselinePoint, mark=o, color=cyan, forget plot] coordinates {(9.29715e-5, 0.00903229)};

\addplot[baselinePoint, mark=*, color=purple] coordinates {(1.47116e-6, 0.0328946)};
\addlegendentry{LSODA}
\addplot[baselinePoint, mark=o, color=purple, forget plot] coordinates {(8.53015e-5, 0.0193986)};

\addplot[baselinePoint, mark=*, color=orange] coordinates {(1.84484e-5, 0.350575)};
\addlegendentry{DOP853}
\addplot[baselinePoint, mark=o, color=orange, forget plot] coordinates {(1.87790e-3, 0.323532)};

\addplot[baselinePoint, mark=*, color=black!30!sRed] coordinates {(1.92166e-4, 0.052807)};
\addlegendentry{ImpEuler}
\addplot[baselinePoint, mark=o, color=black!30!sRed, forget plot] coordinates {(1.70459e-3, 0.00805579)};

\addplot[paretoLine, color=sRed, forget plot] coordinates {(4.09318e-8, 0.0952671) (1.55793e-6, 0.0146212)};
\addplot[paretoLine, color=sPurple, forget plot] coordinates {(5.5198e-8, 0.132425) (7.5074e-8, 0.0370395)};

\node[anchor=south west, draw=none, font=\footnotesize] at(axis description cs:0,1.17)
{%
\begin{tabular}{@{}c@{\hspace{4pt}}l@{}}
$\bullet$ & Best MAE\\
$\circ$   & Best Time
\end{tabular}%
};

\end{axis}

\end{tikzpicture}
    \captionof{figure}{Speed--accuracy trade-off for the single-inverter RMS benchmark. Note that the time is plotted on a logarithmic scale. }
    \label{fig:inv_speed_accuracy}
    
    \vspace{10pt}
    
    \begin{tikzpicture}

\begin{axis}[
    title={},
    xlabel={Average number of advances},
    ylabel={Average runtime (s)},
    ylabel style={xshift=-10pt},
    ymode=log,
    grid=both,
    xmin=0, xmax=2500,
    xtick distance={400},
    ymin=7e-3, ymax=0.7,
    clip=false,
    legend columns=3, 
    legend style={
        at={(axis description cs:1,1.05)}, 
        anchor=south east,
        nodes={scale=1, anchor=west},
        draw=none,
        column sep=0.1cm
    }
]

\tikzset{
  paretoLine/.style={line width=1.1pt, opacity=0.8}
}

\node[anchor=south west, draw=none, font=\footnotesize] at (axis description cs:-0,1.17)
{%
\begin{tabular}{@{}c@{\hspace{4pt}}l@{}}
$\bullet$ & Best MAE\\
$\circ$   & Best Time
\end{tabular}%
};


\addplot[only marks, mark=square*, color=red, mark size=3pt] coordinates {(458.7, 0.0952671)};
\addlegendentry{RAPTOR--6}
\addplot[only marks, mark=square, color=red, mark size=3pt, forget plot] coordinates {(38.0, 0.0146212)};
\addplot[paretoLine, color=red, forget plot] coordinates {(458.7, 0.0952671) (38.0, 0.0146212)};

\addplot[only marks, mark=triangle*, color=magenta, mark size=3pt] coordinates {(97.3, 0.1324248)};
\addlegendentry{RAPTOR--8}
\addplot[only marks, mark=triangle, color=magenta, mark size=3pt, forget plot] coordinates {(83.2, 0.0370395)};
\addplot[paretoLine, color=magenta, forget plot] coordinates {(97.3, 0.1324248) (83.2, 0.0370395)};

\addplot[only marks, mark=*, color=black, mark size=2pt] coordinates {(1999.0, 0.0716698)};
\addlegendentry{Radau}
\addplot[only marks, mark=o, color=black, mark size=2pt, forget plot] coordinates {(1999.0, 0.0268074)};

\addplot[only marks, mark=*, color=gray, mark size=2pt] coordinates {(1999.0, 0.0560258)};
\addlegendentry{BDF}
\addplot[only marks, mark=o, color=gray, mark size=2pt, forget plot] coordinates {(1999.0, 0.0290554)};

\addplot[only marks, mark=*, color=blue, mark size=2pt] coordinates {(47.7, 0.0124818)};
\addlegendentry{Kvaerno5}
\addplot[only marks, mark=o, color=blue, mark size=2pt, forget plot] coordinates {(22.1, 0.0075130)};

\addplot[only marks, mark=*, color=cyan, mark size=2pt] coordinates {(146.6, 0.0198416)};
\addlegendentry{Kvaerno4}
\addplot[only marks, mark=o, color=cyan, mark size=2pt, forget plot] coordinates {(46.9, 0.0091891)};


\addplot[only marks, mark=*, color=purple, mark size=2pt] coordinates {(1999.0, 0.0392848)};
\addlegendentry{LSODA}
\addplot[only marks, mark=o, color=purple, mark size=2pt, forget plot] coordinates {(1999.0, 0.0199716)};

\addplot[only marks, mark=*, color=orange, mark size=2pt] coordinates {(1999.0, 0.3960050)};
\addlegendentry{DOP853}
\addplot[only marks, mark=o, color=orange, mark size=2pt, forget plot] coordinates {(1999.0, 0.3319290)};




\addplot[only marks, mark=*, color=black!30!red, mark size=2pt] coordinates {(2326.7, 0.0492495)};
\addlegendentry{ImpEuler}
\addplot[only marks, mark=o, color=black!30!red, mark size=2pt, forget plot] coordinates {(240.6, 0.0084782)};

\end{axis}
\end{tikzpicture}
    \captionof{figure}{Accepted advances versus runtime for the single-inverter RMS benchmark. Note that the time is plotted on a logarithmic scale. }
    \label{fig:inv_steps_time}
\end{minipage}
\end{figure}

The single-inverter benchmark employs a classic 11-state \ac{DAE} model with 7 inputs of a grid-following converter. The inverter is connected to a stiff grid, modeled as a stiff voltage source behind an impedance. We include this benchmark as a controlled first test of RAPTOR on coupled electrical, PLL, power-control, current-control, and filtering dynamics before considering the more demanding weak-grid, oscillatory, and networked cases.

Fig.~\ref{fig:inv_steps_time} relates this performance to the number of numerical advances. The \emph{Best Time} \ac{RAPTOR}--6 configuration completes the simulation in approximately \(0.014\)~s using only \(38\) accepted intervals while maintaining error on the order of \(10^{-6}\). Kvaerno4, Kvaerno5, and implicit Euler also require relatively few accepted timesteps and can achieve shorter runtimes; however, Fig.~\ref{fig:inv_speed_accuracy} shows that these configurations have substantially larger trajectory errors. In the more accurate operating regimes, the remaining classical solvers require hundreds to thousands of accepted timesteps. Thus, RAPTOR provides a favorable accuracy--runtime trade-off by combining low trajectory error with comparatively few accepted advances on this benchmark.

\subsubsection{Analysis for Weak Grid Operation}
\label{sec:res_sensitivity}

We next examine how solver behavior changes as the grid-following inverter is connected to progressively weaker power systems, by reducing the short-circuit ratio (SCR) from $\mathrm{SCR}=5.0$ to $\mathrm{SCR}=1.2$. This test is particularly relevant for converter-dominated systems because a lower SCR means that the grid is less stiff at the point of connection. As a result, the inverter has a stronger influence on the terminal voltage, and the interaction between the control loops and the network becomes tighter. Numerically, this complicates the implicit solve: a change in one variable causes larger changes in the others, so Newton iterations are more sensitive and the solver may need smaller timesteps to converge accurately.

Fig.~\ref{fig:scr_sweep_plot} depicts how the number of accepted advances changes with a reduction of grid strength, i.e. SCR. As the grid weakens, BDF increases from 94 to 123 accepted steps, while Radau increases from 42 to 62, indicating that both solvers must advance more cautiously. In contrast, RAPTOR requires only 8-10 advances and shows a different trend. The number of accepted advances slightly decreases with a reduction in SCR, with only a slight increase for very weak grids. This highlights RAPTOR's ability to capture sensitivity and variability better through solving entire intervals.

\begin{figure}[h]
    \centering
    \begin{tikzpicture}
\begin{axis}[
    xlabel={Short-Circuit Ratio (SCR)},
    ylabel={Accepted Advances},
    x dir=reverse,
    xtick={5.0,3.0,2.0,1.2},
    ymode=log,
    xmin=1.0, xmax=5.2,
    ymin=0, ymax=130,
    grid=major,
    grid style={dashed, gray!30},
    legend style={at={(0.5,1.05)}, anchor=south,  legend columns=4, draw = none},
    thick
]


\addplot[
    color=teal!80!black,
    mark=otimes*,
    dashed,
    line width=1.2pt
] coordinates {
    (5.0, 10) (4.0, 10) (3.0, 9) (2.5, 9) (2.0, 9) (1.5, 8) (1.2, 9)
};
\addlegendentry{RAPTOR}

\addplot[
    color=red!80!black,
    mark=triangle*,
    dashed,
    line width=1.2pt
] coordinates {
    (5.0, 94) (4.0, 95) (3.0, 100) (2.5, 103) (2.0, 107) (1.5, 111) (1.2, 123)
};
\addlegendentry{BDF}

\addplot[
    color=purple!80!black,
    mark=*,
    dotted,
    line width=1.2pt
] coordinates {
    (5.0, 42) (4.0, 44) (3.0, 44) (2.5, 46) (2.0, 49) (1.5, 53) (1.2, 62)
};
\addlegendentry{Radau}

\addplot[
    color=orange!90!black,
    mark=diamond*,
    dashdotted,
    line width=1.2pt
] coordinates {
    (5.0, 21) (4.0, 21) (3.0, 22) (2.5, 23) (2.0, 24) (1.5, 26) (1.2, 29)
};
\addlegendentry{Kvaerno5}

\end{axis}
\end{tikzpicture}
    \caption{Sensitivity of the accepted step or interval count to grid strength, measured by the short-circuit ratio (SCR).}
    \label{fig:scr_sweep_plot}
\end{figure}

\subsubsection{Model Description}

The dynamics of the grid-following converter are captured by the eleven-dimensional state vector
\[
\mathbf{x} = \bigl[\,
\xi_{\mathrm{pll}},\,\theta_{\mathrm{pll}},\,i_d,\,i_q,\,
\xi_{i_d},\,\xi_{i_q},\,\xi_P,\,\xi_Q,\,
P_{\mathrm{filt}},\,Q_{\mathrm{filt}},\,V_{\mathrm{filt}}
\bigr]^\mathsf{T}
\]
the seven-dimensional input vector
\[
\mathbf{u} = \bigl[\,
P_{\mathrm{ref}},\,Q_{\mathrm{ref}},\,V_{\mathrm{ref}},\,\omega_{\mathrm{ref}},\,
v_{ga},\,v_{gb},\,v_{dc}
\bigr]^\mathsf{T}
\]
and the fourteen-dimensional parameter vector
\begin{align}
\mathbf{p} = \bigl[\,
&k_{p_{dp}},\,k_{i_{dp}},\,k_{p_{dq}},\,k_{i_{dq}}, \notag \\
&k_{p_{id}},\,k_{i_{id}},\,k_{p_{iq}},\,k_{i_{iq}},\\
k_{p_{\mathrm{pll}}},\,k_{i_{\mathrm{pll}}}, 
&\omega,\,L,\,R,\,\omega_{\mathrm{filt}}
\bigr ]^\mathsf{T}. \notag
\end{align}

Measured phase-to-ground voltages \(v_{ga},v_{gb}\) are first transformed into the synchronous \(dq\) frame by
\begin{align}
v_{gd} &= v_{ga}\cos\theta_{\mathrm{pll}} + v_{gb}\sin\theta_{\mathrm{pll}},\\
v_{gq} &= -\,v_{ga}\sin\theta_{\mathrm{pll}} + v_{gb}\cos\theta_{\mathrm{pll}}.
\end{align}

These signals drive a phase-locked loop (PLL) whose PI controller uses gains \(k_{p_{\mathrm{pll}}},k_{i_{\mathrm{pll}}}\) to process the quadrature component \(v_{gq}\). The resulting frequency correction \(\omega_{\mathrm{pll}} = k_{p_{\mathrm{pll}}}\,v_{gq} + k_{i_{\mathrm{pll}}}\,\xi_{\mathrm{pll}}\) is added to the reference \(\omega_{\mathrm{ref}}\) to yield the total angular speed \(\omega = \omega_{\mathrm{ref}} + \omega_{\mathrm{pll}}\). The PLL integrator and angle update are
\[
\dot\xi_{\mathrm{pll}} = v_{gq}, \qquad \dot\theta_{\mathrm{pll}} = \omega.
\]

The outer control loop regulates active and reactive power. Power errors are defined by
\[
dP = P_{\mathrm{ref}} - P_{\mathrm{filt}}, \qquad dQ = -\,Q_{\mathrm{ref}} + Q_{\mathrm{filt}},
\]
and a secondary PI stage with gains \(k_{p_{dp}},k_{i_{dp}}\) and \(k_{p_{dq}},k_{i_{dq}}\) generates current references
\begin{align}
i_{d,\mathrm{ref}} &= k_{p_{dp}}\,dP + k_{i_{dp}}\,\xi_P,\\
i_{q,\mathrm{ref}} &= k_{p_{dq}}\,dQ + k_{i_{dq}}\,\xi_Q,
\end{align}
while the integrator states obey
\[
\dot\xi_P = dP, \qquad \dot\xi_Q = dQ.
\]

In the inner current control loop, the \(d\)- and \(q\)-axis current errors are fed to PI regulators with gains \(k_{p_{id}},k_{i_{id}}\) and \(k_{p_{iq}},k_{i_{iq}}\). These produce converter voltage commands
\begin{align}
v_{md} &= k_{p_{id}}(i_{d,\mathrm{ref}} - i_d) + k_{i_{id}}\,\xi_{i_d} 
         -\frac{\omega}{\omega_{\mathrm{ref}}}\,L\,i_q,\\
v_{mq} &= k_{p_{iq}}(i_{q,\mathrm{ref}} - i_q) + k_{i_{iq}}\,\xi_{i_q}
         +\frac{\omega}{\omega_{\mathrm{ref}}}\,L\,i_d,
\end{align}
with integrator dynamics
\[
\dot\xi_{i_d} = i_{d,\mathrm{ref}} - i_d, \qquad
\dot\xi_{i_q} = i_{q,\mathrm{ref}} - i_q.
\]

The electrical network seen by the converter is modeled as a series \(RL\) impedance in the synchronous frame. The current dynamics are
\begin{align}
\dot i_d &= -\frac{R}{L} i_d + \omega\,i_q + \frac{1}{L}\bigl(v_{md}-v_{gd}\bigr), \\
\dot i_q &= -\frac{R}{L} i_q - \omega\,i_d + \frac{1}{L}\bigl(v_{mq}-v_{gq}\bigr).
\end{align}

Finally, the converter’s modulated voltage magnitude \(v_m = \sqrt{v_{md}^2 + v_{mq}^2}\) and current magnitude \(i_m = \sqrt{i_d^2 + i_q^2}\) yield instantaneous power
\[
P = v_{md}\,i_d + v_{mq}\,i_q, \qquad
Q = v_{mq}\,i_d - v_{md}\,i_q.
\]
These signals are filtered by first-order dynamics
\[
\dot P_{\mathrm{filt}} = (P - P_{\mathrm{filt}})\,\omega_{\mathrm{filt}}, \qquad
\dot Q_{\mathrm{filt}} = (Q - Q_{\mathrm{filt}})\,\omega_{\mathrm{filt}}, \qquad
\dot V_{\mathrm{filt}} = (v_m - V_{\mathrm{filt}})\,\omega_{\mathrm{filt}}.
\]

Together, these continuous dynamics describe the nested PLL, power, and current control loops, the converter’s electrical behavior, and the measurement filtering in a unified state-space framework.

\subsection{M-section Line Model}
\label{app:msection_model}

The M-section line model is a mixture of the T-section and \(\pi\)-section models, which allows most of the potential and magnetic energy of the cable to be stored within the cable section while the cable terminal voltages have faster dynamics. The model is shown in Fig.~\ref{fig:m_section}.

\begin{figure}[h]
    \centering
    
    {
        \begin{circuitikz}[scale=0.5, transform shape, american voltages]

        \foreach \i in {1,2,3} {
            \begin{scope}[shift={(4*\i-4,0)}]
        
                \draw (0,0) to[short] (0,0.5);
                \draw (-0.5,0) to[short] (0.5,0);
                \draw (-0.5,-2) to[short] (0.5,-2);
        
                \draw (-0.5,0)  to[R, l_={$G_\i$}, font=\Large] (-0.5,-2);
                \draw (0.5,0)   to[C,  l={$C_\i$}, font=\Large] (0.5,-2);
        
                \draw (0,-2) node[ground] {};
            \end{scope}
        }
        \foreach \i in {1,2} {
            \begin{scope}[shift={(4*\i-4,0)}]
                \draw (0,.5) to[R,-, l={$R'$}, font=\Large] (2,0.5) to[L,-, l={$L'$}, font=\Large] (4,.5);
            \end{scope}
        }

        \draw  (0,0.5) to[short] (-2.5,0.5)
                to[open, v=${v}^m_{\alpha\beta}$, font=\large, *-*] (-2.5,-2) node[ground] {};

        \draw  (8,0.5) to[short] (10.5,0.5)
                to[open, v=${v}^k_{\alpha\beta}$, font=\large, *-*] (10.5,-2) node[ground] {};

        \node[draw=blue,text=blue, circle, inner sep=1pt, font=\large] at (-2.5,1) {m};        
        \node[draw=blue,text=blue, circle, minimum size=.5cm, inner sep=1pt, font=\large] at (4,1) {c};        
        \node[draw=blue,text=blue, circle, inner sep=1pt, font=\Large] at (10.5,1) {k};        
        \end{circuitikz}
        } 
    \caption{$m$-section line model.}
    \label{fig:m_section}
\end{figure}

The dynamics of the cascaded three-section model are captured by the ten-dimensional state vector
\[
\mathbf{x}
= \begin{bmatrix}
  v_{a}^{m} & v_{b}^{m} & i_{a}^{m\!c} & i_{b}^{m\!c} & 
  v_{a}^{c} & v_{b}^{c} & i_{a}^{c\!k} & i_{b}^{c\!k} & 
  v_{a}^{k} & v_{b}^{k}
\end{bmatrix}^\top,
\]
the four-dimensional input vector
\[
\mathbf{u}
= \begin{bmatrix}
  i_{a}^{m} & i_{b}^{m} & i_{a}^{k} & i_{b}^{k}
\end{bmatrix}^\top,
\]
and the six-dimensional parameter vector
\[
\mathbf{p}
= \begin{bmatrix}
  L & R & C & G & \omega & N
\end{bmatrix}^\top.
\]

Here \(\eta_{\mathrm{sec}}\in(0,1)\) partitions the total shunt capacitance \(C\) and conductance \(G\) across the three \(\pi\)-sections via
\[
\begin{aligned}
C_1 &= \tfrac{1-N}{2}\,C, \quad &C_2 &= N\,C, \quad &C_3 &= \tfrac{1-N}{2}\,C,\\
G_1 &= \tfrac{1-N}{2}\,G, \quad &G_2 &= N\,G, \quad &G_3 &= \tfrac{1-N}{2}\,G,
\end{aligned}
\]
and each series branch has
\[
R' = \tfrac{R}{2}, \qquad L' = \tfrac{L}{2}.
\]

\subsubsection*{Section \(m\) (sending end)}
\begin{align}
\dot v_{a}^{m}
&= -\frac{G_1}{C_1}\,v_{a}^{m}
  +\omega\,v_{b}^{m}
  +\frac{1}{C_1}\bigl(i_{a}^{m}-i_{a}^{m\!c}\bigr),\\
\dot v_{b}^{m}
&= -\frac{G_1}{C_1}\,v_{b}^{m}
  -\omega\,v_{a}^{m}
  +\frac{1}{C_1}\bigl(i_{b}^{m}-i_{b}^{m\!c}\bigr),\\
\dot i_{a}^{m\!c}
&= -\frac{R'}{L'}\,i_{a}^{m\!c}
  +\omega\,i_{b}^{m\!c}
  +\frac{1}{L'}\bigl(v_{a}^{m}-v_{a}^{c}\bigr),\\
\dot i_{b}^{m\!c}
&= -\frac{R'}{L'}\,i_{b}^{m\!c}
  -\omega\,i_{a}^{m\!c}
  +\frac{1}{L'}\bigl(v_{b}^{m}-v_{b}^{c}\bigr).
\end{align}

\subsubsection*{Section \(c\) (central section)}
\begin{align}
\dot v_{a}^{c}
&= -\frac{G_2}{C_2}\,v_{a}^{c}
  +\omega\,v_{b}^{c}
  +\frac{1}{C_2}\bigl(i_{a}^{m\!c}-i_{a}^{c\!k}\bigr),\\
\dot v_{b}^{c}
&= -\frac{G_2}{C_2}\,v_{b}^{c}
  -\omega\,v_{a}^{c}
  +\frac{1}{C_2}\bigl(i_{b}^{m\!c}-i_{b}^{c\!k}\bigr),\\
\dot i_{a}^{c\!k}
&= -\frac{R'}{L'}\,i_{a}^{c\!k}
  +\omega\,i_{b}^{c\!k}
  +\frac{1}{L'}\bigl(v_{a}^{c}-v_{a}^{k}\bigr),\\
\dot i_{b}^{c\!k}
&= -\frac{R'}{L'}\,i_{b}^{c\!k}
  -\omega\,i_{a}^{c\!k}
  +\frac{1}{L'}\bigl(v_{b}^{c}-v_{b}^{k}\bigr).
\end{align}

\subsubsection*{Section \(k\) (receiving end)}
\begin{align}
\dot v_{a}^{k}
&= -\frac{G_3}{C_3}\,v_{a}^{k}
  +\omega\,v_{b}^{k}
  +\frac{1}{C_3}\bigl(i_{a}^{k}+i_{a}^{c\!k}\bigr),\\
\dot v_{b}^{k}
&= -\frac{G_3}{C_3}\,v_{b}^{k}
  -\omega\,v_{a}^{k}
  +\frac{1}{C_3}\bigl(i_{b}^{k}+i_{b}^{c\!k}\bigr).
\end{align}

\[
\mathbf{y}
= \begin{bmatrix}
  -\,i_{a}^{m\!c} & -\,i_{b}^{m\!c} & -\,i_{a}^{c\!k} & -\,i_{b}^{c\!k}
\end{bmatrix}^\top.
\]

\subsection{Grid-Following Converter EMT Model}
\label{app:emt_gfl_model}

\subsubsection{Model Description}

The dynamics of the three-phase grid-following converter with LCL filter and PLL are captured by the fifteen-dimensional state vector
\begin{align}
    \mathbf{x} =
\bigl[
  i_{c,a},\,i_{c,b},\,i_{c,c},\,
  i_{g,a},\,i_{g,b},\,i_{g,c},\,
  u_{f,a},\,u_{f,b},\,u_{f,c},\, \\
  \xi_{\mathrm{pll}},\,\theta_{\mathrm{pll}},\,
  \xi_P,\,\xi_Q,\,
  \xi_{id},\,\xi_{iq}
\bigr]^\mathsf{T},
\end{align}

the four-dimensional input vector
\[
\mathbf{u} =
\bigl[
  P^\star,\,Q^\star,\,
  E_{\mathrm{abs}},\,\phi_e
\bigr]^\mathsf{T},
\]
and the thirteen-dimensional parameter vector
\begin{align}
\mathbf{p} = \bigl[\,
&L_c,\,L_g,\,C_f,\,R_c,\,R_f,\,R_{lg},\notag\\
&K^{\mathrm{pc}}_p,\,K^{\mathrm{pc}}_i,\,
 K^{\mathrm{cc}}_p,\,K^{\mathrm{cc}}_i,\,
 K^{\mathrm{pll}}_p,\,K^{\mathrm{pll}}_i,\,
 \omega_N
\bigr]^\mathsf{T}.
\end{align}

For compactness we define three-phase vectors
\[
\mathbf{i}_c =
\begin{bmatrix}
  i_{c,a}\\ i_{c,b}\\ i_{c,c}
\end{bmatrix},\quad
\mathbf{i}_g =
\begin{bmatrix}
  i_{g,a}\\ i_{g,b}\\ i_{g,c}
\end{bmatrix},\quad
\mathbf{v} =
\begin{bmatrix}
  v_a\\ v_b\\ v_c
\end{bmatrix},\quad
\mathbf{u}_f =
\begin{bmatrix}
  u_{f,a}\\ u_{f,b}\\ u_{f,c}
\end{bmatrix},
\]
and similarly for the internal node voltages
\begin{align}
    \mathbf{u}_c = [u_{c,a},u_{c,b},u_{c,c}]^\mathsf{T},\quad
\mathbf{u}_g = [u_{g,a},u_{g,b},u_{g,c}]^\mathsf{T},\quad \\
\mathbf{u}_{\mathrm{conv}} = [u_a,u_b,u_c]^\mathsf{T}.
\end{align}

\subsubsection{Stiff Grid Source}

The three-phase stiff grid voltage is imposed as
\begin{equation}
\label{eq:stiff_source_app}
\mathbf{e}(t) =
E_{\mathrm{abs}}
\begin{bmatrix}
\cos(\omega_N t + \phi_e) \\
\cos(\omega_N t + \phi_e - \tfrac{2\pi}{3}) \\
\cos(\omega_N t + \phi_e + \tfrac{2\pi}{3})
\end{bmatrix}.
\end{equation}
A temporary amplitude reduction can be represented by
\[
E_{\mathrm{abs}}(t) =
\begin{cases}
0.8\,E_{\mathrm{abs}}, & t\in(t_{\mathrm{fault}},t_{\mathrm{clear}}),\\
E_{\mathrm{abs}}, & \text{otherwise}.
\end{cases}
\]

\subsubsection{EMT Network Dynamics (LCL Filter)}

The converter is connected to the grid through an LCL filter with converter-side inductance \(L_c\), grid-side inductance \(L_g\), damping resistors \(R_c\) and \(R_f\), grid resistance \(R_{lg}\), and a shunt capacitor \(C_f\) at the PCC. For each phase, the differential equations of the inductors and capacitor can be written in vector form as
\begin{align}
L_c\,\dot{\mathbf{i}}_c &= \mathbf{u}_{\mathrm{conv}} - \mathbf{u}_c, \label{eq:lc_dyn_app}\\
L_g\,\dot{\mathbf{i}}_g &= \mathbf{v} - \mathbf{u}_g, \label{eq:lg_dyn_app}\\
C_f\,\dot{\mathbf{u}}_f &= \frac{\mathbf{v} - \mathbf{u}_f}{R_f}. \label{eq:cf_dyn_app}
\end{align}
The resistive and algebraic constraints follow from Ohm’s law and KCL at the internal nodes:
\begin{align}
\mathbf{0} &= \mathbf{i}_c + \frac{\mathbf{u}_c - \mathbf{v}}{R_c}, \label{eq:uc_alg_app}\\
\mathbf{0} &= \mathbf{i}_g + \frac{\mathbf{u}_g - \mathbf{e}(t)}{R_{lg}}, \label{eq:ug_alg_app}\\
\mathbf{0} &= \frac{\mathbf{v} - \mathbf{u}_c}{R_c}
          + \frac{\mathbf{v} - \mathbf{u}_f}{R_f}
          + \mathbf{i}_g. \label{eq:pcc_alg_app}
\end{align}

\subsubsection{PLL Dynamics}

The phase-locked loop is modeled in the synchronous \(dq\) frame aligned with the PCC voltage. Its states are the integrator \(\xi_{\mathrm{pll}}\) and the estimated grid angle \(\theta_{\mathrm{pll}}\). With \(v_q\) the \(q\)-axis PCC voltage, the PLL dynamics are
\begin{align}
\dot\xi_{\mathrm{pll}} &= v_q, \label{eq:pll_int_app}\\
\dot\theta_{\mathrm{pll}} &= K^{\mathrm{pll}}_p\,v_q + K^{\mathrm{pll}}_i\,\xi_{\mathrm{pll}} + \omega_N. \label{eq:pll_angle_app}
\end{align}

\subsubsection{Outer Power Control}

The injected active and reactive powers at the PCC are computed from the \(dq\)-axis voltages and currents as
\begin{align}
P &= v_d\,i_d + v_q\,i_q,\\
Q &= v_q\,i_d - v_d\,i_q.
\end{align}
The outer control loop regulates these quantities to the references \(P^\star\) and \(Q^\star\). Power errors are
\begin{align}
dP &= P^\star - P,\\
dQ &= Q - Q^\star,
\end{align}
and a pair of PI controllers with integrator states \(\xi_P\) and \(\xi_Q\) generate the current references
\begin{align}
\dot\xi_P &= dP,\\
i_d^\star &= K^{\mathrm{pc}}_p\,dP + K^{\mathrm{pc}}_i\,\xi_P,\\[0.4em]
\dot\xi_Q &= dQ,\\
i_q^\star &= K^{\mathrm{pc}}_p\,dQ + K^{\mathrm{pc}}_i\,\xi_Q.
\end{align}

\subsubsection{Inner Current Control}

In the inner loop, the \(d\)- and \(q\)-axis current errors are
\[
e_{id} = i_d^\star - i_d, \qquad e_{iq} = i_q^\star - i_q,
\]
and are processed by PI controllers with integrator states \(\xi_{id}\) and \(\xi_{iq}\):
\begin{align}
\dot\xi_{id} &= e_{id},\\
u_d &= K^{\mathrm{cc}}_p\,e_{id} + K^{\mathrm{cc}}_i\,\xi_{id},\\[0.4em]
\dot\xi_{iq} &= e_{iq},\\
u_q &= K^{\mathrm{cc}}_p\,e_{iq} + K^{\mathrm{cc}}_i\,\xi_{iq}.
\end{align}

\subsubsection{Park Transformations and Converter Voltage Synthesis}

Using \(\theta_{\mathrm{pll}}\) as the synchronous angle, the abc-to-\(dq\) Park transformation for the PCC voltages is
\begin{equation}
\theta_a=\theta_{\mathrm{pll}},\qquad
\theta_b=\theta_{\mathrm{pll}}-\frac{2\pi}{3},\qquad
\theta_c=\theta_{\mathrm{pll}}+\frac{2\pi}{3}.
\end{equation}

The PCC voltages are transformed to the synchronous $dq$ frame as
\begin{equation}
\begin{bmatrix}
v_d\\
v_q
\end{bmatrix}
=
\frac{2}{3}
\sum_{p\in\{a,b,c\}}
v_p
\begin{bmatrix}
\cos\theta_p\\
-\sin\theta_p
\end{bmatrix}.
\label{eq:park_voltage}
\end{equation}

Similarly, the grid currents are transformed according to
\begin{equation}
\begin{bmatrix}
i_d\\
i_q
\end{bmatrix}
=
\frac{2}{3}
\sum_{p\in\{a,b,c\}}
i_{g,p}
\begin{bmatrix}
\cos\theta_p\\
-\sin\theta_p
\end{bmatrix}.
\label{eq:park_current}
\end{equation}
The inverse Park transformation maps the \(dq\)-axis voltage commands to converter phase voltages:
\begin{align}
u_a &= \cos\theta_{\mathrm{pll}}\,u_d - \sin\theta_{\mathrm{pll}}\,u_q,\\
u_b &= \cos(\theta_{\mathrm{pll}} - \tfrac{2\pi}{3})\,u_d
       - \sin(\theta_{\mathrm{pll}} - \tfrac{2\pi}{3})\,u_q,\\
u_c &= \cos(\theta_{\mathrm{pll}} + \tfrac{2\pi}{3})\,u_d
       - \sin(\theta_{\mathrm{pll}} + \tfrac{2\pi}{3})\,u_q.
\end{align}

\subsubsection{DAE Formulation}

Collecting all terms, the model can be written compactly as a nonlinear differential--algebraic system
\begin{align}
\dot{\mathbf{x}} &= \mathbf{f}(\mathbf{x},\mathbf{u},\mathbf{p},t),\\
\mathbf{0} &= \mathbf{g}(\mathbf{x},\mathbf{u},\mathbf{p},t),
\end{align}
where \(\mathbf{f}\) is defined by the differential equations \eqref{eq:lc_dyn_app}--\eqref{eq:cf_dyn_app}, \eqref{eq:pll_int_app}--\eqref{eq:pll_angle_app}, and the PI controller dynamics for \(\xi_P,\xi_Q,\xi_{id},\xi_{iq}\), while \(\mathbf{g}\) collects the algebraic constraints \eqref{eq:uc_alg_app}--\eqref{eq:pcc_alg_app}, the power expressions, and the abc--\(dq\) mappings.

\subsection{Synchronous Machine--Infinite Bus EMT Model}
\label{app:emt_sm_model}

\subsubsection{Model Description}

The dynamics of the synchronous machine connected to an infinite bus through a three-phase transmission element are captured by the nineteen-dimensional state vector
\begin{align}
  \mathbf{x} = \bigl[
\,i_a,\,i_b,\,i_c,\,
\psi_d,\,\psi_q,\,\psi_f,\,\psi_{q1},\,\psi_o,\,
v_a,\,v_b,\,v_c,\, \\
v_d,\,v_q,\,v_o,\,
i_d,\,i_q,\,i_o,\,
i_f,\,i_{q1}
\bigr]^\mathsf{T},  
\end{align}
where \(\mathbf{i}=[i_a,i_b,i_c]^\top\) are stator phase currents, \(\psi_d,\psi_q\) are rotor \(dq\)-axis stator flux linkages, \(\psi_f\) and \(\psi_{q1}\) are field and damper winding flux linkages, \(\psi_o\) is the zero-sequence flux, \(\mathbf{v}=[v_a,v_b,v_c]^\top\) are the terminal phase voltages, and \((v_d,v_q,v_o)\) and \((i_d,i_q,i_o)\) are the synchronous-frame voltage and current components.

The infinite bus acts as a stiff Thevenin source
\begin{equation}
\mathbf{e}(t) =
\sqrt{2}\,\lvert E_\infty\rvert
\begin{bmatrix}
\cos(\omega_N t + \phi_E)\\
\cos(\omega_N t + \phi_E - 2\pi/3)\\
\cos(\omega_N t + \phi_E + 2\pi/3)
\end{bmatrix},
\end{equation}
with a temporary outage represented by setting \(\mathbf{e}(t)=\mathbf{0}\) over prescribed time intervals.

\subsubsection*{Network (Line) Current Dynamics}

Each stator phase current flows through a per-phase series impedance \((R_e,L_e)\). For \(p\in\{a,b,c\}\),
\begin{equation}
\dot i_p = \omega_N\!\left(-\frac{R_e}{L_e}i_p + \frac{1}{L_e}v_p - \frac{1}{L_e}e_p(t)\right).
\end{equation}

\subsubsection*{Synchronous Machine Flux Dynamics}

The machine’s \(dq0\) flux derivatives follow directly from the Park-transformed stator equations and the field and damper winding circuits:
\begin{align}
\dot\psi_d &= \omega_N\!\left(-R_a i_d - \psi_q - v_d\right),\\
\dot\psi_q &= \omega_N\!\left(-R_a i_q + \psi_d - v_q\right),\\
\dot\psi_f &= \omega_N\!\left(-R_f i_f + V_f\right),\\
\dot\psi_{q1} &= \omega_N\!\left(-R_{q1} i_{q1}\right),\\
\dot\psi_o &= \omega_N\!\left(-R_a i_o - v_o\right).
\end{align}

\subsubsection*{abc--\(dq0\) Transformations}

Let \(\theta(t)=\theta_0+\omega_N t\) denote the electrical rotor angle. The terminal voltages must satisfy the Park transformation constraints:
\begin{align}
0 &= 
\frac{\sqrt{2}}{3}\!\left(
\cos\theta\,v_a +
\cos(\theta-2\pi/3)\,v_b +
\cos(\theta+2\pi/3)\,v_c
\right) - v_d,\\[0.3em]
0 &= 
-\frac{\sqrt{2}}{3}\!\left(
\sin\theta\,v_a +
\sin(\theta-2\pi/3)\,v_b +
\sin(\theta+2\pi/3)\,v_c
\right) - v_q,\\[0.3em]
0 &= \tfrac{1}{3}(v_a+v_b+v_c) - v_o.
\end{align}

Similarly, the stator currents must satisfy
\begin{align}
0 &= 
\frac{\sqrt{2}}{3}\!\left(
\cos\theta\,i_a +
\cos(\theta-2\pi/3)\,i_b +
\cos(\theta+2\pi/3)\,i_c
\right) - i_d,\\[0.3em]
0 &= 
-\frac{\sqrt{2}}{3}\!\left(
\sin\theta\,i_a +
\sin(\theta-2\pi/3)\,i_b +
\sin(\theta+2\pi/3)\,i_c
\right) - i_q,\\[0.3em]
0 &= \tfrac{1}{3}(i_a+i_b+i_c) - i_o.
\end{align}

\subsubsection*{Machine Algebraic Equations}

Flux linkages and currents obey the magnetic coupling relations
\begin{align}
0 &= L_{dd}\,i_d + L_{df}\,i_f - \psi_d,\\
0 &= L_{qq}\,i_q + L_{qq1}\,i_{q1} - \psi_q,\\
0 &= L_{ff}\,i_f + L_{df}\,i_d - \psi_f,\\
0 &= L_{q1q1}\,i_{q1} + L_{qq1}\,i_q - \psi_{q1},\\
0 &= L_{oo}\,i_o - \psi_o.
\end{align}

\subsubsection*{DAE Formulation}

Collecting the differential and algebraic equations above, the synchronous machine--infinite bus model satisfies the nonlinear DAE system
\begin{align}
\dot{\mathbf{x}} &= \mathbf{f}(\mathbf{x},t),\\
\mathbf{0} &= \mathbf{g}(\mathbf{x},t),
\end{align}
where the functions \(\mathbf{f}\) and \(\mathbf{g}\) contain, respectively, the line-current dynamics, synchronous-machine flux dynamics, and the abc--\(dq0\) algebraic constraints together with the magnetic coupling equations.

\subsection{Networked IEEE RMS Benchmark Systems}
\label{app:ieee_rms_models}

The networked RMS benchmarks are constructed from the IEEE 9-, 14-,
39-, 57-, and 118-bus transmission systems. The original AC network of
each test system is retained, including all buses, branches,
transformers, shunt elements, generation buses, and loads. The
steady-state operating point is determined from an AC power flow and
provides the initial condition for the dynamic simulation.

The IEEE network data specify steady-state electrical quantities rather
than a unique set of dynamic parameters. We therefore augment the
steady-state networks with a common dynamic benchmark parameterization.
The same component models and parameterization rules are used for all
five systems. The resulting benchmarks should consequently be
interpreted as dynamic RMS systems constructed on the standard IEEE
networks, rather than as historical dynamic-data sets associated with
the original IEEE cases.

Throughout this appendix, quantities are expressed on a
$S_{\mathrm{base}}=100$~MVA system base and the nominal electrical
frequency is
\begin{equation}
    f_b=60~\mathrm{Hz},
    \qquad
    \omega_b=2\pi f_b .
\end{equation}

Each system contains three classes of dynamic devices:
sixth-order synchronous machines with excitation and
governor--turbine dynamics, grid-following inverter-based resources
with PLL and power-control dynamics, and fifth-order induction-motor
aggregates. The remainder of the demand is represented by static
impedance. The complete differential state is therefore
\begin{equation}
\mathbf{x}
=
\begin{bmatrix}
\mathbf{x}_{\mathrm{SG}}^\mathsf{T} &
\mathbf{x}_{\mathrm{IBR}}^\mathsf{T} &
\mathbf{x}_{\mathrm{IM}}^\mathsf{T}
\end{bmatrix}^{\mathsf T}.
\end{equation}

For $N_{\mathrm{SG}}$ synchronous machines,
$N_{\mathrm{IBR}}$ grid-following inverters, and
$N_{\mathrm{IM}}$ induction motors, the number of differential states is
\begin{equation}
\label{eq:app_ieee_state_count}
n_x
=
10N_{\mathrm{SG}}
+
8N_{\mathrm{IBR}}
+
5N_{\mathrm{IM}}.
\end{equation}

The resulting composition of the five benchmark systems is summarized
in Table~\ref{tab:app_ieee_composition}.

\begin{table*}[t]
\centering
\caption{Composition of the IEEE networked RMS benchmarks.}
\label{tab:app_ieee_composition}
\resizebox{\textwidth}{!}{%
\begin{tabular}{lccccccccll}
\toprule
\textbf{System}
&
$\boldsymbol{n_b}$
&
\textbf{Branches}
&
\textbf{Gen. records}
&
$\boldsymbol{N_{\mathrm{SG}}}$
&
$\boldsymbol{N_{\mathrm{IBR}}}$
&
$\boldsymbol{N_{\mathrm{IM}}}$
&
$\boldsymbol{n_x}$
&
$\boldsymbol{P_{\mathrm{IM}}/P_L}$
&
\textbf{IBR buses}
&
\textbf{Motor buses}
\\
\midrule
IEEE 9
& 9 & 9 & 3 & 3 & 1 & 3 & 53
& 25.00\%
& 1
& 5, 7, 9
\\
IEEE 14
& 14 & 20 & 5 & 4 & 2 & 6 & 86
& 17.86\%
& 1, 2
& 2, 3, 6, 9, 13, 14
\\
IEEE 39
& 39 & 46 & 10 & 8 & 3 & 6 & 134
& 11.72\%
& 30, 34, 37
& 3, 4, 8, 16, 20, 39
\\
IEEE 57
& 57 & 80 & 7 & 6 & 2 & 6 & 106
& 9.49\%
& 3, 12
& 1, 6, 8, 9, 12, 16
\\
IEEE 118
& 118 & 186 & 54 & 44 & 11 & 6 & 558
& 5.12\%
& 12, 25, 31, 46, 49, 54, 59, 61, 87, 103, 111
& 15, 42, 54, 59, 80, 90
\\
\bottomrule
\end{tabular}
}
\end{table*}

A target fraction
\begin{equation}
\label{eq:app_ibr_share}
    \eta_{\mathrm{IBR}}=0.20
\end{equation}
of the dispatched system active power is represented by grid-following
IBRs. Since the steady-state IEEE cases do not specify generation
technology, the allocation is performed at existing generator buses.
Active generating units are ordered by dispatched active power and the
smallest units are assigned to IBR operation first until the target
system-wide IBR share is reached. A marginal unit is split when required
to obtain exactly $\eta_{\mathrm{IBR}}=20\%$. Zero-active-power
synchronous condensers remain synchronous devices.

For the IEEE 9-bus system, bus 1 is the marginal plant and its IBR
fraction is $0.8893$. For IEEE 14, bus 2 is entirely inverter based and
bus 1 has an IBR fraction of $0.0623$. For IEEE 39, buses 30 and 34 are
entirely inverter based, while the IBR fraction at bus 37 is $0.9288$.
For IEEE 57, bus 3 is entirely inverter based and the IBR fraction at
bus 12 is $0.6959$. For IEEE 118, all listed IBR plants except bus 25
are entirely represented by the grid-following model; bus 25 is the
marginal plant with IBR fraction $0.5313$.

At selected load buses, the target active-power fraction assigned to
the induction-motor aggregate is
\begin{equation}
\label{eq:app_motor_target}
    \eta_{\mathrm{IM}}=0.25.
\end{equation}
At most six dynamic motor buses are used. Candidate buses satisfy
\begin{equation}
    P_{L,k}>0.03~\mathrm{pu},
    \qquad
    Q_{L,k}>0.005~\mathrm{pu},
\end{equation}
and are ordered by active demand. The motor fraction is additionally
limited when necessary to avoid assigning more reactive demand to the
motor than is available at the original operating point. Consequently,
the resulting system-wide dynamic-motor active-power fractions are
those reported in Table~\ref{tab:app_ieee_composition}. The remaining
load is retained as static impedance.

The complete networked model is a semi-explicit index-1 DAE,
\begin{subequations}
\label{eq:app_ieee_dae}
\begin{align}
\dot{\mathbf{x}}
&=
\mathbf{f}
\left(
t,\mathbf{x},\mathbf{V},\mathbf{\sigma}
\right),
\\
\mathbf{0}
&=
\mathbf{A}_{\sigma}\mathbf{V}
-
\mathbf{b}(\mathbf{x}),
\end{align}
\end{subequations}
where
\(
\mathbf{V}\in\mathbb{C}^{n_b}
\)
contains the complex positive-sequence bus voltages and $\mathbf{\sigma}$
denotes the active network or disturbance discrete state.

The second equation in \eqref{eq:app_ieee_dae} is an algebraic
Kirchhoff-current constraint on the complete network. It does not
represent a reduction in the number of buses. Since
$\mathbf{A}_{\sigma}$ is nonsingular in the considered operating
region, the algebraic voltage solution is
\begin{equation}
\label{eq:app_exact_voltage}
\boxed{
\mathbf{V}
=
\mathbf{A}_{\sigma}^{-1}
\mathbf{b}(\mathbf{x})
}.
\end{equation}
Thus, the bus voltages can be eliminated algebraically while retaining
the complete physical network and all differential component states.

The initial operating point is obtained from the conventional AC
power-flow equations. Let
\begin{equation}
V_k
=
|V_k|e^{j\theta_k},
\end{equation}
and let $\mathbf{Y}_{\mathrm{bus}}$ denote the nodal admittance matrix
of the corresponding IEEE network. The complex power injected at bus
$k$ is
\begin{equation}
\label{eq:app_power_flow}
S_k
=
P_k+jQ_k
=
V_k
\left(
\sum_{\ell=1}^{n_b}
Y_{k\ell}V_\ell
\right)^{*}.
\end{equation}
The slack-bus angle, PV-bus voltage magnitudes, specified active-power
injections, and PQ-bus active and reactive powers define the usual
power-flow constraints. The converged solution
$\mathbf{V}_0$ is subsequently used to initialize each dynamic device
such that its initial electrical injection is consistent with the
power-flow solution.

Each synchronous generator is represented by
\begin{equation}
\label{eq:app_sg_state}
\mathbf{x}_g =
\begin{bmatrix}
\delta_g \\
\omega_g \\
E'_{q,g} \\
E'_{d,g} \\
E''_{q,g} \\
E''_{d,g} \\
E_{\mathrm{fd},g} \\
V_{m,g} \\
P_{m,g} \\
P_{v,g}
\end{bmatrix}
\in \mathbb{R}^{10}.
\end{equation}
Here $\delta_g$ is the rotor electrical angle,
$\omega_g$ the rotor speed in per unit,
$E'_{q,g}$ and $E'_{d,g}$ the transient internal voltages,
$E''_{q,g}$ and $E''_{d,g}$ the subtransient internal voltages,
$E_{\mathrm{fd},g}$ the excitation voltage,
$V_{m,g}$ the measured terminal-voltage magnitude,
$P_{m,g}$ the mechanical power, and $P_{v,g}$ the
governor/valve state.

For a generator connected at bus $k$, define
\begin{equation}
\gamma_g
=
\delta_g-\frac{\pi}{2},
\end{equation}
and express its terminal voltage in the rotor reference frame as
\begin{equation}
\label{eq:app_sg_vdq}
V_{d,g}+jV_{q,g}
=
V_k e^{-j\gamma_g}.
\end{equation}
The subtransient stator currents are
\begin{align}
I_{d,g}
&=
\frac{E''_{q,g}-V_{q,g}}{x''_g},
\\
I_{q,g}
&=
\frac{V_{d,g}-E''_{d,g}}{x''_g}.
\end{align}
The current injected into the network is
\begin{equation}
\label{eq:app_sg_current}
I_g
=
(I_{d,g}+jI_{q,g})e^{j\gamma_g},
\end{equation}
and the corresponding electrical active power is
\begin{equation}
\label{eq:app_sg_pe}
P_{e,g}
=
\Re\!\left\{
V_k I_g^{*}
\right\}.
\end{equation}

The electromechanical rotor equations are
\begin{align}
\dot{\delta}_g
&=
\omega_b(\omega_g-1),
\\
\dot{\omega}_g
&=
\frac{
P_{m,g}
-
P_{e,g}
-
D_g(\omega_g-1)
}{
2H_g
}.
\end{align}
The transient and subtransient voltage dynamics are
\begin{align}
T'_{do,g}\dot E'_{q,g}
&=
E_{\mathrm{fd},g}
-
E'_{q,g}
-
(x_{d,g}-x'_{d,g})I_{d,g},
\\
T'_{qo,g}\dot E'_{d,g}
&=
-
E'_{d,g}
+
(x_{q,g}-x'_{q,g})I_{q,g},
\\
T''_{do,g}\dot E''_{q,g}
&=
E'_{q,g}
-
E''_{q,g}
-
(x'_{d,g}-x''_g)I_{d,g},
\\
T''_{qo,g}\dot E''_{d,g}
&=
E'_{d,g}
-
E''_{d,g}
+
(x'_{q,g}-x''_g)I_{q,g}.
\end{align}

The terminal-voltage measurement is represented by
\begin{equation}
T_{r,g}\dot V_{m,g}
=
|V_k|-V_{m,g}.
\end{equation}
The excitation command is
\begin{equation}
\label{eq:app_avr_command}
E_{\mathrm{fd},g}^{\star}
=
\operatorname{clip}
\left(
K_{A,g}
[V_{\mathrm{ref},g}-V_{m,g}],
-100,
100
\right),
\end{equation}
followed by
\begin{equation}
T_{A,g}\dot E_{\mathrm{fd},g}
=
E_{\mathrm{fd},g}^{\star}
-
E_{\mathrm{fd},g}.
\end{equation}
The wide bound in \eqref{eq:app_avr_command} acts as a soft field
runaway guard and remains inactive close to the initialized operating
point.

The governor and turbine are represented by
\begin{align}
T_{g,g}\dot P_{v,g}
&=
P_{\mathrm{ref},g}
-
\frac{\omega_g-1}{R_g}
-
P_{v,g},
\\
T_{t,g}\dot P_{m,g}
&=
P_{v,g}-P_{m,g}.
\end{align}

The synchronous-machine parameters common to the benchmark are
\begin{equation}
\begin{aligned}
x_d &= 1.80,
&
x_q &= 1.70,
&
x'_d &= 0.30,
&
x'_q &= 0.55,
\\
T'_{qo} &= 0.40~\mathrm{s},
&
T''_{do} &= 0.030~\mathrm{s},
&
T''_{qo} &= 0.050~\mathrm{s},
&
D &= 1.0,
\\
K_A &= 20,
&
T_A &= 0.050~\mathrm{s},
&
T_r &= 0.020~\mathrm{s},
&
R_g &= 0.05,
\\
T_g &= 0.20~\mathrm{s},
&
T_t &= 0.50~\mathrm{s}. &&
\end{aligned}
\end{equation}
To avoid an artificially identical machine population, three machine
parameters vary deterministically with the dispatched-unit index
$r=0,1,\ldots$:
\begin{align}
H_r
&=
3.5
+
0.4\left[(r\bmod 4)+1\right]
\quad \mathrm{s},
\\
T'_{do,r}
&=
6.0
+
0.7(r\bmod 4)
\quad \mathrm{s},
\\
x''_r
&=
0.20
+
0.01(r\bmod 3).
\end{align}
Consequently,
\begin{equation}
H_r\in\{3.9,4.3,4.7,5.1\}\ \mathrm{s},
\end{equation}
\begin{equation}
T'_{do,r}\in\{6.0,6.7,7.4,8.1\}\ \mathrm{s},
\end{equation}
and
\begin{equation}
x''_r\in\{0.20,0.21,0.22\}.
\end{equation}

At initialization, the machine states are selected to satisfy the
power-flow current injection. If $S_g$ and $V_k$ denote the assigned
initial complex power and terminal voltage, respectively,
\begin{equation}
I_{g,0}
=
\left(
\frac{S_g}{V_k}
\right)^{*}.
\end{equation}
The initial internal subtransient voltage is
\begin{equation}
E''_{g,0}
=
V_k
+
jx''_g I_{g,0},
\end{equation}
from which $\delta_g$, $E''_{q,g}$, $E''_{d,g}$,
$E'_{q,g}$, $E'_{d,g}$, and $E_{\mathrm{fd},g}$ are obtained.
The mechanical states are initialized according to
\begin{equation}
P_{m,g}(0)=P_{v,g}(0)=\Re\{S_g\},
\end{equation}
and
\begin{equation}
V_{\mathrm{ref},g}
=
|V_k|
+
\frac{E_{\mathrm{fd},g}(0)}{K_A},
\end{equation}
so that the machine starts from the power-flow equilibrium.

Each grid-following IBR is represented by the state vector
\begin{equation}
\label{eq:app_ibr_state}
\mathbf{x}_i
=
\begin{bmatrix}
\theta_i &
\xi_{\mathrm{pll},i} &
P_{f,i} &
Q_{f,i} &
\xi_{P,i} &
\xi_{Q,i} &
i_{d,i} &
i_{q,i}
\end{bmatrix}^{\mathsf T}.
\end{equation}
The states describe the estimated grid angle, PLL integral state,
filtered active and reactive powers, outer-loop PI integral states,
and the $dq$ current injection.

For an inverter connected to bus $k$, its rotating-frame terminal
voltage is
\begin{equation}
\label{eq:app_ibr_vdq}
v_{d,i}+jv_{q,i}
=
V_k e^{-j\theta_i}.
\end{equation}
The injected active and reactive powers are
\begin{align}
P_i
&=
v_{d,i}i_{d,i}
+
v_{q,i}i_{q,i},
\\
Q_i
&=
v_{q,i}i_{d,i}
-
v_{d,i}i_{q,i}.
\end{align}

The normalized PLL error is
\begin{equation}
\label{eq:app_pll_error}
e_{\mathrm{pll},i}
=
\frac{
v_{q,i}
}{
\max(|V_k|,V_\epsilon)
},
\end{equation}
with
\begin{align}
\dot{\xi}_{\mathrm{pll},i}
&=
e_{\mathrm{pll},i},
\\
\dot{\theta}_i
&=
k_{p,\mathrm{pll}}
e_{\mathrm{pll},i}
+
k_{i,\mathrm{pll}}
\xi_{\mathrm{pll},i}.
\end{align}

The measured active and reactive powers are filtered according to
\begin{align}
T_{pq}\dot P_{f,i}
&=
P_i-P_{f,i},
\\
T_{pq}\dot Q_{f,i}
&=
Q_i-Q_{f,i}.
\end{align}
Defining
\begin{equation}
e_{P,i}
=
P_{\mathrm{ref},i}-P_{f,i},
\qquad
e_{Q,i}
=
Q_{\mathrm{ref},i}-Q_{f,i},
\end{equation}
the outer PI integrators satisfy
\begin{align}
\dot{\xi}_{P,i}
&=
e_{P,i},
\\
\dot{\xi}_{Q,i}
&=
e_{Q,i}.
\end{align}

The voltage used to form the current request is regularized as
\begin{equation}
v_{d,\epsilon}
=
\max(|v_{d,i}|,V_\epsilon),
\end{equation}
where
\begin{equation}
V_\epsilon=0.20~\mathrm{pu}.
\end{equation}
The unconstrained current references are then
\begin{align}
i_{d,i}^{\star,\mathrm{raw}}
&=
\frac{P_{\mathrm{ref},i}}{v_{d,\epsilon}}
+
k_{p,P}e_{P,i}
+
k_{i,P}\xi_{P,i},
\\
i_{q,i}^{\star,\mathrm{raw}}
&=
-
\frac{Q_{\mathrm{ref},i}}{v_{d,\epsilon}}
-
k_{p,Q}e_{Q,i}
-
k_{i,Q}\xi_{Q,i}.
\end{align}

A circular converter-current limit is applied to these commands.
Define
\begin{equation}
I_i^{\star,\mathrm{raw}}
=
\sqrt{
(i_{d,i}^{\star,\mathrm{raw}})^2
+
(i_{q,i}^{\star,\mathrm{raw}})^2
},
\end{equation}
and
\begin{equation}
\kappa_i
=
\min
\left(
1,
\frac{I_{\max,i}}
     {\max(I_i^{\star,\mathrm{raw}},\epsilon)}
\right).
\end{equation}
The limited commands are
\begin{equation}
\label{eq:app_ibr_limit}
\begin{bmatrix}
i_{d,i}^{\star}\\
i_{q,i}^{\star}
\end{bmatrix}
=
\kappa_i
\begin{bmatrix}
i_{d,i}^{\star,\mathrm{raw}}\\
i_{q,i}^{\star,\mathrm{raw}}
\end{bmatrix},
\end{equation}
which guarantees
\begin{equation}
\sqrt{
(i_{d,i}^{\star})^2
+
(i_{q,i}^{\star})^2
}
\le I_{\max,i}.
\end{equation}

The inner converter/current response is represented by a first-order
lag,
\begin{align}
T_i\dot i_{d,i}
&=
i_{d,i}^{\star}
-
i_{d,i},
\\
T_i\dot i_{q,i}
&=
i_{q,i}^{\star}
-
i_{q,i}.
\end{align}
The current injected into the network is
\begin{equation}
\label{eq:app_ibr_injection}
I_{\mathrm{IBR},i}
=
(i_{d,i}+ji_{q,i})e^{j\theta_i}.
\end{equation}

The common IBR parameters are
\begin{equation}
\begin{aligned}
k_{p,\mathrm{pll}} &= 35,
&
k_{i,\mathrm{pll}} &= 600,
&
T_{pq} &= 0.020~\mathrm{s},
\\
k_{p,P} &= 0.40,
&
k_{i,P} &= 15,
&
k_{p,Q} &= 0.40,
\\
k_{i,Q} &= 15,
&
T_i &= 0.004~\mathrm{s},
&
V_\epsilon &= 0.20~\mathrm{pu}.
\end{aligned}
\end{equation}
The initial inverter angle is aligned with the power-flow voltage,
\begin{equation}
\theta_i(0)=\angle V_k,
\end{equation}
and the initial $dq$ current is obtained from
\begin{equation}
I_{i,0}
=
\left(
\frac{S_i}{V_k}
\right)^{*},
\qquad
I_{dq,i,0}
=
I_{i,0}e^{-j\theta_i(0)}.
\end{equation}
The references are initialized as
\begin{equation}
P_{\mathrm{ref},i}=P_i(0),
\qquad
Q_{\mathrm{ref},i}=Q_i(0),
\end{equation}
with
\begin{equation}
P_{f,i}(0)=P_i(0),
\qquad
Q_{f,i}(0)=Q_i(0).
\end{equation}
The current capability is scaled to the initialized device injection:
\begin{equation}
\label{eq:app_imax}
I_{\max,i}
=
1.30
\max
\left(
|I_{dq,i,0}|,0.10
\right).
\end{equation}
Thus, an initialized inverter with appreciable power injection has
approximately $30\%$ current headroom relative to its initial operating
current.

Each dynamic induction-motor aggregate is represented by
\begin{equation}
\label{eq:app_motor_state}
\mathbf{x}_m
=
\begin{bmatrix}
E'_{q,m} &
E'_{d,m} &
E''_{q,m} &
E''_{d,m} &
s_m
\end{bmatrix}^{\mathsf T},
\end{equation}
where $s_m$ is rotor slip. For the positive-sequence terminal voltage
\begin{equation}
V_k=V_d+jV_q,
\end{equation}
define
\begin{equation}
\Delta_m
=
r_s^2+(L'')^2.
\end{equation}
The motor currents are
\begin{align}
i_{d,m}
&=
\frac{
r_s(V_d+E''_{d,m})
+
L''(V_q+E''_{q,m})
}{
\Delta_m
},
\\
i_{q,m}
&=
\frac{
r_s(V_q+E''_{q,m})
-
L''(V_d+E''_{d,m})
}{
\Delta_m
}.
\end{align}

The transient electrical dynamics satisfy
\begin{align}
T'\dot E'_{q,m}
&=
-E'_{q,m}
-
(L_s-L')i_{d,m}
-
\omega_b s_m T'E'_{d,m},
\\
T'\dot E'_{d,m}
&=
-E'_{d,m}
+
(L_s-L')i_{q,m}
+
\omega_b s_m T'E'_{q,m}.
\end{align}
Define
\begin{equation}
a_m
=
\frac{
T''(L_s-L')
+
T'(L'-L'')
}{
T'T''
}.
\end{equation}
The subtransient dynamics are
\begin{align}
\dot E''_{q,m}
&=
\frac{T'-T''}{T'T''}E'_{q,m}
-
a_m i_{d,m}
-
\frac{E''_{q,m}}{T''}
-
\omega_b s_m E''_{d,m},
\\
\dot E''_{d,m}
&=
\frac{T'-T''}{T'T''}E'_{d,m}
+
a_m i_{q,m}
-
\frac{E''_{d,m}}{T''}
+
\omega_b s_m E''_{q,m}.
\end{align}

The mechanical rotor speed is
\begin{equation}
\omega_m=1-s_m.
\end{equation}
The electrical air-gap power is
\begin{equation}
P_{\mathrm{gap},m}
=
\Re
\left\{
V_kI_m^{*}
\right\}
-
r_s|I_m|^2,
\end{equation}
where
\begin{equation}
I_m=i_{d,m}+ji_{q,m}.
\end{equation}
The electromagnetic torque is
\begin{equation}
T_{e,m}
=
\frac{
P_{\mathrm{gap},m}
}{
\max(\omega_m,0.05)
}.
\end{equation}
The mechanical load torque follows the speed-dependent characteristic
\begin{equation}
T_{m,m}
=
T_{m0,m}
\mu_m(t)
\left[
\max(\omega_m,0.05)
\right]^{E_{\mathrm{trq}}},
\end{equation}
and the slip equation is
\begin{equation}
\label{eq:app_motor_slip}
\dot s_m
=
\frac{
T_{m,m}-T_{e,m}
}{
2H_m
}.
\end{equation}

The common induction-motor parameters are
\begin{equation}
\begin{aligned}
r_s &= 0.020,
&
L_s &= 2.50,
&
L' &= 0.18,
&
L'' &= 0.12,
\\
T' &= 0.15~\mathrm{s},
&
T'' &= 0.025~\mathrm{s},
&
H_m &= 0.50~\mathrm{s},
&
E_{\mathrm{trq}} &= 2.0 .
\end{aligned}
\end{equation}
The initial slip is
\begin{equation}
s_m(0)=0.04,
\qquad
\omega_m(0)=0.96.
\end{equation}
For the prescribed initial slip and power-flow voltage, the four
electrical states are obtained from their steady-state equations,
\begin{equation}
\dot E'_q
=
\dot E'_d
=
\dot E''_q
=
\dot E''_d
=
0.
\end{equation}
The initial electromagnetic torque is then used to determine
$T_{m0,m}$ such that
\begin{equation}
\dot s_m(0)=0.
\end{equation}

For each selected motor bus, the motor model is first evaluated with
unit scaling. If its corresponding active power is
$P_{m,\mathrm{unit}}$, the motor scale is selected to target
\begin{equation}
P_{m,\mathrm{target}}
=
0.25 P_{L,k},
\end{equation}
subject to the available reactive-power demand at that bus. The
remaining complex load is represented by static impedance.

For a static load with initial complex power
\begin{equation}
S_{Z,k}
=
P_{Z,k}+jQ_{Z,k}
\end{equation}
and initial bus voltage $V_{k,0}$, the equivalent shunt admittance is
\begin{equation}
\label{eq:app_static_load}
Y_{Z,k}
=
\frac{
S_{Z,k}^{*}
}{
|V_{k,0}|^2
}.
\end{equation}
Its current is therefore
\begin{equation}
I_{Z,k}=Y_{Z,k}V_k,
\end{equation}
so its active and reactive consumption vary with voltage magnitude
rather than remaining constant after a disturbance.

All components are coupled through the complete positive-sequence AC
network. Let
\begin{equation}
\mathbf{Y}_{\mathrm{bus}}
\in
\mathbb{C}^{n_b\times n_b}
\end{equation}
denote the bus-admittance matrix assembled from the IEEE branches,
transformers, charging susceptances, and bus shunts.

Each synchronous generator admits a Norton representation. Define its
subtransient internal voltage in network coordinates as
\begin{equation}
E''_g
=
(E''_{d,g}+jE''_{q,g})
e^{j(\delta_g-\pi/2)}.
\end{equation}
Its shunt contribution is
\begin{equation}
Y_{\mathrm{SG},g}
=
\frac{1}{jx''_g},
\end{equation}
and its Norton source is
\begin{equation}
b_g
=
\frac{E''_g}{jx''_g}.
\end{equation}

The GFL inverter behaves as a controlled current source,
\begin{equation}
b_i
=
(i_{d,i}+ji_{q,i})e^{j\theta_i}.
\end{equation}

For induction motor $m$, define
\begin{equation}
E''_m
=
E''_{d,m}+jE''_{q,m}.
\end{equation}
Its network admittance is
\begin{equation}
Y_{\mathrm{IM},m}
=
\frac{\kappa_m}
     {r_s+jL''},
\end{equation}
where $\kappa_m$ denotes the motor power scaling at its bus, and its
Norton source is
\begin{equation}
b_m
=
-
\frac{
\kappa_mE''_m
}{
r_s+jL''
}.
\end{equation}

Collecting all static and dynamic admittances yields
\begin{equation}
\label{eq:app_anet}
\mathbf{A}
=
\mathbf{Y}_{\mathrm{bus}}
+
\operatorname{diag}
\left(
\mathbf{Y}_{Z}
+
\mathbf{Y}_{\mathrm{SG}}
+
\mathbf{Y}_{\mathrm{IM}}
\right),
\end{equation}
while the source at bus $k$ is
\begin{align}
b_k(\mathbf{x})
={}&
\sum_{g\in\mathcal{G}_k}
\frac{
(E''_{d,g}+jE''_{q,g})
e^{j(\delta_g-\pi/2)}
}{
jx''_g
}
\nonumber\\
&
+
\sum_{i\in\mathcal{I}_k}
(i_{d,i}+ji_{q,i})e^{j\theta_i}
\nonumber\\
&
-
\sum_{m\in\mathcal{M}_k}
\frac{
\kappa_m
(E''_{d,m}+jE''_{q,m})
}{
r_s+jL''
}.
\end{align}
Here $\mathcal{G}_k$, $\mathcal{I}_k$, and $\mathcal{M}_k$ denote the
sets of synchronous machines, inverters, and motors connected at bus
$k$, respectively.

Kirchhoff's current law for all buses can therefore be written
compactly as
\begin{equation}
\label{eq:app_full_network}
\boxed{
\mathbf{A}\mathbf{V}
=
\mathbf{b}(\mathbf{x})
}.
\end{equation}
Equation~\eqref{eq:app_full_network} is linear in the algebraic bus
voltages, although $\mathbf{b}(\mathbf{x})$ depends nonlinearly on the
dynamic states. Hence,
\begin{equation}
\mathbf{V}
=
\mathbf{A}^{-1}
\mathbf{b}(\mathbf{x}),
\end{equation}
and substitution into the component equations gives the equivalent
reduced differential representation
\begin{equation}
\label{eq:app_reduced_ode}
\dot{\mathbf{x}}
=
\mathbf{F}(\mathbf{x},t)
=
\mathbf{f}
\left(
\mathbf{x},
\mathbf{A}^{-1}
\mathbf{b}(\mathbf{x}),
t
\right).
\end{equation}
This elimination is algebraically exact for the specified model and
does not constitute a Kron reduction of the IEEE network.

Two disturbance families are used. In the mixed-setpoint scenario,
three successive disturbances excite the motor, inverter, and
synchronous-machine control dynamics. With disturbance-scale parameter
$d$ and nominal value
\begin{equation}
d=1,
\end{equation}
the induction-motor mechanical-torque multiplier is
\begin{equation}
\mu_m(t)
=
\begin{cases}
1,
& t<0.20~\mathrm{s},
\\
1+0.05d,
& t\ge0.20~\mathrm{s}.
\end{cases}
\end{equation}
The inverter active-power reference multiplier is
\begin{equation}
\mu_{\mathrm{IBR}}(t)
=
\begin{cases}
1,
& t<0.55~\mathrm{s},
\\
1+0.03d,
& t\ge0.55~\mathrm{s},
\end{cases}
\end{equation}
such that
\begin{equation}
P_{\mathrm{ref},i}(t)
=
\mu_{\mathrm{IBR}}(t)
P_{\mathrm{ref},i}^{0}.
\end{equation}
Finally, the synchronous-machine voltage reference is
\begin{equation}
V_{\mathrm{ref},g}(t)
=
\begin{cases}
V_{\mathrm{ref},g}^{0},
& t<0.80~\mathrm{s},
\\
V_{\mathrm{ref},g}^{0}+0.01d,
& t\ge0.80~\mathrm{s}.
\end{cases}
\end{equation}

A separate fault-ride-through scenario introduces a balanced
positive-sequence three-phase fault at the bus with the largest active
load. The fault is represented by the finite shunt impedance
\begin{equation}
Z_f
=
0.01+j0.10~\mathrm{pu},
\end{equation}
or equivalently
\begin{equation}
Y_f
=
\frac{1}{0.01+j0.10}.
\end{equation}
The fault is active over
\begin{equation}
0.20~\mathrm{s}
\le t
<
0.28~\mathrm{s}.
\end{equation}
During this interval, the network matrix becomes
\begin{equation}
\mathbf{A}_{\mathrm{fault}}
=
\mathbf{A}
+
\operatorname{diag}
(\mathbf{y}_{f}),
\end{equation}
where $\mathbf{y}_f$ is zero at all buses except the faulted bus, where
its value is $Y_f$. The fault is cleared at $t=0.28$~s, after which the
pre-fault network admittance is restored.

In the fault-ride-through scenario, the IBR active-power references are
also increased at
\begin{equation}
t=0.65~\mathrm{s}
\end{equation}
according to
\begin{equation}
P_{\mathrm{ref},i}(t)
=
\begin{cases}
P_{\mathrm{ref},i}^{0},
& t<0.65~\mathrm{s},
\\
(1+0.02d)P_{\mathrm{ref},i}^{0},
& t\ge0.65~\mathrm{s}.
\end{cases}
\end{equation}

The differential states remain continuous across network events,
whereas the algebraic voltages may change instantaneously because the
active algebraic constraint changes from
\(
\mathbf{A}\mathbf{V}=\mathbf{b}(\mathbf{x})
\)
to
\(
\mathbf{A}_{\mathrm{fault}}\mathbf{V}
=
\mathbf{b}(\mathbf{x})
\)
and back. Setpoint disturbances change the forcing of the dynamic
equations without changing the network topology.

Collecting the complete component models, every benchmark therefore has
the common structure
\begin{subequations}
\label{eq:app_ieee_complete}
\begin{align}
\dot{\mathbf{x}}_{\mathrm{SG}}
&=
\mathbf{f}_{\mathrm{SG}}
\left(
\mathbf{x}_{\mathrm{SG}},
\mathbf{V},
t
\right),
\\
\dot{\mathbf{x}}_{\mathrm{IBR}}
&=
\mathbf{f}_{\mathrm{IBR}}
\left(
\mathbf{x}_{\mathrm{IBR}},
\mathbf{V},
t
\right),
\\
\dot{\mathbf{x}}_{\mathrm{IM}}
&=
\mathbf{f}_{\mathrm{IM}}
\left(
\mathbf{x}_{\mathrm{IM}},
\mathbf{V},
t
\right),
\\
\mathbf{0}
&=
\mathbf{A}_{\sigma}\mathbf{V}
-
\mathbf{b}(\mathbf{x}).
\end{align}
\end{subequations}

The five systems therefore differ primarily in network size,
generation dispatch, dynamic-device placement, and the resulting state
dimension, while retaining the same component-level equations and
parameterization. The benchmarks span $53$ differential states in the
IEEE 9-bus system to $558$ differential states in the IEEE 118-bus
system and combine machine electromechanical dynamics, subtransient
electrical dynamics, excitation and governor controls, fast
grid-following inverter control loops, nonlinear converter current
limits, induction-motor dynamics, voltage-dependent loads, algebraic
network coupling, and discrete disturbance events within one common
RMS transient-simulation framework.

Figure~\ref{fig:case118_raptor_reference} illustrates a representative transient response for the IEEE~118-bus system, the largest network considered. RAPTOR closely follows the high-accuracy reference across machine, voltage, motor, and inverter states under the three disturbance events while requiring only 32 accepted intervals, whose endpoints are indicated by the red markers.
\begin{figure*}[t]
    \centering
    \includestandalone[width=0.98\textwidth]
    {Plots/appendix_case118}
    \caption{IEEE 118-bus transient response: high-accuracy reference
    (solid black) and RAPTOR (dashed red). Red bullets mark the endpoints
    of the 32 accepted RAPTOR intervals. Vertical dashed lines indicate
    the motor-torque, inverter active-power-reference, and AVR-reference
    events.}
    \label{fig:case118_raptor_reference}
\end{figure*}

\begin{table*}[t]
\centering
\caption{Sequential advancement and nonlinear-solve effort for the selected
accuracy-qualified IEEE RMS benchmark configurations. ``Corr.'' denotes
RAPTOR modified-Newton (Chord) coefficient corrections. The LU columns report
the measured number of matrix factorizations.}
\label{tab:ieee_solver_effort}
\scriptsize
\setlength{\tabcolsep}{3.6pt}
\begin{tabular}{lccccc|cc|cc|c}
\toprule
&
\multicolumn{5}{c|}{\textbf{RAPTOR}}
&
\multicolumn{2}{c|}{\textbf{BDF}}
&
\multicolumn{2}{c|}{\textbf{Radau}}
&
\textbf{LSODA}
\\
\textbf{System}
& \textbf{Intervals}
& \textbf{Corr.}
& \textbf{Corr./int.}
& \textbf{Jac.}
& \textbf{LU}
& \textbf{Adv.}
& \textbf{LU}
& \textbf{Adv.}
& \textbf{LU}
& \textbf{Adv.}
\\
\midrule
IEEE 9
& 28 & 44 & 1.57 & 3 & 64
& 193 & 63
& 123 & 286
& 246
\\
IEEE 14
& 22 & 37 & 1.68 & 4 & 68
& 192 & 62
& 126 & 356
& 233
\\
IEEE 39
& 27 & 57 & 2.11 & 4 & 88
& 234 & 72
& 145 & 334
& 342
\\
IEEE 57
& 23 & 43 & 1.87 & 5 & 80
& 199 & 68
& 150 & 378
& 258
\\
IEEE 118
& 32 & 51 & 1.59 & 1 & 32
& 195 & 62
& 153 & 396
& 226
\\
\bottomrule
\end{tabular}
\end{table*}
Table~\ref{tab:ieee_solver_effort} quantifies the computational mechanism illustrated by Fig.~\ref{fig:case118_raptor_reference}. Across the five networked benchmarks, RAPTOR requires only \(22\)--\(32\) accepted intervals, compared with \(192\)--\(234\) advances for BDF, \(123\)--\(153\) for Radau, and \(226\)--\(342\) for LSODA. Despite covering substantially more physical time per interval, the RAPTOR nonlinear solve requires only \(1.57\)--\(2.11\) modified-Newton coefficient corrections per accepted interval. Jacobian reuse further limits the number of Jacobian constructions to \(1\)--\(5\) over the complete simulation. The factorization reduction is particularly pronounced relative to Radau, reaching \(396/32=12.4\times\) on IEEE~118. Relative to BDF, the factorization counts are comparable on the smaller systems, showing that RAPTOR's speedup there is primarily associated with fewer sequential advances and inexpensive convergence of the longer interval solves rather than fewer factorizations alone.

\subsection{Benchmark Parameters and Reproducibility}
\label{app:reproducibility}

This appendix collects the benchmark-specific physical parameters, controller
gains, initialization settings, and internal RAPTOR numerical parameters used
in the numerical experiments.

\subsection{Standalone RMS Grid-Following Inverter}

The standalone RMS inverter uses a nominal frequency of \(50\)~Hz,
\(\omega_N=2\pi50\)~rad/s, and the parameter values in
Table~\ref{tab:app_rms_inv_params}.

\begin{table}[t]
\centering
\caption{Standalone RMS grid-following inverter parameters.}
\label{tab:app_rms_inv_params}
\scriptsize
\setlength{\tabcolsep}{4pt}
\begin{tabular}{lll}
\toprule
\textbf{Parameter} & \textbf{Description} & \textbf{Value} \\
\midrule
\(P_{\mathrm{ref}}\) & Active-power reference & \(1.0\) pu \\
\(Q_{\mathrm{ref}}\) & Reactive-power reference & \(0.0\) pu \\
\(V_{\mathrm{ref}}\) & Voltage reference & \(1.0\) pu \\
\(\omega_{\mathrm{ref}}\) & Frequency reference & \(2\pi50\) rad/s \\
\(v_{\mathrm{dc}}\) & DC-link voltage & \(1.0\) pu \\
\midrule
\(k_{p_{dp}}\) & Active-power proportional gain & \(1.0\) \\
\(k_{i_{dp}}\) & Active-power integral gain & \(50\) \\
\(k_{p_{dq}}\) & Reactive-power proportional gain & \(1.0\) \\
\(k_{i_{dq}}\) & Reactive-power integral gain & \(50\) \\
\(k_{p_{id}}\) & \(d\)-axis current proportional gain & \(1.0\) \\
\(k_{i_{id}}\) & \(d\)-axis current integral gain & \(500\) \\
\(k_{p_{iq}}\) & \(q\)-axis current proportional gain & \(1.0\) \\
\(k_{i_{iq}}\) & \(q\)-axis current integral gain & \(500\) \\
\(k_{p_{\mathrm{pll}}}\) & PLL proportional gain & \(30\) \\
\(k_{i_{\mathrm{pll}}}\) & PLL integral gain & \(900\) \\
\midrule
\(R\) & Series resistance & \(0.01\) pu \\
\(L\) & Series inductance & \(0.1/(2\pi50)\) \\
\(\omega_{\mathrm{filt}}\) & Measurement-filter rate & \(25~\mathrm{s}^{-1}\) \\
\bottomrule
\end{tabular}
\end{table}

The nominal initial state is
\[
\mathbf{x}_0=
[0,\,0,\,1,\,0,\,0,\,0,\,0,\,0,\,1,\,0,\,1]^\mathsf{T}.
\]
The additional initial conditions perturb \(i_d\) by \(+0.2\) pu,
\(i_q\) by \(+0.1\) pu, \(P_{\mathrm{filt}}\) to \(0.8\) pu,
\(V_{\mathrm{filt}}\) to \(1.05\) pu, and
\(\theta_{\mathrm{pll}}\) by \(+0.05\) rad.

For the weak-grid sensitivity experiment, the controller parameters remain
unchanged and the network impedance is varied with SCR according to
\[
L(\mathrm{SCR})=\frac{1}{2\pi50\,\mathrm{SCR}},
\qquad
R(\mathrm{SCR})=\frac{0.05}{\mathrm{SCR}},
\]
for
\[
\mathrm{SCR}\in\{5,\,3,\,2,\,1.2\}.
\]

The common internal RAPTOR settings used for this model are summarized in
Table~\ref{tab:app_rms_raptor}.

\begin{table}[t]
\centering
\caption{Internal RAPTOR settings for the standalone RMS inverter.}
\label{tab:app_rms_raptor}
\scriptsize
\setlength{\tabcolsep}{4pt}
\begin{tabular}{lll}
\toprule
\textbf{Parameter} & \textbf{Meaning} & \textbf{Value} \\
\midrule
\(M\) & Collocation points & \(8\) \\
\(\Delta t_{\min}\) & Minimum interval length & \(5\times10^{-5}\) s \\
\(s_0\) & Initial-interval safety multiplier & \(1.0\) \\
\(\nu_{\max}\) & Maximum rejected retries & \(5\) \\
\(k_P\) & PI proportional exponent coefficient & \(0.7\) \\
\(k_I\) & PI integral exponent coefficient & \(0.3\) \\
\(\gamma_{\min}\) & Minimum interval factor & \(0.2\) \\
\(\eta_{\mathrm{saf}}\) & PI safety factor & \(0.9\) \\
\(\gamma_{\max}\) & Maximum interval factor & \(2.0\) \\
\(\lambda_{\mathrm{cont}}\) & Continuation regularization & \(10^{-3}\) \\
Chord rtol & Nonlinear-solver relative tolerance & \(10^{-3}\) \\
Chord atol & Nonlinear-solver absolute tolerance & \(10^{-3}\) \\
Max. Chord steps & Nonlinear iterations per attempt & \(300\) \\
Random seed & RBF sampling seed & \(0\) \\
\bottomrule
\end{tabular}
\end{table}

The PI interval controller uses
\[
\Delta t_{k+1}
=
0.9\,\Delta t_k
\operatorname{clip}
\left[
\mathrm{err}_k^{-0.7/r}
\mathrm{err}_{k-1}^{-0.3/r},
\,0.2,\,2.0
\right],
\qquad
r=\nu+1.
\]

\subsection{Cascaded M-Section Transmission Line}

The three-section line uses the physical parameters in
Table~\ref{tab:app_msection_params}.

\begin{table}[t]
\centering
\caption{Cascaded M-section line parameters.}
\label{tab:app_msection_params}
\scriptsize
\setlength{\tabcolsep}{4pt}
\begin{tabular}{lll}
\toprule
\textbf{Parameter} & \textbf{Description} & \textbf{Value} \\
\midrule
\(R\) & Total series resistance & \(0.005\) \\
\(L\) & Total series inductance & \(0.1/(2\pi50)\) \\
\(C\) & Total shunt capacitance & \(0.0066\) \\
\(G\) & Total shunt conductance & \(0.2\) \\
\(\omega\) & Synchronous angular frequency & \(2\pi50\) rad/s \\
\(\eta_{\mathrm{sec}}\) & Central-section partition factor & \(0.999\) \\
\(I_{\mathrm{in}}\) & Boundary-input amplitude & \(0.02\) \\
\(f_{\mathrm{in}}\) & Boundary-input frequency & \(50\) Hz \\
\bottomrule
\end{tabular}
\end{table}

The resulting section parameters are
\[
R'=\frac{R}{2}=0.0025,
\qquad
L'=\frac{L}{2}=\frac{0.05}{2\pi50},
\]
\[
C_1=C_3=\frac{1-N}{2}C=3.3\times10^{-6},
\qquad
C_2=NC=6.5934\times10^{-3},
\]
and
\[
G_1=G_3=\frac{1-N}{2}G=10^{-4},
\qquad
G_2=NG=0.1998.
\]

The unperturbed initial state is \(\mathbf{x}_0=\mathbf{0}\). Randomized
initial conditions are formed by independent zero-mean Gaussian perturbations
with standard deviation \(0.04\) for both voltage and current states. The
initial-condition random seed is \(7\).

\begin{table}[t]
\centering
\caption{Internal RAPTOR settings for the M-section benchmark.}
\label{tab:app_msection_raptor}
\scriptsize
\setlength{\tabcolsep}{4pt}
\begin{tabular}{lll}
\toprule
\textbf{Parameter} & \textbf{Meaning} & \textbf{Value} \\
\midrule
\(M\) & Collocation points & \(10\) \\
\(N\) & Gaussian RBFs in tuned configuration & \(8\) \\
\(\Delta t_{\min}\) & Minimum interval & \(5\times10^{-5}\) s \\
\(s_0\) & Initial-interval safety multiplier & \(1.0\) \\
\(\nu_{\max}\) & Maximum retries & \(8\) \\
\(k_P\) & PI proportional coefficient & \(0.7\) \\
\(k_I\) & PI integral coefficient & \(0.3\) \\
\(\gamma_{\min}\) & Minimum interval factor & \(0.2\) \\
\(\eta_{\mathrm{saf}}\) & PI safety factor & \(0.9\) \\
\(\gamma_{\max}\) & Maximum interval factor & \(2.0\) \\
Rejection factor & Maximum rejected-interval reduction & \(0.5\) \\
Ridge parameter & Linear collocation regularization & \(10^{-10}\) \\
Random seed & Initial-condition seed & \(7\) \\
Timing repetitions & Repeated executions & \(3\) \\
\bottomrule
\end{tabular}
\end{table}

\subsection{Networked IEEE RMS DAE Benchmarks}

The IEEE 9-, 14-, 39-, 57-, and 118-bus systems use a \(100\)-MVA system
base and \(60\)-Hz nominal frequency. The dynamic-device composition is given
in Table~\ref{tab:app_ieee_comp}.

\begin{table*}[t]
\centering
\caption{Composition of the networked IEEE RMS benchmarks.}
\label{tab:app_ieee_comp}
\scriptsize
\resizebox{\textwidth}{!}{%
\begin{tabular}{lcccccccll}
\toprule
System & Buses & Branches & SG & IBR & IM & \(n_x\)
& \(P_{\mathrm{IM}}/P_L\) & IBR buses & Motor buses \\
\midrule
IEEE 9
& 9 & 9 & 3 & 1 & 3 & 53
& 25.00\% & 1 & 5,7,9 \\
IEEE 14
& 14 & 20 & 4 & 2 & 6 & 86
& 17.86\% & 1,2 & 2,3,6,9,13,14 \\
IEEE 39
& 39 & 46 & 8 & 3 & 6 & 134
& 11.72\% & 30,34,37 & 3,4,8,16,20,39 \\
IEEE 57
& 57 & 80 & 6 & 2 & 6 & 106
& 9.49\% & 3,12 & 1,6,8,9,12,16 \\
IEEE 118
& 118 & 186 & 44 & 11 & 6 & 558
& 5.12\% & 12,25,31,46,49,54,59,61,87,103,111
& 15,42,54,59,80,90 \\
\bottomrule
\end{tabular}}
\end{table*}

The target inverter share is
\[
\eta_{\mathrm{IBR}}=0.20,
\]
and the target induction-motor fraction at a selected dynamic load bus is
\[
\eta_{\mathrm{IM}}=0.25.
\]
At most six motor buses are used. Candidate motor buses satisfy
\[
P_L>0.03\ {\rm pu},
\qquad
Q_L>0.005\ {\rm pu}.
\]

The fractional IBR assignments at marginal plants are
\[
\begin{array}{c|ccccc}
\text{System} & 9 & 14 & 39 & 57 & 118\\
\hline
\text{Marginal bus} & 1 & 1 & 37 & 12 & 25\\
\text{IBR fraction} & 0.8893 & 0.0623 & 0.9288 & 0.6959 & 0.5313 .
\end{array}
\]

\begin{table}[t]
\centering
\caption{Synchronous-machine, AVR, and governor parameters.}
\label{tab:app_ieee_sg}
\scriptsize
\setlength{\tabcolsep}{4pt}
\begin{tabular}{lll}
\toprule
\textbf{Parameter} & \textbf{Description} & \textbf{Value} \\
\midrule
\(x_d\) & \(d\)-axis synchronous reactance & \(1.80\) \\
\(x_q\) & \(q\)-axis synchronous reactance & \(1.70\) \\
\(x'_d\) & \(d\)-axis transient reactance & \(0.30\) \\
\(x'_q\) & \(q\)-axis transient reactance & \(0.55\) \\
\(T'_{qo}\) & \(q\)-axis transient time constant & \(0.40\) s \\
\(T''_{do}\) & \(d\)-axis subtransient time constant & \(0.030\) s \\
\(T''_{qo}\) & \(q\)-axis subtransient time constant & \(0.050\) s \\
\(D\) & Damping coefficient & \(1.0\) \\
\(K_A\) & AVR gain & \(20\) \\
\(T_A\) & AVR time constant & \(0.050\) s \\
\(T_r\) & Voltage-measurement time constant & \(0.020\) s \\
\(R_g\) & Governor droop & \(0.05\) \\
\(T_g\) & Governor time constant & \(0.20\) s \\
\(T_t\) & Turbine time constant & \(0.50\) s \\
\bottomrule
\end{tabular}
\end{table}

Machine-to-machine diversity is introduced deterministically using
\[
H_r=3.5+0.4[(r\bmod4)+1]\ {\rm s},
\]
\[
T'_{do,r}=6.0+0.7(r\bmod4)\ {\rm s},
\qquad
x''_r=0.20+0.01(r\bmod3),
\]
giving
\[
H_r\in\{3.9,4.3,4.7,5.1\}\ {\rm s},
\]
\[
T'_{do,r}\in\{6.0,6.7,7.4,8.1\}\ {\rm s},
\qquad
x''_r\in\{0.20,0.21,0.22\}.
\]

\begin{table}[t]
\centering
\caption{Grid-following IBR parameters for the IEEE RMS benchmarks.}
\label{tab:app_ieee_ibr}
\scriptsize
\setlength{\tabcolsep}{4pt}
\begin{tabular}{lll}
\toprule
\textbf{Parameter} & \textbf{Description} & \textbf{Value} \\
\midrule
\(k_{p,\mathrm{pll}}\) & PLL proportional gain & \(35\) \\
\(k_{i,\mathrm{pll}}\) & PLL integral gain & \(600\) \\
\(T_{pq}\) & Power-filter time constant & \(0.020\) s \\
\(k_{p,P}\) & Active-power proportional gain & \(0.40\) \\
\(k_{i,P}\) & Active-power integral gain & \(15\) \\
\(k_{p,Q}\) & Reactive-power proportional gain & \(0.40\) \\
\(k_{i,Q}\) & Reactive-power integral gain & \(15\) \\
\(T_i\) & Inner current-response time constant & \(0.004\) s \\
\(V_\epsilon\) & Voltage regularization floor & \(0.20\) pu \\
\bottomrule
\end{tabular}
\end{table}

The inverter current limit is initialized as
\[
I_{\max,i}
=
1.30\max\!\left(|I_{dq,i}(0)|,0.10\right).
\]

\begin{table}[t]
\centering
\caption{Induction-motor parameters for the IEEE RMS benchmarks.}
\label{tab:app_ieee_im}
\scriptsize
\setlength{\tabcolsep}{4pt}
\begin{tabular}{lll}
\toprule
\textbf{Parameter} & \textbf{Description} & \textbf{Value} \\
\midrule
\(r_s\) & Stator resistance & \(0.020\) \\
\(L_s\) & Synchronous inductance & \(2.50\) \\
\(L'\) & Transient inductance & \(0.18\) \\
\(L''\) & Subtransient inductance & \(0.12\) \\
\(T'\) & Transient time constant & \(0.15\) s \\
\(T''\) & Subtransient time constant & \(0.025\) s \\
\(H_m\) & Motor inertia constant & \(0.50\) s \\
\(E_{\mathrm{trq}}\) & Mechanical torque exponent & \(2.0\) \\
\(s_m(0)\) & Initial slip & \(0.04\) \\
\(\omega_m(0)\) & Initial motor speed & \(0.96\) pu \\
\bottomrule
\end{tabular}
\end{table}

The internal RAPTOR parameters used in the IEEE configuration sweep are
summarized in Table~\ref{tab:app_ieee_raptor}.

\begin{table}[t]
\centering
\caption{Internal RAPTOR parameters for the IEEE RMS DAE experiments.}
\label{tab:app_ieee_raptor}
\scriptsize
\setlength{\tabcolsep}{3.5pt}
\begin{tabular}{lll}
\toprule
\textbf{Parameter} & \textbf{Meaning} & \textbf{Value} \\
\midrule
\(N\) & Gaussian RBFs per state & \(7\) \\
\(M\) & Collocation points & \(7\) \\
\(N_{\mathrm{audit}}\) & Off-grid audit points & \(9\) \\
\(\Delta t_0\) & Initial interval & \(10^{-2}\) s \\
\(\Delta t_{\min}\) & Minimum interval & \(2\times10^{-5}\) s \\
\(\mathrm{atol}_{R}\) & Fixed residual absolute scale & \(2\times10^{-7}\) \\
Newton tolerance & Coefficient-correction tolerance & \(3\times10^{-2}\) \\
Maximum corrections & Chord/Newton safety cap & \(5\) \\
Jacobian refresh & Internal refresh parameter & \(2\) \\
GMRES rtol & Iterative linear-solve tolerance & \(10^{-4}\) \\
\(\eta_{\mathrm{saf}}\) & Interval safety factor & \(0.93\) \\
\(\gamma_{\min}\) & Minimum growth factor & \(0.35\) \\
\(\gamma_{\max}\) & Maximum growth factor & \(2.5\) \\
\(\gamma_{\mathrm{rej}}\) & Rejected-interval factor & \(0.50\) \\
Random seed & Gaussian-width seed & \(0\) \\
\bottomrule
\end{tabular}
\end{table}

The tested Jacobian-reuse periods are
\[
n_{\mathrm{reuse}}\in\{8,12\}
\]
accepted intervals, and both one-point and two-point coefficient
initialization are tested. Gaussian centers are
\[
\beta_j=\frac{j+\tfrac12}{N},
\qquad j=0,\ldots,N-1,
\]
and their positive shape parameters are initialized using random seed \(0\).

\subsection{EMT Grid-Following Inverter}

The production EMT grid-following inverter uses a \(50\)-Hz base frequency,
with the electrical quantities in the source parameter file converted to the
continuous-time model according to
\[
\omega_N=2\pi50.
\]

\begin{table}[t]
\centering
\caption{EMT grid-following inverter electrical and controller parameters.}
\label{tab:app_emt_gfl}
\scriptsize
\setlength{\tabcolsep}{3.5pt}
\begin{tabular}{lll}
\toprule
\textbf{Parameter} & \textbf{Description} & \textbf{Value} \\
\midrule
\(L_g\) & Grid-side inductance & \(0.1/(2\pi50)\) \\
\(R_{lg}\) & Grid-side resistance & \(0.005\) \\
\(L_c\) & Converter-side inductance & \(0.1/(2\pi50)\) \\
\(R_c\) & Converter-side resistance & \(0.005\) \\
\(C_f\) & Filter capacitance & \(0.0011840835645/(2\pi50)\) \\
\(R_f\) & Damping resistance & \(0.0757\) \\
\midrule
\(K_p^{\mathrm{cc}}\) & Current-controller proportional gain & \(0.1\) \\
\(K_i^{\mathrm{cc}}\) & Current-controller integral gain & \(20\) \\
\(K_p^{\mathrm{pc}}\) & Power-controller proportional gain & \(0.5\) \\
\(K_i^{\mathrm{pc}}\) & Power-controller integral gain & \(30\) \\
\(K_p^{\mathrm{pll}}\) & PLL proportional gain & \(25\) \\
\(K_i^{\mathrm{pll}}\) & PLL integral gain & \(300\) \\
\bottomrule
\end{tabular}
\end{table}

Numerically,
\[
L_c=L_g=3.18310\times10^{-4},
\qquad
C_f=3.76996\times10^{-6}.
\]

The initialization uses
\[
|V_{\mathrm{PCC}}(0)|=1,\qquad
\angle V_{\mathrm{PCC}}(0)=0,
\]
\[
|I_g(0)|=0,\qquad
\angle I_g(0)=0,
\]
with initial event states
\[
s_e(0)=1,\qquad P^\star(0)=0,\qquad Q^\star(0)=0.
\]

For the voltage-sag trajectory used in the reported waveform comparison,
\[
s_e=
\begin{cases}
0.8, & 1.1\le t<1.2~\mathrm{s},\\
1.0, & \text{otherwise},
\end{cases}
\]
and a forced event-localization interval of \(10^{-5}\)~s is used immediately
around the event.

\begin{table}[t]
\centering
\caption{Internal RAPTOR parameters for the EMT grid-following inverter.}
\label{tab:app_emt_gfl_raptor}
\scriptsize
\setlength{\tabcolsep}{3.5pt}
\begin{tabular}{lll}
\toprule
\textbf{Parameter} & \textbf{Meaning} & \textbf{Value} \\
\midrule
\(N\) & Gaussian RBFs per differential state & \(6\) \\
\(M\) & Collocation points & \(6\) \\
\(\lambda_{\mathrm{alg}}\) & Algebraic-residual weight & \(1.0\) \\
\(\Delta t_{\min}\) & Minimum interval & \(10^{-3}\) s \\
\(s_0\) & Initial-interval safety multiplier & \(1.0\) \\
\(\nu_{\max}\) & Maximum retry count & \(20\) \\
\(k_P\) & PI proportional coefficient & \(0.7\) \\
\(k_I\) & PI integral coefficient & \(0.3\) \\
\(\gamma_{\min}\) & Minimum interval factor & \(0.2\) \\
\(\eta_{\mathrm{saf}}\) & PI safety factor & \(0.9\) \\
\(\gamma_{\max}\) & Maximum interval factor & \(2.0\) \\
\(\epsilon_{\mathrm{PI}}\) & PI error floor & \(10^{-8}\) \\
\(\tau_{\mathrm{relax}}\) & Time-scale limiter factor & \(1.0\) \\
Random seed & RBF sampling seed & \(0\) \\
Maximum Chord steps & Nonlinear solve cap & \(40\) \\
Algebraic projection steps & Post-event projection cap & \(15\) \\
Chord rtol & Interval nonlinear solve & \(10^{-8}\) \\
Chord atol & Interval nonlinear solve & \(10^{-8}\) \\
Projection rtol & Algebraic projection & \(10^{-10}\) \\
Projection atol & Algebraic projection & \(10^{-10}\) \\
\bottomrule
\end{tabular}
\end{table}

The interval controller uses the endpoint differential residual,
\[
\mathrm{err}_k
=
\left[
\frac{1}{n_x}
\sum_i
\left(
\frac{r_{k,i}}
{\pi_{\mathrm{atol}}+
 \pi_{\mathrm{rtol}}|x_{k,i}|}
\right)^2
\right]^{1/2},
\]
with \(k_P=0.7\) and \(k_I=0.3\), and proposes
\[
\Delta t_{k+1}
=
0.9\,\Delta t_k
\operatorname{clip}
\left[
\mathrm{err}_k^{-0.7/r}
\mathrm{err}_{k-1}^{-0.3/r},
0.2,2.0
\right],
\qquad r=\nu+1.
\]

\subsection{Synchronous-Machine EMT Benchmark}

The synchronous-machine EMT benchmark uses the operating point
\[
P_0=0.5~\mathrm{pu},
\qquad
Q_0=0.1~\mathrm{pu},
\qquad
V_0=1.0~\mathrm{pu},
\]
and a \(50\)-Hz nominal frequency. Its primitive and derived parameters are
listed in Table~\ref{tab:app_emt_sm}.

\begin{table}[t]
\centering
\caption{Synchronous-machine EMT benchmark parameters.}
\label{tab:app_emt_sm}
\scriptsize
\setlength{\tabcolsep}{3.3pt}
\begin{tabular}{lll}
\toprule
\textbf{Parameter} & \textbf{Description} & \textbf{Value} \\
\midrule
\(X_e\) & External line reactance & \(0.20\) \\
\(R_e\) & External line resistance & \(0.01\) \\
\(L_e\) & External line inductive parameter & \(0.20\) \\
\(R_a\) & Stator resistance & \(0.005\) \\
\(X_d\) & \(d\)-axis reactance & \(2.4\) \\
\(X_q\) & \(q\)-axis reactance & \(2.4\) \\
\(X_l\) & Leakage reactance & \(0.2\) \\
\(X'_d\) & \(d\)-axis transient reactance & \(0.40\) \\
\(X'_q\) & \(q\)-axis transient reactance & \(0.25\) \\
\(L_{oo}\) & Zero-sequence inductance & \(0.1\) \\
\(T'_{do}\) & Open-circuit \(d\)-axis time constant & \(7.0\) s \\
\(T'_{qo}\) & Open-circuit \(q\)-axis time constant & \(0.3\) s \\
\midrule
\(L_{dd}\) & \(d\)-axis self inductance & \(2.4\) \\
\(L_{df}\) & \(d\)-axis mutual inductance & \(2.2\) \\
\(L_{ff}\) & Field self inductance & \(2.42\) \\
\(R_f\) & Field resistance & \(1.10044\times10^{-3}\) \\
\(L_{qq}\) & \(q\)-axis self inductance & \(2.4\) \\
\(L_{qq1}\) & \(q\)-axis mutual inductance & \(2.2\) \\
\(L_{q1q1}\) & Damper self inductance & \(2.25116\) \\
\(R_{q1}\) & Damper resistance & \(2.38856\times10^{-2}\) \\
\bottomrule
\end{tabular}
\end{table}

The infinite-bus phasor resulting from the initialized operating point is
\[
|E_\infty|=0.9800133,
\qquad
\angle E_\infty=-0.101192~\mathrm{rad},
\]
and the initialized rotor electrical angle is
\[
\theta_0=0.767791~\mathrm{rad}.
\]

The switching event sets the infinite-bus source multiplier from
\[
s_e=1 \quad\longrightarrow\quad s_e=0
\]
at
\[
t_e=0.05~\mathrm{s},
\]
and the post-event localization interval is \(10^{-6}\)~s.

\begin{table}[t]
\centering
\caption{Internal RAPTOR parameters for the synchronous-machine EMT benchmark.}
\label{tab:app_emt_sm_raptor}
\scriptsize
\setlength{\tabcolsep}{3.3pt}
\begin{tabular}{lll}
\toprule
\textbf{Parameter} & \textbf{Meaning} & \textbf{Value} \\
\midrule
\(N\) & Gaussian RBFs per differential state & \(6\) \\
\(M\) & Collocation points & \(6\) \\
\(\lambda_{\mathrm{alg}}\) & Algebraic-residual weight & \(10\) \\
\(\Delta t_{\min}\) & Minimum interval & \(5\times10^{-4}\) s \\
\(s_0\) & Initial-interval multiplier & \(10^{-3}\) \\
\(\nu_{\max}\) & Maximum retries & \(10\) \\
\(k_P\) & PI proportional coefficient & \(0.7\) \\
\(k_I\) & PI integral coefficient & \(0.3\) \\
\(\gamma_{\min}\) & Minimum interval factor & \(0.2\) \\
\(\eta_{\mathrm{saf}}\) & PI safety factor & \(0.9\) \\
\(\gamma_{\max}\) & Maximum interval factor & \(2.1\) \\
\(\epsilon_{\mathrm{PI}}\) & PI error floor & \(10^{-8}\) \\
\(\tau_{\mathrm{relax}}\) & Time-scale limiter & \(1.0\) \\
Random seed & RBF seed & \(0\) \\
Maximum Chord steps & Nonlinear solve cap & \(20\) \\
Algebraic projection steps & Projection cap & \(10\) \\
Event localization interval & Post-event interval & \(10^{-6}\) s \\
\bottomrule
\end{tabular}
\end{table}

For the Gaussian RBF representation in this EMT benchmark, the centers are
uniformly distributed on \([0,1]\), and the sampled shape parameters use
random seed \(0\). The adaptive controller again uses
\[
k_P=0.7,\qquad k_I=0.3,
\]
with acceptance whenever the scaled endpoint residual satisfies
\[
\mathrm{err}_k<1.
\]

\bibliographystyle{IEEEtran}
\bibliography{bibliography}